\pdfoutput=1
\RequirePackage{ifpdf}

 \ifpdf
 \documentclass[final,5p,times,twocolumn]{elsarticle}
\else
 \documentclass[final,5p,times,twocolumn]{elsarticle}
 \fi

\let \oldref=\ref
\renewcommand{\ref}[1]{{\color{red}\oldref{#1}}}

\input{tpcnimpackages}

\newcommand{\gas}{\mbox{Ne--CO$_2$--N$_2$}\xspace}
\newcommand{\mixture}{\mbox{85.7--9.5--4.8}\xspace}

\graphicspath{{figcalcom/}{figchamb/}{figcool/}{figDCS/}{figelect/}{figfcage/}{figgas/}{figinfra/}{figintro/}{figlaser/}{figtpc/}{figoverv/}}

\journal{Nuclear Instruments and Methods}

\begin{document}
\begin{frontmatter}

\title{The ALICE TPC, a large 3-dimensional tracking device with fast readout for ultra-high multiplicity events}

\long\def\symbolfootnote[#1]#2{\begingroup%
\def\thefootnote{\fnsymbol{footnote}}\footnote[#1]{#2}\endgroup} 

\author[label0]{J.~Alme}
\author[label3]{Y.~Andres}
\author[label6]{H.~Appelsh\"{a}user}
\author[label0]{S.~Bablok}
\author[label6]{N.~Bialas}
\author[label0]{R.~Bolgen}
\author[label5]{U.~Bonnes}
\author[label7]{R.~Bramm}
\author[label5,label7,label12,label13]{P.~Braun-Munzinger}
\author[label3]{R.~Campagnolo}
\author[label9]{P.~Christiansen}
\author[label9]{A.~Dobrin}
\author[label3]{C.~Engster}
\author[label0]{D.~Fehlker}
\author[label7]{P.~Foka}
\author[label7]{U.~Frankenfeld}
\author[label4]{J.J.~Gaardh{\o}je}
\author[label7]{C.~Garabatos}
\author[label8]{P.~Gl\"{a}ssel}
\author[label3]{C.~Gonzalez~Gutierrez}
\author[label9]{P.~Gros}
\author[label9]{H.-A.~Gustafsson}
\author[label1]{H.~Helstrup}
\author[label3]{M.~Hoch}
\author[label7]{M.~Ivanov}
\author[label2]{R.~Janik}
\author[label3]{A.~Junique}
\author[label5]{A.~Kalweit}
\author[label14]{R.~Keidel}
\author[label6]{S.~Kniege}
\author[label10]{M.~Kowalski}
\author[label0]{D.T.~Larsen}
\author[label3]{Y. Lesenechal}
\author[label3]{P.~Lenoir}
\author[label4]{N.~Lindegaard}
\author[label3]{C.~Lippmann}

\author[label3]{M.~Mager}
\author[label3]{M.~Mast}
\author[label10]{A.~Matyja}
\author[label0]{M.~Munkejord}
\author[label3]{L.~Musa}
\author[label4]{B.S.~Nielsen}
\author[label11]{V.~Nikolic}
\author[label5]{H.~Oeschler}
\author[label4]{E.K.~Olsen}
\author[label9]{A.~Oskarsson}
\author[label9]{L.~Osterman}
\author[label2]{M.~Pikna}
\author[label3]{A.~Rehman}
\author[label4]{G.~Renault}
\author[label6]{R.~Renfordt}
\author[label3]{S.~Rossegger}
\author[label0]{D.~R\"{o}hrich}
\author[label1]{K.~R{\o}ed} 
\author[label0]{M.~Richter}
\author[label6]{G.~Rueshmann}
\author[label10]{A.~Rybicki}
\author[label7]{H.~Sann\symbolfootnote[2]{Deceased}}
\author[label7]{H.-R. Schmidt}
\author[label2]{M.~Siska}
\author[label2]{B.~Sit\'{a}r}
\author[label4]{C.~Soegaard}
\author[label8]{H.-K.~Soltveit}
\author[label7]{D.~Soyk}
\author[label8]{J.~Stachel}
\author[label7]{H.~Stelzer}
\author[label9]{E.~Stenlund}
\author[label6]{R.~Stock}
\author[label2]{P.~Strme\v{n}}
\author[label2]{I.~Szarka}
\author[label0]{K.~Ullaland}
\author[label7]{D.~Vranic}
\author[label3]{R.~Veenhof}
\author[label4]{J.~Westergaard}
\author[label8]{J.~Wiechula}
\author[label8]{B.~Windelband}

\address[label0]{Department of Physics, University of Bergen, Bergen, Norway}
\address[label1]{Faculty of Engineering, Bergen University College, Bergen, Norway}
\address[label2]{Faculty of Mathematics, Physics and Informatics, Comenius University, Bratislava, Slovakia}
\address[label3]{European Organization for Nuclear Research (CERN), Geneva}
\address[label4]{Niels Bohr Institute, University of Copenhagen, Copenhagen, Denmark}
\address[label5]{Institut f\"{u}r Kernphysik, Technische Universit\"{a}t Darmstadt, Darmstadt, Germany}
\address[label6]{Institut f\"{u}r Kernphysik, Johann-Wolfgang-Goethe Universit\"{a}t Frankfurt, Frankfurt, Germany}
\address[label7]{GSI Helmholtzzentrum f\"{u}r Schwerionenforschung GmbH, Darmstadt, Germany}
\address[label8]{Physikalisches Institut, Ruprecht-Karls-Universit\"{a}t Heidelberg, Heidelberg, Germany}
\address[label9]{Division of Experimental High Energy Physics, University of Lund, Lund, Sweden}
\address[label10]{The Henryk Niewodniczanski Institute of Nuclear Physics, Polish Academy of Sciences, Cracow, Poland}
\address[label11]{Rudjer Bo\v{s}kovi\'{c} Institute, Zagreb, Croatia}
\address[label12]{ExtreMe Matter Institute, EMMI, GSI, Darmstadt, Germany}
\address[label13]{Frankfurt Institute for Advanced Studies, J.W. Goethe University, Frankfurt, Germany}
\address[label14]{Zentrum f\"{u}r Technologietransfer und Telekommunikation (ZTT), Fachhochschule Worms, Worms, Germany}

\begin{abstract}
The design, construction, and commissioning of the ALICE Time-Projection
Chamber (TPC) is described. It is the main device for pattern recognition, 
tracking, and identification of charged particles in the ALICE experiment at the CERN LHC.
 The TPC is cylindrical in shape with a volume 
close to 90~\meter$^3$ and is operated in a 0.5 T solenoidal magnetic field
parallel to its axis. 

In this paper we describe in detail the design considerations for this
detector for operation in the extreme multiplicity environment of central
Pb--Pb collisions at LHC energy. The implementation of the resulting
requirements into hardware (field cage, read-out chambers, electronics),
infrastructure (gas and cooling system, laser-calibration system), and
software led to many technical innovations which are described along with a
presentation of all the major components of the detector, as currently
realized. We also report on the performance achieved after completion of the
first round of stand-alone calibration runs and demonstrate results close to
those specified in the TPC Technical Design Report.
\end{abstract}


\end{frontmatter}

\section{Introduction}

The ALICE~\cite{TP:alice,ALICEjinst} Time-Projection Chamber (TPC)~\cite{TDR:tpc}  
is the main device, in the ALICE `central barrel', 
for tracking of charged particles and particle identification.

The main goal of the ALICE experiment at the CERN Large Hadron Collider (LHC) is the investigation of
 Pb--Pb collisions at a center-of-mass energy of 5.5 TeV per nucleon
 pair.  Tracking of charged particles in such an environment  can only be
 performed with a detector which can cope with unprecedented  densities
 of charged particles: the maximum expected rapidity density in Pb--Pb
 collisions at LHC energy is about 3\,000~\cite{lhc_predictions}. Furthermore, a comprehensive
 experiment needs to  cover full azimuth and provide a significant
 acceptance in pseudo-rapidity $\eta = -\ln{\tan{\theta/2}}$ with
 $\theta$ the polar angle. In addition, the detector should provide
 excellent momentum and energy-loss resolution and run at extremely high
 rates ($\unit{>300}{\hertz}$ for Pb--Pb central collisions, $\unit{>1.4}{\kilo\hertz}$ for
 proton--proton collisions).

\begin{figure*}
\center
\includegraphics[width=0.8\linewidth]{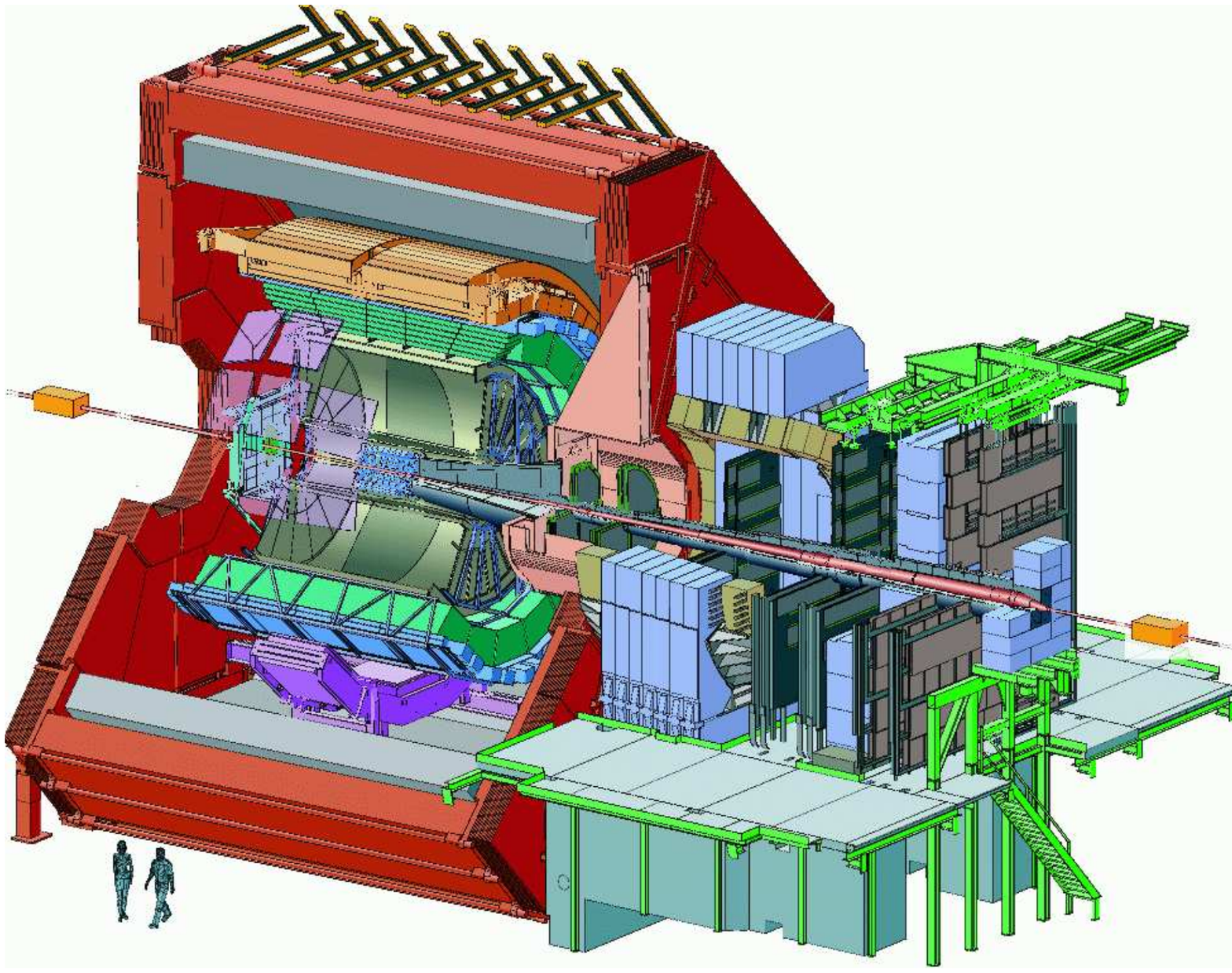} 
\caption{ALICE schematic layout \cite{ALICEjinst}.}
\label{intro:ALICElayout}
\end{figure*}

The resulting detector choice was a large-volume TPC with overall `conventional' lay-out but with nearly all other design parameters beyond the state of the art. This manuscript describes in detail the resulting detector and outlines the path from design considerations to construction and commissioning.

In outline the ALICE TPC consists of a hollow cylinder whose axis is
aligned with the beams from the LHC and is
parallel to the ALICE detector's solenoidal magnetic
field. The active volume has an inner radius of about $\unit{85}{\centi\metre}$, an outer radius
of about $\unit{250}{\centi\metre}$, and an overall length along the beam direction of 
$\unit{500}{\centi\metre}$. A
conducting electrode at the center of the cylinder, charged to $\unit{100}{\kilo\volt}$,
provides, together with a voltage dividing network at the surface of the outer
and inner cylinder, a precise axial electric field of $\unit{400}{\volt\per\centi\metre}$.
The detector is filled with a counting gas consisting of a Ne--CO$_2$--N$_2$ mixture at
atmospheric pressure. Charged particles traversing the detector ionize the
gas. The ionization electrons drift, under the influence of the electric
field, to the endplates of the cylinder, where their arrival point in the
cylinder plane is precisely measured. Together with an accurate measurement of
the arrival time (relative to some external reference such as the collision
time of the beams from the LHC) the complete trajectory in space of all
charged particles traversing the TPC can be determined with precision.

The ALICE set-up is shown  in  Fig.~\ref{intro:ALICElayout}. 
The TPC surrounds the Inner Tracking System (ITS) which is optimized 
for the determination of the primary and secondary vertices and precision 
tracking of low-momentum particles.  
On the outside the Transition Radiation Detector (TRD) is designed for
electron identification. The outermost Time-Of-Flight (TOF) array provides
pion, kaon, and proton identification. In addition, there are three single-arm 
detectors: the Photon Spectrometer (PHOS), the Electro-Magnetic CALorimeter (EMCAL) 
and an array of RICH counters optimized for High-Momentum Particle IDentification (HMPID).

The $\unit{0.5}{\tesla}$ magnetic field in the central barrel is provided 
by the L3 solenoidal magnet previously used by the L3 experiment.

The ALICE TPC was designed to cope with the highest conceivable charged
particle multiplicities 
predicted, at the time of the Technical Proposal (TP), 
for central \mbox{Pb--Pb} collisions at LHC energy ~\cite{TP:alice, ppr1, ppr2},
i.e.~rapidity densities approaching d$N_\text{ch}$/d$y = 8\,000$ 
at center-of-mass energy of $\unit{5.5}{\tera e\volt}$.\footnote{More recent
  estimates \cite{lhc_predictions} put this number at
  d$N_\text{ch}$/d$y < 3\,000$.} 
Its acceptance covers $2\pi$
in azimuthal angle and a pseudo-rapidity interval $\lvert\eta\rvert < 0.9$. Including secondaries, the above 
charged particle rapidity density 
could amount to 20\,000 tracks in one interaction in the TPC acceptance. 

Furthermore, the design of the readout chambers, electronics, and data handling
allows inspection of up to several hundred such events per second with a maximum
interaction rate of $\unit{8}{\kilo\hertz}$ for \mbox{Pb--Pb} collisions, implying special precautions
to minimize the effects of space-charge built-up in
the drift volume of the TPC on the track reconstruction.

To realize a detector which performs efficiently in such an environment required the development of many new components and procedures.
 A summary of the design parameters is presented in 
Tabs.~\ref{overv:overview1}--\ref{overv:overview3}.  
 A summary and system overview can be found in \cite{ALICEjinst}.

\begin{table*}
\begin{center}
\caption{General parameters of the ALICE TPC.}
\label{overv:overview1}
\vspace{2mm}
\begin{tabular}{|l|c|}\hline
Pseudo-rapidity coverage        & $-0.9<\eta<0.9$ for full radial track length\\
                                & $-1.5<\eta<1.5$ for $1/3$ radial track length\\
Azimuthal coverage              & $\unit{360}{\degree}$ \\
Radial position (active volume)          & $848<r<\unit{2\,466}{\milli\metre}$ \\
Radial size of vessel (outer dimensions) & $610<r<\unit{2\,780}{\milli\metre}$  \\
Radial size of vessel (gas volume)       & $788<r<\unit{2\,580}{\milli\metre}$  \\
Length (active volume)          & $\unit{2 \times 2\,497}{\milli\metre}$ \\
Segmentation in $\varphi$       & $\unit{20}{\degree}$\\
Segmentation in r               & 2 chambers per sector\\
Total number of readout chambers & $2\times2\times 18 = 72$ \\
\hline
\hline
Inner readout chamber geometry  &  trapezoidal, $ 848<r<\unit{1\,321}{\milli\metre}$ active area\\
\quad pad size                  & $\unit{4\times7.5}{\milli\metre\squared}$ \quad ($r\varphi\times r$)\\
\quad pad rows                  & 63 \\
\quad total pads                & 5\,504 \\ \hline
Outer readout chamber geometry  &  trapezoidal, $ 1\,346<r<\unit{2\,466}{\milli\metre}$ active area\\
\quad pad size                  & $6\times10$ and $\unit{6\times15}{\milli\metre\squared}$
         \quad ($r\varphi\times r$)\\
\quad pad rows                  & $\phantom{0\,0}64+\phantom{0\,0}32=\phantom{0\,0}96$\quad (small and large pads) \\
\quad total pads                & $5\,952 + 4\,032 = 9\,984$\quad (small and large pads) \\
\hline
\hline
Detector gas                    & Ne--CO$_2$--N$_2$   \quad [85.7--9.5--4.8] \\
Gas volume                      & $\unit{90}{\metre\cubed}$      \\
Drift voltage                   & $\unit{100}{\kilo\volt}$ \\
Anode voltage (nominal)         & $\unit{1\,350}{\volt}$ (IROC) \\
                                & $\unit{1\,570}{\volt}$ (OROC) \\
Gain (nominal)                  & $7\,000-8\,000$ \\
Drift field                     & $\unit{400}{\volt\per\centi\metre}$ \\
Drift velocity (NTP)            & $\unit{2.65}{\centi\metre\per\micro\second}$ \\
Drift time (NTP)                & $\unit{94}{\micro\second}$\\
Diffusion (longitudinal and transversal) & $\unit{220}{\micro\metre\per\sqrt{\centi\metre}}$ \\
\hline
\hline
Material budget (including counting gas)       & $X/X_0 = 3.5$\% near $\eta = 0$ \\
\hline

\end{tabular}
\end{center}
\end{table*}

\begin{table*}
\begin{center}
\caption{ALICE TPC electronics parameters. }
\label{overv:overview2}
\vspace{2mm}
\begin{tabular}{|l|c|}\hline

\hline
Front-End Cards (FECs)          & 121 per sector $\times$ 36 = 4356 \\
Readout partitions              & 6 per sector, 18 to 25 FECs each \\
Total readout control units     & 216 \\
Total pads --- readout channels & 557\,568\\
\hline
Pad occupancy (for d$N$/d$y = 8\,000$) & 40 to 15\%\quad inner / outer radius\\
Pad occupancy (for pp) & 5 to $2\times 10^{-4}$ \quad inner / outer radius\\
\hline
Event size \hspace{7mm} (for d$N$/d$y = 8\,000)$ & $\approx\unit{70}{\mega\byte}$\\
Event size \hspace{7mm} (for pp) & $\unit{0.1-0.2}{\mega\byte}$ \\
Total bandwidth                 & $\unit{35}{\giga\byte\per\second}$\\
Maximum trigger rate  & $\unit{300}{\hertz}$ Pb--Pb central events\\
                                & $\unit{1.4}{\kilo\hertz}$ proton--proton events\\
\hline
ADC                             & $\unit{10}{\bit}$ \\
\quad sampling frequency        & $\unit{5-10}{\mega\hertz}$ \\
\quad time samples              & $500-1\,000$   \\
Conversion gain                 & $\unit{6}{ADC\; counts\per\femto\coulomb}$ \\
\hline

\end{tabular}
\end{center}
\end{table*}

\begin{table*}
\begin{center}
\caption{Expected resolution parameters.}
\label{overv:overview3}
\vspace{2mm}
\begin{tabular}{|l|c|}\hline

Position resolution ($\sigma$) in $r\varphi$  & 1100 to $\unit{800}{\micro\metre}$\quad inner / outer radii\\
\phantom{Position resolution ($\sigma$)} in $z$     & 1\,250 to $\unit{1\,100}{\micro\metre}$ \\
\hline
d$E$/d$x$ resolution, isolated tracks & 5.0\%     \\
\phantom{d$E$/d$x$ resolution,} d$N$/d$y = 8\,000$ &  6.8\% \\
\hline
\end{tabular}
\end{center}
\end{table*}

In this paper we describe
the major components of the detector as currently realized and report on the
performance achieved after completion of the first round of calibration runs.

The first major challenge was the design and construction of the field cage,
whose overall thickness should not exceed 5\% of a radiation length while
providing, over a volume of nearly $\unit{90}{\metre\cubed}$, an axial electric field of 
$\unit{400}{\volt\per\centi\metre}$ with distortions in the $10^{-4}$ range.
The realization of this device is described in Sec.~\ref{fcage}.

The readout chambers are installed at the two endplates of the cylinder. Their
design is based on the 
Multi-Wire Proportional Chamber (MWPC)
technique with pad
readout. To ensure low diffusion of the drifting electrons and a large ion
mobility, Ne was chosen as the main component of the counting
gas. Furthermore, the size of the readout pads had to be adapted to the
expected large multiplicities, implying pad sizes as small as
$\unit{4\times7.5}{\milli\metre\squared}$ in the
innermost region.  As a consequence, the readout chambers have to be operated
safely at gains near $10^4$. In Sec.~\ref{chamb} we describe
the technical implementation and report on the first operating experience of
these detectors.

In Sec.~\ref{elect}  we discuss the design and implementation of the
electronics chain. Because of the high granularity (557\,568 readout
channels) special emphasis was placed on very low power consumption. To cope
with the large dynamic range needed to track particles from very low to high
momenta, and to provide low noise performance combined with efficient
baseline restoration and zero-suppression, the signals from the
preamplifier/shaper chip were fed into a 10-bit, $\unit{10}{\msps}$ ADC
integrated into a digital chip. We report on the implementation and running experience of the
electronics chain as realized in the ALICE TPC.      

Successful operation of a  very large detector like the ALICE TPC depends on 
a considerable amount of infrastructure and services, along with sophisticated
gas and cooling systems. One of the major challenges in this context was to
provide a temperature stability of less than 0.1~K across the full volume
of the TPC. This requirement originates from the strong temperature dependence
of the drift velocity in the Ne--CO$_2$--N$_2$ mixture at realistically
accessible electric 
fields. Furthermore, it is essential to control the O$_2$ content of the
counting gas below a level of 5 ppm to keep to a minimum the absorption of
electrons over  the long drift length.  The approach to solve these and many
other technological challenges is described in the sections on
cooling, gas system, infrastructure and services, and 
Detector-Control System (DCS).

Calibration and commissioning of the ALICE TPC relied, before the availability
of any collisions from the LHC, on three different methods: a set of external
UV laser beams was used to characterize field distortions and to determine the
magnitude of the correction from $E\times B$ effects on the drifting electrons
originating from the residual non-parallelism of the electric and magnetic
field inside the drift volume. Furthermore, radioactive krypton was inserted
through the gas system into the detector to provide efficient and precise
amplitude calibration of all 557\,568 readout channels. Finally, extensive
measurements with cosmic rays were performed to determine tracking
efficiencies, energy loss, and momentum resolution of the detector. The
methods used and results obtained  during these calibrations are described in detail in
Secs.~\ref{calcom} and~\ref{perform}. They demonstrate that detector performance is close to that
specified in the original technical design report~\cite{TDR:tpc}.

\begin{figure}
\centering
\includegraphics[height=\linewidth,angle=90,clip]{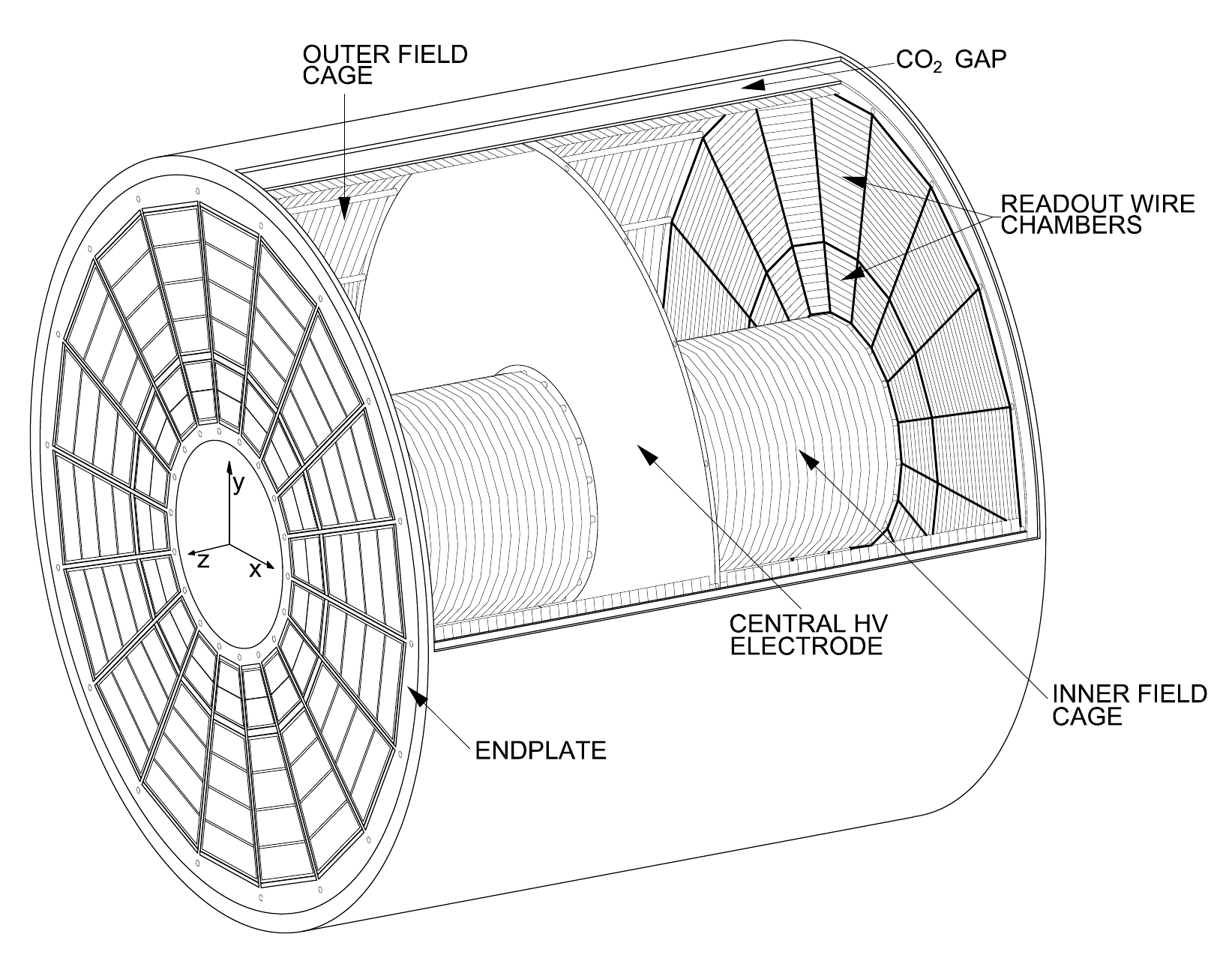} 
\caption{3D view of the TPC field cage.  The high voltage electrode is
  located at  the center  of the drift  volume. The endplates  with 18
  sectors and 36 readout chambers on each end are shown.}
\label{fcage:fieldcage}
\end{figure}

\section{Field cage}
\label{fcage}

The purpose  of the  field cage is  to define a  uniform electrostatic
field in  the gas  volume in order  to transport  ionization electrons
from their point of creation  to the readout chambers on the endplates
without  significant distortions.   The field  cage provides  a stable
mechanical structure for precise positioning of the chambers and other
detector  elements  while  being  as  thin as  possible  in  terms  of
radiation  lengths  presented to  the  tracks  entering  the TPC  (see
Fig.~\ref{fcage:fieldcage}).  In addition,  the walls  of  the field
cage provide  a gas-tight  envelope and ensure  appropriate electrical
isolation of the field cage from the rest of the experiment.

\begin{figure}
\centering
\includegraphics[width=\linewidth,clip]{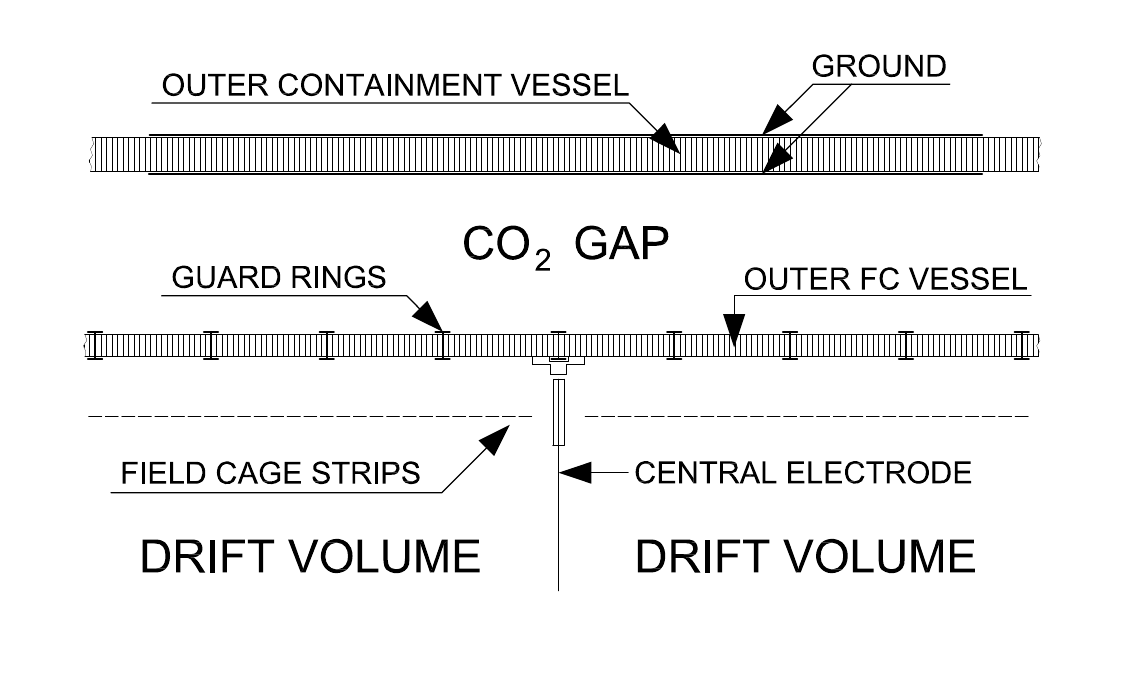} 
\caption{Detail  view  of  the  outer  field  cage  near  the  central
  electrode.}
\label{fcage:fieldcagedetail}
\end{figure}

It is  a classical TPC field  cage with the high  voltage electrode in
the middle of  the detector.  Electrons drift to both  end plates in a
uniform electric field that runs parallel to the axis of the cylinder.
The TPC is filled with a mixture of neon, carbon dioxide, and nitrogen
because  the  multiple  coulomb  scattering  in this  gas  mixture  is
relatively low,  it has good  diffusion characteristics, and it  has a
high positive  ion mobility that helps  to clear positive  ions out of
the drift  volume in a  short amount of time  (see Sec.~\ref{gas}).
However, to also have  fast electron drift velocities requires putting
$\unit{100}{\kilo\volt}$ on the central electrode.
The isolation of  the high voltage
field cage from the rest of  the experiment is ensured by using CO$_2$
filled gas  gaps between  the containment vessels  and the  field cage
vessels; see Fig.~\ref{fcage:fieldcagedetail}.

The design  of the ALICE  field cage is  similar to the design  of the
field cage used in the NA49 experiment~\cite{NA49}.  An important part
of the design  is the requirement to prevent  charge build-up, and possible
breakdown,  on  solid insulator  surfaces  between the  field-defining
strips and so  the use of these insulators  is minimized or completely
avoided.

The ALICE field cage consists of two parts; a field cage vessel with a
set of  coarsely segmented  guard rings and  a finely  segmented field
cage  which is  located inside the field  cage vessel.
The guard rings on the field  cage vessel help to avoid large electric
fields due to charge build-up on  the surface of the vessel. The rings
have a $\unit{92}{\milli\metre}$ gap between them and this corresponds
to a relatively low field  gradient of  $\unit{46.7}{\volt\per\milli\metre}$
on the  insulating surface  between the
rings.  The  guard rings are  made of $\unit{13}{\milli\metre}$ wide
strips  of aluminum
tape and they are placed on  both sides of the containment vessel with
a pitch of  $\unit{105}{\milli\metre}$.
Small holes were drilled through
the  walls of  the  vessel  to allow  for  electrical contact  between
corresponding rings  and filled with Al foil  feed-throughs and sealed
with  epoxy. The  potentials for  the guard  rings are  defined  by an
independent chain of $24\times\unit{500}{\mega\ohm}$ resistors (per end). The
first of these  resistors is connected to the  rim of the high-voltage
electrode.   The  last  one  is  connected to  ground  through  a
$\unit{100}{\kilo\ohm}$ resistor, across  which the  voltage drop  is  measured for
monitoring  purposes.   The field  gradient  between  the guard  rings
matches the  field gradient on  the finely segmented field  cage which
lies inside the guard ring vessel.

The finely  segmented field  cage is made  of 165 free  standing mylar
strips.  In  principle, there is  space for 166 strips  but mechanical
considerations near the central electrode prevents the installation of
the first  strip and so  it is left  out. (The resistor chain  for the
field cage  includes 166  resistors and so  in this  way the
missing strip is  included, see below.)  The mylar  strips do not come
into contact with the field  cage vessel or the gas containment vessel
but,    instead,     are    wrapped    around    a     set    of    18
Macrolon\textsuperscript{\textregistered}  rods   that  are  regularly
spaced    around    the    circumference    of     the    TPC.     The
Macrolon\textsuperscript{\textregistered} rods are located at a radius
of $\unit{2\,542.5}{\milli\metre}$ on the outer  field cage and 
$\unit{815}{\milli\metre}$ on the  inner field
cage. The field  cage strips are made of $\unit{13}{\milli\metre}$
wide aluminized mylar.
They are  stretched over the Macrolon\textsuperscript{\textregistered}
rods with a pitch of $\unit{15}{\milli\metre}$.  This leaves a 
$\unit{2}{\milli\metre}$ insulation gap between
each pair of strips and creates  a voltage gradient of
$\unit{300}{\volt\per\milli\metre}$ across
the gap.   The neon gas mixture  is the only  insulator that separates
the strips  of the field cage  except for the region  where the strips
touch the  rods. The resistor chains  for the field  cages are located
inside one  of the Macrolon\textsuperscript{\textregistered}  rods for
the outer field cage, one on each end, and also inside one rod for the
inner field cage, one on each end.

The combination of the fine  segmentation of the field cage strips and
the   coarsely  segmented  guard   rings  is   a  robust   and  stable
electrostatic design when $\unit{100}{\kilo\volt}$
is applied to the central electrode.
This design  minimizes the electric field  distortions that would
occur inside the  drift volume of the TPC if  the electric field lines
were  to go  directly to  ground from  the field  cage.  Electrostatic
calculations demonstrate  that the field shape  distortions inside the
drift  volume are  below $10^{-4}$  at a  distance of
$\unit{15}{\milli\metre}$ from the strips~\cite{Vranic1997}.

\subsection{Vessels}
\label{fcage:vessels}

\begin{figure}[t]
\centering
\includegraphics[width=\linewidth,clip]{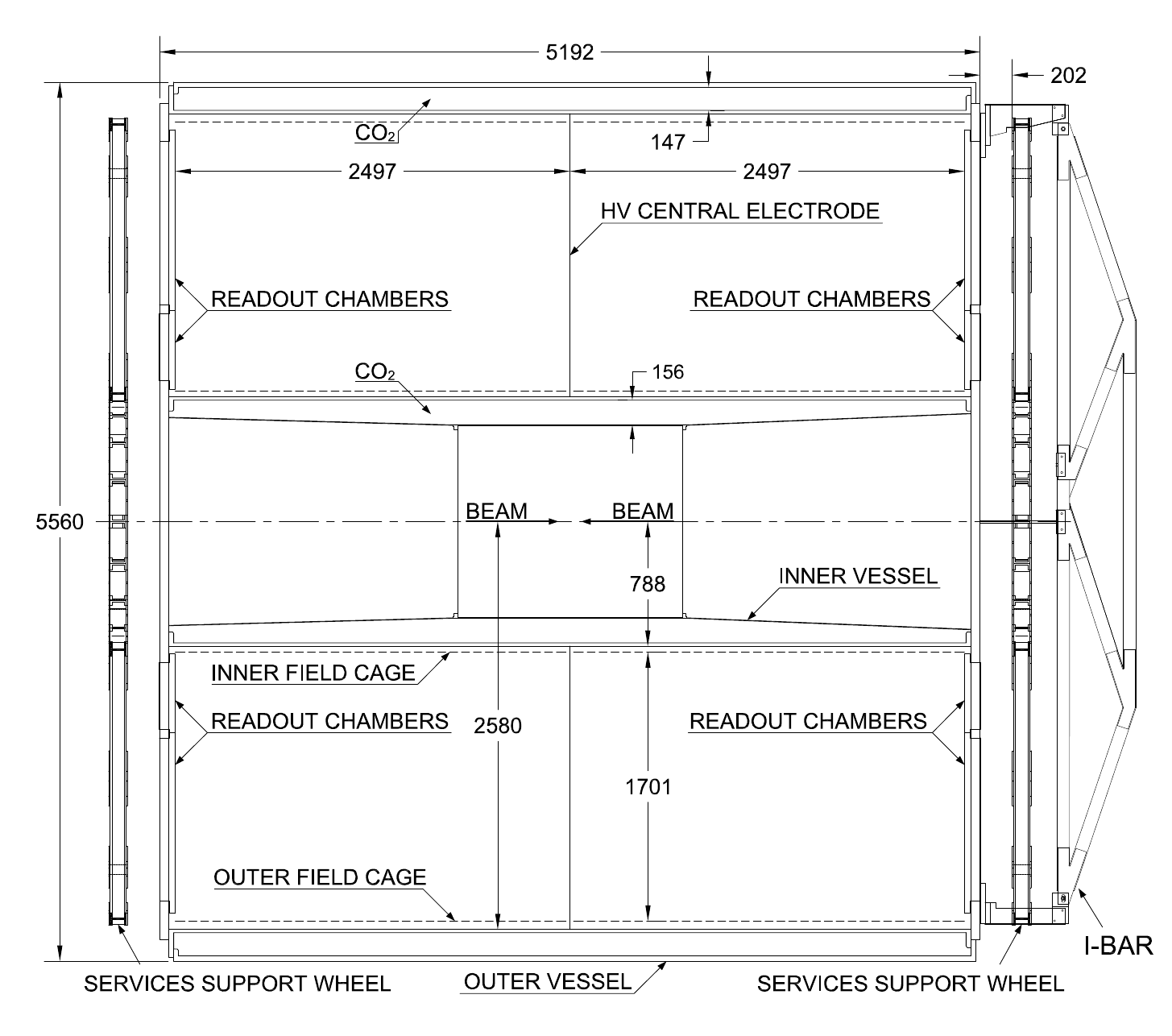} 
\caption{Cross-sectional side view of the TPC with relevant dimensions
  (in mm). The  service support wheels and one of  the I-bars are also
  shown.}
\label{fcage:TPCside}
\end{figure}

Four  cylinders  are required  to  make  the complete  field  cage
structure; two  field-cage vessels (one  inner and one outer)  and two
containment vessels  (inner and  outer).  The cylinders  are composite
structures   made    with   a   Nomex\textsuperscript{\textregistered}
honeycomb  core sandwiched between  prepreg sheets  (epoxy fiberglass)
and Tedlar\textsuperscript{\textregistered} foils, to provide a light,
rigid, and gas-tight  structure. A cross-sectional view of the  TPC with some
relevant dimensions is shown in Fig.~\ref{fcage:TPCside}.

The inner and  outer field-cage vessels (see Fig.~\ref{fcage:TPCside})
define the gas volume of the TPC.  They have radii of 
788 and $\unit{2580}{\milli\metre}$,
respectively.          The          thickness          of          the
Nomex\textsuperscript{\textregistered}  honeycomb core  is
$\unit{20}{\milli\metre}$  for
both vessels. Clamps are glued to the inside walls of the cylinders to
provide support  for the outer rim  of the high  voltage electrode and
all the rods.

The  containment vessels  surround  the field  cage  vessels and  they
provide  gas tight  and grounded  enclosures  at the  inner and  outer
diameters of  the TPC. To  maintain a good  ground, both walls  of the
cylindrical composite  structures are covered with
$\unit{50}{\micro\metre}$ thick
aluminum foil.   The containment vessels are separated  from the field
cage  vessels by  an insulating  gap and  these gaps  are continuously
flushed  with  CO$_2$ to  isolate  the  field  cage voltage  from  the
grounded containment walls.  The distance between the outer field cage
vessel   and   the  outer   containment   vessel   is  
$\unit{147}{\milli\metre}$.    The
Nomex\textsuperscript{\textregistered} core for the containment vessel
is $\unit{30}{\milli\metre}$ thick.
The inner containment vessel is made of three parts:
a central drum which surrounds the inner tracking system (ITS) and two
cones  that support  the drum; see  Fig.~\ref{fcage:TPCside}.  The
central drum  is cylindrical  in shape ($\unit{1420}{\milli\metre}$
long and  $\unit{610}{\milli\metre}$ in
radius), and has a Nomex\textsuperscript{\textregistered} core that is
$\unit{5}{\milli\metre}$ thick. The
support  cones are  made of  $\unit{3}{\milli\metre}$ thick  aluminum
and  they  span the
distance from the central drum to each endplate.  They provide support
for  the  ITS  while  leaving  room for  a  muon
absorber on one end and services  for the inner tracking system on the
other.   The attachment  between the  central  drum and  the cones  is
sealed with  a $\unit{2}{\milli\metre}$ thick flat neoprene  rubber  
ring.  The CO$_2$ gap
between the inner  containment vessel and the inner  field cage vessel
is $\unit{156}{\milli\metre}$ thick at the centerline  of the detector
and decreases to $\unit{80}{\milli\metre}$ near the endplates.

The thickness of the critical components of the field cage vessels 
are listed in Tab.~\ref{fcage:X} in units of radiation length.

\begin{table}[t]
\begin{center}
\caption{The thickness of the inner and outer field cage components are 
  listed in radiation lengths.  The total thickness presented to a
  particle entering the TPC at $\eta=0$  is about 1\% $X_0$. 
}
\begin{tabular}{|l|c|}
\hline
Part                     & $X/X_0$ [\%] \\
\hline
\hline
Central drum             & 0.470 \\
Inner CO$_2$ gap         & 0.085 \\
Inner field cage vessel  & 0.401 \\
Inner field cage strip   & 0.012 \\
\hline
Inner field cage total   & 0.968 \\
\hline
\hline
Drift gas                & 0.607 \\
\hline
\hline
Outer field cage strip   & 0.012 \\
Outer field cage vessel  & 0.401 \\
Outer CO$_2$ gap         & 0.081 \\
Outer containment vessel & 1.330 \\
\hline
Outer field cage total   & 1.824 \\
\hline
\end{tabular}
\label{fcage:X}
\end{center}
\end{table}

\subsection{Central electrode}
\label{fcage:CE}

The central electrode is made  of a stretched $\unit{23}{\micro\metre}$
thick mylar
foil which is  aluminized on both sides and held flat  by an inner and
outer aluminum  rim.  Three foils  were glued together by  laying them
side by  side and gluing $\unit{50}{\milli\metre}$
wide aluminized mylar  bands over the
junctions.   The resulting $6\times\unit{6}{\metre\squared}$ 
foil was  stretched with
pneumatic jacks and  glued onto a set of inner  and outer rims.  After
curing, a  second set of rims  was lowered into position  and glued to
the foil.

\subsection{Rods}
\label{fcage:rods}

A total of 72 rods are positioned axially on the internal walls of the
inner  and outer  field-cage vessels  and in  the corners  between the
readout chambers. Their main role is to hold the field cage strips
for the inner and outer field  cages. The rods are made of several 178
and  $\unit{209}{\milli\metre}$ long pieces  of
Macrolon\textsuperscript{\textregistered}
tube (a  special sort  of Plexiglas) which  have been  glued together.
The  final rod  assemblies have  an outer  (inner) diameter  of
$\unit{44}{\milli\metre}$ ($\unit{36}{\milli\metre}$).
Their outer  surfaces were  machined with  $\unit{2}{\milli\metre}$
wide and $\unit{2.5}{\milli\metre}$ deep grooves,  at a pitch of
$\unit{15}{\milli\metre}$,  to increase the distance
along the  insulator surface between  the strips. An aluminum  ring at
each glue junction helps to minimize and redistribute the accumulation
of charge  along the  rods.  The gluing  operation was performed  on a
precision jig in  order to achieve a uniform spacing  of the strips to
within $\unit{100}{\micro\metre}$.  The rods for  the outer field cage,
  except for
their grooves, are coated with  copper to avoid charge accumulation on
their  exposed surface.  The  rods  are  held  in position  with
holding clamps which are glued to  the walls of the field cages with a
$\unit{500}{\milli\metre}$ spacing between the clamps.

\subsubsection{Resistor rods}
\label{fcage:resistorrods}

\begin{figure}
\centering
\includegraphics[width=\linewidth,clip]{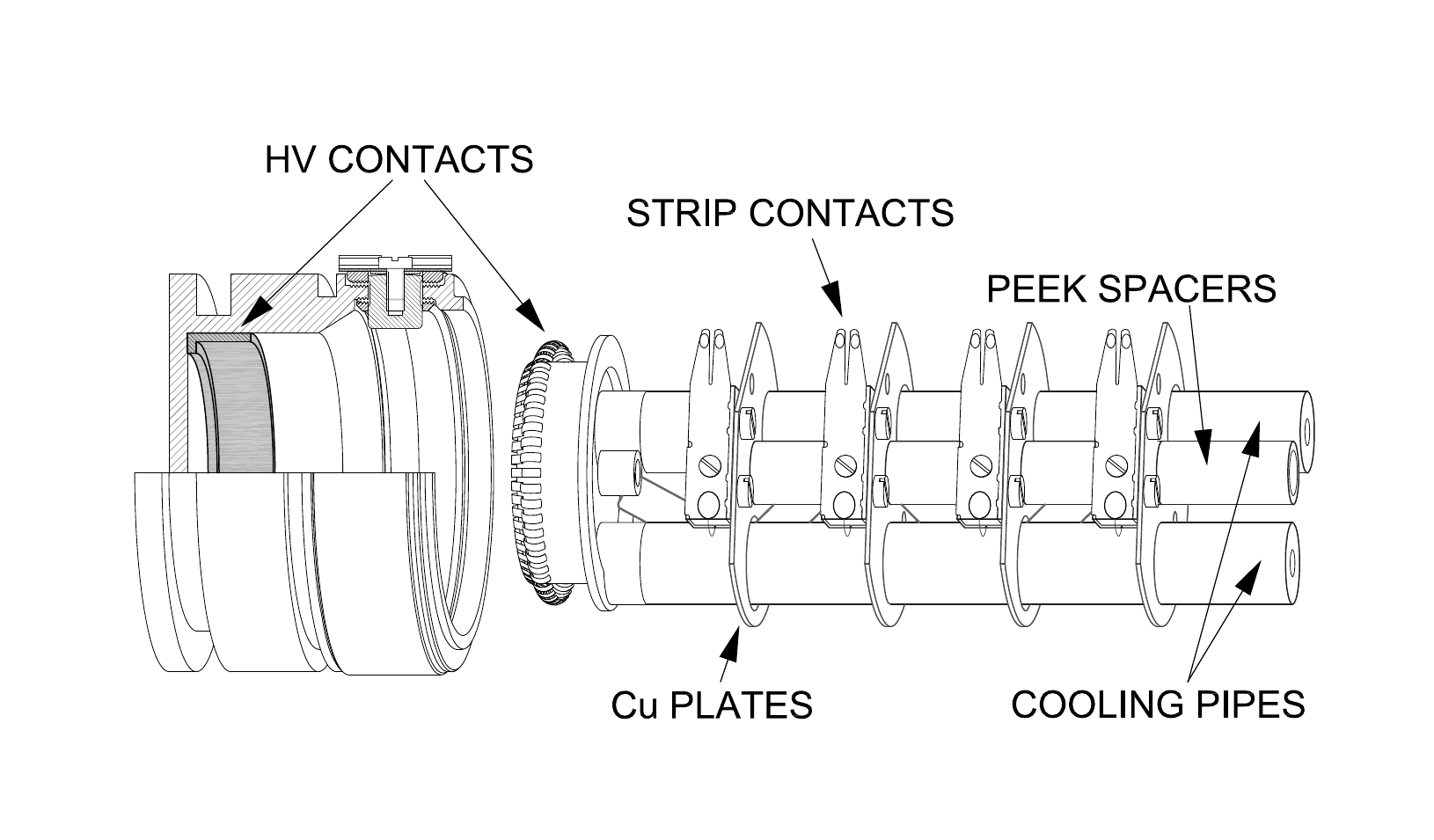} 
\caption{Detailed view  of the high-voltage  end of the  resistor rod,
  showing   the   cooling   pipes,   the  central   PEEK   pipe,   the
  heat-dissipating copper plates, the  contacts to the strips, and the
  high-voltage contact, which matches the contact at the housing rod.}
\label{fcage:rrhv}
\centering
\includegraphics[width=\linewidth,clip]{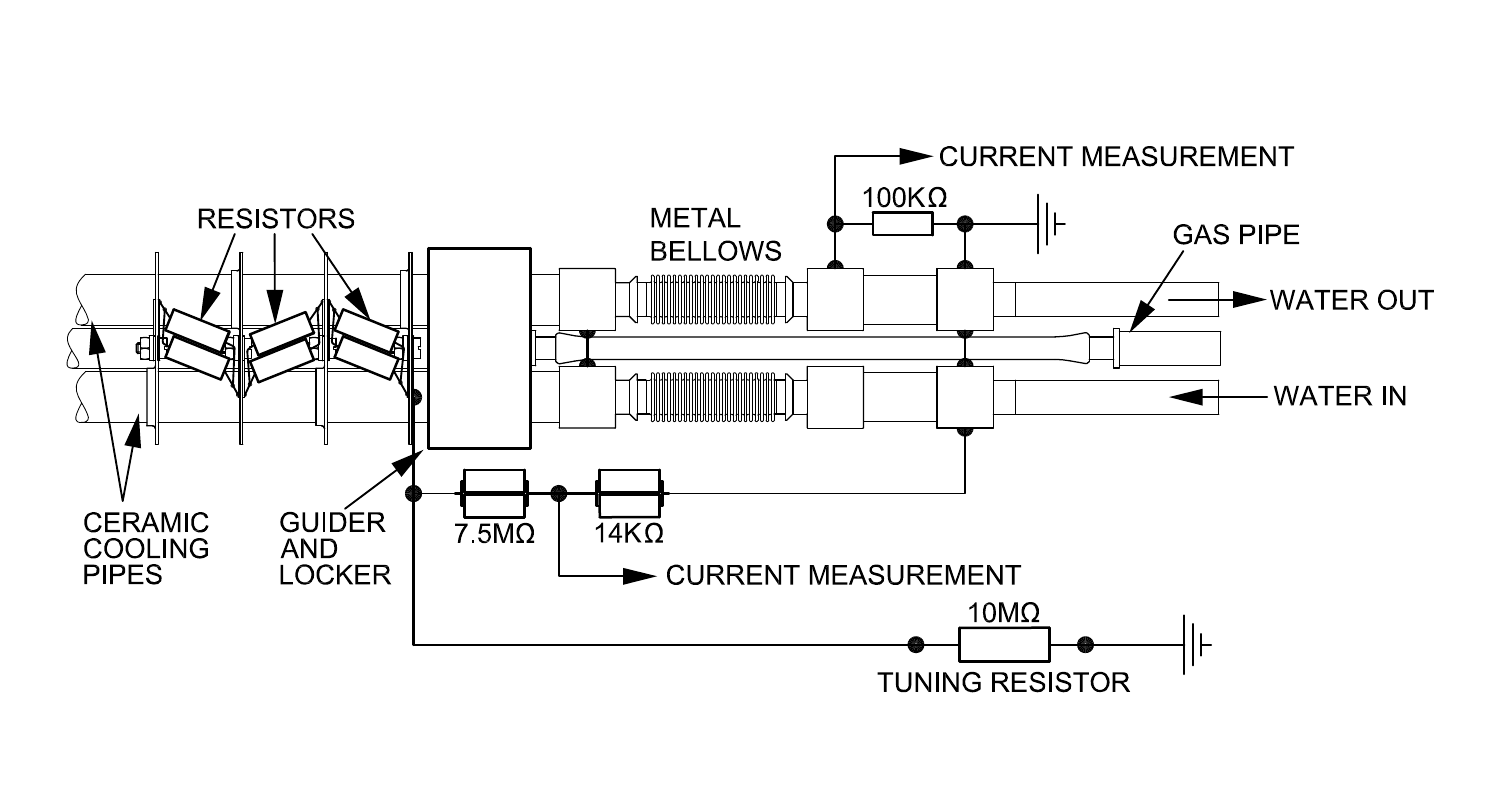} 
\caption{Schematic of the mechanical and electrical arrangement of the
  ground side of  the resistor rod. The currents  through the resistor
  chain and through the cooling water are measured independently.}
\label{fcage:rrgnd}
\end{figure}

The voltage dividers are  integrated with the so-called resistor rods,
and they are inserted into four  of the rods of the field cage: inner,
outer,  and on  both  sides.  The  resistor  rods contain  a chain  of
resistors  which define  the  potential  on each  strip  of the  field
cage. The innovative design of these rods allows for water cooling and
serviceability. The power dissipated by  the resistors is removed by a
water-cooling circuit that runs back and forth through the rods. A set
of contacts  ensures a  good connection between  the resistors  in the
chain  and  to  each   strip.   Provisions  are  made  for  insertion,
contacting, locking, and removal of  the rods for service.  Details of
both ends of a resistor  rod can be seen in Figs.~\ref{fcage:rrhv}
and~\ref{fcage:rrgnd}.

For each resistor rod, a set  of 165 copper plates, 
$\unit{0.5}{\milli\metre}$ thick, are
held        together         by        short        sections        of
PEEK\textsuperscript{\textregistered}
 (polyacryletheretherketone, a thermoplastic) 
tubing  which are glued together,
 thus defining the
$\unit{15}{\milli\metre}$
pitch  for the  strips.  The  resulting tube  is 
$\unit{2.5}{\metre}$ long  and the
central hole  is used  to flush  the system with  drift gas  since the
Macrolon\textsuperscript{\textregistered}   rod  is   not  necessarily
gas-tight. In addition to the central hole, two more holes are drilled
into the  copper plates through  which two ceramic pipes,  
$\unit{3}{\milli\metre}$ inner and  $\unit{9}{\milli\metre}$ outer
diameter,  are  inserted.  The  copper  plates  are
connected to these ceramic pipes  with thermally conductive glue in an
alternating  pattern.  The pipes  are  bridged  together  at the  high
voltage end by  a stainless steel tube so  that de-ionized water flows
into one of  the ceramic pipes and returns through  the other pipe. In
this manner, the power dissipated by the voltage divider is coupled to
the copper plates and is removed by the cooling water.

A $\unit{7.5}{\mega\ohm}$ resistor is connected between each  of the
Cu plates. The first resistor,  from the central membrane to  the
first Cu plate, has a value of $\unit{15}{\mega\ohm}$ to compensate
for the missing first strips
in  the field  cage.   In all  cases,  the resistors  are soldered  to
washers  which are  then screwed  to the  plates.  This  results  in a
resistor chain with a total  resistance of $\unit{1\,245}{\mega\ohm}$
inside the TPC drift volume plus an additional
$\unit{4.286}{\mega\ohm}$ at the end of the
chain to allow  for precisely tuning the voltage on  the last strip of
the  field  cage;  see  Fig.~\ref{fcage:rrgnd}.  A  small  piece  of
PEEK\textsuperscript{\textregistered}  material  is  screwed  to  each
copper  plate,  onto which  a  flexible,  gold-plated stainless  steel
electrical contact is again screwed.  A $\unit{50}{\micro\metre}$
Au--W wire is used
to  make  the connection  between  the  plate  and the  contact,  thus
minimizing the amount of heat transmitted to the contacts and into the
drift volume.   At the high-voltage  end, a connector consisting  of a
crown of  flexible contacts provides the electrical  connection to the
corresponding     part     in     the     field-cage     rod;   see
Fig.~\ref{fcage:rrhv}.

The ground end of the resistor rod, shown in Fig.~\ref{fcage:rrgnd},
is  equipped  with  various  resistors for  properly  terminating  the
assembly to ground, tuning the  potential of the last strip, measuring
the  current through  the  resistor chain  and  measuring the  current
through the  cooling water.  The contacts on  the field-cage  rods are
made  of  gold-plated  brass and  are  glued  into  holes in  the  rod
wall.  The hooks to  which the  strips are  attached are  screwed onto
these contacts.

With  $\unit{100}{\kilo\volt}$ on  the  central electrode,  the  total
current  flowing through one of  the resistor chains is
$\unit{91}{\micro\ampere}$; this  is a sum of $\unit{80}{\micro\ampere}$
flowing  through   the  field   cage   resistor  chain,
$\unit{8.4}{\micro\ampere}$ through  the guard ring chain  and
$\unit{2.5}{\micro\ampere}$ flowing
through the cooling water for the rod.

\subsubsection{High-voltage cable rod}
\label{fcage:cablerod}

The  cable that  provides high  voltage  to the  central electrode  is
inserted         into         one         of         the         outer
Macrolon\textsuperscript{\textregistered} rods.  The ground shield for
the cable has been removed over the entire length of the rod
($\unit{250}{\centi\metre}$)
and replaced  by a  semi-conductive carbon loaded  polyethylene sleeve
that provides a  smooth voltage gradient inside the  rod.  The contact
for  the cable is  similar to  the resistor-rod  contact, and  again a
special cable  connects the  rod's contact to  the rim of  the central
electrode. A rod  with a spare contact is installed  on the other side
of the TPC. A flange in  the endplate ensures gas tightness of the rod
and mechanical support for the cable.

\subsubsection{Laser rods}
\label{fcage:laserrods}

Six  outer  rods  per  side  are  devoted  to  the  laser  calibration
system. The  laser rods are  spaced uniformly around the  perimeter of
the  TPC. The  corresponding flanges  for  the rods  include a  quartz
window  for introducing  a laser  beam into  the rod.  The  rod itself
holds, in its interior, a set  of mirrors which deflect the light into
the drift volume of the TPC  through openings machined in the rods for
this purpose. The  laser calibration system is described  in detail in
Sec.~\ref{laser}.

\subsubsection{Gas rods}
\label{fcage:gasrods}

Ten rods  from the outer field cage,  and 17 rods for  the inner field
cage, are  empty and so these  rods plus the  partially obscured laser
rods are used to circulate gas through the TPC.  The rods are machined
with an array of $\unit{1}{\milli\metre}$ holes which have a 
$\unit{15}{\milli\metre}$ pitch. The inner rods
are used for  the gas inlet, and  the outer rods are used  for the gas
outlet and this is  the only way that gas goes in  and out of the TPC.
In  this  manner, the  gas  flows  radially  through the  system  thus
minimizing the forces exerted on the central electrode.

\subsection{Strips}
\label{fcage:strips}

\begin{figure}[t]
\centering
\includegraphics[width=\linewidth,clip]{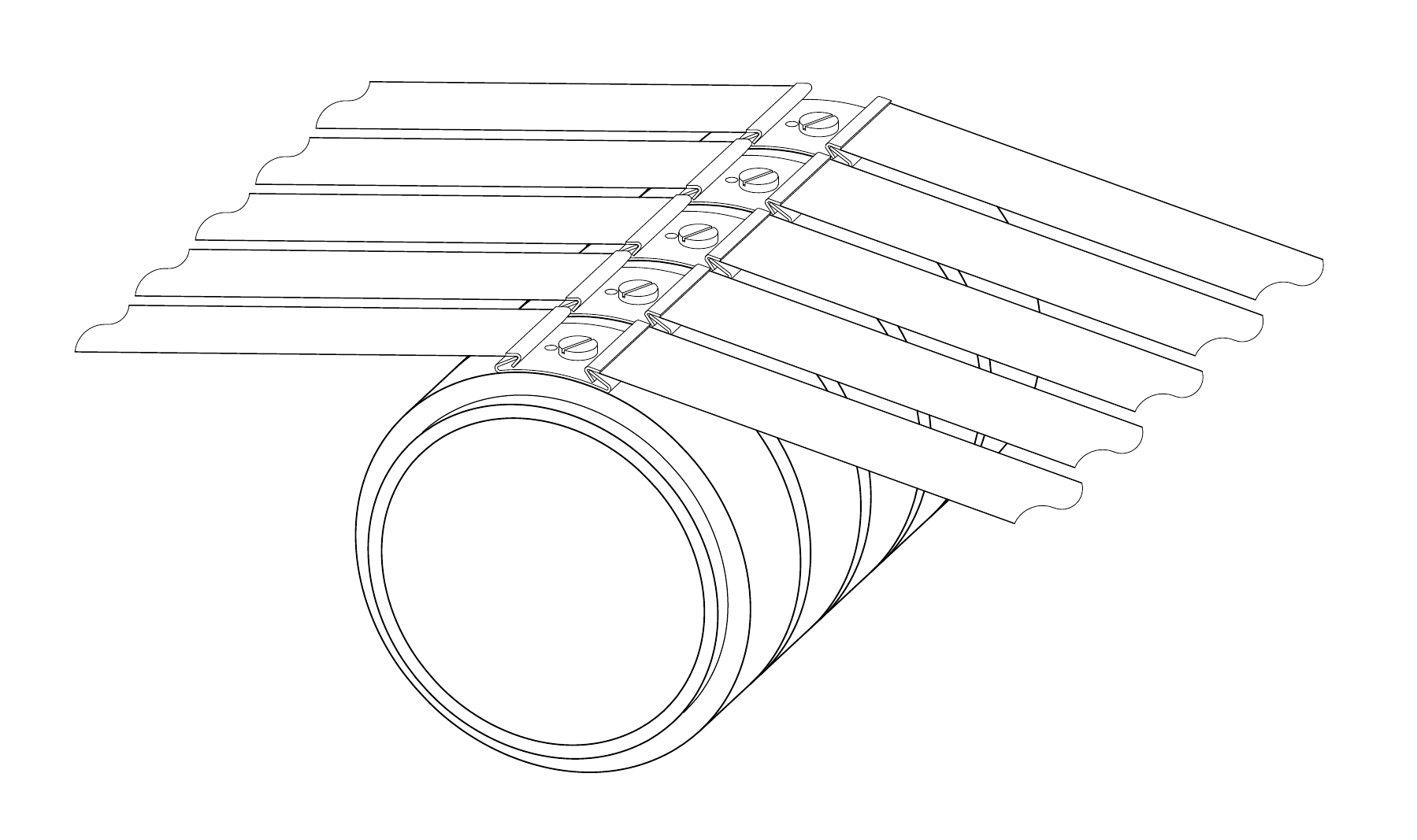} 
\caption{The field cage strips connect to the resistor rods with Cu-Be
  hooks. Metallic screws  hold the hooks onto the  rod and provide the
  electrical connection to contacts inside the rod.}
\label{fcage:rrhooks}
\end{figure}

\begin{figure}[t]
\centering
\includegraphics[width=\linewidth,clip]{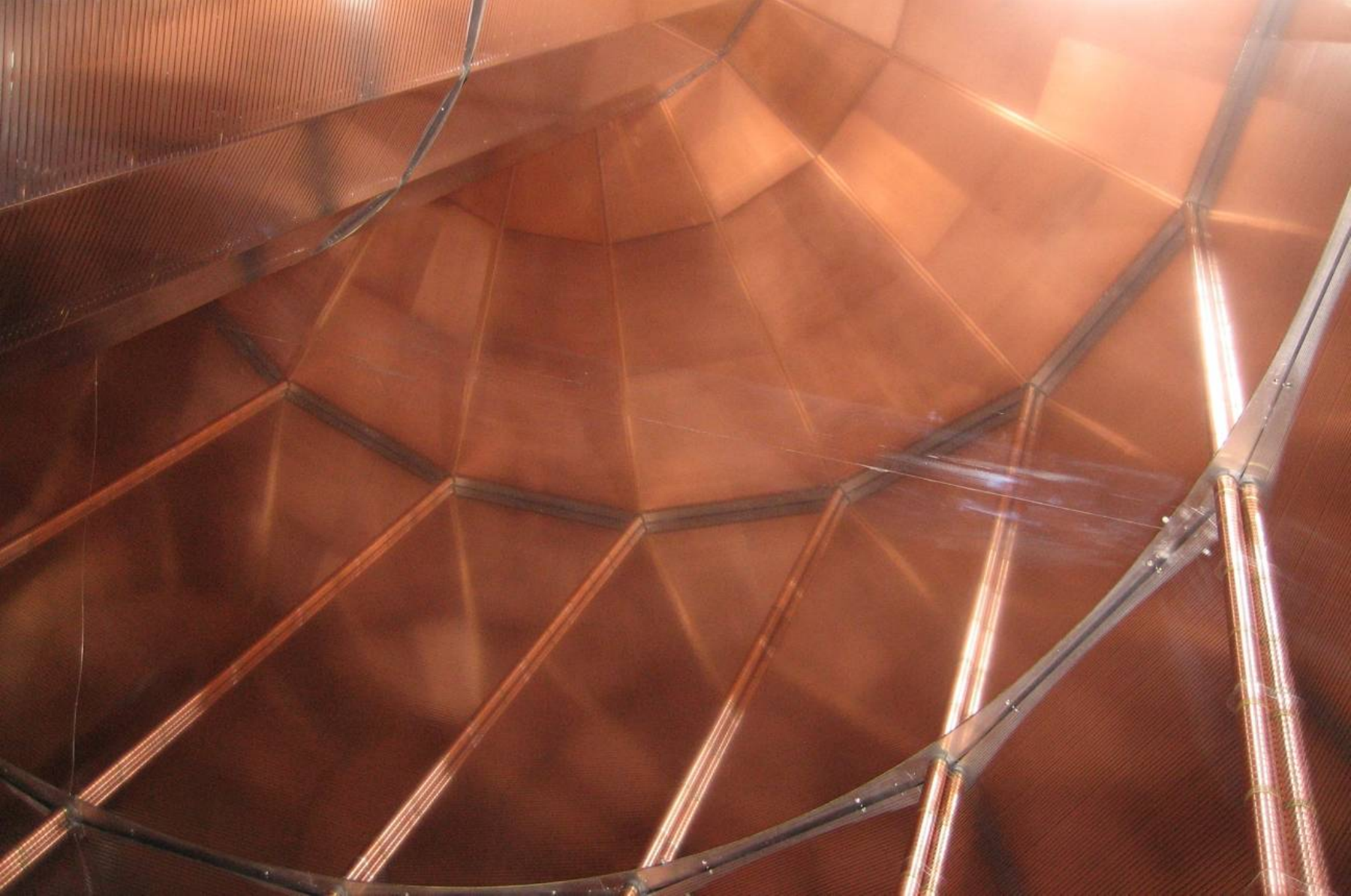} 
\caption{A view inside the field  cage where the strips and supporting
  rods are visible. The central electrode reflects a view of the field
  cage and the readout chambers.  The subdivision of the pad planes of
  the OROCs into four boards  can be seen. The skirt electrodes around
  the OROCs are also visible.}
\label{fcage:photofc}
\end{figure}

The   field-defining   strips   are   made  from   aluminized   mylar,
$\unit{25}{\micro\metre}$ thick and  $\unit{13}{\milli\metre}$ wide.
Under a tension  of $\unit{3.5}{\newton}$, they
are cut to the right length  ($\unit{5.246}{\metre}$ and
$\unit{16.018}{\metre}$ for the inner and
outer strips, respectively).  A custom-made tool was then used to fold
Cu-Be foil around the end of each strip to produce a hook.  The strips
were then strung around the rods and connected to similar hooks on the
resistor  rods, as  shown in  Fig.~\ref{fcage:rrhooks}.   A photograph
showing  the  interior  of  the  finalized  field  cage  is  shown  in
Fig.~\ref{fcage:photofc}.

\subsection{Skirts}
\label{fcage:skirts}

The strips of the inner field cage  run close to the inner edge of the
readout chambers  thus enabling a good  match of the  drift field with
the potentials on  the cover electrodes of the  inner readout chambers
(see  Sec.~\ref{chamb}).  The  voltages  on the  cover electrodes  are
tunable and this helps to ensure a good match. However, there is a gap
between the outer  readout chambers and the strips  of the outer field
cage which is too large to be left unfilled.  The electric field would
be distorted  if it were  left exposed and  so a
$\unit{38}{\milli\metre}$ wide  skirt is
inserted into  the gap.  The skirts  are parallel to  the endplate and
are electrically interconnected  so they can be set  to an appropriate
potential  to minimize the  distortions of  the field.   A temperature
sensor (PT1000) is  glued on the back side of  each skirt sector, thus
allowing for temperature measurements inside the volume of the TPC.

\subsection{Endplates}
\label{fcage:endplates}

The function of the endplates is  to align the cylinders for the field
cage vessels  and to hold the  readout chambers in  position. The four
cylinders  are screwed  to the  flanges  that connect  the field  cage
vessels  and the  containment  vessels, and  are  made gas-tight  with
O-rings. The  aluminum structure of  the endplate is 
$\unit{60}{\milli\metre}$  thick and the spokes are
$\unit{30}{\milli\metre}$ wide.  The cut-outs for the readout chambers are
equipped with provisions for the alignment of the chambers relative to
the central electrode and are  independent of the endplate itself (see
Sec.~\ref{chamb}).  Gas tightness is achieved  by a sealing foil and a
double  O-ring;  one on  the  chamber and  one  on  the endplate.  The
endplates also  provide feed-throughs and  flanges for gas,  laser and
electrical connections.

\subsection{I-bars}
\label{fcage:ibars}

The TPC is  installed at an angle of 0.79 degrees  with respect to the
horizontal due to the inclination  of the LHC accelerator at the ALICE
collision hall.  This  puts a gravity load on  the endplates and leads
to a  displacement of the inner  field cage with respect  to the outer
field cage.   The elastic deformation  of the endplates is  removed by
pulling  on  the   inner  field  cage  with  a   pair  of  I-bars.  In
Fig.~\ref{fcage:TPCside}, the  I-bars are shown attached  on the right
hand side  of the TPC and were  designed so that they  do not obstruct
the area  around the beam-pipe. The  I bars are attached  to the outer
ring of the endplate and can push or pull on the inner field cage ring
in order  to re-align  the field cages.  During assembly in  the ALICE
detector, it  was necessary  to pull  on the inner  field cage  with a
force  of  $\unit{3}{\kilo\newton}$ and  an  alignment  of
about $\unit{150}{\micro\metre}$ was  actually achieved.

\section{Readout chambers}
\label{chamb}

\subsection{Design considerations}
\label{chamb:Design_considerations}

 Large-scale TPCs have been employed and proven to work in collider experiments before~\cite{Wieman1998}, but none of them had to cope with the particle densities and rates anticipated for the ALICE experiment~\cite{ppr1,ppr2}. 

For the design of the Read-Out Chambers (ROCs), this leads to requirements that go beyond an optimization in terms of momentum and d$E$/d$x$ resolution. In particular, the optimization of rate capability in a high-track density environment has been the key input for the design considerations. 
 
The ALICE TPC has adopted MWPCs with cathode pad readout. In preparation of the TPC TDR~\cite{TDR:tpc} alternative readout concepts had also been considered, such as Ring Cathode Chambers (RCCs)~\cite{Janik2009} or Gas Electron Multipliers (GEMs)~\cite{Sauli2004} as amplification structures. However, those concepts seemed, though conceptually convincing, not yet in an R\&D state to be readily adopted for a large detector project, which had to be realized within a relatively short time span.

\subsection{Mechanical structure}
\label{chamb:Mechanical_structure}

The azimuthal segmentation of the readout plane is common with the subsequent ALICE detectors TRD and TOF, i.e. 18 trapezoidal sectors, each covering 20$^{\circ}$ in azimuth. The radial dependence of the track density leads to different requirements for the readout-chamber design as a function of radius. Consequently, there are two different types of readout chambers, leading to a radial segmentation of the readout plane into Inner and Outer ReadOut Chamber (IROC and OROC, respectively). In addition, this segmentation eases the assembly and handling of the chambers as compared to a single large one, covering the full radial extension of the TPC.

\begin{figure}\centering
\includegraphics[width=\linewidth,clip]{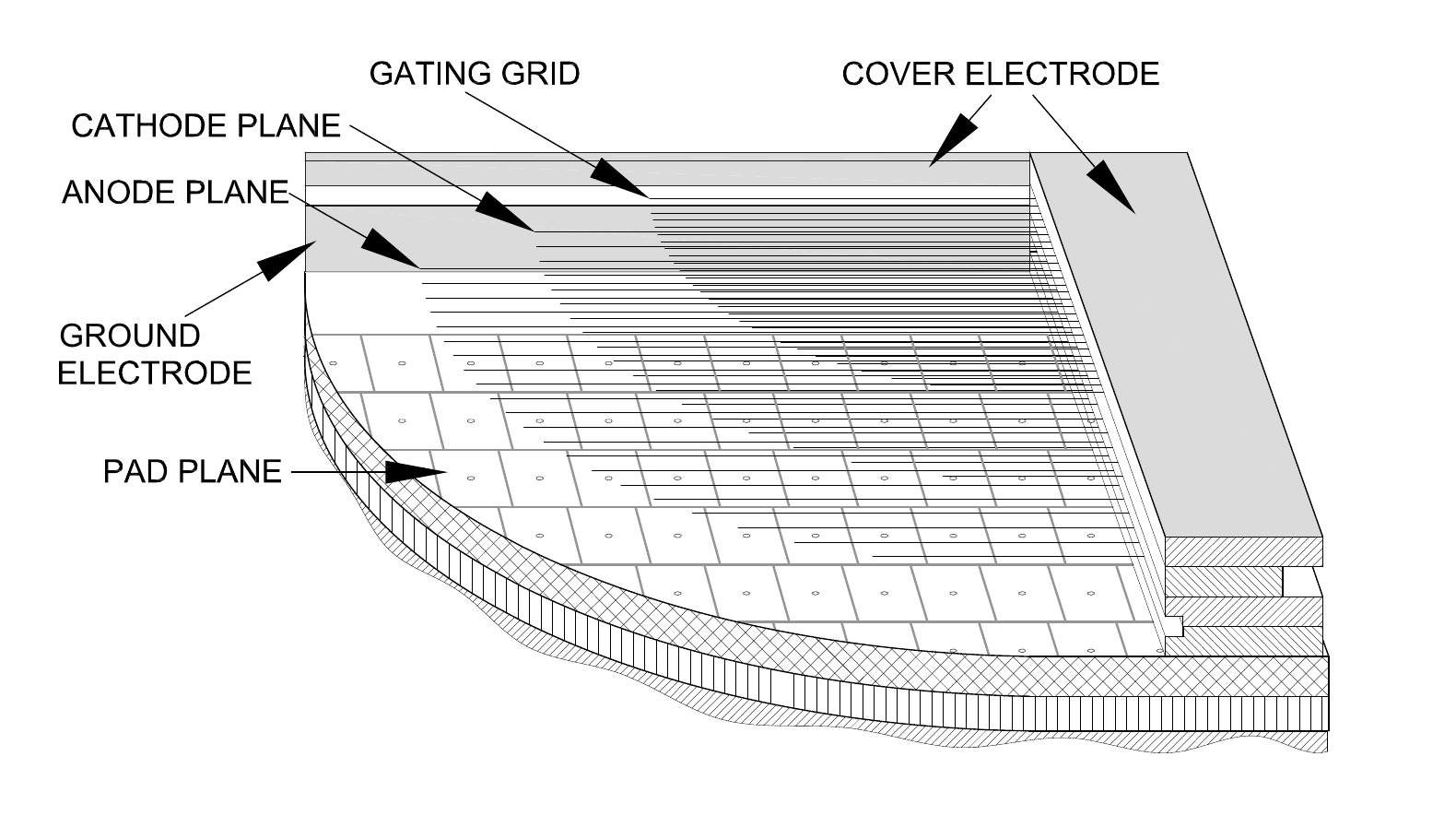} 
\caption{Cross section through a readout chamber showing the pad plane, the wire planes and the cover electrode.}
\label{chamb:wire_3D}
\end{figure}

The dead space between neighboring readout chambers is minimized by a special mounting technique (described in Sec.~\ref{chamb:Precommissioning}) by which the readout chambers are attached to the endplate from the inside of the drift volume. The dead space between two adjacent chambers in the azimuthal direction is 27~mm. This includes the width of the wire frames of 12~mm on each chamber (see Fig.~\ref{chamb:wire_3D}) and a gap of 3~mm between two chambers. The total active area of the ALICE~TPC readout chambers is 32.5~m$^{2}$.
The inner and outer chambers are radially aligned, again matching the acceptance of the external detectors. The effective active radial length (taking edge effects into account) varies from 84.1~cm to 132.1~cm (134.6~cm to 246.6~cm) for the inner (outer) readout chambers.
The mechanical structure of the readout chamber itself consists of four main components: the wire planes, the pad plane, made of a multilayer Printed Circuit Board (PCB), an additional 3 mm Stesalit insulation plate, and a trapezoidal aluminum frame.

\subsubsection{Wires}
The wire length is given by the overall detector layout and varies from 27~cm to 44~cm in the inner chambers, and from 45~cm to 84~cm in the outer chambers.

At constant potential, the gas gain increases with decreasing anode-wire diameter. Thus, a small anode-wire diameter is preferred. Owing to their superior strength, gold-plated tungsten  is preferable to copper--beryllium (an alloy of 98\% Cu and 2\% Be) for the thin anode wires. 
However, for the thicker cathode and gating grid wires this dense material would require unaffordable tensions on the thin wire ledges. Therefore, copper--beryllium is used.

However, electrostatic and gravitational forces cause the anode wires to sag, leading to gas-gain variations along the wire. The electrostatic sag is approximately proportional to the square of the length of the wire, and inversely proportional to the stretching force, while the gravitational sag depends on the density of the wire material. Therefore, the wires need to be mechanically strong enough to withstand the required stretching forces. We have chosen for the anode wires a diameter of 
20~$\micro$m and a stretching force of 0.45~N. The cathode and gating grid wires have a diameter of
75~$\micro$m and a stretching force of 0.6 and 1.2~N for inner and outer chamber, respectively. The wire tension has been measured during production for all wires (see Sec.~\ref{chamb:Steps}). The measured values ensure a wire sag around 50~$\micro$m and thus are below the specified limit of 70 $\micro$m \cite{TDR:tpc}.

\subsubsection{Wire planes}
\begin{figure}
\centering
\includegraphics[width=\linewidth,clip]{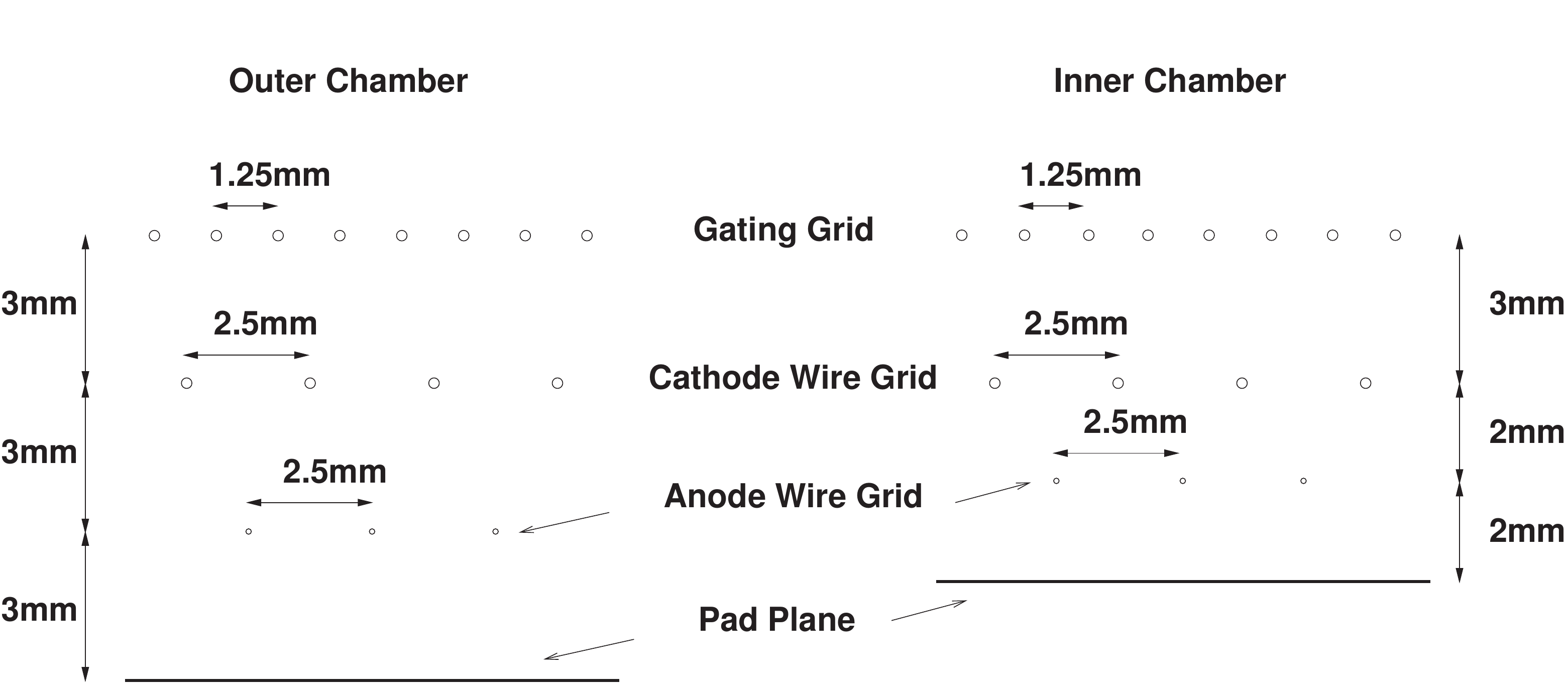} 
\caption{Wire geometries of the outer (left) and inner (right) readout chambers.}
\label{chamb:wire_structure}
\end{figure}

The ALICE-TPC readout chambers employ a commonly used scheme of wire planes, i.e. a grid of anode wires above the pad plane, a cathode-wire grid, and a gating grid. All wires run in the azimuthal direction. Since the design constraints are different for the inner and outer chambers (see below), their wire geometry is different, as shown in Fig.~\ref{chamb:wire_structure}. The gap between the anode-wire grid and the pad plane is 3\,mm for the outer chambers, and only 2\,mm for the inner chambers. The same is true for the distance between the anode-wire grid and the cathode-wire grid. The gating grid is located 3\,mm above the cathode-wire grid in both types of chamber. The anode-wire grid and the gating grid are staggered with respect to the 
cathode-wire grid. Henceforth we abbreviate the wire geometry of the inner chamber by (2-2-3), and that of the outer chamber by (3-3-3). 

\subsubsection{Anode-wire grid}
Because of the expected high particle multiplicity and the relatively large gas gains required for the readout chambers (see below) a small anode-wire pitch was chosen for the ALICE TPC to minimize the accumulated charge per unit length of the anode wire and hence the risk of rate-induced gas-gain variations. This led to the choice of a 2.5~mm pitch for the anode wires. There are no field wires since they would reduce the signal coupling to the pads, as they pick up a significant fraction of the signal. The absence of field wires also considerably reduces the mechanical forces on the wire frames. However, a chamber without field wires requires a somewhat higher voltage to achieve the required gas gain and a higher geometrical precision in the positioning of the wires. 

\subsubsection{Cathode-wire grid} 
The cathode-wire grid separates the drift volume from the amplification region. A large number of the ions produced in the amplification avalanche are collected at the cathode wires without causing a noticeable reduction in electron transmission. The cathode wire pitch is 2.5~mm.
Electrostatic calculations substantiating the above layout numerically are described in detail in Ref.~\cite{TDR:tpc}.

\subsubsection{Gating-wire grid}
The gating grid is located above the cathode-wire grid, with alternating wires connected together electrically. In the open gate mode, all the gating grid wires are held at the same potential $V_{\rm G}$, admitting electrons from the drift volume to enter the amplification region. In the absence of a valid trigger, the gating grid is biased with a bipolar field $V_{\rm G}\pm\Delta V$ (see Sec.~\ref{infra:Gate_pulser}), which prevents electrons from the drift volume to get to the amplification region. This considerably reduces the integral charge deposit on the anode wires. In addition, the closed gate stops ions created in the avalanche processes of previous events from drifting back into the drift volume. This is important because escaping ions accumulate in the drift volume and can cause severe distortions of the drift field \cite{blumrolandi}. The goal is therefore to avoid increasing the ion charge density above that created by primary ionization. The resulting requirement is that the ion leakage from the amplification region has to be less than $10^{-4}$. To achieve an electron transparency close to 100\% in the open mode while trapping ions and electrons in the closed mode, the offset and bias potentials of the gating grid are carefully adjusted. On the other hand, any ionization produced by particles traversing the gap between the gating grid and pad plane will unavoidably be amplified at the anode wires and thus contribute to the integral charge accumulation. To minimize this effect, the gap between the gating and cathode-wire grid is only 3~mm, sufficient to trap the ions within a typical gate opening time of 100~$\micro$s. To keep the alternating bias voltages low, the pitch between the gating grid wires is 1.25~mm.

\subsubsection{Cover and edge geometry}
\begin{figure}[t]
\centering
\includegraphics[width=\linewidth,clip]{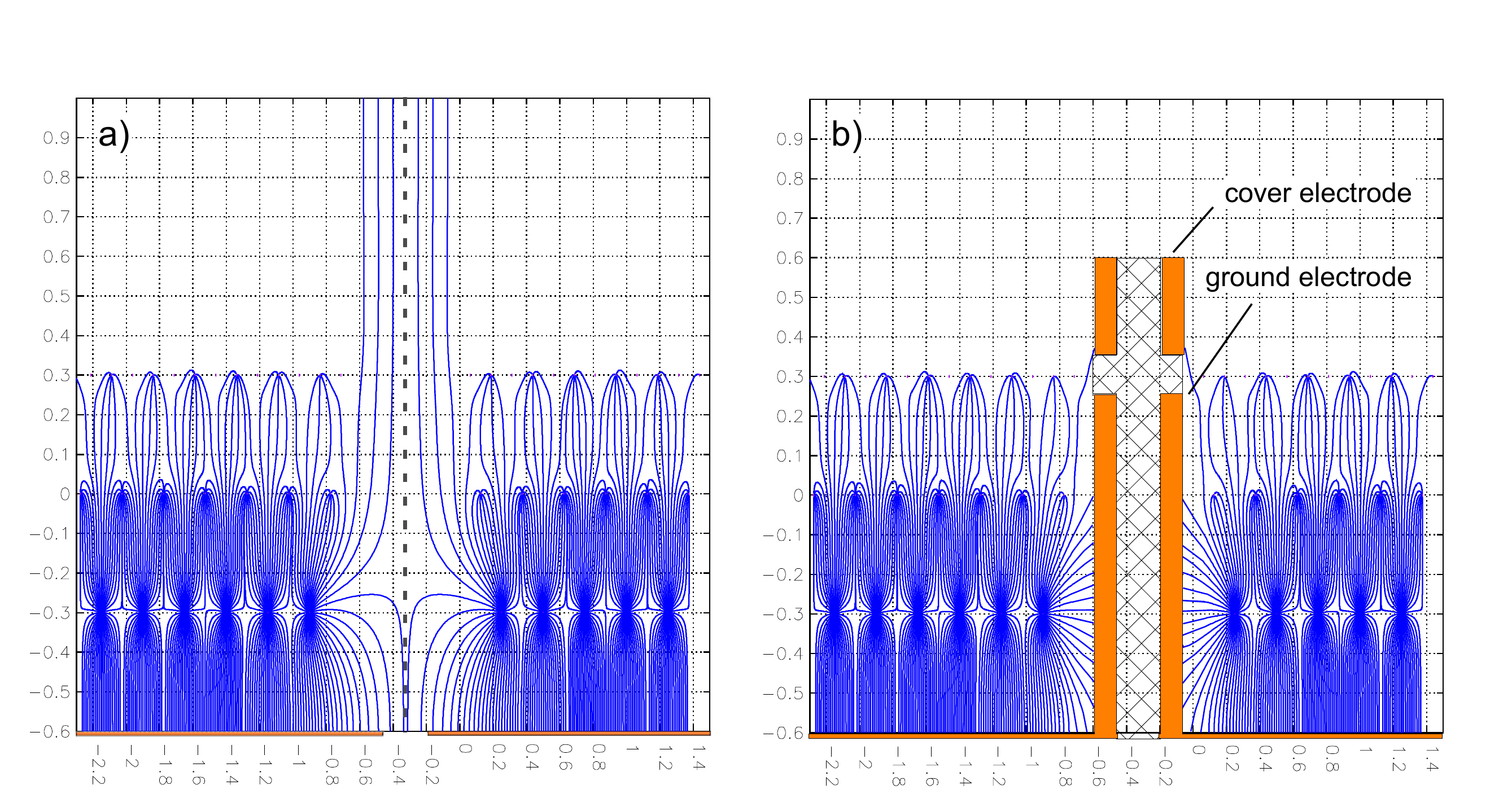} 
\caption{Drift lines for positive ions at the border of two readout chambers with gate closed and standard  wire configuration before (a) and after (b) the optimization of the electrostatic configuration at the borders of the chambers.}
\label{chamb:ggleak}
\end{figure}

The standard wire configuration (see Fig.~\ref{chamb:wire_structure}) has a discontinuity at the transition to the next chamber in the radial direction. Electrostatic simulations, as shown in Fig.~\ref{chamb:ggleak}a for the standard wire configuration, revealed a substantial inefficiency of the ion gate. 

The drift lines of positive ions originating from the amplification zone around the anode wire are shown. A sizable number of positive ions could leak back into the drift zone for this particular configuration. In order to improve the electrostatic configuration additional electrodes, i.e. ground and cover strips, were introduced (see Fig.~\ref{chamb:wire_3D}). The voltage of the cover strip, which frames the whole chamber, can be tuned to maximize the homogeneity of the drift field in the amplification zone. The ground strip, together with the HV of the cover strip forces all drift lines to end on either the cover or ground strip. In addition, two thicker edge anode wires ($\unit{75}{\micro\metre}$) were introduced. Their HV can be set independently thus providing a lower gain in the edge region. The corresponding field lines from electrostatic simulations are shown in Fig.~\ref{chamb:ggleak}b. Measurements of the ion-back flow for this configuration are given in Sec.~\ref{chamb:Tests}.

\subsubsection{Pad plane, connectors and flexible cables}
The readout pad structure has been optimized for signal-to-noise ratio and position resolution at the desired gas gain. A detailed account for the considerations leading to the chosen pad layout is given in Ref.~\cite{TDR:tpc}. The adopted pad sizes are given in Tab.~\ref{overv:overview1}.

The pad size increases with radius in two steps following the radially decreasing track density. The pad plane itself is a 3 mm thick halogen-free FR4 printed circuit board. The signal from the pad is routed in three layers of traces and vias to the connector side. The routing of the traces from the pads to their connector pad was realized employing an auto-router and was optimized for minimum trace length and maximum trace-to-trace distance. The boundary conditions for electrical design of the inner (outer) readout chamber pad plane were the line width of 4 (8) mil \footnote{1 mil = 25.4~$\micro$m} and the minimum distance between lines of 13 (31) mil.

The  pad plane connectors are standard for vertical connection of flat flexible cables. They have 23 pins each with a pitch of 1 mm. Six connectors in the radial direction are grouped to connect to the 128 readout channels of one FEC; 4 of them use 2 ground lines and  2 use 1 ground line to connect the ground on the pad plane with the ground of the FEC. The cables themselves are flexible Kapton\texttrademark \space cables, 8.2 cm long.

\subsubsection{Pad plane capacitance measurements}
An important optimization parameter of the pad plane is the minimization of the pad and traces-to-board capacitances.  One way to reduce the pad-to-board (ground) capacitances is to make the traces as short as possible. Typically,  traces from the border pads to their connectors are the longest ones. After optimization, i.e. basically overwriting auto-routers choice `by hand', the capacitances were generally below 9\,pF and as low as 6\,pF for the shortest connections. 

\subsubsection{Al-body}
\begin{figure}[ht]
\centering
\includegraphics[width=\linewidth,clip]{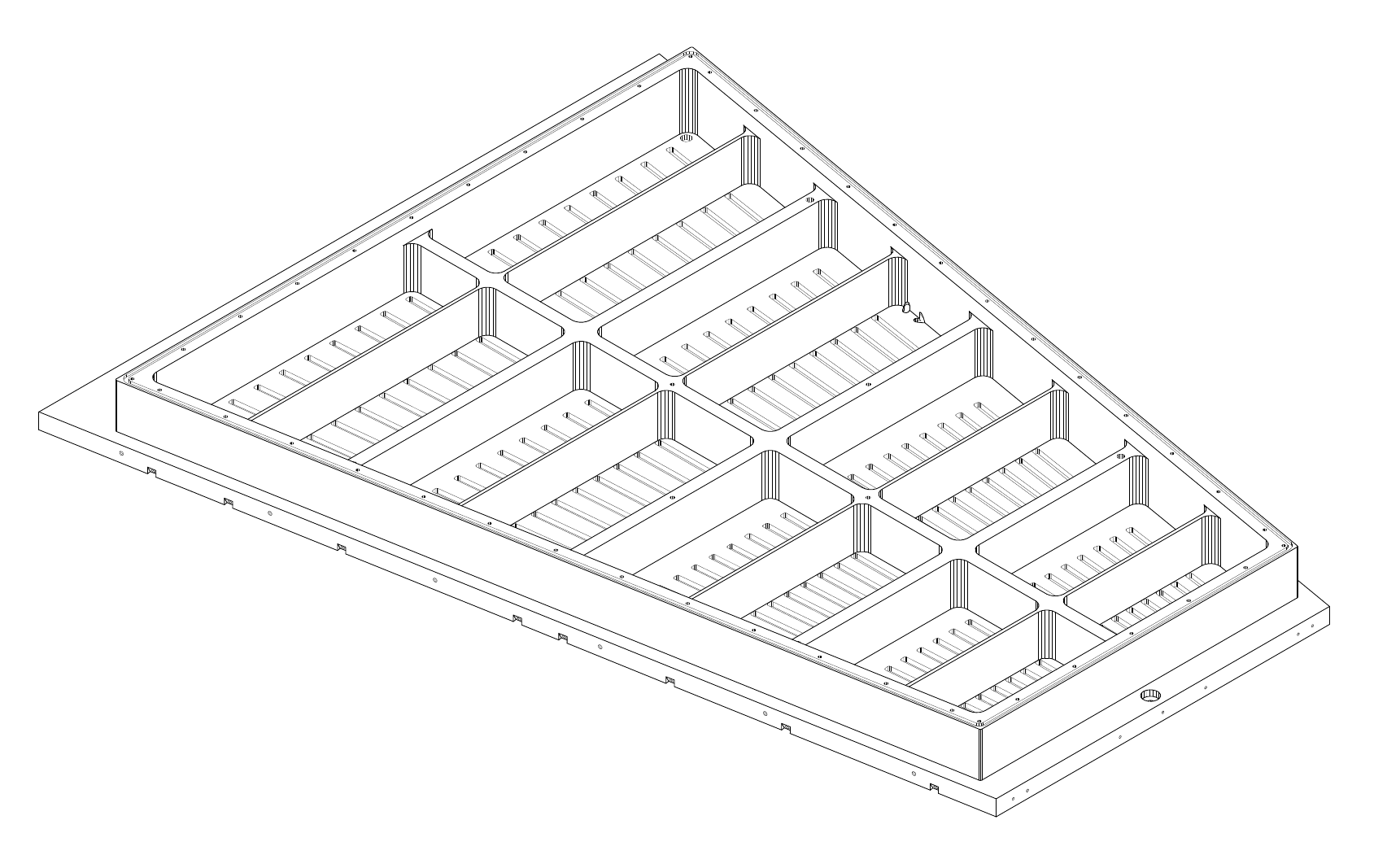} 
\caption{Drawing of the Al-body of an outer readout chamber. Shown is the FEC side with the cut-outs for the flexible cables.}
\label{chamb:ALUBODY-OROC-3D}
\end{figure}

Figure~\ref{chamb:ALUBODY-OROC-3D} shows the aluminum body for an IROC, which holds the pad and wire planes. Its stability against deformation from the forces of the wire planes was optimized employing Finite-Element (FEM) calculations.

The Al-body has cut-outs to allow for the connection of the FECs to the connectors on the backside of the pad-plane.  A cooling pipe was introduced into the Al-body to remove residual heat not taken away by the main cooling of the FECs or heat transmitted to the Al-body via the flat Kapton cables (see Sec.~\ref{cool}). The mechanical deformation of the readout chambers under the forces of the stretched wires was estimated via FEM calculation. The input to the FEM calculations is based  on the technical specification of the TPC readout chambers and on the material parameters specified by the producer. The calculations yield the mechanical stress, the stiffness as well as the deformation of the overall structure: the maximal overall deformation of an inner module is 10~$\micro$m and 25~$\micro$m for an outer module. These values, as well as the corresponding values for the stress, are below the values, which are considered to be critical, i.e. would influence noticeably the performance of the chamber.

\subsection{Tests with prototype chambers}
\label{chamb:Tests}

Tests were performed both with several small custom-built chambers to investigate specific properties of TPC components and with real-size prototypes to verify the design before mass production of the chambers. The tests, with both small and real-size prototypes, are described in detail in Ref.~\cite{Frankenfeld2002, Stelzer2003, Christiansen2005}. In summary it was verified that the

\begin{itemize}
\item  {\sl gating efficiency}, measured both with a radioactive source and the laser is better than  $\approx 0.7 \times 10^{-4}$, i.e. of the same order than the inverse of the envisioned gain of $2 \times 10^{4}$;
\item  {\sl cross talk} in the flat cables is of the order of   0.5--1.0\%, i.e. it is thus not expected that the tracking performance is deteriorated in a significant way;
\item  {\sl gas gain}, estimated as a function of anode voltage, is of the order of $3 \times10^{4}$ at 1280~V (for the original mixture without N$_2$) and thus sufficiently high to achieve a signal/noise ratio S/N=20;
\item average current  is stable ({\sl long term stability}) during the irradiation of a chamber with a source corresponding to one year of \mbox{Pb--Pb} (d$N_{\mathrm{ch}}$/d$y = 8000$) running at 400~Hz trigger rate;
\item chamber performance does not suffer from  {\sl aging} or  {\sl electron attachment}  induced by out-gassing of construction materials;
\item  chambers are stable at {\sl high beam rate} and perform according to the design values for position and energy resolution.
\end{itemize}

\subsubsection{Description of production steps}
\label{chamb:Steps}

The work to produce readout chambers is split into work packages defined such that they can be carried out in parallel. 
One work package was the preparation of the module body, which included the insertion and gluing of the cooling loop,
the insulation plate and the pad plane onto the Al-body. 

\begin{figure}[t]
\centering
\includegraphics[width=\linewidth,clip]{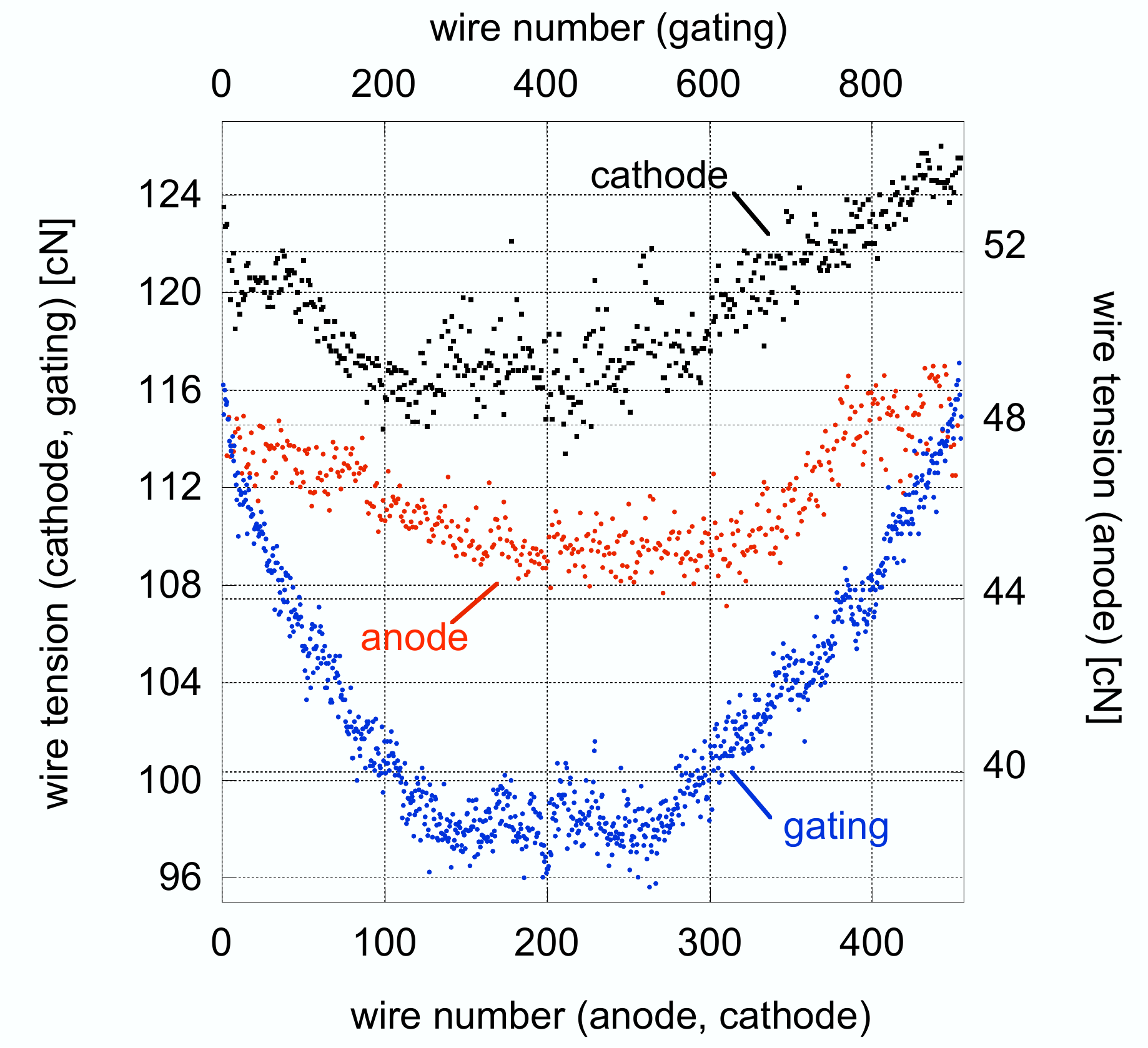} 
\caption{Measured wire tension for all types of wires (anode, cathode, gating) for one of the outer readout chambers.}
\label{chamb:wire_tension_oroc_9}
\end{figure}

After geometrical tests and cleaning, the module bodies were ready to receive the wire planes. The work package with the longest irreducible time span was the winding and gluing of the three wire planes (5 days); which defined the maximum chamber production rate, i.e. 1 chamber/week. The total production time per chamber was 30 days. A third work package was the testing of the final chamber which included a measurement of the wire tension (see Fig.~\ref{chamb:wire_tension_oroc_9}) and the connectivity of the wire planes, as well as leak and performance tests, which are described below.  The wire tension, e.g. for the anode wires, varies by $\pm$5\%. The wire sag at the nominal voltage (1500 V)  is for the longest wires (90 cm) about 50 $\micro$m, which changes the gain by about 4\%. Such gain variations are fully equilibrated by the krypton calibration (see Sec.~\ref{sec:krypton}).

The production time for all 80 chambers was nominally 400 days, which, adding 25\% contingency, amounts to an total effective production time of $\approx 2.5$ years.  In fact, the production of the multiwire proportional readout chambers started in March 2001 and finished in May 2004, i.e. took a little more than three years.

\subsubsection{Quality assurance and tests}
\label{chamb:QA}

All chambers were tested during and after production in order to validate them for the final assembly into the TPC. These tests included:
\begin{itemize}
\item gas tightness;
\item pad plane deformation;
\item performance: measurement of the gain as a function of high voltage;
\item continuous operation under irradiation;
\item uniformity of response: irradiation scans over the active area addressing the gain homogeneity.
\end{itemize}

For these tests a dedicated setup was used, consisting of a gas box with a short field cage and auxiliary sensors, into which a readout chamber could be mounted. The installed sensors allowed us to monitor the O$_{2}$ and H$_{2}$O content of the gas, temperature and pressure, as well as the currents of anode, cathode and gating grid wires.

\paragraph{Leak tests}
\label{chamb:Leak_tests}

The leak rate of each chamber was estimated from the O$_{2}$ contamination at the chamber outlet when flushing the chamber with a certain flow of fresh gas as described in Ref.~\cite{Stelzer2003}. Typically, the leak rate is 0.2~ml/h ($5.5 \times 10^{-5}$ mbar~l/s) at a flow rate f$ \approx$ 0.023 m$^{3}$/h. The acceptable O$_{2}$ contamination for the whole TPC is less than 5 ppm~\cite{TDR:tpc}, for a gas regeneration rate of 15~m$^{3}$/h. This translates into a leak rate of 0.5~l/h. If this leak was entirely due to the $2 \times 36$ readout chambers, each of them would be allowed to contribute with a maximum leak rate of 5~ml/h.

\paragraph{Long term stability tests}
\label{chamb:Stability}
Each chamber is subjected to a long term stability test. For this, the anode voltage for each of the chambers is set to a value corresponding to a gain of $3 \times 10^4$. A collimated iron source is placed at a fixed position for a full two-days irradiation test. The currents, X-ray fluxes, pulse height spectra and ambient pressure and temperature are continuously (every 15 min) recorded. The chamber is validated if no visible deterioration of its performance is observed. 

\begin{figure}
\centering
\includegraphics[width=\linewidth]{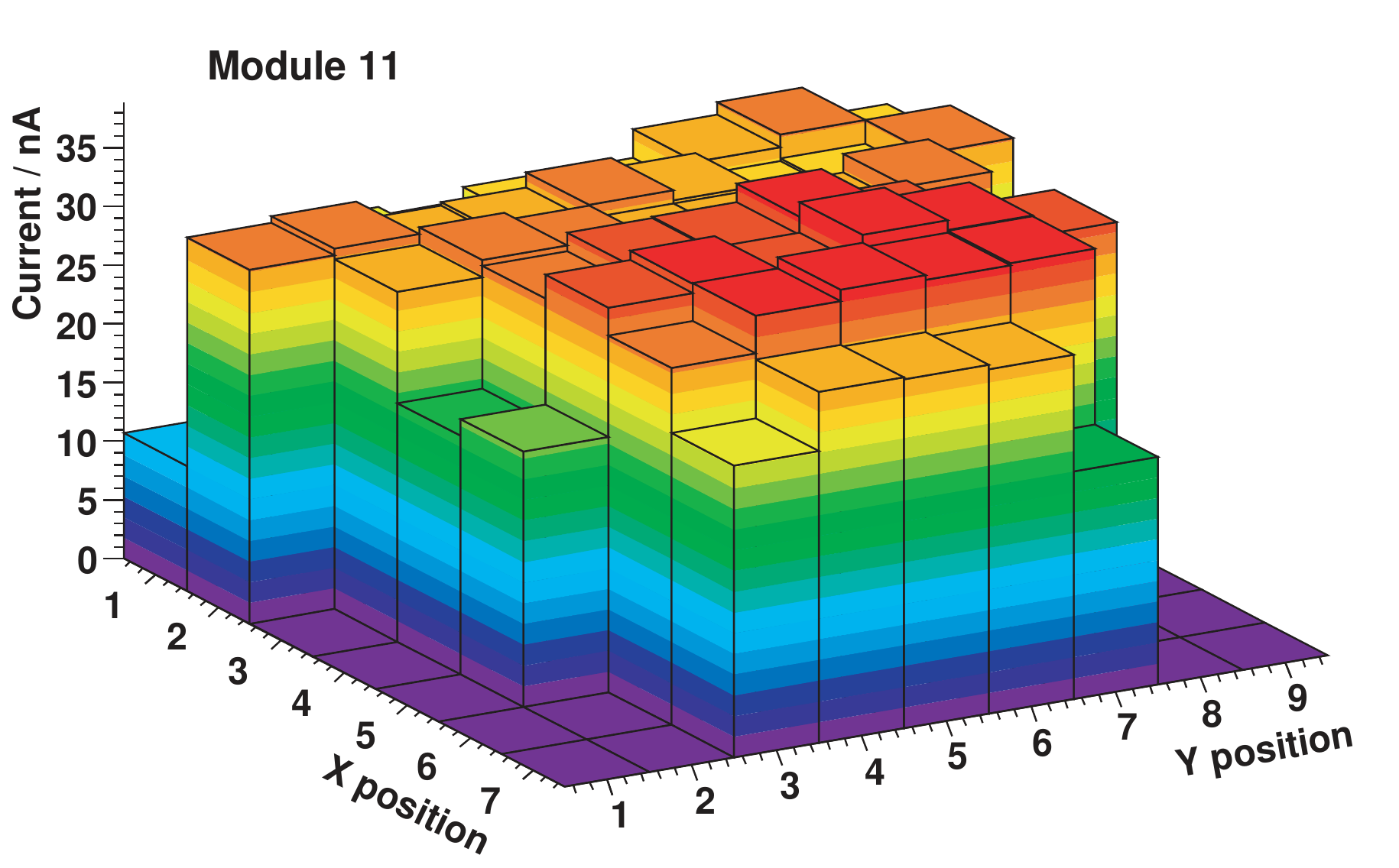}
\caption{Scan performed on OROC after the long term irradiation test, which was done at the central position.}
\label{chamb:homogenity_test}
\end{figure}

\paragraph{Gain homogeneity tests}
\label{chamb:Homogeneity}
After the long term test, a scan over the active area of the chamber is performed. Keeping the same voltages, the $^{55}$Fe source is consecutively placed in various  predefined positions. Currents were recorded for each position to map the gain uniformity of the chamber. Misalignment of wire planes or sags due to insufficient wire tension would result in observable patterns on such a gain scan. Figure~\ref{chamb:homogenity_test} shows the scan performed on one of the chambers. The spots on the corners fall partially outside the active area of the chamber and present therefore a
lower efficiency. Owing to the tight electrode geometry, high gains and the absence of field wires, a gain uniformity of the order of 10--20 \% was measured. However, no evidence of loose wires was observed.

From the 20 IROCs tested this way, 17 showed a stable and uniform ($\Delta$G/G $< \pm 20$ \% ) performance. Three of them did not pass the validation tests. Two chambers showed large (order of $\micro$A) dark or leakage currents at voltages below the operational ones. It is suspected that the reason for this behavior was a bad pad plane: either dirty or with a rough surface. These two chambers were therefore discarded for installation into the TPC. In one chamber an anode wire broke after several minutes under nominal voltage. The anode was burned at some 5~mm from the holding ledge. This was traced to faulty wire material. After this incident it was decided to inspect the wire quality employing electron microscopy before winding any anode wire plane.

\begin{figure}
\centering
\includegraphics[width=\linewidth]{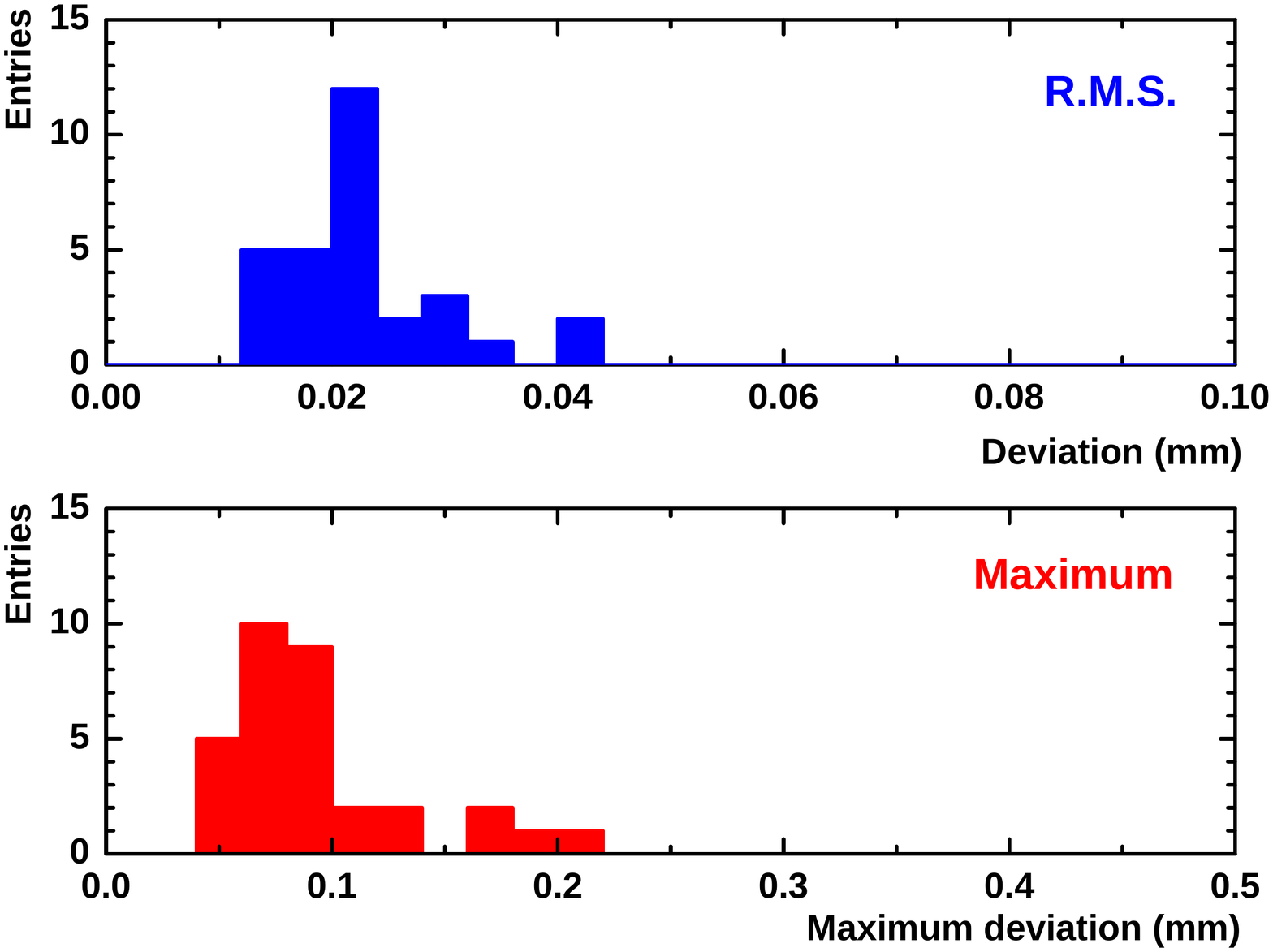}
\caption{Top: RMS of the deviation of the pad plane references points from the mean for an IROC chamber. Bottom: Maximum distance between any two measured reference points of an IROC chamber. }
\label{chamb:PadDeformation_IROC}
\end{figure}

\begin{figure}[t]
\centering
\includegraphics[width=\linewidth]{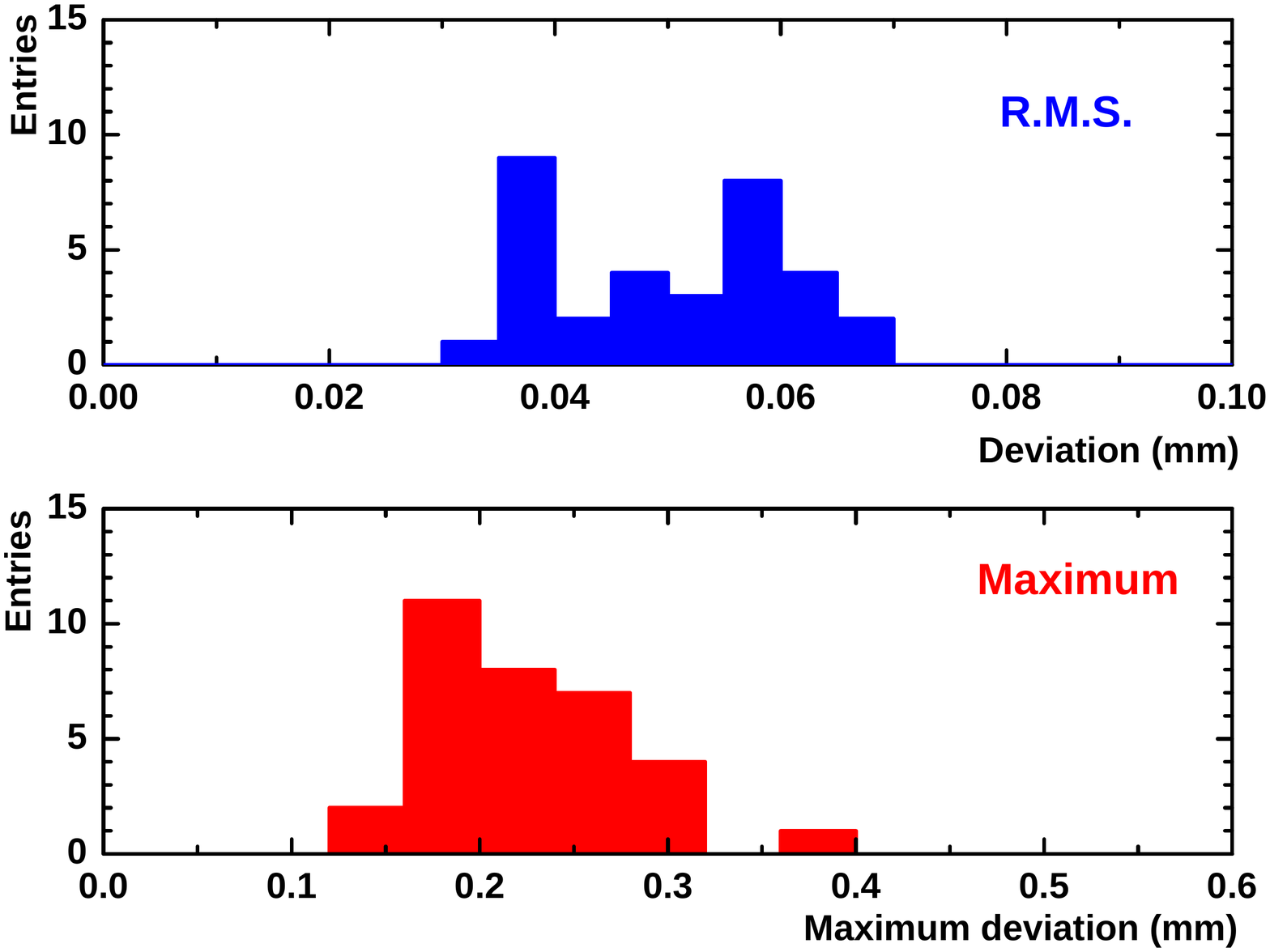} 
\caption{Top: RMS of the deviation of the pad plane references points from the mean for an OROC chamber. Bottom: Maximum distance between any two measured reference points of an OROC chamber. }
\label{chamb:PadDeformation_OROC}
\end{figure}

\paragraph{Pad plane deformation}
\label{chamb:Deformation}
The spatial homogeneity of the chamber gain depends on the distance between anode wire and the cathode (wire) planes, i.e. on the wire sag due to gravitational and electrostatic forces and on the planarity of the cathode pad plane. To ensure an acceptable contribution of the pad plane inhomogeneity to the gain variation, the pad plane deformation should be comparable to the average wire sag. After gluing the pad planes onto the Al-body the homogeneity of the pad plane has been measured for each chamber on a $xy$-table. 24 and 28 reference points were surveyed for IROC and OROC, respectively. The results are depicted in Figs.~\ref{chamb:PadDeformation_IROC} and ~\ref{chamb:PadDeformation_OROC}. The RMS value of the pad plane deformation is of the same magnitude as the average wire sag ($\approx 50~\micro$m)  and thus contributes with a value of less than 5\% to the gain variation. The maximum deviations are significantly higher and can contribute with values of up to 15\% to the gain inhomogeneity.

\subsection{Chamber mounting and pre-commissioning}
\label{chamb:Precommissioning}

\begin{figure}
\centering
\includegraphics[width=0.8\linewidth,clip]{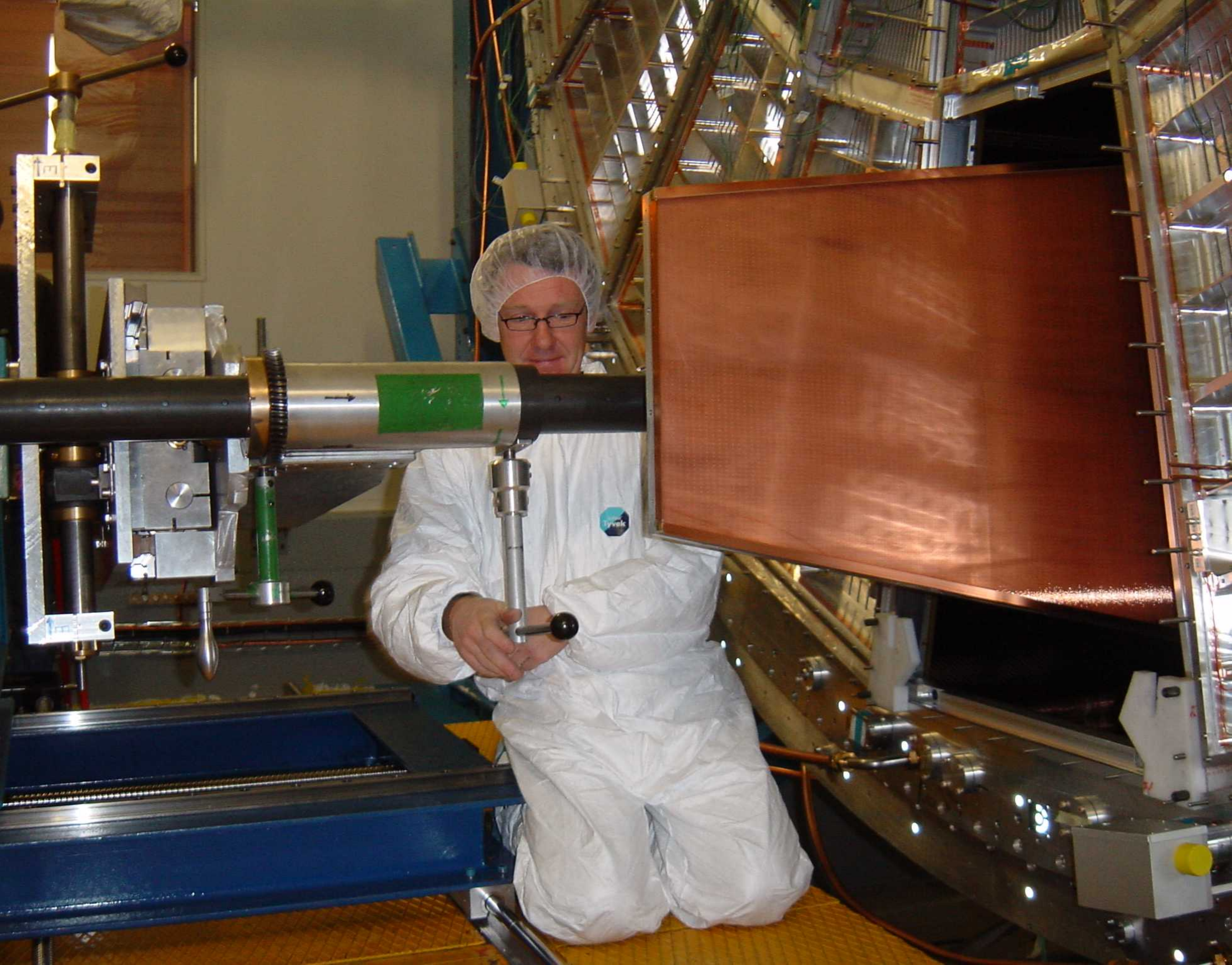} 
\caption{Insertion of an OROC through the endplate. The tilt, polar and azimuthal angles of the chambers can be adjusted via handles and a transmission system.}
\label{chamb:insertion}
\end{figure}

\begin{figure}
\centering
\includegraphics[width=0.8\linewidth,clip]{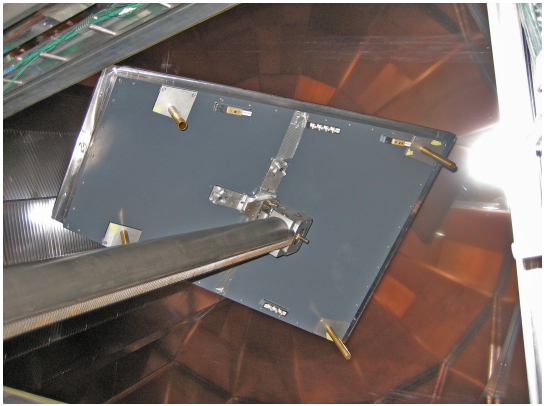} 
\caption{Rotation of an IROC inside the field cage. To prevent dirt falling into the field cage the FEC side of the chambers is closed with a cover.}
\label{chamb:retraction}
\end{figure}

The chambers are attached to the endplate from the inside to minimize dead space between neighboring chambers. This required a special mounting technique, by which the chambers are attached to a long manipulator arm, which allows the rotation and tilting of the chambers. This mounting technology had already been used by the ALEPH collaboration, from which we inherited the manipulator device. The chambers, attached to the tip of the manipulator arm, are first adjusted in angles such that they can be moved through the endplate (see Fig.~\ref{chamb:insertion}), thereafter they are turned (see Fig.~\ref{chamb:retraction}) and retracted into their final position.

Five types of measurements were done for all sectors:
\begin{itemize}
\item pedestal and noise measurements; 
\item calibration pulser measurements (study the shaping properties of the electronics);
\item measurements with the TPC laser system (for alignment purposes);
\item optical measurement of the readout-chamber position relative to inner field cage vessel;
\item gain and drift-time measurements (using a cosmics trigger).
\end{itemize}

The results of the first four measurements as well as the gain measurement, which were repeated after the installation in the L3 magnet, are reported in Sec.~\ref{perform}.

\section{Front-end electronics and readout}
\label{elect}

\subsection{General specifications}

Charged particles traversing the TPC volume ionize the 
gas along their path, liberating electrons that drift towards the endplate of the chamber. 
The signal amplification is provided through avalanche effect in the 
vicinity of the anode wires of the readout chambers. The electrons and positive ions 
created in the avalanche, which move respectively towards the anode wire and the surrounding 
electrodes, induce a positive current signal on the pad plane. The current signal of a single avalanche, 
which is characterized by a fast rise time (less than $\unit{1}{\nano\second}$) and a long tail (of the order of 
$\unit{50}{\micro\second}$), carries a charge that, in the ALICE TPC, can be as low as a few $\femto\coulomb$. 
It is then delivered on the detector  impedance which, to a very good approximation, is a pure 
capacitance of the order of a few $\pico\farad$. The shape of the signal tail, which is due to the motion 
of the positive ions, is rather complex and depends on the details of the chamber and pad
geometry~\cite{Mota2004500,Ross09}. 
This tail, causing pile-up effects, sets the main limitation to the maximum track density at which 
a MWPC can be operated. 

The readout of the signal is done by 557\,568 pads that form the cathode pad plane
of the readout chambers. The signals from the pads are passed to 4\,356 Front-End Cards (FECs),
located $\unit{7}{\centi\metre}$ away from the pad plane, via flexible Kapton cables. In the FECs a
custom-made charge-sensitive shaping amplifier, named PASA (PreAmplifier ShAper), transforms the charge signal induced in
the pads into a differential semi-Gaussian voltage signal that is fed to the input of the ALTRO 
(ALice Tpc Read Out) chip. Each ALTRO contains 16 channels operating concurrently that digitize and process 
the input signals. Upon arrival of a first-level trigger, the data stream corresponding  to the detector drift time
($\unit{\lesssim 100}{\micro\second}$) is stored in a memory. When a second-level trigger 
(accept or reject) is received, 
the latest event data stream is either frozen in the data memory, until its complete
readout takes place, or discarded. The readout can take place at any time at a speed 
of up to $\unit{200}{\mega\byte\per\second}$ through a 40-bit-wide backplane bus linking the 
FECs to the Readout Control Unit (RCU), which interfaces them to the Data AQuisition (DAQ),
the Trigger and the Detector Control System (DCS).    

The main requirements for the readout electronics and the way they are derived from the detector performance 
requirements are discussed in Sec.~5.1.1 of~\cite{TDR:tpc} and listed in Tab.~\ref{table:specs}.

\begin{table}[t]
   \centering
   \caption{Readout electronics requirements.}
   \label{table:specs}
   \begin{tabular}{|c|c|}
     \hline
     number of channels           & 557\,568 \\
     dynamic range                & $900:1$  \\
     noise (ENC)                  & $<\unit{1000}{e}$ (rms) \\
     conversion gain              & $\unit{12}{\milli\volt\per\femto\coulomb}$ \\
     crosstalk                    & $<1\%$ \\
     shaping time                 & $\approx\unit{200}{\nano\second}$ \\
     sampling rate                & $\unit{5-10}{\mega\hertz}$        \\
     tail correction after $\unit{1}{\micro\second}$ & $0.1\%$        \\
     maximum readout rate (Pb--Pb) & $\unit{300}{\hertz}$              \\
     maximum readout rate (pp)   & $\unit{1.4}{\kilo\hertz}$           \\
     power consumption            & $<\unit{100}{\milli\watt\per channel}$ \\      
     \hline
   \end{tabular}
\end{table}

One of the tightest requirements is defined by the extremely high pulse rate with which the 
ALICE TPC Front-End Electronics (FEE) has to cope. Indeed, the FEE has been designed to cope with a 
signal occupancy as high as 50\%. Furthermore the extremely large raw data volume
($\unit{750}{\mega\byte\per event}$) asks for zero suppression already in the FEE in order to fit 
events at the foreseen event rate into the DAQ bandwidth (216 links at $\unit{160}{\mega\byte\per\second}$).
For example, for a trigger rate of $\unit{1}{\kilo\hertz}$ as planned for pp collisions, this leads
to a raw data throughput of $\unit{750}{\giga\byte\per\second}$, which is beyond the present data handling
capabilities. It should be noticed that in a high occupancy
environment, in order to preserve the full resolution on the signal features (charge and arrival time), a very 
accurate cancellation of the signal tail and correction of the baseline have to be performed before the zero suppression.

Besides the optimization for the maximum rate the detector can be operated at, its dead time has to be
minimized. This is done by introducing a derandomizing Multiple Event Buffer (MEB) to eliminate the contribution
due to the random  nature of the trigger arrival times. Simulations showed
that four entries were a good trade-off between cost/size and effect (refer to Sec.~5.1.7 of~\cite{TDR:tpc}).

The front-end electronics system has to satisfy many other constraints, while meeting the required 
performance specifications. Mainly, the readout electronics needs to fit into the overall detector structure, 
in particular into the available space, which has important consequences for the requirements on reliability, 
power, and cooling.

The radiation load on the TPC is rather low (less than $\unit{1}{\kilo\rad}$ over 10 years).
Thus standard radiation-soft technologies are suitable for the implementation of 
this electronics. However, some special care has to be taken to protect the system
against Single Event Upset (SEU), see Sec.~\ref{elect:radiation}.

\begin{figure}
\centering
\includegraphics[width=0.96\linewidth,clip]{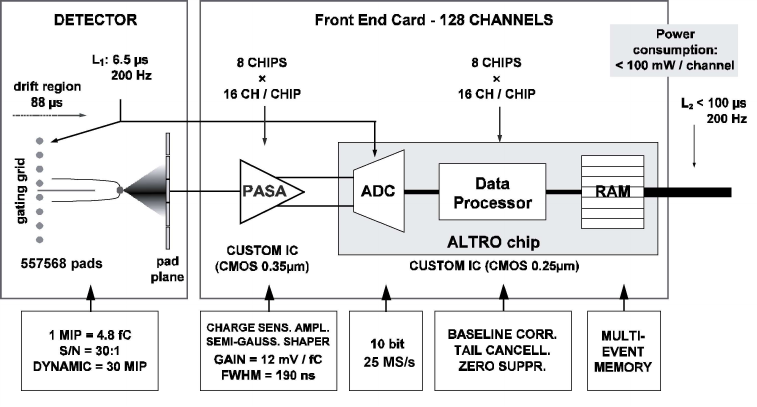} 
\caption{An overview of the ALICE TPC front end electronics.}
\label{elect:fig_feechain}
\end{figure}

\subsubsection{System overview}
\label{elect:overview}

A single readout channel is comprised of four basic functional units 
(Fig.~\ref{elect:fig_feechain}):
1) a charge sensitive amplifier/shaper (PASA); 
2) a 10-bit $\unit{25}{\msps}$ (mega samples per second) low power ADC; 
3) a digital circuit that contains a 
   shortening filter for the tail cancellation, the baseline subtraction 
   and zero suppression circuits; 
4) the MEB.

The charge collected at the TPC pads is amplified and integrated by a 
low input impedance amplifier. It is based on a charge sensitive amplifier 
followed by a semi-Gaussian pulse shaper. These analogue functions are realized 
by a custom integrated circuit, PASA (Sec.~\ref{elect:PASA}), which contains 16 channels. 
The circuit has a conversion gain of $\unit{12}{\milli\volt\per\femto\coulomb}$, an output dynamic range of
$\unit{2}{\volt}$,
and produces a differential semi-Gaussian pulse with a shaping time (FWHM) of $\unit{190}{\nano\second}$. 

The output signals of the PASA chip are digitized by a 10-bit pipelined $\unit{25}{\msps}$ ADC (one per channel) 
operated at a sampling rate in the range of $5$ to $\unit{10}{\mega\hertz}$. The digitized signal is then 
processed by a set of circuits that perform the baseline subtraction, tail 
cancellation, zero-suppression, formatting and buffering. The ADC and the 
digital circuits are contained in a single, custom-made
chip named ALTRO (Sec.~\ref{elect:ALTRO}). 

\begin{figure}[b]
\centering
\includegraphics[width=0.96\linewidth,clip]{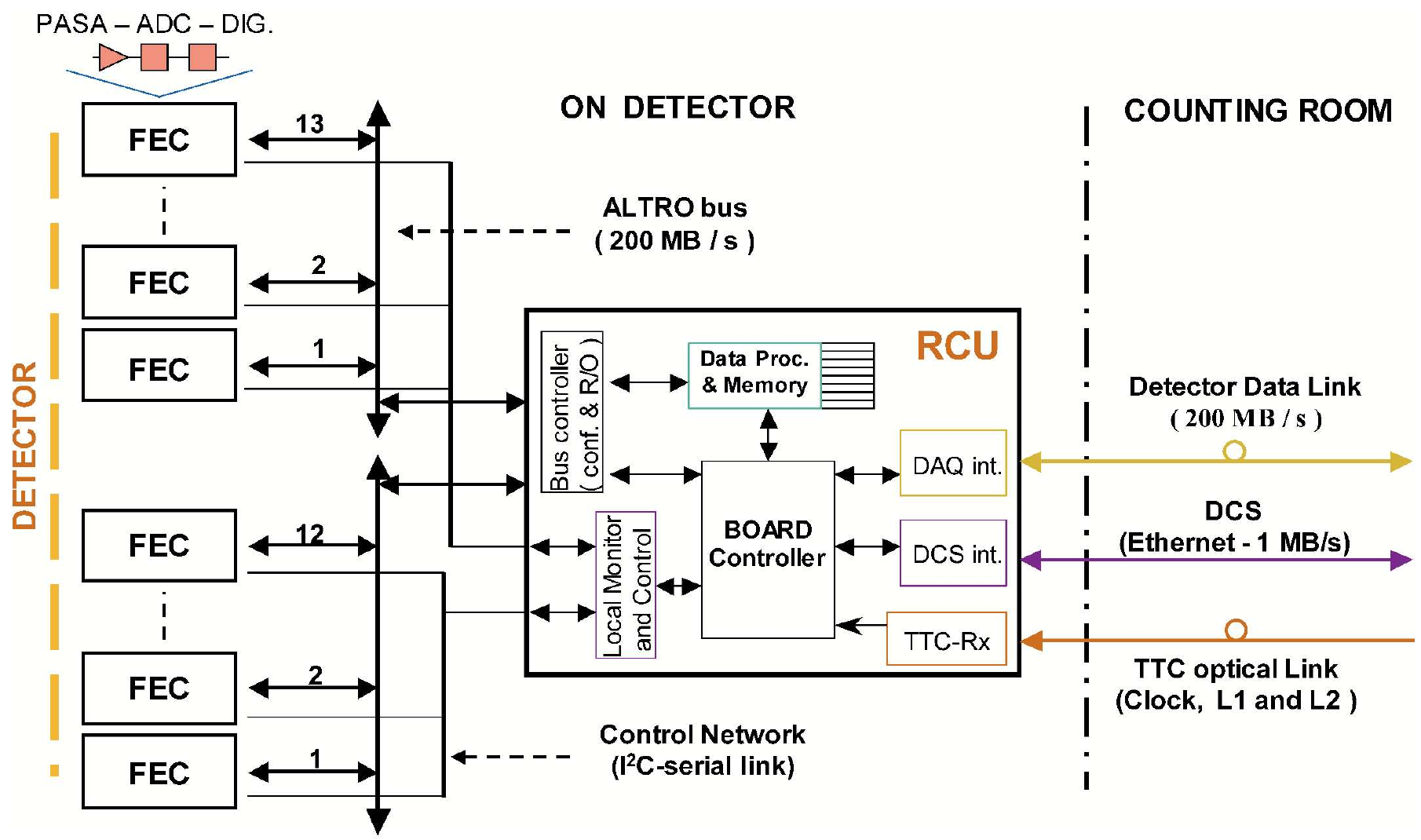} 
\caption{Block diagram of the basic readout partition. 
The overall TPC readout consists of 216 readout partitions.}
\label{elect:fig_system}
\end{figure}

The complete readout chain is contained in FECs (Sec.~\ref{elect:FEC}), which are plugged 
in crates that are supported by the service support wheel.
They are mechanically decoupled with respect to the detector by Kapton cables (see Sec.~\ref{infra:Service_support_wheel}).
Each FEC houses 8 PASAs and 8 ALTRO chips, 
128 channels in total. Another important component of the FEC is the Board Controller (BC), 
which implements a number of key functions for the readout and system monitoring.
The FECs are connected to the cathode plane by means of six $\unit{8.2}{\centi\metre}$ long flexible cables. 
As illustrated in Fig.~\ref{elect:fig_system}, FECs are grouped in readout partitions 
controlled by RCUs (Sec.~\ref{elect:RCU}). 
The number of FECs per partition 
varies according to its radial position on the detector, due to the trapezoidal shape of the 
sectors and the different pad sizes. 
Within a readout partition, 
the FECs are organized in two branches, connected to the RCU via separate backplanes. 

The FECs communicate with the RCU by means of two independent buses: the ALTRO bus and the 
front-end control bus, both based on low-voltage 
signaling technology (Gunning Transistor Logic, GTL). The configuration of the FECs, the distribution of the clock and 
trigger signals and the readout of trigger related data are performed via the ALTRO bus. 
The RCU uses the front-end control bus for all operations related to the safety and monitoring 
of the readout partition. 
Each of the $2\times 18$ TPC sectors is equipped with 6 readout partitions with respectively (from the innermost  
to the outermost) 18, 25, 18, 20, 20, and 20 FECs, accounting for a total of 4\,356 FECs and 216 RCUs.
From the readout and control point of view, each partition represents an independent system.

\subsection{PASA}
\label{elect:PASA}

\begin{figure}
\centering
\includegraphics[width=0.96\linewidth,clip]{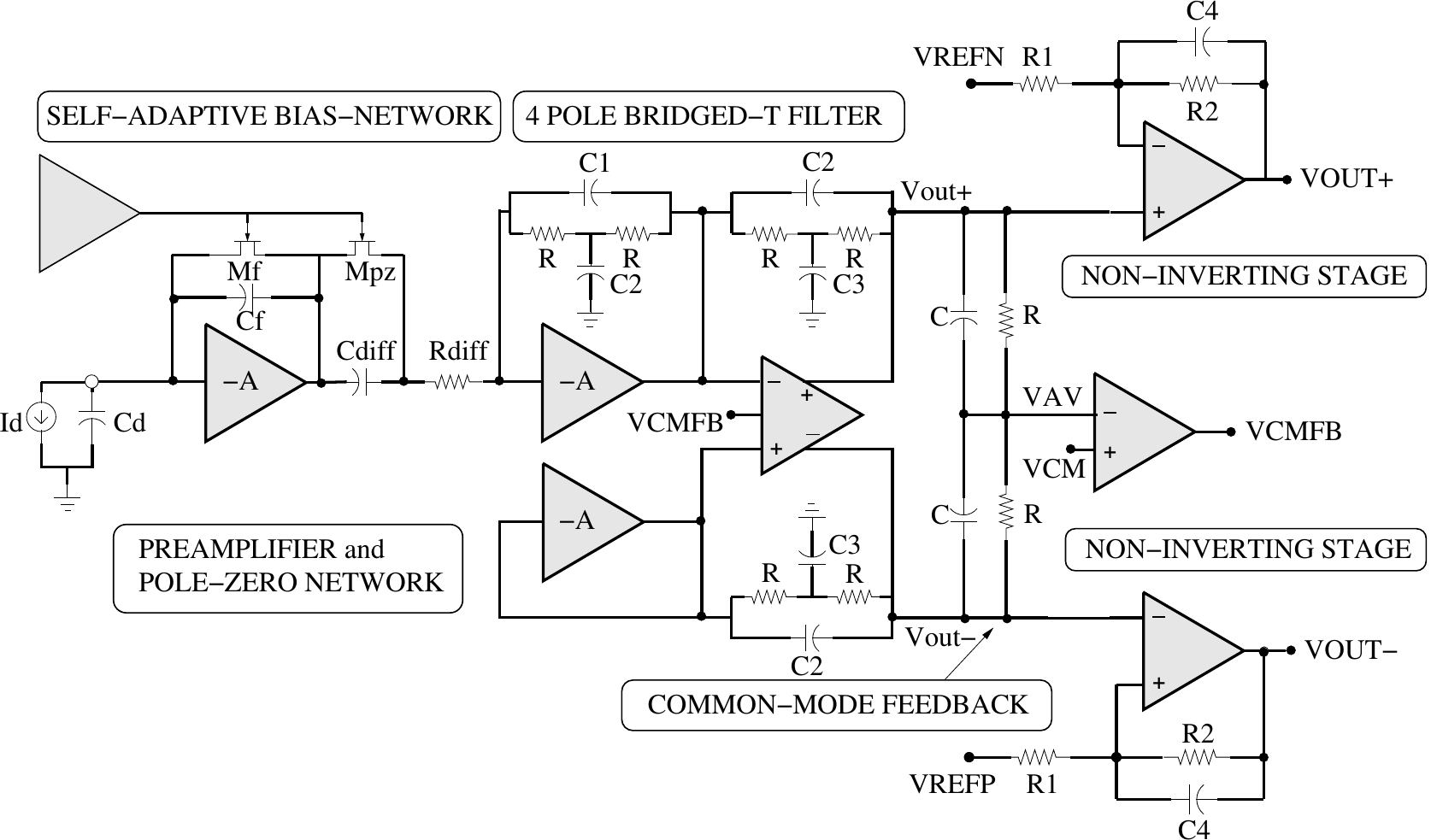} 
\caption{A simplified block diagram of the PreAmplifier-ShAper (PASA) signal processing
chain.} 
\label{elect:PASA_arch}
\end{figure}

The PASA~\cite{elect:Soltveitpaper} integrates 16 identical Charge
Sensitive Amplifiers (CSAs) followed by a pole-zero cancellation network and a shaping amplifier. 
A simplified block diagram of the signal processing chain is shown in Fig.~\ref{elect:PASA_arch}.
The positive polarity CSA, with a capacitive and resistive feedback connected in parallel, is followed by a
pole-zero cancellation  network with a self-adaptive bias network, a $CR$-filter,
two $(RC)^2$-bridged-T filters, a common-mode feed-back network and two quasi-differential gain-2 amplifiers.
The circuit is optimized for a detector capacitance of $\unit{18-25}{\pico\farad}$. 

The amplifier topology is based on a single-ended
folded cascode amplifier followed by a source follower.
As seen in Fig.~\ref{elect:PASA_arch}, an NMOS transistor,
which is operated in the subthreshold region to implement a large resistor, is connected in parallel to the
feedback capacitor $C_\text{f}$. The purpose of this resistor is to avoid saturation of
the CSA by continuously discharging the feedback capacitance. 
This resistor contributes to the parallel noise at the
CSA input. A value of $\unit{10}{\mega\ohm}$ is chosen as trade off between noise and count rate. 
Still, the relatively long discharge time constant of the CSA may cause signal pile-up. 
For this reason, the CSA is followed by a pole-zero cancellation network ($M_\text{pz}$ and $C_\text{diff}$), 
which is combined with the $C_\text{diff}R_\text{diff}$-filter stage.
The signal is then amplified and further shaped by two second order bridged-T filters to optimize the 
signal-to-noise ratio and to limit the signal bandwidth.
In the last PASA stage the signal levels are adjusted to match the input of the ALTRO chip.

\begin{figure}[t]
\centering
\includegraphics[width=0.96\linewidth,clip]{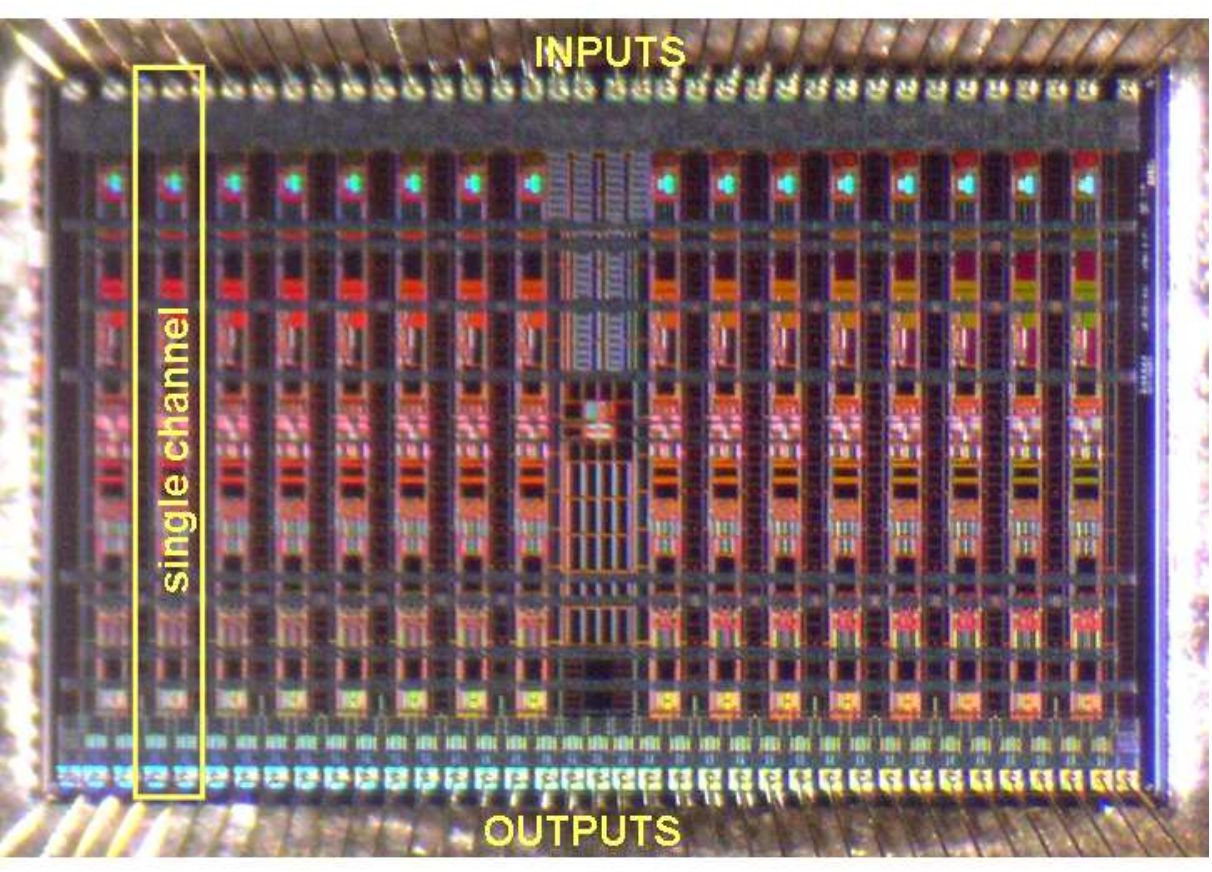}
\caption{Layout of the PASA chip.}
\label{elect:PASA_layout}
\end{figure}

The chip was manufactured in the C35B3C1 $\unit{0.35}{\micro\metre}$ CMOS technology featured by 
Austriamicrosystems. It has a width of $\unit{5.3}{\milli\metre}$ and a length of $\unit{3.4}{\milli\metre}$
and gives in total an area of $\unit{18}{\milli\metre\squared}$. 
A picture of the produced PASA is shown in Fig.~\ref{elect:PASA_layout}. 
The input and output pins are distributed along the length of the chip. 
The 16 channels, divided in 2 groups of 8 channels each, are placed on each side 
of the bias network, located at the center of the chip. 
The de-coupling capacitors together with pads connected to ground are placed 
adjacent to each channel, creating a physical distance between the channels 
in order to reduce the crosstalk. In addition each channel is surrounded 
by a guard ring connected to the substrate which isolates them from each other 
and further reduces the crosstalk.

As detailed in Sec.~\ref{elect:testing}, the performance of about 48\,000 PASA was tested;
98\% of the chips was fully functional. The general performance of the chip is listed in Tab.~\ref{elect:PASA_perf}.

\begin{table}
   \centering
   \caption{PASA key performance figures. All values are given
     for a detector capacitance of $\unit{12}{\pico\farad}$.}
   \label{elect:PASA_perf}
   \begin{tabular}{|c|ccc|}
     \hline
     Parameter & Specs & Simulation & Test \\
     \hline
     Noise & $<\unit{1000}{e}$ & $\unit{385}{e}$ & 
     $\approx\unit{385}{e}$\\
     Shaping time [$\nano\second$] & 190 & 212 & $\approx 190$ \\
     Non-linearity & $<1\%$ & $0.19\%$ & $0.2\%$ \\
     Crosstalk & $<0.3\%$ & --- & $<$ $0.1\%$ \\
     Baseline variation [$\milli\volt$]& --- & --- & $\pm80$ \\
     Conv.\ gain [$\milli\volt\per\femto\coulomb$] & 12 & 12.74&
     $\approx 12.8$\\
     Power [$\milli\watt\per\text{ch}$] & $< 20$ & 11 & 11.67\\
     \hline
   \end{tabular}
\end{table}

\subsection{ALTRO}
\label{elect:ALTRO}

\subsubsection{Circuit description}

\begin{figure}[t]
\centering
\includegraphics[width=0.96\linewidth,clip]{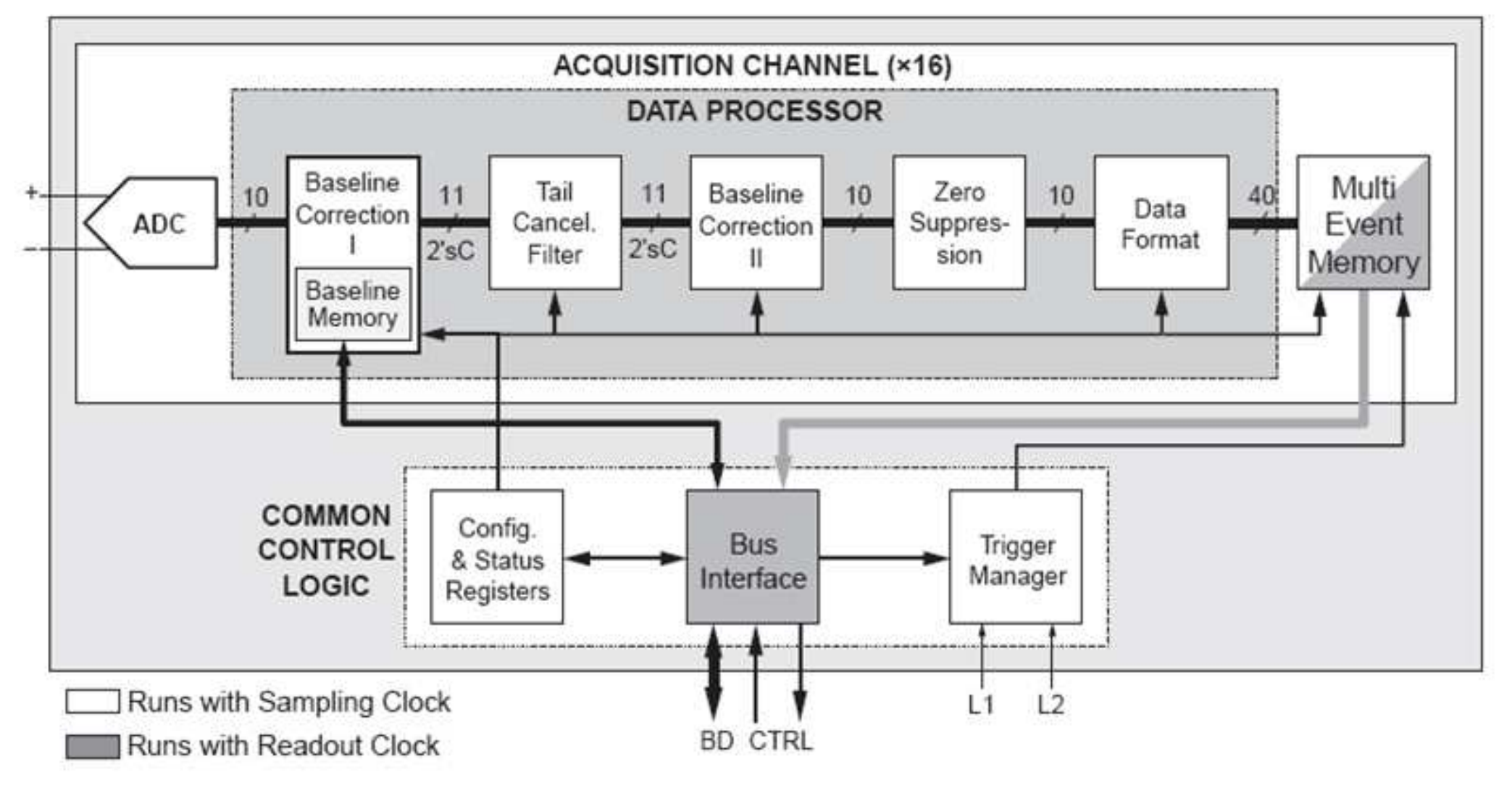}
\caption{ALTRO chip block diagram.}
\label{elect:ALTRO_arch}
\end{figure}

The ALTRO~\cite{EsteveBosch:2003bj}
is a mixed-signal custom integrated circuit containing 16 channels 
operating concurrently and continuously on the analogue signals coming from 
16 independent inputs. 
It is designed to process a train of pulses sitting on a common baseline. 
Figure~\ref{elect:ALTRO_arch} shows a simplified block diagram of the chip.
When a first-level trigger is received, a predefined number of samples is 
processed and temporarily stored in a data memory (acquisition). 
The acquisition is frozen if a positive second-level trigger is received; 
otherwise it is overwritten by the next acquisition.

With reference to Fig.~\ref{elect:ALTRO_arch}, a short description of the main building blocks is 
given following the signal processing path.

\paragraph{ADC}

\begin{figure}
\centering
\includegraphics[width=0.96\linewidth]{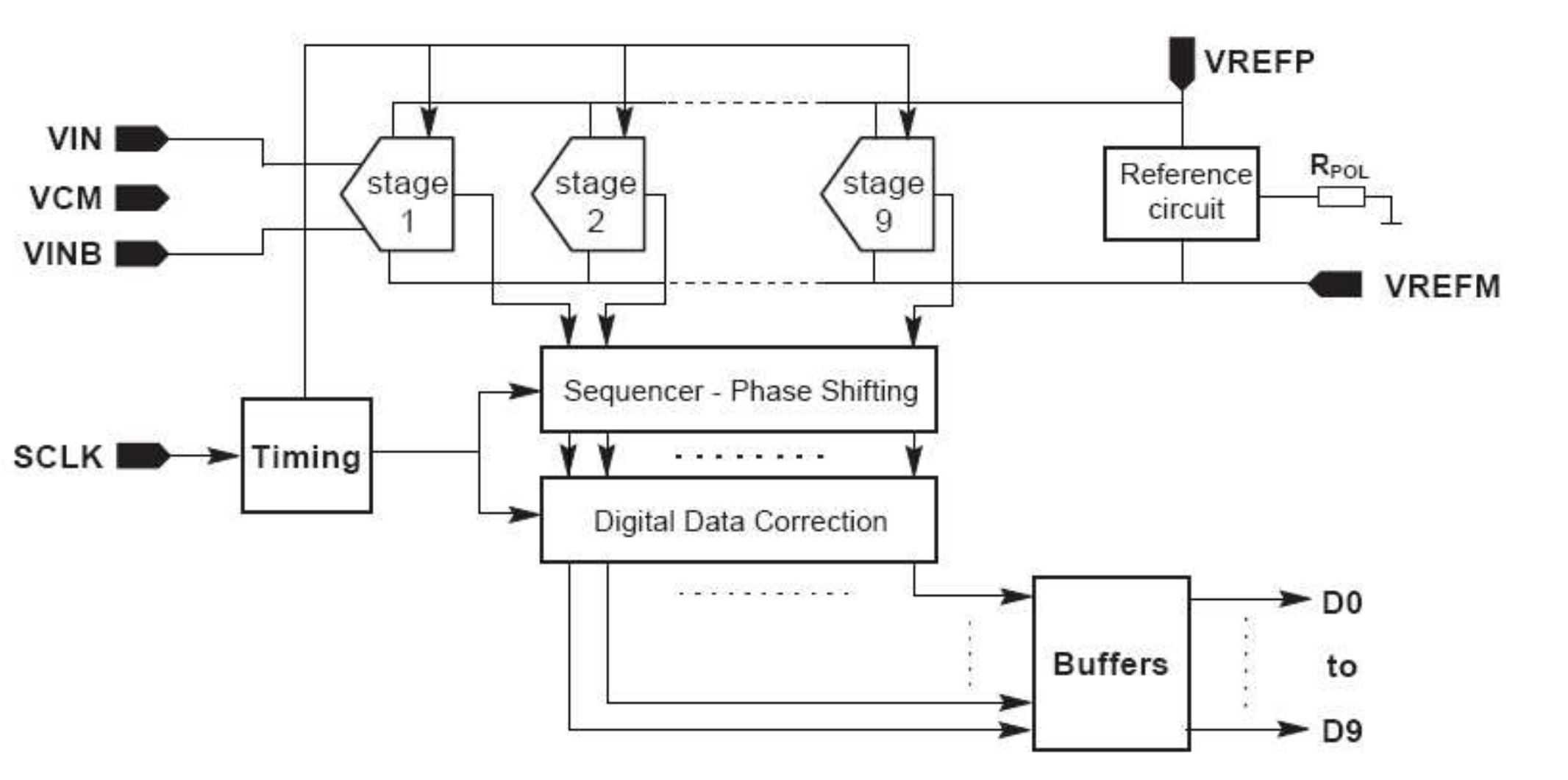}
\caption{Block diagram of the ALTRO ADC.}
\label{elect:ADC_arch}
\end{figure}

The Analogue to Digital conversion is based on the ST Microelectronics 
TSA1001~\cite{elect:STM_ADC}, a CMOS 10-bit pipelined ADC. The block diagram of this 
ADC is presented in Fig.~\ref{elect:ADC_arch}. 
The conversion pipeline consists of 9 stages with an overall latency of 5.5 clock cycles.

The internal construction of the ADC is fully differential, and allows up to $\unit{2}{\volt}$ 
differential swing. A polarization current, provided for each channel by an internal resistor, defines 
the ADC bandwidth and power consumption. The polarization resistor is divided in multiple 
taps such that only one metal layer has to be changed in order to optimize the power 
consumption for the required bandwidth. Two versions of the 
ALTRO chip have been produced. They were optimized for maximum sampling rates of $25$ and $\unit{40}{\mega\hertz}$ 
and power consumptions of $12.5$ and $\unit{43}{\milli\watt\per channel}$, respectively. The ALICE TPC uses the
$\unit{25}{\msps}$ version.

\paragraph{Data Processor}

\begin{figure}
\centering
\includegraphics[width=0.96\linewidth,clip]{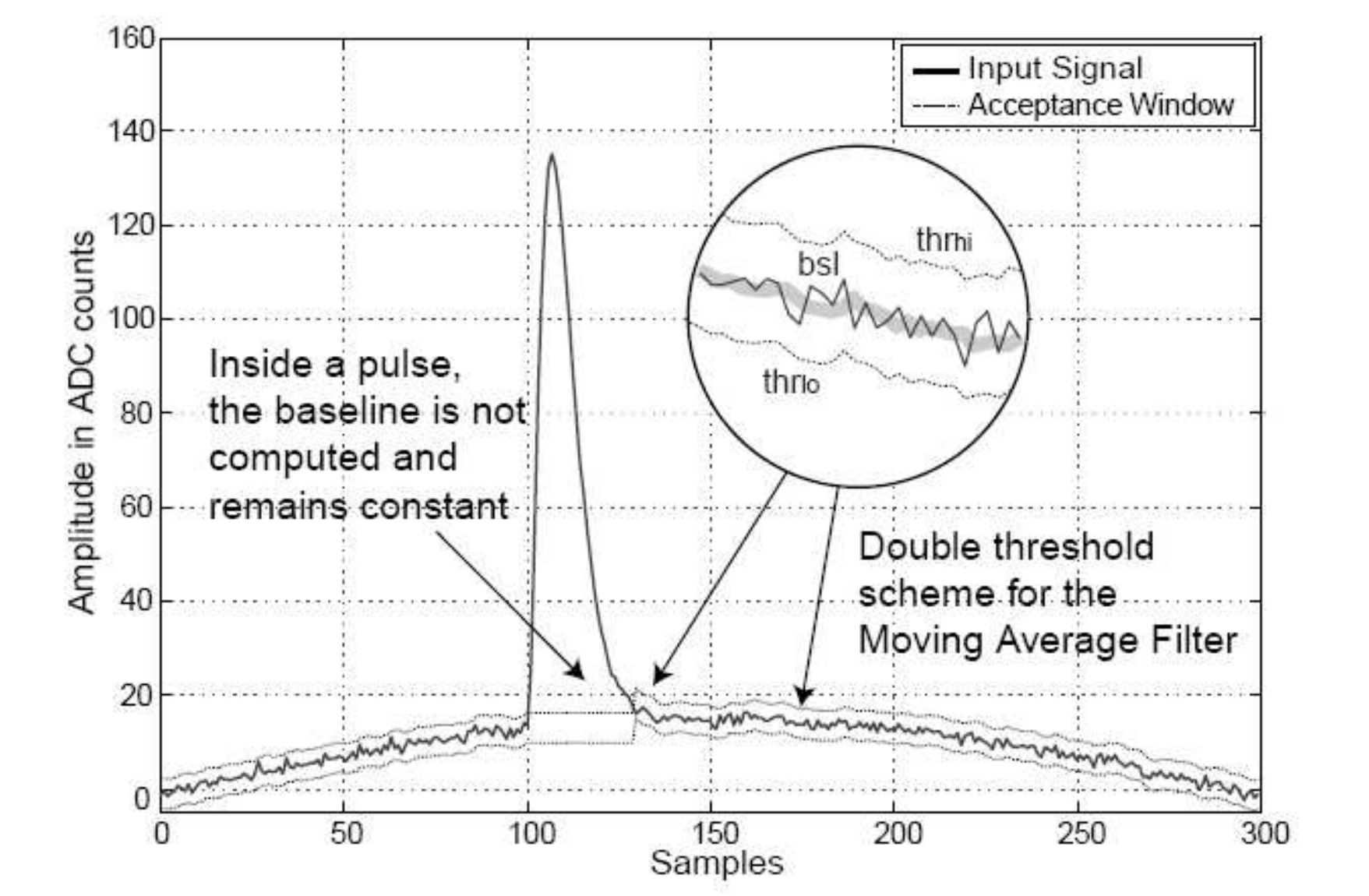}
\caption{Baseline Correction II block operation principle.}
\label{elect:ALTRO_bc2}
\centering
\includegraphics[width=0.96\linewidth,clip]{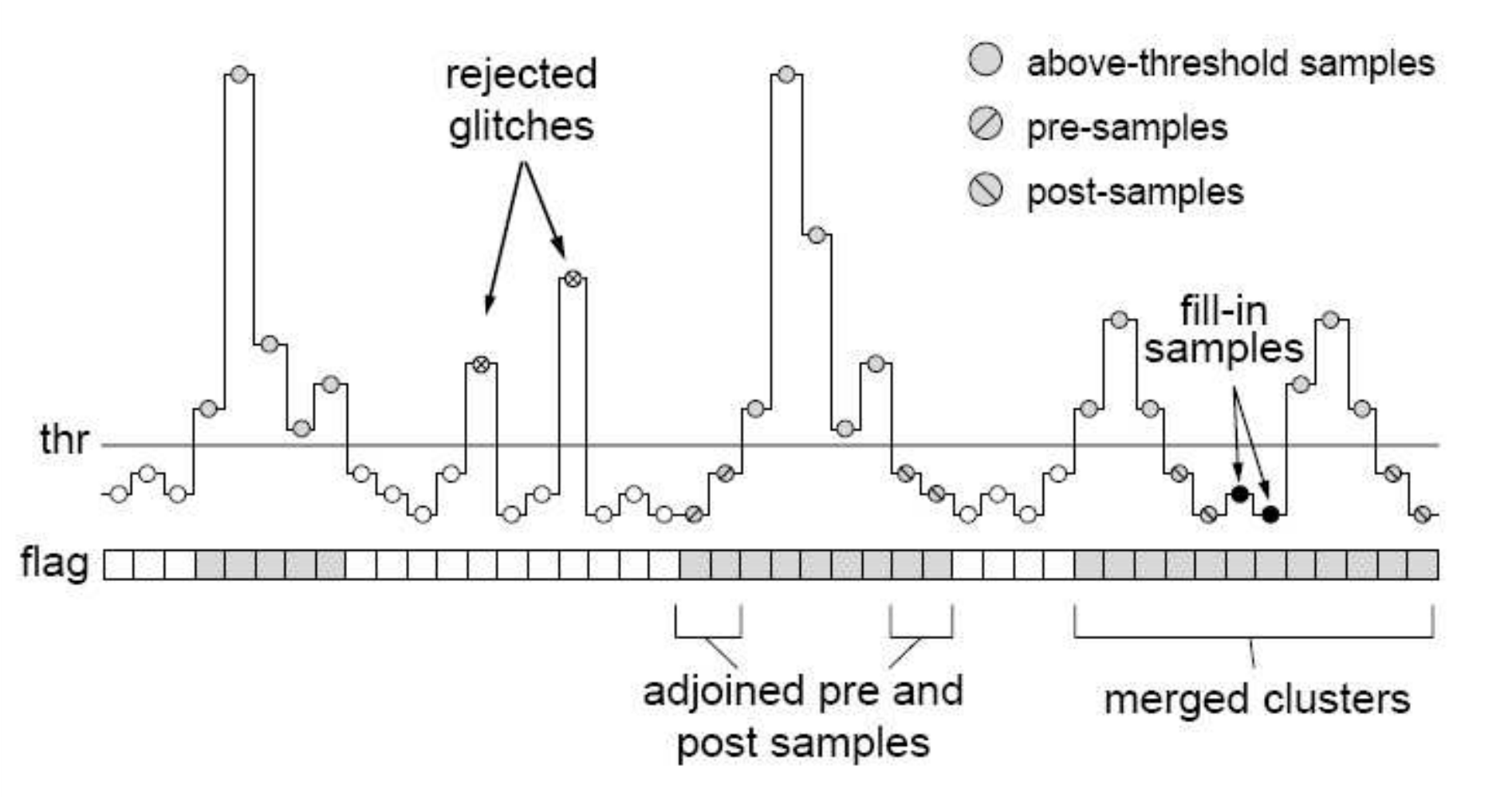}
\caption{Zero Suppression scheme.}
\label{elect:ALTRO_zs}
\end{figure}

The Data Processor conditions the signal 
in several processing stages. The first stage is the Baseline Correction I. 
Its main task is to prepare the signal for the tail cancellation by removing 
low frequency perturbations and trigger correlated effects. 
While the first is implemented by a self-calibration circuit based on a first order infinite impulse response filter,
the latter is achieved by subtracting a pattern stored in a 10-bit wide 1k deep Pedestal MEMory (PMEM).
The next processing block, the Tail Cancellation Filter, which is based on a third order infinite 
impulse response filter, is able to suppress the tail of the pulses within $\unit{1}{\micro\second}$ after the peak, 
with an accuracy of $\unit{1}{\lsb}$~\cite{Mota2004500}. Since the filter coefficients for each channel are 
fully programmable, the circuit is able to cancel a wide 
range of signal tail shapes. This also allows maintaining a constant quality of 
the output signal regardless of the actual detector operation parameters (gas 
and anode voltage), aging effects on the detector, and channel-to-channel fluctuations. 
The subsequent processing block, the Baseline Correction II, 
applies a baseline correction scheme based on a 8-tap moving average filter. 
This scheme removes non-systematic perturbations of the baseline that are 
superimposed to the signal (see Fig.~\ref{elect:ALTRO_bc2}). 
At the output of this block, the signal baseline 
is constant with an accuracy of $\unit{1}{\lsb}$. Such accuracy allows an efficient signal 
compression implemented in the Zero Suppression unit, which discards all data below a 
programmable threshold (see Fig.~\ref{elect:ALTRO_zs}). In addition this unit features the following three functions.
1) A glitch filter checks for a consecutive number of samples above the threshold, 
   confirming the existence of a real pulse, and thus reducing the impulsive noise 
   sensitivity.
2) In order to keep enough information for further extraction, 
   the complete pulse shape must be recorded. Therefore, the possibility to record 
   pre- and post-samples is provided.
3) Finally, the merging of two consecutive data 
   sets that are closer than three samples is performed.
In the Data Format unit, each data packet is formatted with 
its time stamp and size information such that reconstruction is possible 
afterward. The output of the Data Processor is sent to a $\unit{5}{\kilo\byte}$ data memory 
able to store up to 8 acquisitions.

\paragraph{Multi-Event Memory}

The ALTRO channel data memory ($1024\times\unit{40}{\bit}$) is partitioned in either 4 or 8 buffers. 
The size of the memory allows storing 4 complete 1\,000-sample acquisitions with 
non-zero-suppressed data. If the Data Processor is configured to process less 
than 500 samples, the 8-buffer partitioning can be used.
In order to reduce the noise, the basic principle of operation 
is that all bus activity should be stopped during the acquisition time. 
For this reason, the data memory manager interrupts the readout when a trigger 
is received and resumes only once the acquisition has finished. 

Data can be read out from the chip, as standalone circuit, at a maximum speed 
of $\unit{60}{\mega\hertz}$ through a 40-bit wide bus, yielding a total bandwidth 
of $\unit{300}{\mega\byte\per\second}$. 
The readout speed and the ADC sampling frequency are independent. 
In the FEC the ALTRO chips are readout at a frequency of $\unit{40}{\mega\hertz}$. 

\subsubsection{Physical implementation}
The ALTRO chip is manufactured in the ST Microelectronics CMOS 
$\unit{0.25}{\micro\metre}$ (HCMOS-7) technology.
The main physical characteristics of the circuit are summarized in Tab.~\ref{elect:ALTRO_physical}.

\begin{table}
   \centering
   \caption{ALTRO physical characteristics.}
   \label{elect:ALTRO_physical}
   \begin{tabular}{|c|c|}
     \hline
     Process          & ST HCMOS-7 ($\unit{0.25}{\micro\metre}$)      \\
     Area             & $\unit{64}{\milli\metre\squared}$             \\
     Dimensions       & $\unit{7.70\times8.35}{\milli\metre\squared}$ \\
     Transistors      & 6 million                                     \\
     Embedded memory  & $\unit{800}{\kilo\bit}$                       \\
     Supply voltage   & $\unit{2.5}{\volt}$                           \\
     Package          & TQFP-176                                      \\
     \hline
   \end{tabular}
\end{table}

\begin{figure}[t]
\centering
\includegraphics[width=0.96\linewidth,clip]{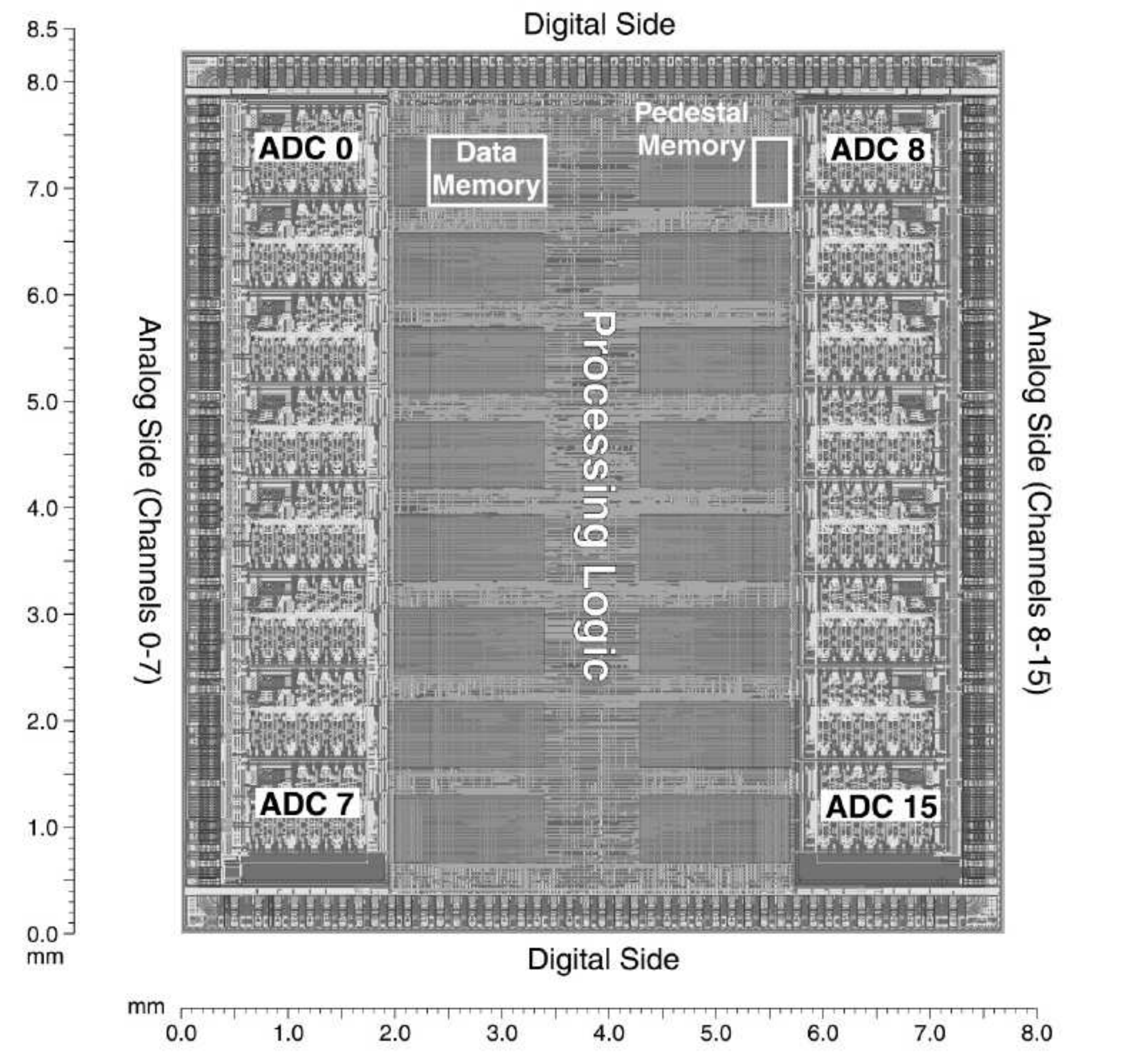}
\caption{ALTRO chip layout.}
\label{elect:ALTRO_layout}
\end{figure}

The integration of the ADC imposes certain restrictions to the layout and 
the pin-out of the chip in order to guarantee a good performance in terms of 
noise and conversion reliability. The 16 ADCs are arranged in two octal ADC macros. 
The pedestal memories are placed close to the macros on the left and right 
side as shown in Fig.~\ref{elect:ALTRO_layout}. The data memories are placed towards 
the center of the chip, distant from the ADCs macros. The placement of the 
memories reflects the regular structure of the 16 concurrent processing channels. 
The processing logic is distributed in the remaining space.
To reduce the effect of digital noise on the ADC, the following strategy was 
applied during the layout phase. Since $95\%$ of the logic works on the sampling clock, 
the phase of the clock signal distributed to all the flip-flops can be adjusted 
such that the switching of all digital nodes occurs outside the aperture time 
of the ADC. Each ADC block contains a passive clock tree balanced with an
accuracy of $\unit{1}{\pico\second}$.

As detailed in Sec.~\ref{elect:testing}, 48\,000 chips have been fabricated 
with a production yield of $84\%$.
A number of tests were implemented to assess the performance of the chip
(see Tab.~\ref{elect:ALTRO_performance}). 

\begin{table}[h]
   \centering
   \caption{ALTRO key performance figures. The three and four letter acronyms stand for:
     Effective Number Of Bits (ENOB),
     Differential Non-Linearity (DNL) 
     Integral Non-Linearity (INL), and
     Spurious-Free Dynamic Range (SFDR).
     $^1$$f_\text{S}=\unit{10}{\mega\hertz}$, $R_\text{pol}=\unit{90}{\kilo\ohm}$, internal;
     $^2$$f_\text{RDO}=\unit{60}{\mega\hertz}$;
     $^3$$f_\text{in}=\unit{960}{\kilo\hertz}$, $\unit{1}{\volt_\text{pp}}$;
     $^4$aggressor: $f_\text{in}=\unit{960}{\kilo\hertz}$, $\unit{1}{\volt_\text{pp}}$,
     victim closed to $\unit{100}{\ohm}$.
   }
\label{elect:ALTRO_performance}
   \begin{tabular}{|c|c|}
     \hline
     Power consumption            & $\unit{320}{\milli\watt}$$^1$           \\
     Max. readout bandwidth       & $\unit{300}{\mega\byte\per\second}$$^2$ \\
     ADC resolution               & $\unit{10}{\bit}$                       \\
     ENOB                         & $\unit{9.7}{\bit}$$^{1,3}$              \\
     $N_{\text{rms}}$ (RMS noise) & $\unit{0.35}{\lsb}$ rms                 \\
     DNL                          & $<\unit{0.2}{\lsb}$ rms$^{1,3}$         \\
     INL                          & $<\unit{0.8}{\lsb}$ abs.$^{1,3}$        \\
     SFDR                         & $\unit{78}{\deci\bel}$$^{1,3}$          \\
     Crosstalk                    & $\unit{0.05}{\lsb}$ rms$^{1,4}$         \\
     \hline
   \end{tabular}
\end{table}

\subsection{Front-End Card (FEC)}
\label{elect:FEC}

The Front-End Card (FEC) contains the complete readout chain for the amplification, 
shaping, digitization, processing and buffering of the TPC signals; it must handle 
the signal dynamic range of about 10 bits with minimal degradation of resolution.

\subsubsection{Circuit description}

\begin{figure}[t]
\centering
\includegraphics[width=0.96\linewidth,clip]{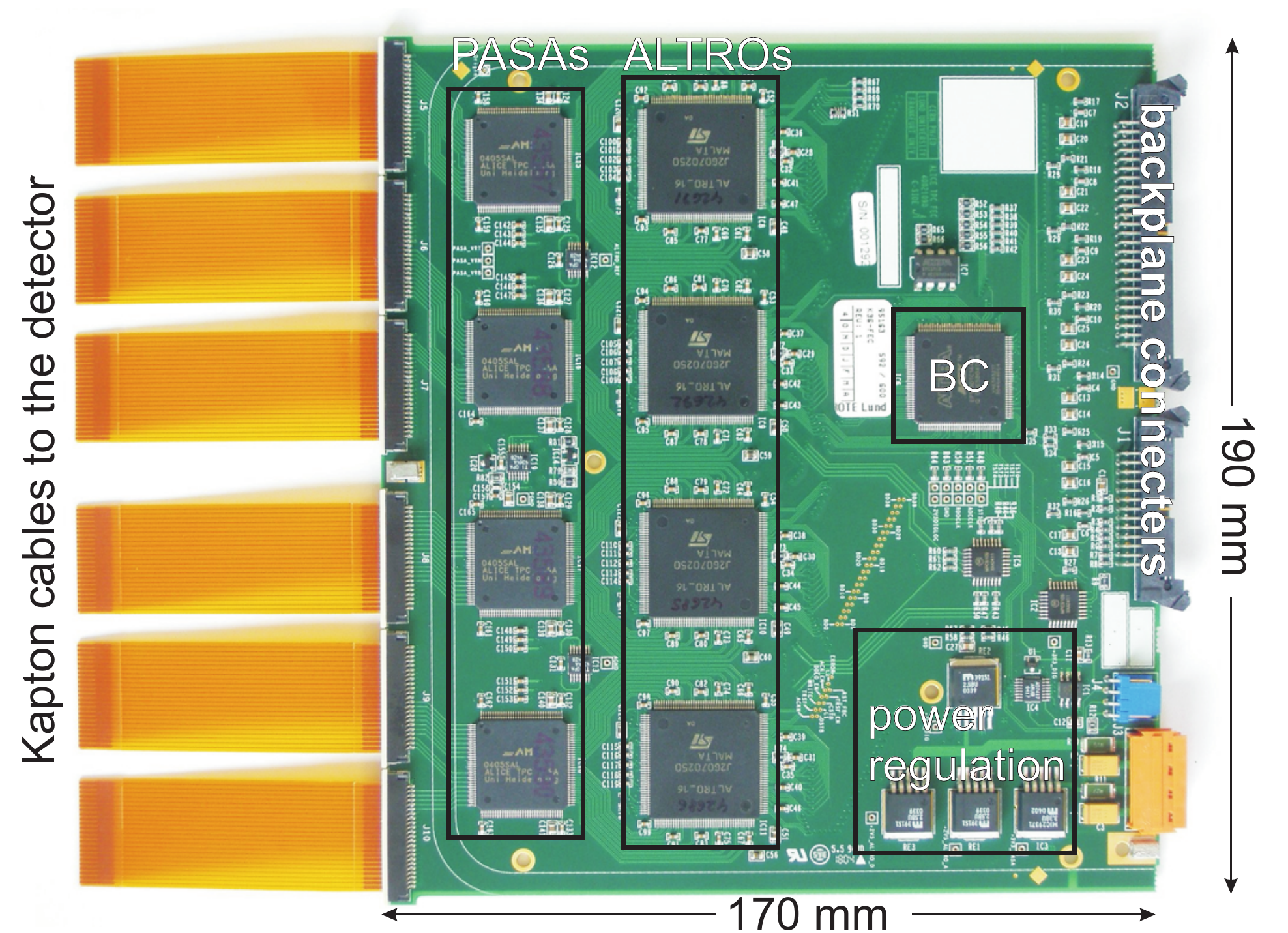} 
\caption{FEC layout. The components are mounted on both sides of the board. 
The figure shows the board topside with 4 PASAs, 4 ALTROs, 1 FPGA, the voltage
 regulators and some other minor components. On the bottom side are placed the 
other 4 PASAs and 4 ALTROs and, close to the readout bus connectors, the GTL 
transceivers.}
\label{elect:FEC_layout}
\end{figure}

With reference to Fig.~\ref{elect:FEC_layout},
hereafter the FEC layout following the signals flow is described. 
The FEC receives 128 analogue signals through 6 flexible Kapton cables 
and the corresponding connectors. The input signals are very fast, with a 
rise time of less than 1~ns. Therefore, to minimize the channel-to-channel 
crosstalk, the 8 PASA circuits have to be very close to the input connectors. 
The analogue to digital conversion and the digital processing are done by the 
ALTROs, which are connected to the corresponding PASAs with differential signals. 
It should be noticed that the PASA and ALTRO can also be interconnected in a 
single-ended mode. However, according to measurements, the noise increases by a factor 
two with respect to the differential-mode interconnection adopted in the FEC.

The data lines of the ALTROs are multiplexed, at the board level, through an LVCMOS bus. 
It features an asynchronous protocol, which is enhanced by a clocked 
block-transfer that provides a bandwidth of up to $\unit{200}{\mega\byte\per\second}$.
The FEC is interfaced 
to the RCU through a bus based on the GTL technology, named ALTRO bus. At the board 
output the bus signals are translated from LVCMOS level to GTL level by bi-directional 
transceivers. The configuration, readout and test of the board are done via the GTL bus. 
Moreover, the FEC contains the BC, implemented in an FPGA, 
which provides the RCU with an independent access to the FEC via an I$^2$C link. 
This secondary access is used to control the state of the voltage regulators 
and monitor the board activity, power supplies and temperature.

The board offers a number of test facilities. As an example a data pattern can be 
written in the ALTRO chip and read out back exercising the complete readout chain. 
The BC allows verifying the bus activities, the presence of the clock 
and the number of triggers received.

The ALTRO chips and the BC work synchronously under the master clock with a 
frequency up to $\unit{40}{\mega\hertz}$. The ALTRO circuits usually perform the same operations
concurrently, under the control of the RCU. However, the latter can also control 
a single channel at a time. This is performed in the configuration phase and for 
test purposes. The RCU broadcasts the trigger information to the individual FECs 
and controls the readout procedure. Both functions are implemented via the GTL bus.

\subsubsection{Physical implementation}

In order to match the position of the connectors on the chamber pad plane, 
the FEC has a width of $\unit{190}{\milli\metre}$. Moreover, in order to fit into the available space its 
height and thickness are $\unit{170}{\milli\metre}$ and $\unit{14}{\milli\metre}$ respectively.

The FEC Printed Circuit Board (PCB) contains four signal layers and four power layers 
(two supply layers with the corresponding ground layers). The power layers have 
essentially the same geometry. The duplication of the power and ground layers
provides the following advantages:
1) it eases the implementation of controlled-impedance lines;
2) it reduces the voltage drop over the power layers; 
3) it reduces the noise produced by ground bouncing.
From the power supply point 
of view the board is divided in three main sections: the PASA section, the ALTRO/ADC 
section and the digital section. Each power layer consists of three different 
power planes. The ALTRO/ADC and the digital planes are supplied with the same input 
voltage ($\unit{+2.5}{\volt}$), and are closed together at the input of the voltage regulators. 
The PASA plane is supplied at $\unit{+3.3}{\volt}$. The three ground planes (PASA ground, ALTRO/ADC 
ground and digital ground) are closed together with a pad, which is located upstream of
the voltage regulators.

The FEC has a maximum power consumption of about $\unit{6}{\watt}$. In order to minimize the heat 
transfer to the detector sensitive volume, the FECs are embedded in two copper plates 
cooled by water (see Sec.~\ref{cool:FEC}).

\subsection{RCU}
\label{elect:RCU}

The readout control unit (RCU) is the central node in the readout and control networks. It acts as
bridge between the different interfaces to the 
TPC (DAQ, Trigger and DCS) and its underlying electronics (FECs). In addition, 
the RCU provides core functionality to configure, trigger, readout, monitor 
and debug the FEE.

The RCU consists of a motherboard, which contains the main control circuit, and two
daughter cards: the SIU (Source Interface Unit) and DCS cards.
The RCU functions, which are mostly implemented on programmable logic devices,
can be re-programmed remotely. This has proven to be an essential feature while commissioning and
debugging the TPC FEE and its interplay with the backend systems. Moreover, the programmability of the 
RCU allows us to incorporate new features that might become relevant when 
operating the detector with high luminosity heavy ion beams.
The SIU, which is the common interface card to the ALICE DAQ, is described in detail in~\cite{elect:DDL}.
The motherboard and DCS cards will be discussed in the following sub-sections. 

\subsubsection{RCU motherboard}
The motherboard hosts the main FPGA, a Xilinx Virtex-II Pro (XC2VP7~\cite{Xilinx:2007a,Xilinx:2007b}),
the interface to the daughter cards, as well as the line drivers for the buses that connect to the FECs.

\begin{figure}
\centering
\includegraphics[width=0.96\linewidth,clip]{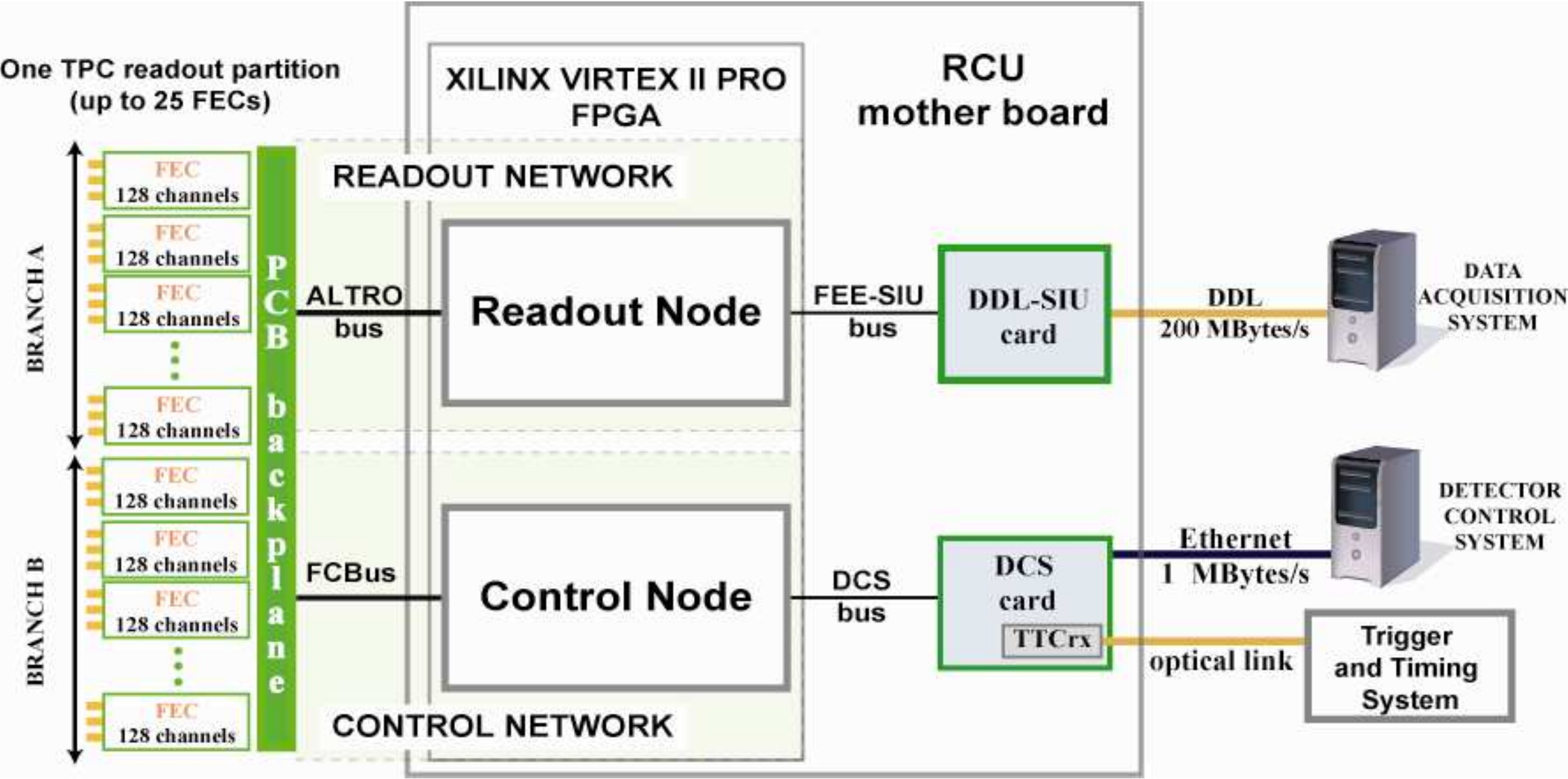} 
\caption{The RCU main FPGA firmware. The Control Node is seen on
the left, while the Readout Node is to the right.} 
\label{elect:RCU_fw_main}
\end{figure}

\begin{figure}
\centering
\includegraphics[width=0.96\linewidth,clip]{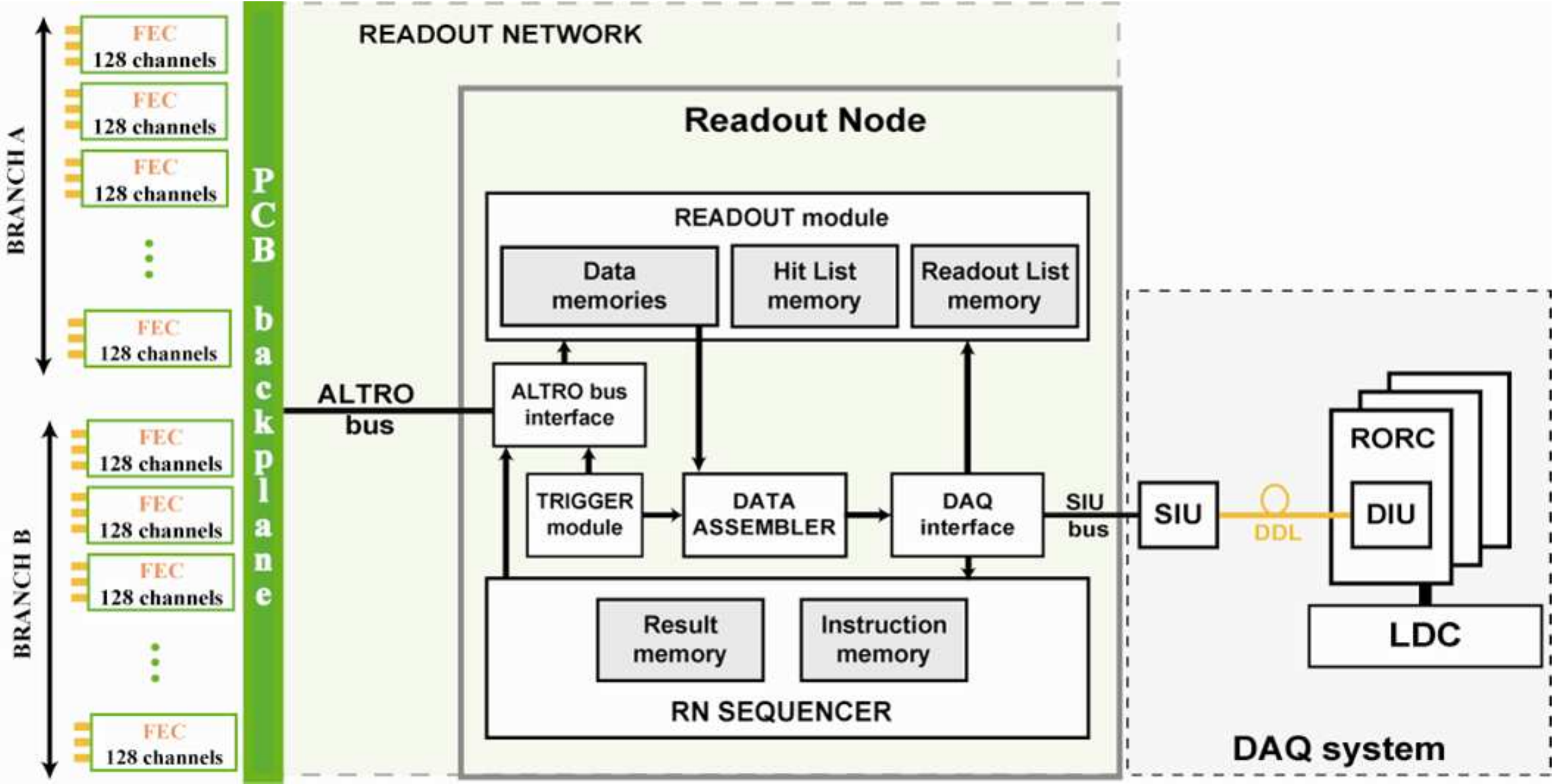}
\caption{The RCU main FPGA firmware. Simplified block diagram of the Readout Node.} 
\label{elect:RCU_fw_rn}
\end{figure}

\begin{figure}
\centering
\includegraphics[width=0.96\linewidth,clip]{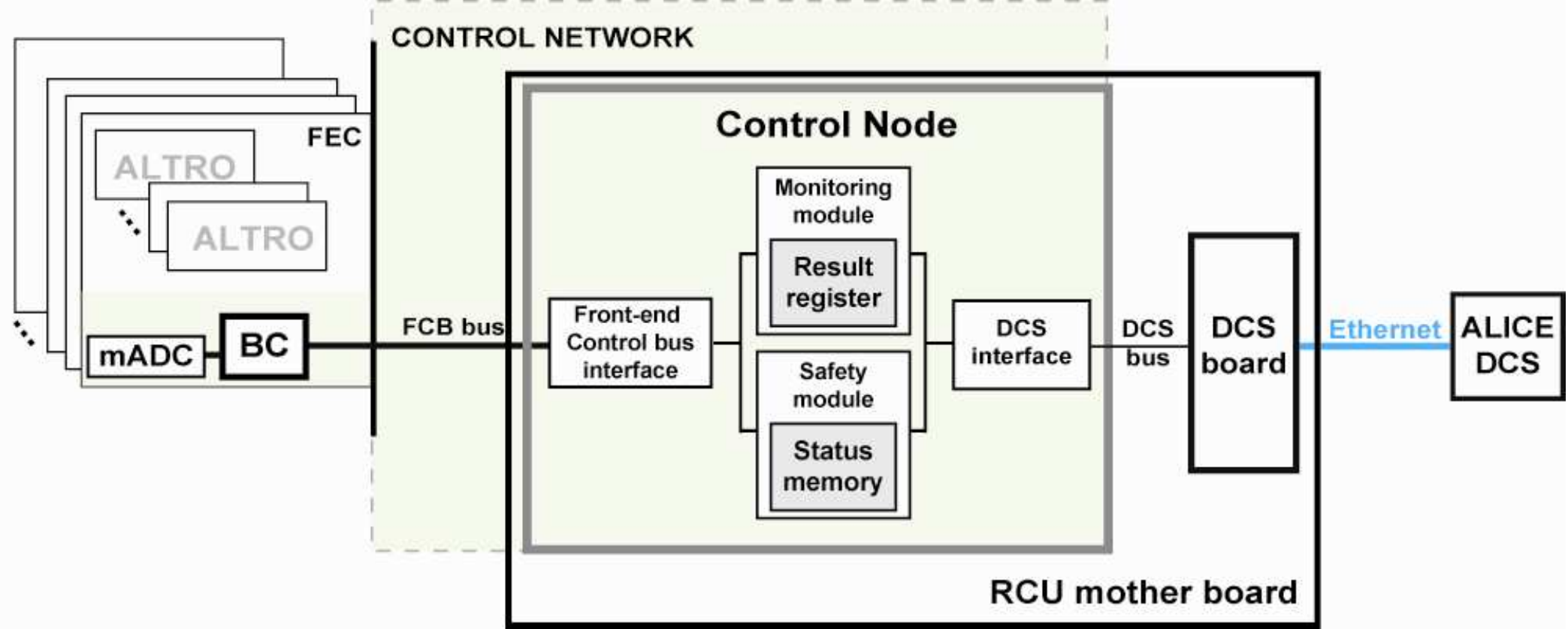} 
\caption{The RCU main FPGA firmware. Simplified block diagram of the Control Node.} 
\label{elect:RCU_fw_cn}
\end{figure}

The firmware for the FPGA is schematically sketched in Figs.~\ref{elect:RCU_fw_main}--\ref{elect:RCU_fw_cn}.   			
The readout node provides a sequencer to readout the ALTRO channels in a programmed order and allows
the program to skip
empty channels to speed up the readout process (called `sparse readout'). Special emphasis was given
to the handling of erroneous bus transactions and their reporting, which is essential to operate a system
with half a million channels in a stable fashion. In the course of implementing a fault-proof system,
the RCU wraps the ALTRO event data (10-bit words) into a simple 32-bit format, adding marker bits
to re-align to single channels in case of data corruption. This allows the system to decode the event
data packet even when part of the system fails.
The control node uses the slow control bus to read and monitor the BC values. 
It also implements an interrupt handler that can turn off FECs within $\unit{100}{\micro\second}$ 
upon a severe error, e.g.\ a temporary short circuit produced by a highly ionizing particle.

The RCU's main FPGA is also responsible for distributing the clock signals (readout and sampling clock) to the FECs. 
They are derived from the global LHC clock ($\unit{\approx 40}{\mega\hertz}$) that is distributed via the 
LHC Timing Trigger and Control (TTC) system \cite{elect:TTC} to the FEE. The sampling clock can be selected to be 
either $2.5$, $5$, $10$ or $\unit{20}{\mega\hertz}$, and the RCU provides the necessary synchronization logic
to keep the sampling clock phases of all RCUs equal. In this context, it should be mentioned that the phase of the 
trigger signal with respect to the sampling clock is measured for each event by the RCU, in LHC bunch crossing periods 
($\unit{25}{\nano\second}$), and is included in the trailer of the event data packet. 

Besides the main FPGA, the RCU motherboard also hosts a flash FPGA (ACTEL ProAsic+~\cite{Actel:2007}) and some 
flash memory. This set of circuits is used verify the configuration of the main FPGA and to possibly
reconfigure it while it is operating, making the RCU a radiation tolerant circuit as further
discussed in Sec.~\ref{elect:radiation}.

\subsubsection{DCS board}
The DCS board provides the interfaces to the ALICE Trigger and DCS systems. It is based on an
Altera Excalibur FPGA (EPXA1) with an embedded processor core
(ARM 922T)~\cite{Excalibur:2002} running a tailored version of Linux~\cite{Krawutschke:2008}. 
The connection to the DCS system is established via a $\unit{10}{\mega\bit\per\second}$
Ethernet network interface, which is electrically adapted to run in a magnetic field. 
This architecture has proven to be extremely flexible and easy to maintain as it is built on widely 
supported hard- and software platforms.

The main application running on the DCS boards is a server application, called FeeServer, 
that provides communication channels to the DCS system for configuration and monitoring. 
It is described in more detail in Sec.~\ref{DCS:Electr}.

\subsection{Trigger subsystem}
\label{elect:trigger}
The ALICE trigger system (CTP) is based on three trigger levels~\cite{elect:CTP}:
\begin{description}
\item[L0:] The `level zero' trigger pulse has a fixed latency of about $\unit{1.2}{\micro\second}$ 
  with respect to the interaction.
\item[L1a/L1r:] Each L0 can be followed by a `level one accept' pulse after a fixed latency of about 
  $\unit{7.7}{\micro\second}$ with respect to the interaction time. If this is not the case, it is referred
  to as `level one reject' and the trigger sequence has finished. If an L1a was issued, an asynchronous
  message containing basic event information (containing the event ID) will follow.
\item[L2a/L2r:] A third level trigger (`level two accept' or `level two reject') completes the 
  trigger sequence by deciding if the triggered event should be transferred from the FEE data buffers to the DAQ. 
  This trigger level is dispatched as an asynchronous message after a minimum time of about $\unit{100}{\micro\second}$,
  which corresponds to the TPC drift time, in order to ensure the completion of the TPC readout.
  It is relevant to mention that the rejection of events that are superimposed within the TPC drift time, 
  can be implemented at this trigger level. 
\end{description}
The synchronous trigger pulses (L0 and L1a) and the trigger messages (L1 message, L2a/L2r) are transmitted
together with the LHC clock signal via the TTC optical fibers.

The TPC data acquisition is started either upon an L0 or upon an L1a, according to the configuration
of the trigger detectors participating in the run. 
The readout process always starts after an L2a.

\begin{figure}[t]
\centering
\includegraphics[width=0.96\linewidth,clip]{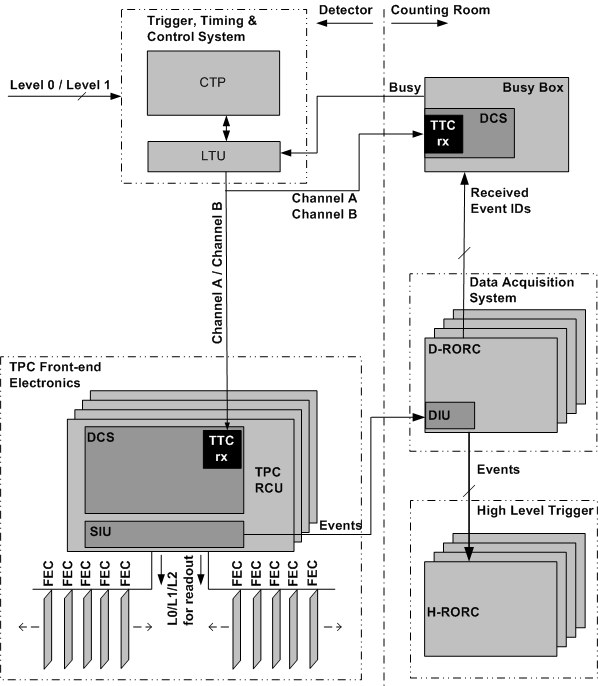} 
\caption{Overview of the trigger subsystem.} 
\label{elect:fee_trigger}
\end{figure}

The dead-time generated by the TPC has two contributions: detector dead time, i.e.\ the drift time, and the FEE dead time 
(event readout time). Whenever the TPC cannot process any further events, a signal (busy signal) is asserted to prevent the 
CTP from issuing subsequent triggers. The busy signal is generated by the BusyBox, which keeps track of the triggers 
issued and events received by the DAQ machines (via separate links to each of the 216 D-RORCs),
as well as of the number of free MEB entries in the ALTROs (refer to Fig.~\ref{elect:fee_trigger}).

\subsection{Radiation tolerance}
\label{elect:radiation}
Exposure to energetic particles can produce instantaneous failures and, over time, the degrading 
of electronic components.
The requirements for the TPC electronics in terms of radiation tolerance are 
twofold. On the one hand the estimated radiation is rather low both in terms 
of flux and dose (see below), which allows for the use of standard, 
radiation-soft technologies. On the other hand it has to be assured that 
the impact of rare (but existing) effects due to radiation are not 
affecting the overall performance or even causing irreversible damages to the system.

Simulations suggest that the total radiation load on the TPC will be less 
then $\unit{13}{\gray}$ over 10 years, and the flux through the
electronics less than about
$\unit{800}{particles\per\centi\metre\squared\second}$~\cite{elect:ALICE-INT-2002-028}.
Nevertheless, for the design and implementation of the TPC FEE, the following three effects were carefully evaluated:
\begin{itemize}
\item \textsl{Total Ionization Dose (TID) effects}. TID is a cumulative effect, 
related to the damage of the semiconductor lattice, which causes slow gradual 
degradation of the device's performance.
\item \textsl{Single Event Upset (SEU)}. When an ionizing particle traverses the 
  sensitive region of a memory cell it can change its logical state (0 to 1 or 
  vice versa).
\item \textsl{Single Event Latch-up (SEL)}. The ionized track of an particle can form
  a conducting path through the substrate of the semiconductor creating a short 
  between the supply rails.
\end{itemize}

Concerning the TID effects, all components were qualified to withstand a dose corresponding 
to 150 years of LHC operation. 
The measures to protect the FEE  components against SEUs and SELs were chosen according to
their potential impact on the system. Two examples, the protection of the ALTRO chips and
the RCU FPGA, are hereunder discussed in more detail.

\subsubsection{SEU}
\paragraph{ALTRO}
In the ALTRO all control state machines are protected against SEU by Hamming coding
that implements an algorithm to recover from single bit and to detect double bit errors. In particular
this prevents electrical conflicts of two ALTRO chips on the I/O bus. As this 
virtually makes them fail-safe, the most severe problem that can still occur 
is a corruption of the configuration registers content. Moreover, data could
also be affected, but the estimated bit error rate does not justify any protection on this level.
Combining the simulated particle flux with the SEU cross-sections measured for the ALTRO chip,
the corresponding mean time between failures (MTBF) have been determined as listed in
Tab.~\ref{elect:MTBF}.

\begin{table}
\begin{center}
\caption{MTBF of the internal elements of the ALTRO chip.
The MTBF values are quoted for a readout partition with 25 FECs. $^1$It refers to the
occurrence of SEUs in a finite state machine (FSM), which is recovered by the Hamming protection.}
\label{elect:MTBF}
\begin{tabular}{|c|c|}
\hline
part      & MTBF \\
\hline
registers & 36 hours \\
PMEM      & 168 minutes \\
MEB       & 42 minutes \\
FSM$^1$   & 58 days \\
\hline
\end{tabular}
\end{center}
\end{table}

\begin{figure}
\centering
\includegraphics[width=0.96\linewidth,clip]{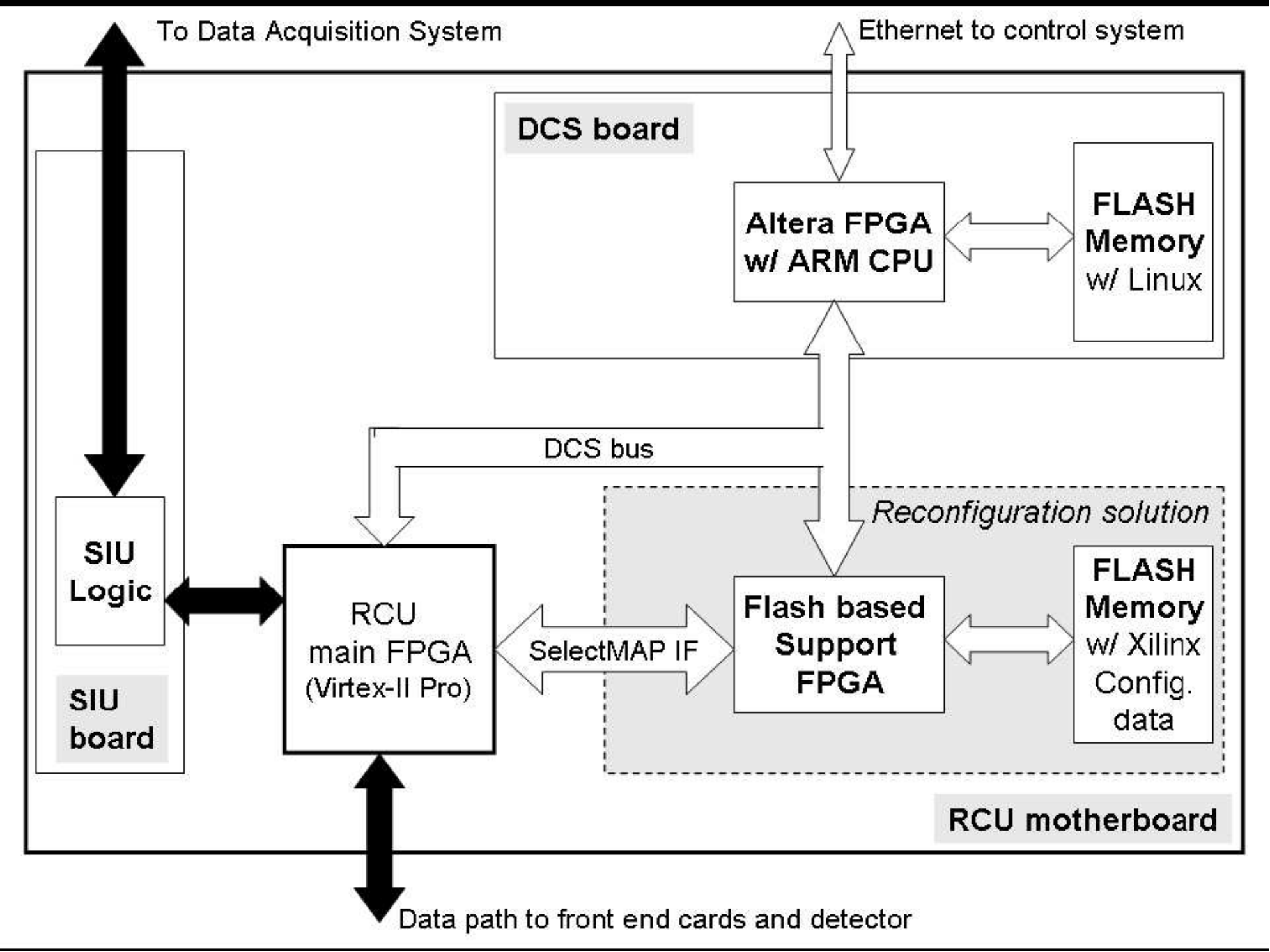} 
\caption{RCU block diagram, emphasizing the circuit for Active
Partial Reconfiguration at the bottom right. The data path is given by black arrows.} 
\label{elect:fee_rad}
\end{figure}

\paragraph{RCU}
In the RCU there is a single device to be protected against SEUs, the main FPGA,
which is based on SRAM technology. In this context, it should be mentioned that the
hardware resources needed for the implementation of the RCU functions are featured
only in large SRAM programmable devices. However,  these devices store their configuration, 
which defines their logical function, inside a radiation sensitive SRAM. 
If an SEU occurs in this configuration memory, it might cause a circuit malfunction,
which however can be corrected by rewriting the originally stored value to 
the memory cell. Partial reconfiguration is a feature offered by several Xilinx FPGAs,
which allows to reconfigure a subset of the configuration memory without interrupting 
the operation of the device. This has been implemented in the RCU system in order to 
detect and correct SEUs~\cite{Roed:2009}. As shown in Fig.~\ref{elect:fee_rad}, the main parts of the 
verification and reconfiguration circuit are an Actel flash-based FPGA (auxiliary FPGA), 
and a flash-memory device.

\subsubsection{SEL}
The radiation tests carried out on the FEE components have never shown the occurrence
of SELs. However, the FECs were designed before the radiation tests were 
completed and, therefore, were protected against the occurrence of short circuits induced
by SEL as a precautionary measure. The BC continuously monitor the FEC's power consumption and asserts 
an interrupt if a programmable threshold is exceeded. Whenever the RCU detects an interrupt 
it turns off the corresponding FEC, removes it from the readout process, and notifies the event
to the DCS.

\subsection{Testing procedure}
\label{elect:testing}
The testing and qualification procedure for the electronics was implemented in two consecutive stages:
first the individual custom made chips (ALTROs and PASAs) and then the assembled FECs.

After having been packaged, all chips had to go through a thorough test that included burn-in testing 
and functional testing based on a semiautomatic set-up using a robot~\cite{TPCint}. 
The digital and analog parts were subject to different acceptance tests. The digital parts were
required to be 100\% functional. The analog parts were allowed a small variation in the most critical
parameters (conversion gain, peaking time and output DC offset).
The requirements and sources of failures are displayed in Tabs.~\ref{elect:yieldsALTRO} 
and \ref{elect:yieldsPASA}, which also include the numbers as obtained for a later production
of the circuits mainly for STAR in 2006.
The yields for ALTROs and PASAs are 84\% and 98\%, respectively. 
The PASA yields were so good, that we were able to apply very
stringent selection criteria as listed in Tab.~\ref{elect:yieldsPASA}.

Once the parts had been qualified, the assembled FEC was tested to reject clear electrical faults (e.g.\ shorts) 
and to characterize the interplay of the components. The corresponding yields are shown in Tab.~\ref{elect:yieldsFEC}.

During all tests the components were marked and their characteristics filled into a database. For the FECs the
ID is also stored in an EEPROM on the card itself.

\begin{table}
\begin{center}
\caption{Failures in the ALTROs. Tests were performed in the listed order and the procedure was
stopped upon the first detected failure.}
\label{elect:yieldsALTRO}
\begin{tabular}{|c|cc|cc|}
\hline
\multirow{2}{*}{failure} & \multicolumn{4}{|c|}{frequency} \\
& \multicolumn{2}{|c|}{ALICE} & \multicolumn{2}{|c|}{STAR} \\
\hline
none         & 41\,297 &  84.1\% & 14\,273 &  86.6\% \\
\hline
power        &  2\,307 &   4.7\% &     674 &   4.1\% \\
register     &  1\,032 &   2.1\% &     190 &   1.2\% \\
PMEM         &     621 &   1.3\% &     199 &   1.2\% \\
MEB          &  2\,203 &   4.5\% &     633 &   3.8\% \\
DSP          &     753 &   1.5\% &     241 &   1.5\% \\
ADC          &     712 &   1.4\% &     228 &   1.4\% \\
misc.        &     153 &   0.3\% &      49 &   0.3\% \\
not tested   &      48 &   0.1\% &       2 &   0.01\% \\
\hline
total        & 49\,127 & 100.0\% & 16\,489 & 100.0\% \\
\hline
\end{tabular}
\end{center}
\begin{center}
\caption{Acceptance levels for the PASAs as chosen for the ALICE TPC. Multiple counting occurs but
in the case of too high power consumption where the testing procedure was aborted.}
\label{elect:yieldsPASA}
\begin{tabular}{|c|cc|cc|}
\hline
not accepted       & \multicolumn{4}{|c|}{frequency}    \\
 & \multicolumn{2}{|c|}{ALICE} & \multicolumn{2}{|c|}{STAR} \\
\hline
none               & 40\,938  &  85.9\% & 20\,954 &  91.9\% \\
\hline
power              &     776  &   1.6\% &      72 &   0.3\% \\
\hline
conv.\ gain (5\%)  &  1\,009  &   2.1\% &     179 &   0.8\% \\
peaking time (6\%) &  1\,408  &   3.0\% &     153 &   0.7\% \\
outp. offset ($\unit{50}{\milli\volt}$)
                   &  5\,428  &  11.4\% &  2\,122 &   9.3\% \\
\hline
total              & 47\,637  & 100.0\% & 22\,795 & 100.0\% \\
\hline
\end{tabular}
\end{center}
\begin{center}
\caption{Failures in the FECs. Multiple counting occurs. The numbers are biased by the `unknown' entry, which
  refers to FECs that fail the test but still have to be debugged/repaired, as the FECs with most
  obvious errors were repaired first.}
\label{elect:yieldsFEC}
\begin{tabular}{|p{5cm}|cc|}
\hline
failure & \multicolumn{2}{|c|}{frequency} \\
\hline
none               & 4\,320 &  90.0\% \\
any		   &    380 &   7.9\% \\
\hline
damaged PASAs      &     80 &   1.7\% \\
damaged ALTROs     &     40 &   0.8\% \\
improper placement or soldering of connectors   &   35 &   0.7\% \\
PCB traces         &      5 &   0.1\% \\
improper soldering of IC pins &  107 &   2.2\% \\
defective, misplaced or missing passive components &   73 &   1.5\% \\
unknown             &   140 &   2.9\% \\
\hline
total              & 4\,800 & 100.0\% \\
\hline
\end{tabular}
\end{center}
\end{table}

\section{Cooling and temperature stabilization system}
\label{cool}
\subsection{Overview}
\label{cool:Overview}

In this section we describe the water based cooling and temperature stabilization system of the ALICE TPC. We substantiate the need for a temperature stabilization system in addition to the cooling (heat removal) of the front-end electronics. The principle of the leakless operation of the cooling system is briefly introduced and the cooling strategy is outlined. In Sec.~\ref{cool:Commissioning} the experience gained during the commissioning of the cooling systems (2006--2008) is reported. The achieved temperature homogeneity during these first runs is also discussed.

\subsection{The necessity for uniform temperatures}
\label{cool:Necessity}

The design goal for the temperature stability and homogeneity within the TPC drift volume is $\Delta$T $<$ 0.1~$^\circ$C \cite{TDR:tpc}. This value is a consequence of our particular gas choice, a mixture of Ne--CO$_2$--N$_2$.

\begin{figure}
\centering
\includegraphics[width=0.96\linewidth]{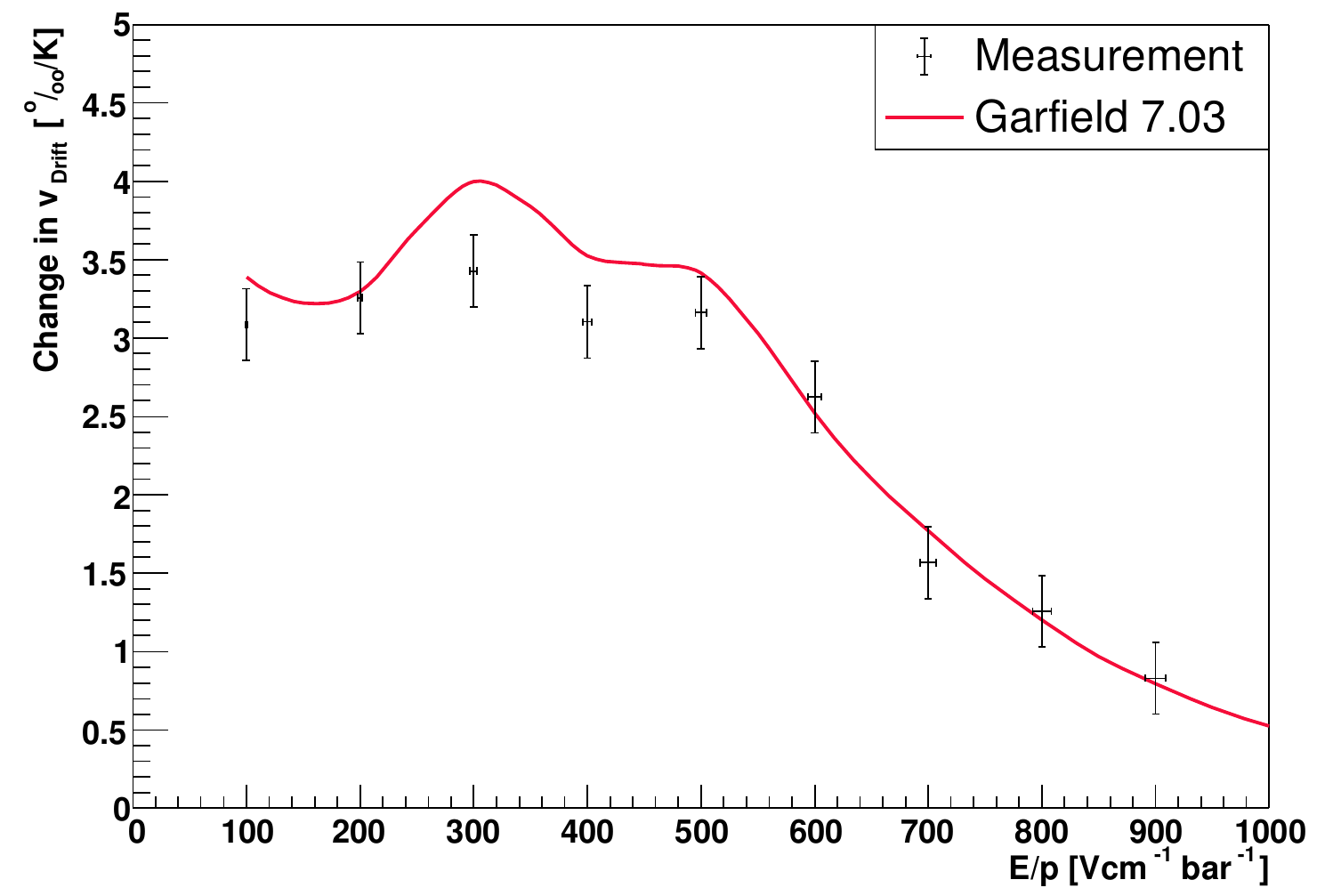}
\caption{Relative change of the drift velocity with temperature as a function of the drift field for the binary mixture Ne--CO$_2$ in proportion 90--10 \cite{Wiechula2005, cool:Wiechula2004}.}
\label{cool:EvsdVdrift}
\end{figure}

Figure~\ref{cool:EvsdVdrift}  shows the relative change in drift velocity with temperature as a function of the drift field \cite{Wiechula2005, cool:Wiechula2004}.  As can be seen from the graph, the change in drift velocity is about $\unit{0.35\%}{\per\kelvin}$ at $\unit{400}{\volt\per\centi\metre}$. This means that a temperature difference larger than 0.1~K over the full drift length  of $\unit{250}{\centi\metre}$  results in longitudinal position variations $\Delta z$  of the order of 1~mm at $v_\text{drift} = \unit{2.65}{\centi\metre\per\micro\second}$.  This exceeds the internal resolution of the readout chambers.

\subsubsection{Heat load and Computational Fluid Dynamics calculations}
\label{cool:Heat_load}

The main heat contribution stems from the FECs,
which are connected via short ($\unit{8.2}{\centi\metre}$) flexible cables to the cathode pad plane of the readout chambers. The $2 \times 18$ sectors, each being equipped with 121 FEC cards, dissipate a total of $\unit{28}{\kilo\watt}$. This heat load has to be removed by the FEC-cooling circuits. The bus bars, providing the low voltage power to the FEC, are integrated into the SSW spokes and dissipate a total of about $\unit{0.54}{\kilo\watt}$.

Another important heat source affecting the TPC gas is the power produced by the four field-cage resistor rods. While the power is relatively small ($\unit{8}{\watt\per rod}$), it would, without countermeasures, be dissipated directly into the gas volume.

Other heat sources are neighboring detectors, namely the Inner Tracking System (ITS) inside and the Transition Radiation Detector (TRD) outside of the TPC.

\begin{figure}
\centering
\includegraphics[width=0.96\linewidth]{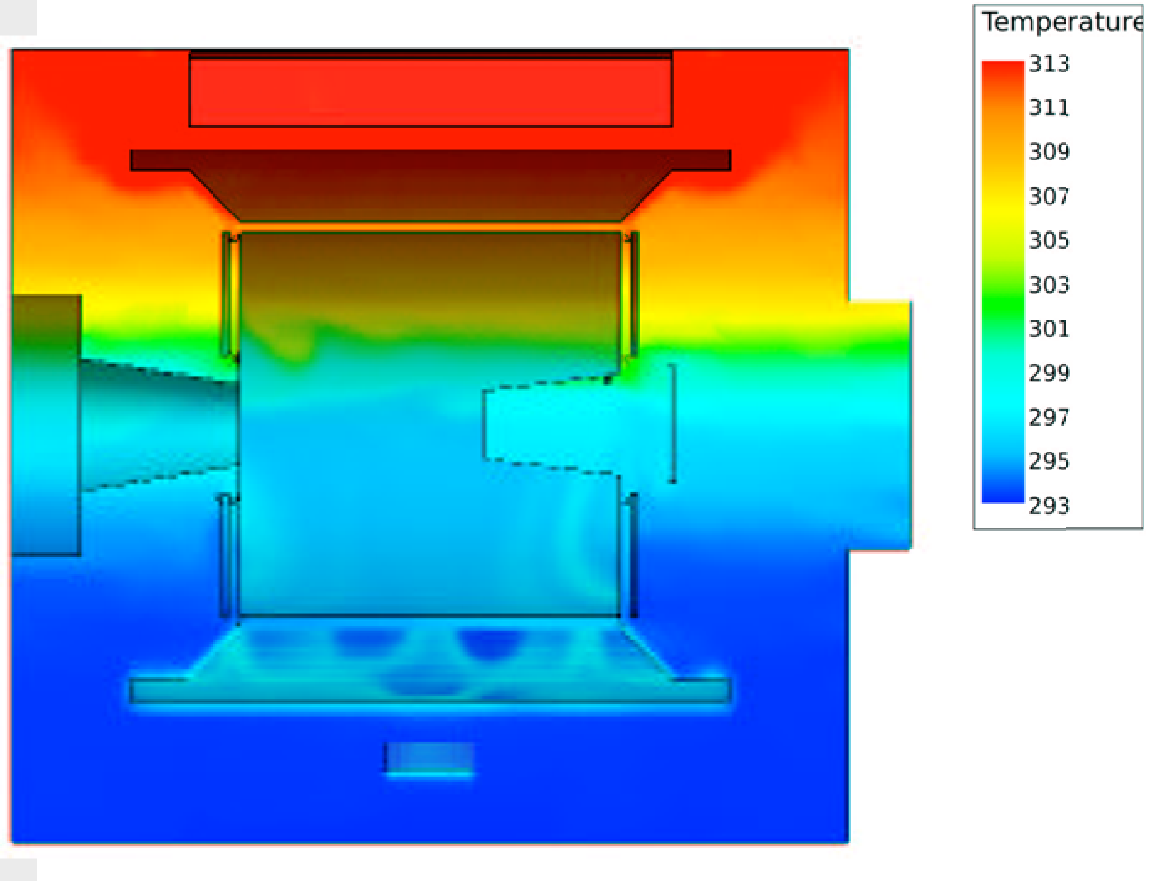}
\caption{Temperature profile in the L3 solenoid magnet resulting from CFD calculations \cite{cool:Mueller2004}.}
\label{cool:CFD}
\end{figure}

In the context of the optimization of the
ventilation scheme inside the ALICE L3 magnet, a Computational Fluid Dynamics (CFD) study has been carried out to estimate the residual heat distribution within the L3 magnet \cite{cool:Mueller2004}. Figure~\ref{cool:CFD} shows the temperature profile resulting from the CFD calculations. A  90\% cooling efficiency of the various detector cooling systems has been assumed for this study leading to a dissipated heat of $\unit{17}{\kilo\watt}$. It was shown that, irrespective of the detailed layout of the ventilation scheme, a temperature gradient of about $\unit{5}{\kelvin}$ develops across the vertical dimension of the TPC. This study clearly demonstrates that the TPC fiducial volume needs additional thermal shielding and compensation of temperature gradients.

\subsection{Principle of underpressure cooling }
\label{cool:Underpressure}

\begin{figure}[t]
\centering
\includegraphics[width=0.96\linewidth]{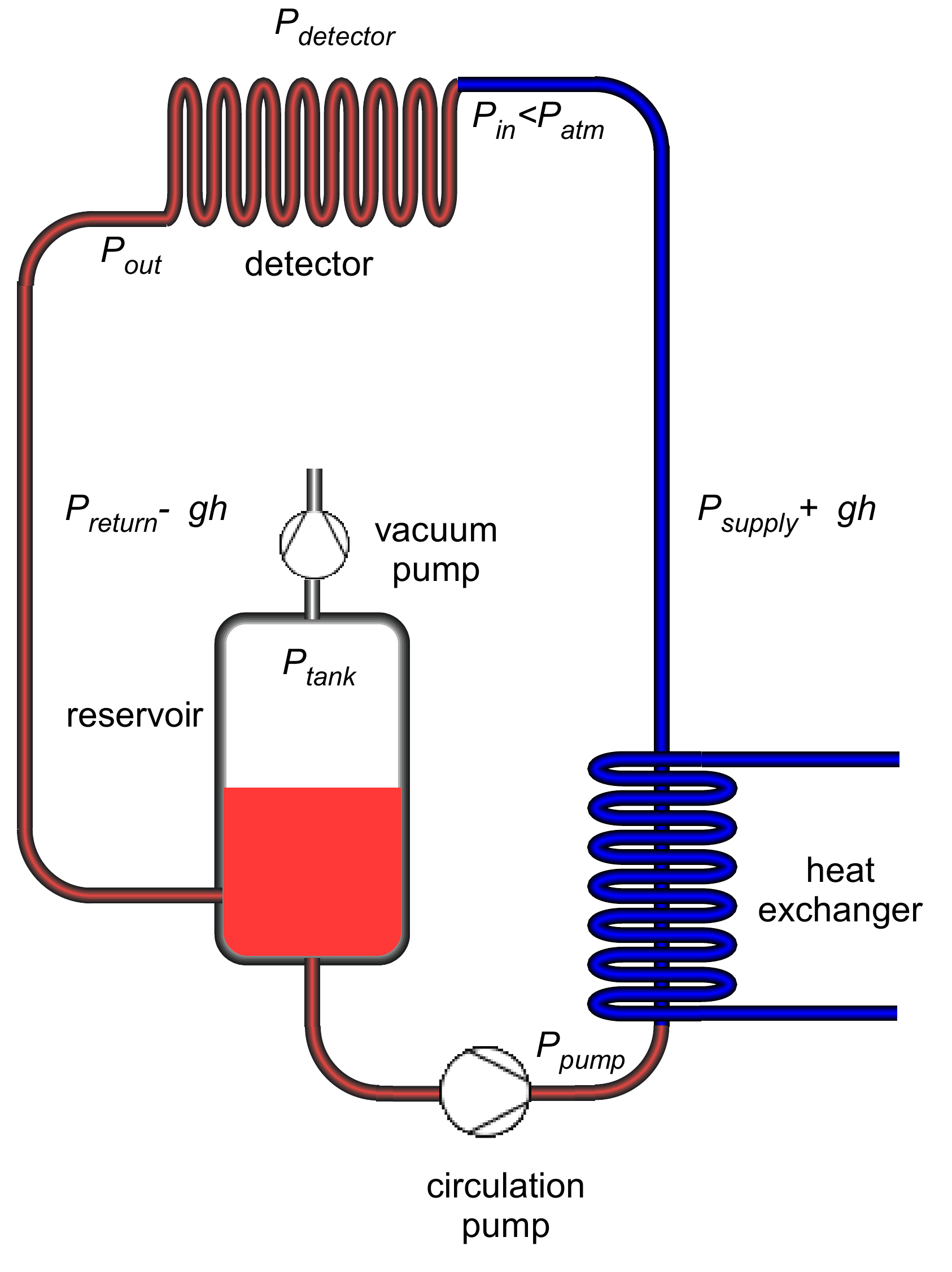}
\caption{Principle of underpressure operation.}
\label{cool:underpressure}
\end{figure}

The principle of underpressure cooling \cite{Bonneau:1999xi} is depicted in Fig.~\ref{cool:underpressure}. The cooling-liquid circuit is a closed circuit, which allows to operate all or part of the cooling lines  below atmospheric pressure. The cooling-liquid tank is kept at underpressure, which, by the proper choice of length and diameter of the return pipes and of the circulation-pump output pressure, ensures that the water pressure inside the detector is below atmospheric pressure. This has the obvious advantage of an active protection against the occurrence of leaks.

In our case, the space constraints due to the extremely dense front-end readout does not allow space-consuming high-pressure certified fittings. Therefore, we chose simple silicon hoses without any special lock mechanism to couple to the copper tube of the card cooling envelope 
(see Sec.~\ref{cool:FEC}).
 Though the connection between the silicon hose and the copper tube has been tested to hold overpressure of  $\unit{2.5}{\bbar}$ over an extended period (24 hours), it is mechanically fragile, e.g., against tears or cuts. These considerations led us to the choice of the sub-atmospheric `leakless' technology for our cooling circuits. 
In the following paragraphs, we will give a short description of this, a more detailed description can be found in \cite{Bonneau:1999xi}.

An apparent disadvantage of the sub-atmospheric `leakless' technology is the limited range of operation ($p < \unit{1}{\bbar}$). This implies that the allowed pressure loss,  $\Delta p_\text{loss}$, in the detector is rather limited. The situation is further aggravated, in our case, due to the height difference  of about $\unit{8}{\metre}$ between the highest inlet and the cooling plant. In addition the height difference between the inlets of sector 4 and 13 located at the top and bottom of the TPC, respectively, is $\unit{5}{\metre}$. While the input pressure at each inlet can be adjusted independently via balancing valves, this has not been foreseen for the return lines. Thus, all detectors `see'
 a combination of the reservoir pressure, the hydrostatic pressure  and the pressure loss in the return pipes. This can result in very low pressure values in some of the return lines, which might cause cavitation phenomena (see Sec.~\ref{cool:Cavitation}).

\subsection{TPC cooling plants}
\label{cool:Plant}

The TPC main plant \cite{Pimenta2003}, schematically depicted in Fig.~\ref{cool:underpressure},  consists of a $\unit{1200}{\litre}$ reservoir, large enough to buffer most of the water in the installation, a pump,  a heat exchanger connected to the CERN mixed water network, a supply manifold and a return manifold. A total of 48 circuits are connected on these manifolds. All 48 circuits can be temperature and flow adjusted. Temperature regulation is done via individual heaters on the supply manifold. These heaters are controlled via a Proportional-Integral-Derivative (PID) controller using a temperature sensor immediately downstream of the heater as a feedback signal. The precision of the regulation is $\unit{0.1}{\kelvin}$. The effective power of the heaters is $\approx\unit{3}{\kilo\watt}$, which allows a temperature swing of about $\unit{2}{\kelvin}$ at a flow of $\unit{20}{\litre\per min}$. To allow a larger flow and/or temperature swing for the chamber body, the heaters for these loops were upgraded to $\unit{10}{\kilo\watt}$ (see Secs.~\ref{cool:Strategy} and~\ref{cool:Commissioning}).

Flow adjustment was done manually by adjusting valves during the commissioning of the system, in order to ensure that all circuits got the right flow despite of the varying pipe lengths and heights between the plant and the detector. The pressure at the inlet of the reservoir is controlled by a so-called back-pressure valve. This valve had not been foreseen in the original design of the plant. However, during commissioning (see Sec.~\ref{cool:Commissioning}) the necessity of being able to vary the return line pressure became evident. 
A PID controller adjusts the pump speed via a frequency inverter in order to keep a constant pressure in the supply manifold independent of the number of circuits in service. Another PID controller adjusts the flow of mixed water via a 3-way valve in order to maintain a constant temperature at the heat-exchanger outlet. All regulation loops and the plant control is done by a Programmable Logical Controller (PLC). A comprehensive description of the TPC cooling plant, its operational parameters as well as the alarm handling is presented in \cite{Pimenta2003}.

A conceptually similar, but much smaller plant ($V_\text{reservoir} = \unit{80}{\litre}$) is used for the resistor-rod cooling. It has, in addition, an ion-exchanger filter to purify the circulation water. The conductivity of both the supply and return water is measured. An alarm is raised if its value exceeds a threshold value. For this plant the pressure of the four return lines was adjusted individually by introducing fixed restrictions.  They compensate the different pressure drops in the return lines due to the hydrostatic pressure given by the different height of the resistor rods.

\subsubsection{Cooling circuits}
\label{cool:Loops}

\begin{figure}[t]
\centering
\includegraphics[width=0.96\linewidth]{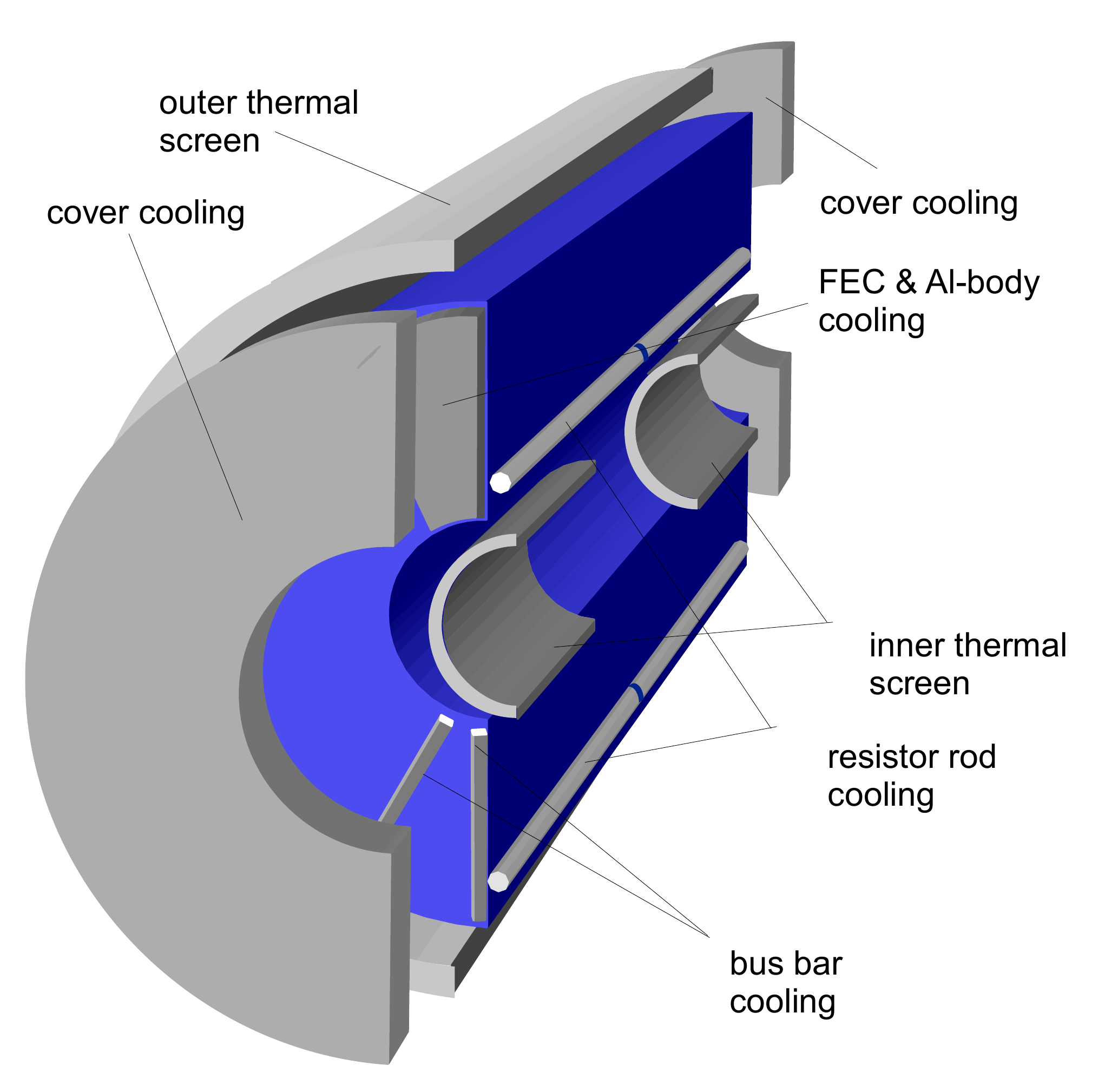}
\caption{Schematic view of the various TPC cooling elements.}
\label{cool:screens}
\end{figure}

A schematic overview of the different cooling units of the TPC, consisting of circuits to remove the heat produced by (i) the front-end electronics, bus bars and resistor rod and (ii) screens to define iso-thermal surfaces,  is shown in Fig.~\ref{cool:screens}.

Cooling and temperature stabilization of the TPC is provided via 60 individual loops which are supplied by three different cooling plants. The main TPC cooling plant supplies:

\begin{itemize}
\item $2 \times 18$ loops for the front-end electronics cooling at the A- and C-side, respectively;
\item $2 \times 2$ loops for the bus bar and  cover cooling. The two loops at each side furnish the top and bottom half of the bus bars and covers, respectively;
\item $2 \times 2$ loops for the chamber body cooling. The two loops at each side supply the top and bottom half of the chamber bodies, respectively;
\item $2 \times 1$ loops for the inner thermal screen, which separates the TPC from the ITS services. Each of the loops is split after the balancing valve and supplies the upper and lower manifold of the screen panel;
\item 1 loop supplies the resistor-rod heat exchanger; 
\end{itemize}

Another plant, the resistor rod-cooling plant, supplies the $2\times2$ loops for the resistor-rod cooling. On each side of the TPC, the inner and the outer resistor rods have their own cooling circuits. A separate plant for the resistor-rod cooling is needed because of the special demand on the purity of the cooling water.

The outer thermal screen, decoupling the TPC and the TRD thermally, is supplied by 9 independent cooling circuits.  The Al-panels of the screen require deionized water, which is provided by the TRD cooling plant.

All loops of the main TPC plant are independent from each other in the sense that the flow and the temperature (within limits) can be regulated independently. The resistor-rod lines have a common temperature set point and individual flow regulation via balancing valves. The outer thermal-screen panels are supplied by a common temperature water flow.  Regulation, e.g.\ between top and bottom panels, is possible only via different water flow settings.

\begin{figure}[t]
\centering
\includegraphics[width=0.96\linewidth]{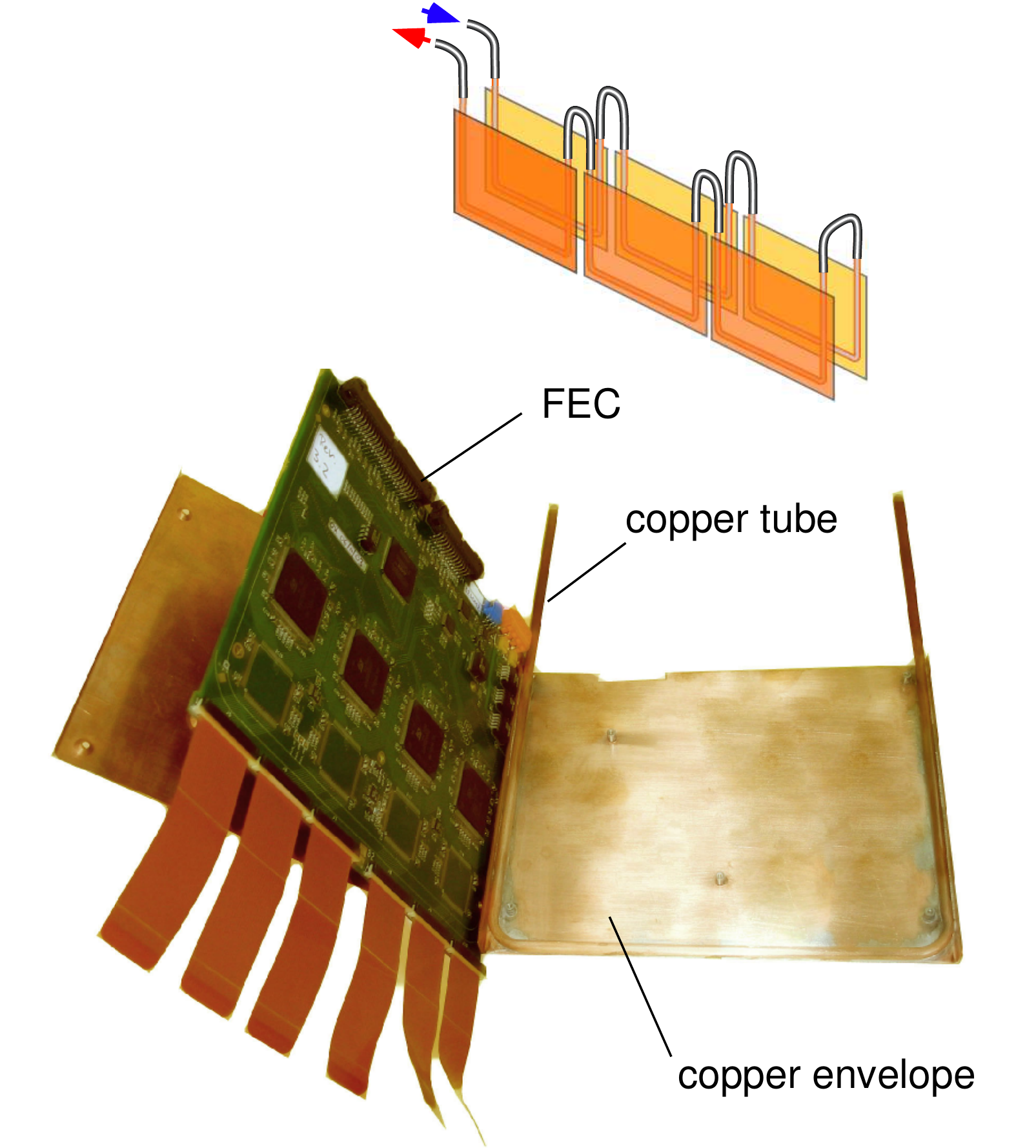}
\caption{Photograph of a FEC with its cooling envelope. The top part of the figure shows a sketch of the routing of the cooling pipes connecting 6 FECs.}
\label{cool:cooling_envelope_dis}
\end{figure}

\paragraph{FEC and Al-body cooling}
\label{cool:FEC}
Cooling to the FECs is provided sector by sector. As shown in Fig.~\ref{cool:cooling_envelope_dis} (top), 6 FECs are grouped together and are connected to the sector manifold. It has been estimated \cite{Pimenta2003} that a flow of about  0.5 l/min per group of  6 FECs,  i.e.\ 10 l/min in total per sector, is sufficient to cool the electronics. A photograph of the FEC and its cooling envelope is shown in Fig.~\ref{cool:cooling_envelope_dis} (bottom); the copper envelope is flipped open and exhibits the FEC inside. As can be seen, a 5~mm copper tube for the cooling water is soldered to one of the copper plates. The two plates are held together via 6 screws, which at the same time serve as heat bridges between the two plates. The main heat load of the FECs is taken up by the cooled copper envelopes. However, to stabilize the temperature and to absorb residual heat transfer, e.g., by the Kapton cables, auxiliary cooling circuits have been integrated into the Al-body of the chambers (see Sec.~\ref{chamb}).

The busbars, suppling the low voltage to the FEC, run along the spokes of the service support wheel. To remove the resistive heat of the busbars,  additional, water-cooled copper bars, which are in thermal contact with the busbars, are integrated into the service support wheel spokes. The busbars and the TPC covers are supplied by the same manifold.

\begin{figure}[t]
\centering
\includegraphics[width=0.96\linewidth]{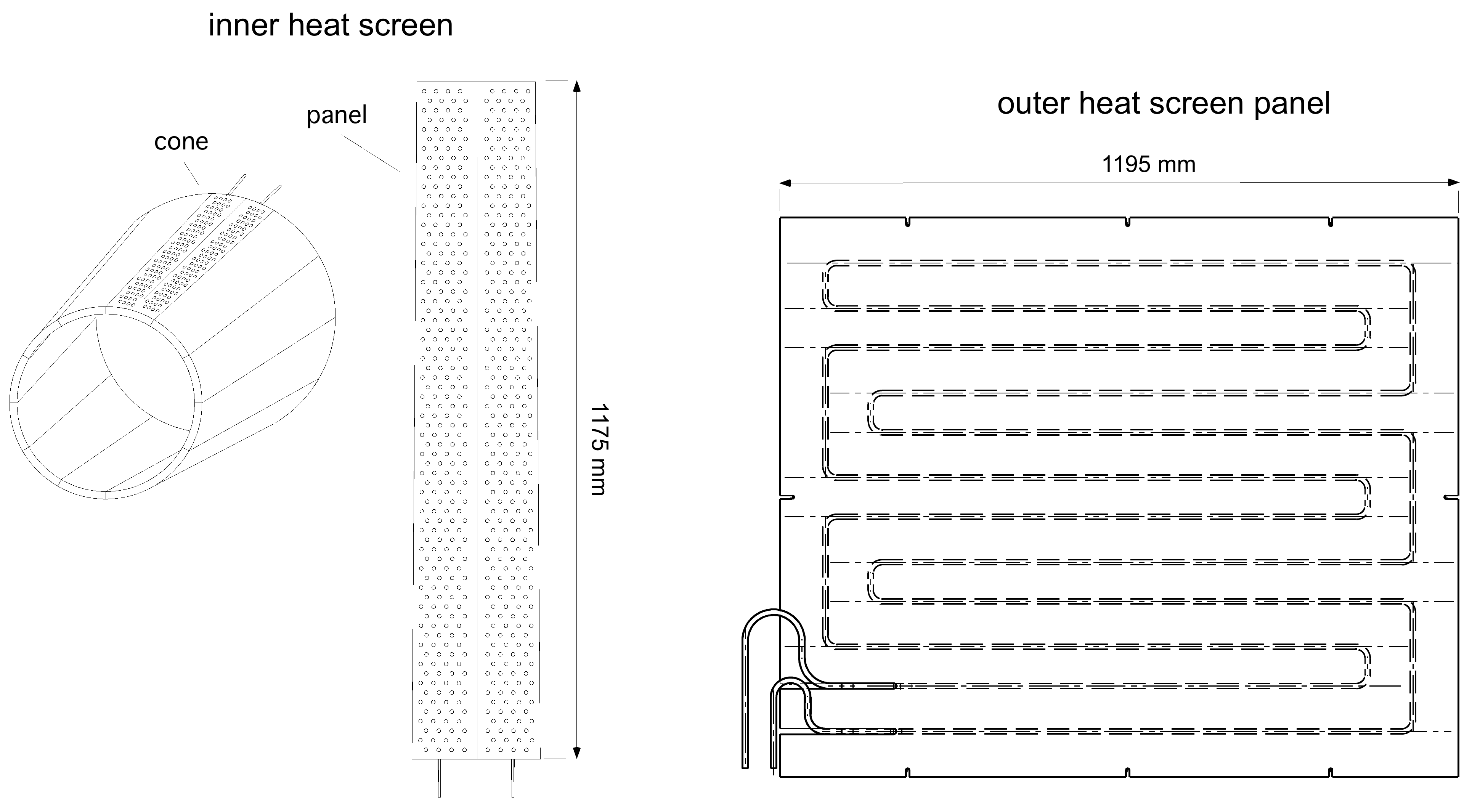}
\caption{Inner (left) and outer (right) heat screen panel.}
\label{cool:heat_screen}
\end{figure}

\paragraph{Thermal screens}
\label{cool:Screens}
The TPC vessel is, to a certain extent, shielded  against outside heat sources or temperature variations by thermal screens. Figure~\ref{cool:heat_screen} (right) shows an example of one of the outer heat-screen panels, which is located between TPC and TRD.  The panels are installed into the space frame structure, which also defines the size of an individual panel. The outer heat screen design follows the space frame structure, i.e.\  18 `super-panels' in $\varphi$ are subdivided into 5 basic panels in z-direction. A design restriction for the panel was the required short radiation length , which led to the choice of 0.5 mm thick Al-sheets and Al-tubes of 10~mm outer and 6~mm inner diameter, respectively. The corresponding average radiation length X/X$_0$ is $\approx 1.2 \%$. The upper nine super-panels are supplied by 6 loops (2 loops per 3 sectors), the lower panels are supplied by 3 loops (1 loop per 3 sectors).

The inner thermal screen shields against the heat of the ITS services. It does not cover the central drum of the TPC and is hence outside of the acceptance of the TRD. On each side of the TPC the screen consists of 12 double-wall stainless steel panels. An example of a panel is show in Fig.~\ref{cool:heat_screen} (left). 

A third thermal screen is defined by the IROC and OROC covers: each readout chamber is covered by a  1 mm thick copper sheet, onto which $3 \times 1$~mm cooling coils are soldered. The covers are fixed to the service support wheel.

\begin{figure}
\centering
\includegraphics[width=0.96\linewidth]{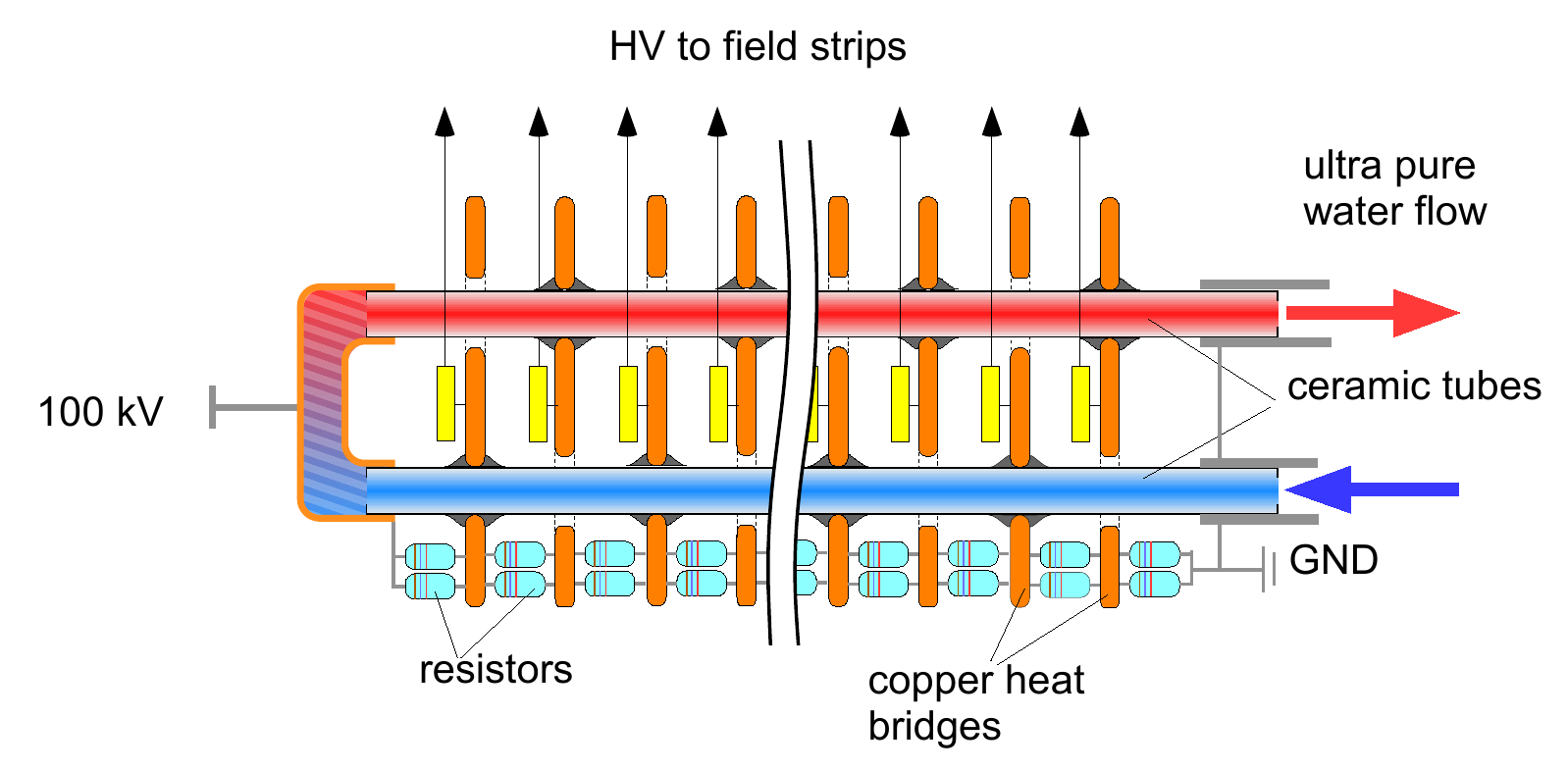}
\caption{Schematic overview of a resistor-rod cooling loop.}
\label{cool:resistor_rod}
\end{figure}

\paragraph{Resistor rod cooling}
\label{cool:Rod}
Though not large in quantity ($4 \times 8$~ W), the heat produced by the field-cage resistor chains is dissipated directly into the fiducial volume of the TPC. To avoid this direct heating, a cooling scheme for the resistor rods has been introduced. A schematic drawing is shown in Fig.~\ref{cool:resistor_rod}. The cooling of the 166 pairs of 15~M$\Omega$ resistors is provided by water flowing through ceramic tubes of 9~mm outer and 3~mm inner diameter, respectively. Copper plates serve both as heat bridges to transfer the heat  from the resistors to the ceramic tubes and as terminal for the field strips. To avoid electrical breakthrough between the high-voltage terminal connecting to the central membrane and the cooling water, the water column is at 100 kV and ground potential at the two ends of the ceramic tube, respectively. This requires ultra-pure water with a resistivity close to $\unit{18}{\mega\ohm\metre}$ in order to keep the current through the water around $\unit{3}{\micro\ampere}$. This value is sufficiently small compared to the resistor rod current of nominally $\unit{80.05}{\micro\ampere}$ at $\unit{100}{\kilo\volt}$.

The water flow through each of the four resistor-rod cooling tubes is only about $\unit{0.5}{\litre\per min}$,  which, however,  corresponds to an almost 15-fold volume exchange per minute of the ceramic cooling tubes. This relatively small flow requires that the temperature of the cooling water is defined precisely at the TPC: owing to the long distance from the cooling plant to the TPC the water with this small amount of flow is likely to pick up the ambient temperature on its way to the TPC, which might be variable and different from the desired gas operating temperature. We have therefore installed four heat exchangers at the input of each of the resistor rod cooling tubes. The primary heat-exchanger circuit is supplied with a high flow ($>\unit{10}{\litre\per min}$) and is thus largely insensitive to ambient temperature changes.

\subsection{Cooling strategy}
\label{cool:Strategy}

The stringent requirement on the TPC temperature stability and homogeneity necessitate an elaborate cooling approach.  The strategy to stabilize the TPC temperature was validated experimentally in a small test setup and is described in Refs.~\cite{cool:Popescu2005, Frankenfeld2005_cooling}. Basically, it was demonstrated that thermal neutrality of the FECs can be achieved by `undercooling', i.e.~the cooling water is injected into the FEC cooling loops several degrees below the desired TPC operating gas temperature. Furthermore, it was shown that the body cooling loops are required to establish stable equilibrium values of the gas operating temperature.  

\subsection{Commissioning of the cooling system}
\label{cool:Commissioning}

\subsubsection{Test with mock-up sectors}
\label{cool:Dummy_Sectors}

\begin{figure}[t]
\centering
\includegraphics[width=0.96\linewidth]{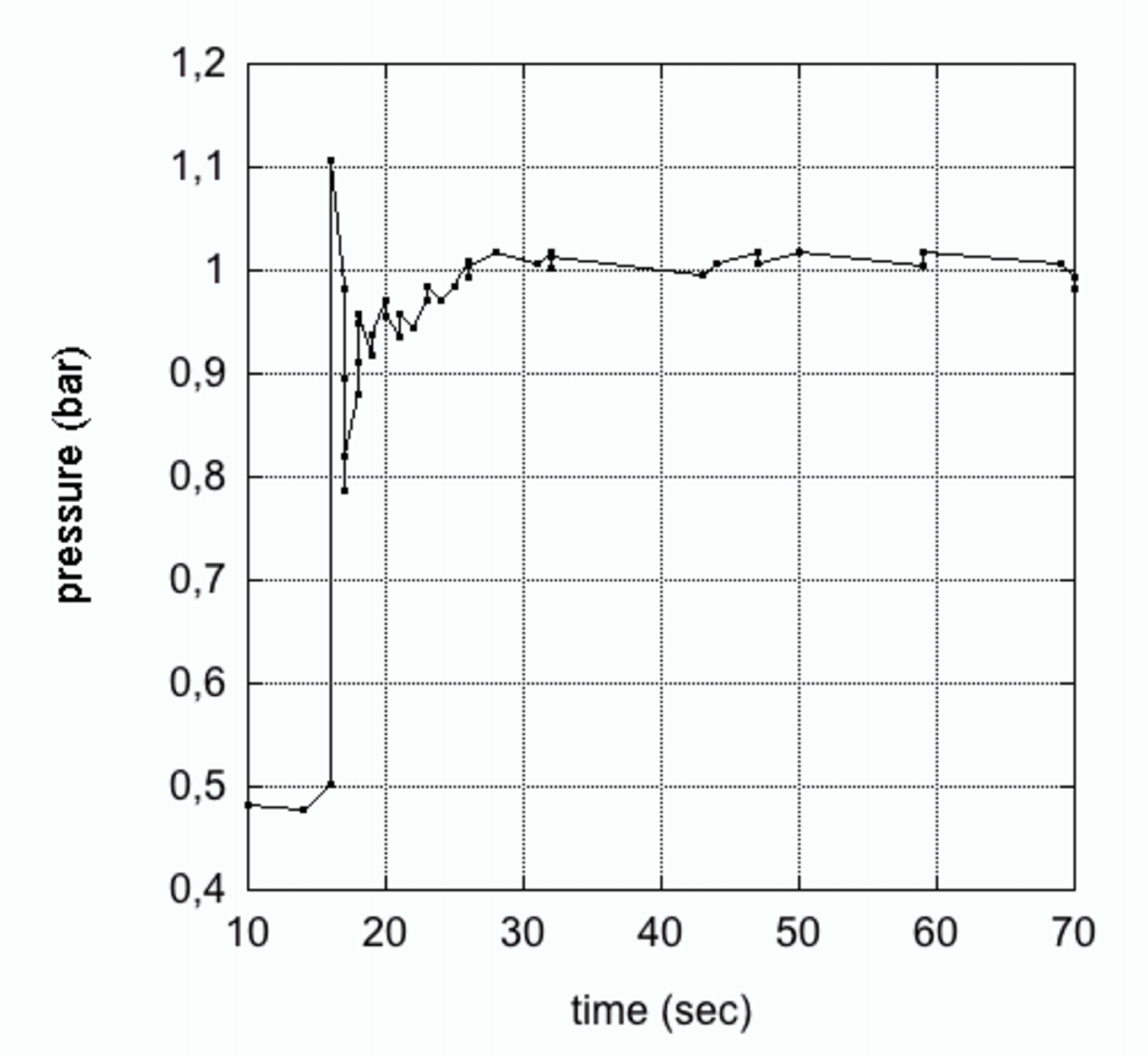}
\caption{Pressure as function of time after a forced leak at t=15~s in one sector. The occurrence of the leak results in a short, immediate pressure spike at which, however, no water is spilled. This appears only after about 10~s, when the pressure raises permanently above atmospheric. }
\label{cool:cooling_commissioning}
\end{figure}

The reaction of the leakless plant to the sudden appearance of a major leak, e.g. the inadvertent removal of a silicon hose with water circulation on, has been tested employing specially made mock-up sectors. The two mock-up sectors used have approximately the same properties as a real sector in terms of silicon-hose tubing, i.e.\ the pressure distribution of the water is similar to a real sector. The most important result of these tests with mock-up sectors is summarized in Fig.~\ref{cool:cooling_commissioning}: on incidence of a large leak (e.g., a silicon hose open) at $t=\unit{15}{\second}$ the system is able to keep the detector cooling sub-atmospheric only for a limited time ($<\unit{15}{\second}$). Thereafter the pressure, measured at the inlet of the detector, rises above atmospheric pressure and water spills out. The reason the system goes beyond $\unit{1\,000}{\milli\bbar}$ is due to the fact that air bubbles sucked into the system clog the return line, increasing the resistance for the water flow. Therefore, we have installed pressure sensors at all inlets. The data from the pressure sensors are fed to a dedicated PLC, which, on detecting a  pressure value higher than a preset value, sends a signal to the PLC of the cooling plant upon which the plant circulations is stopped, water is collected in the tank and underpressure is re-established in all loops.

\subsubsection{Startup procedures and operation}
\label{cool:Startup}

At the very beginning of the cooling-system commissioning it was found that water circulation in many loops could not be established without violating the paradigm of the sub-atmospheric cooling, i.e.\ that the pressure in the detector should not exceed the atmospheric value. The reason for this has been traced back to the routing of both the supply and return lines which were laid out with siphons. In a tube with an air-water mixture, as it exists during the startup of the plant when the tubes are initially air-filled, any siphon reduces the over- or underpressure relative to atmospheric. Hence, in the presence of many or large siphons the limited pressure range of an under-pressure system is not sufficient to start the circulation, i.e.\ to overcome the initial hydrostatic pressure imbalance. Only a careful rerouting of the supply and return lines avoiding siphons a much as possible finally allowed startup of the circulation in all loops. In addition, many of the loops reach circulation only if the pressure in the tank is lowered to at least $\unit{350}{\milli\bbar}$.

While the startup requires a tank pressure as low as possible, once a steady circulation has been established the pressure in the tank can be raised to $\unit{550}{\milli\bbar}$ or higher. The necessity to raise the tank pressure to a level of at least $\unit{550}{\milli\bbar}$ is discussed below in Sec.~\ref{cool:Cavitation}. During the commissioning runs in 2008/9 the cooling systems for resistor rods and FECs have been in operation over an extended period (several month) without problems. 

\subsubsection{Cavitation problem}
\label{cool:Cavitation}

As already mentioned above, a low value of the reservoir pressure ($\unit{350}{\milli\bbar}$), together with a large negative hydrostatic pressure difference ($\unit{850}{\milli\bbar}$ for the highest sectors) in the return line, gives, by \emph{calculation}, negative pressure values which produce instabilities. Physically, this leads to turbulences in the return line and to a phenomenon called cavitation. Cavitation is defined as the phenomenon of formation of vapor bubbles of a flowing liquid in a region where the pressure of the liquid falls below its vapor pressure ($\unit{23}{\milli\bbar}$ at $\unit{22}{\degreecelsius}$).  The collapsing bubbles produce shock waves, which might be strong enough to entail significant damage. In fact, at low-pressure operation a `knocking' noise had been observed in the return lines. To avoid possible damage to the tubing the plant has to be operated at tank pressures above $\unit{550}{\milli\bbar}$, a value at which the dynamic effects seem to be reduced and audible signs of cavitation cease. To be able to switch quickly from low- to high-pressure values of the tank a `back-pressure' valve has been installed at the inlet of the reservoir which regulates the pressure in the return lines. 

For the resistor-rod cooling plant, cavitation constitutes a particularly severe problem owing to the ceramic cooling tubes which are inside the TPC gas volume. Individually-adjusted  restrictions, introduced in the return lines, raise the pressure  to a level above the cavitation threat. To control the pressure distribution in the ceramic tubes, pressure sensors were installed both at the inlet and outlet of the resistor-rod cooling tubes.

\subsection{Temperature monitoring system}
\label{cool:Monitoring}

To monitor the temperature distribution of the TPC, 496 PT1000 sensors are mounted both inside and outside of the gas volume of the TPC. In addition to sensors covering the outside of the Inner and Outer Field Cage containment vessels, several sensors are mounted onto each IROC and OROC. For each sector, sensors measure the cooling water inlet and outlet temperature. Several sensors ($2 \times 18$) are attached to a circular skirt inside the gas volume.  Additional temperature sensors on the front-end electronic cards (one sensor for each of the 4356 FECs) complete the monitoring system. A comprehensive description of the temperature monitoring system including its calibration and its readout system is given in \cite{Frankenfeld2005_temp_system}.

\subsubsection{Temperature profile and homogenization}
\label{cool:profile}

\begin{figure}[t]
\centering
\includegraphics[width=0.96\linewidth]{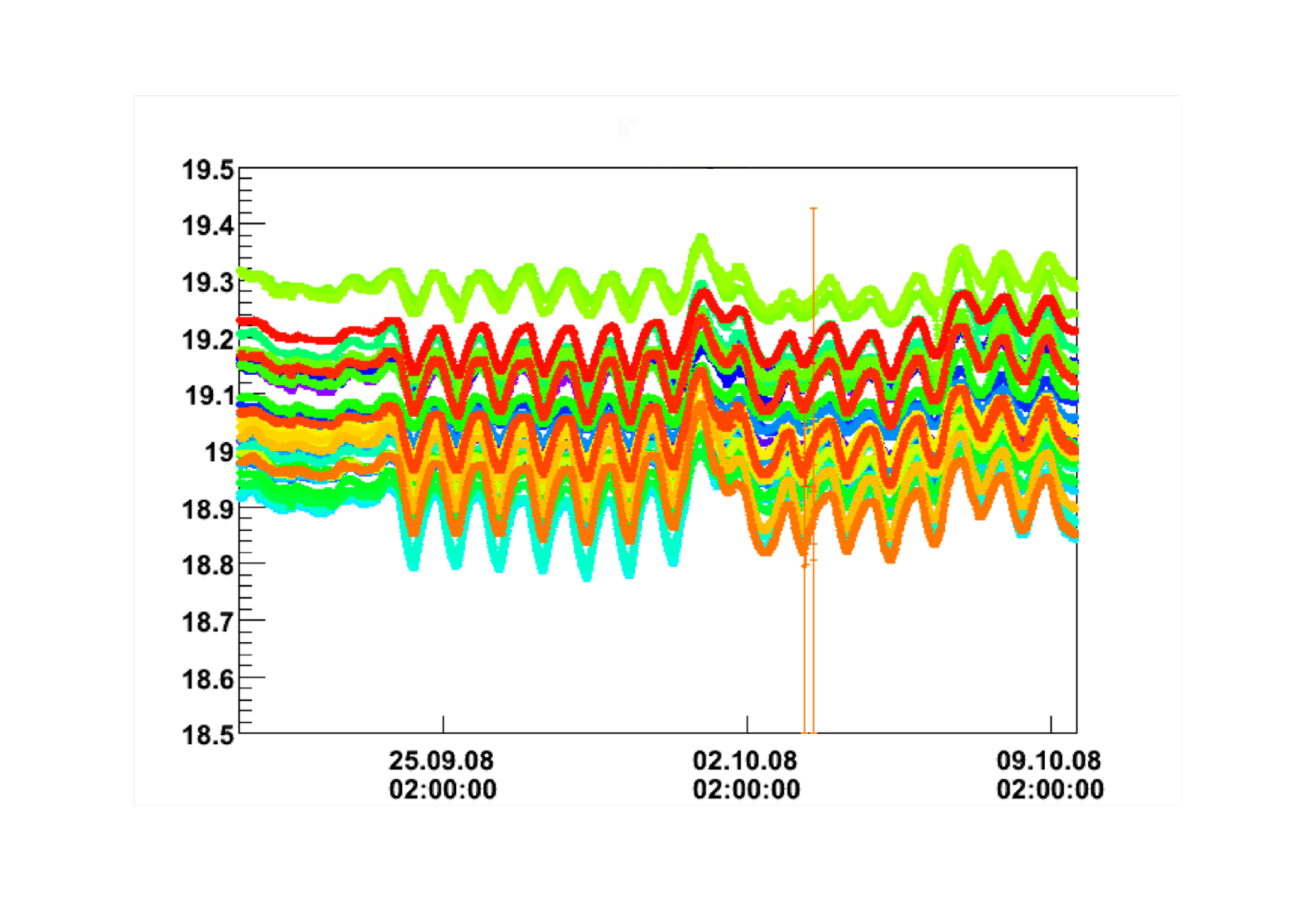}
\caption{Temperatures measured with the skirt PT1000 sensors as a function of time. Each of the curves represents one of the skirt sensors.}
\label{cool:temperature_TOF}
\end{figure}

Figure~\ref{cool:temperature_TOF} shows the temperatures as measured with the skirt sensors, which are located inside the gas volume,  as a function of time. The temperature data were taken during the detector commissioning run in fall 2008. The plot demonstrates the sensitivity of the TPC gas volume to the heat load from other detectors. During the beginning of the measurement period the TOF detector, surrounding the TPC,  had been continuously running. Later it was switched off each night for 8 hrs. The influence on the temperature  inside the TPC is clearly visible by temperature excursions of the order of $\unit{0.2}{\kelvin}$. Even though the TOF detector has been on for $2/3$ of the day no stable temperature (i.e.\ equilibrium) is reached. Measurements have shown that the temperature relaxation times are of the order of 16 hours due to the large mass involved.

\begin{figure}
\centering
\includegraphics[width=0.8\linewidth]{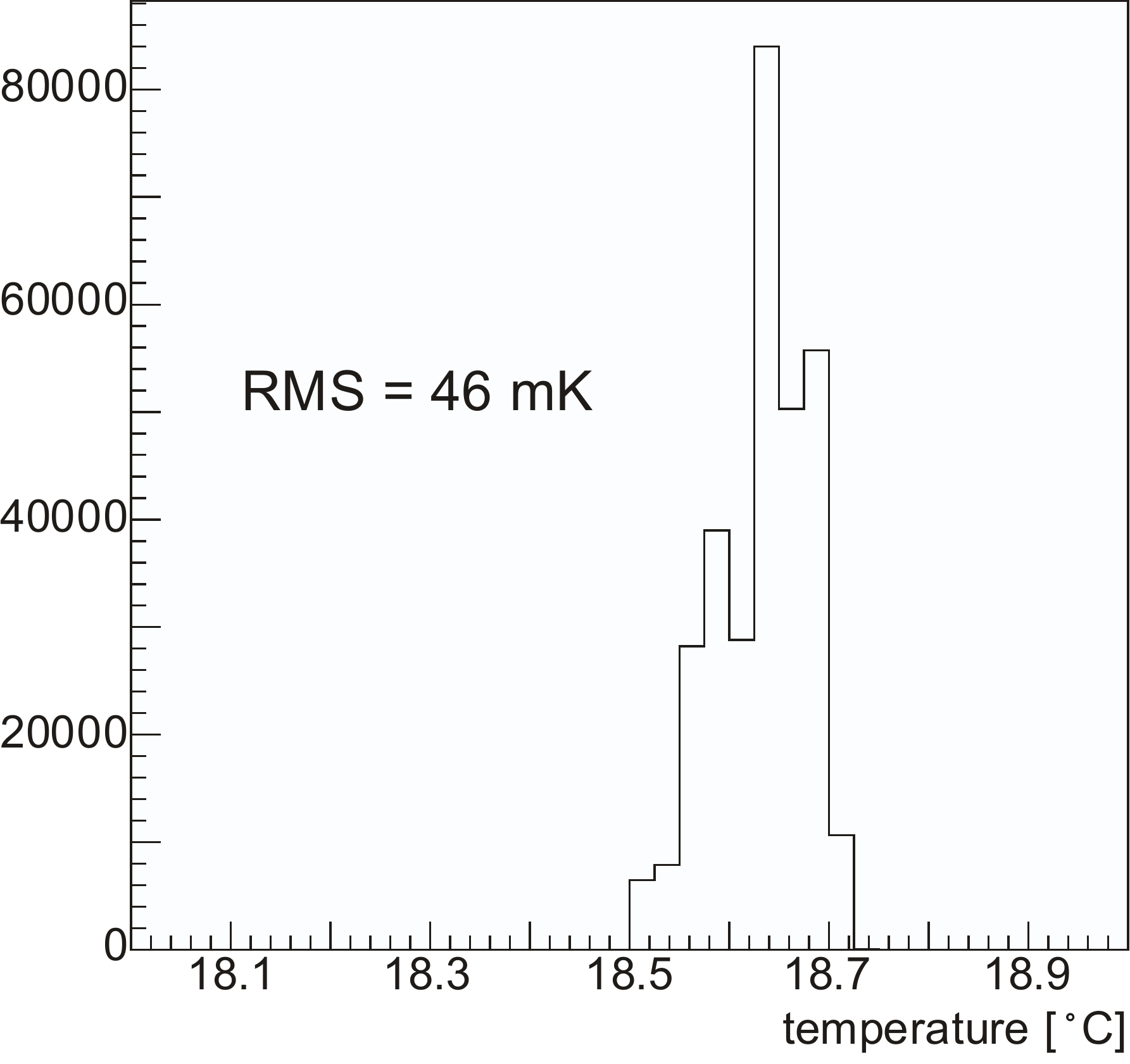}
\caption{Temperatures distribution measured with the skirt PT1000 sensors.}
\label{cool:temperature_skirt_hist}
\end{figure}

The histogram of the skirt temperature sensor distribution is shown in Fig.~\ref{cool:temperature_skirt_hist}. The temperatures were sampled over a period of 24 hrs with stable environmental conditions. The RMS value of the histogram is below $\unit{50}{\milli\kelvin}$ showing that the desired temperature homogenization of the TPC gas volume below $\unit{0.1}{\kelvin}$ is within reach.
It should, however, be noted that the skirt temperature probes represent the $x$-$y$ gradient close to the readout plane. The optimization of the distribution close to the sectors is comparatively straightforward: it involves the proper 'overcooling' of the readout cards and a slight top--bottom difference in the chamber body cooling. The feedback information  from the sensors, both on the readout chamber and on the skirts, is relatively fast, i.e.\ within several hours an equilibrium value is established. The temperature homogenization over the full TPC volume via flow and temperature adjustment of the thermal screens is considerably more involved:  since the sensors are outside of the TPC vessel and are, in addition, insulated from the gas volume by the CO$_2$ volume they thus reveal limited information about the temperature inside the TPC. A better understanding of the gradients over the full TPC volume requires an analysis of straight, laser induced tracks (see Sec.~\ref{laser}).

Above we have described a complex cooling and temperature homogenization system, which involved altogether 60 individual, adjustable cooling circuits.  Conceptually, the circuits have the tasks, to remove heat,  i.e.\ cool the front-end electronics, and to define iso-thermal surfaces around the TPC  in order to provide an as much as possible gradient-free gas volume. 
Overall, we have been able to reduce the temperature gradients inside  the TPC to a RMS $<$ 0.05 $^\circ$C, as it is required for a full exploitation of the TPC internal position resolution.

\section{Gas and gas system}
\label{gas}

The TPC is essentially a 90\,m$^3$ volume filled with gas, where the gas is the detecting medium. The detector performance depends crucially on the gas choice, stability and quality, 
since these influence
the charge transport in the drift volume and the amplification processes in the readout chambers. The choice of the gas composition is constrained by a set of performance requirements and boundary conditions, and in turn the selected gas mixture determines the performance of the detector and conditions various aspects of its design, from the shaping time of the front-end electronics to the temperature uniformity of the gas in the detector. In particular, the gas system is designed to fulfill the requirements derived from the gas choice and the expected performance of the detector.
In the next section we discuss the selection procedure that led to the choice of a Ne--CO$_2$--N$_2$ gas mixture and the implications of this choice. We then describe the gas system that injects, circulates and cleans this gas.

\subsection{Gas choice}
\label{gas:Gas_choice}

\begin{figure}[t]
\centering
\includegraphics[width=\linewidth,clip]{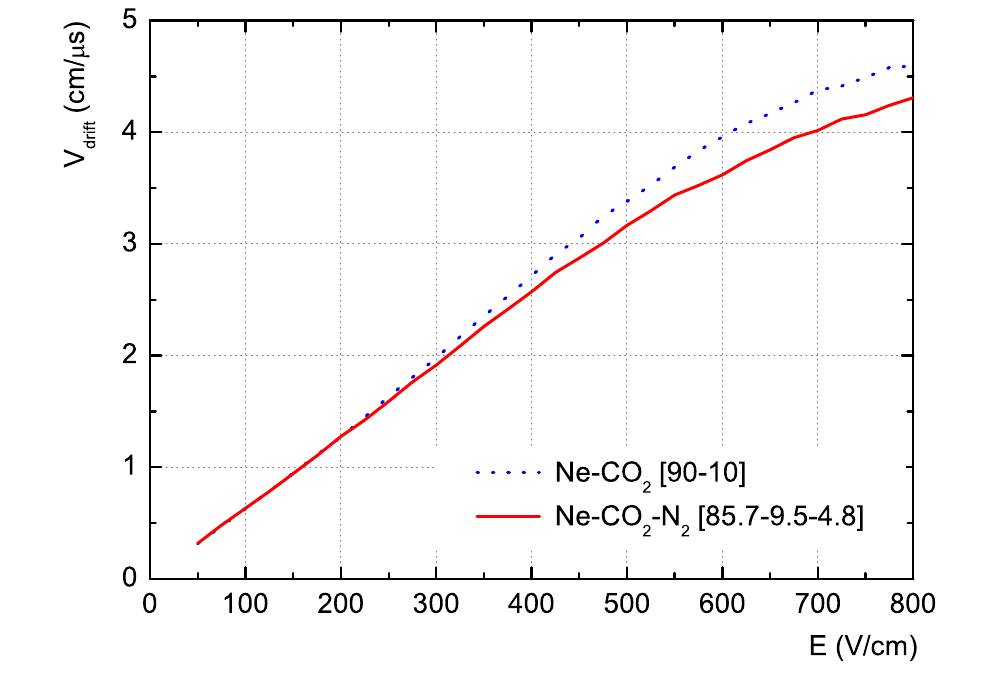} 
\includegraphics[width=\linewidth,clip]{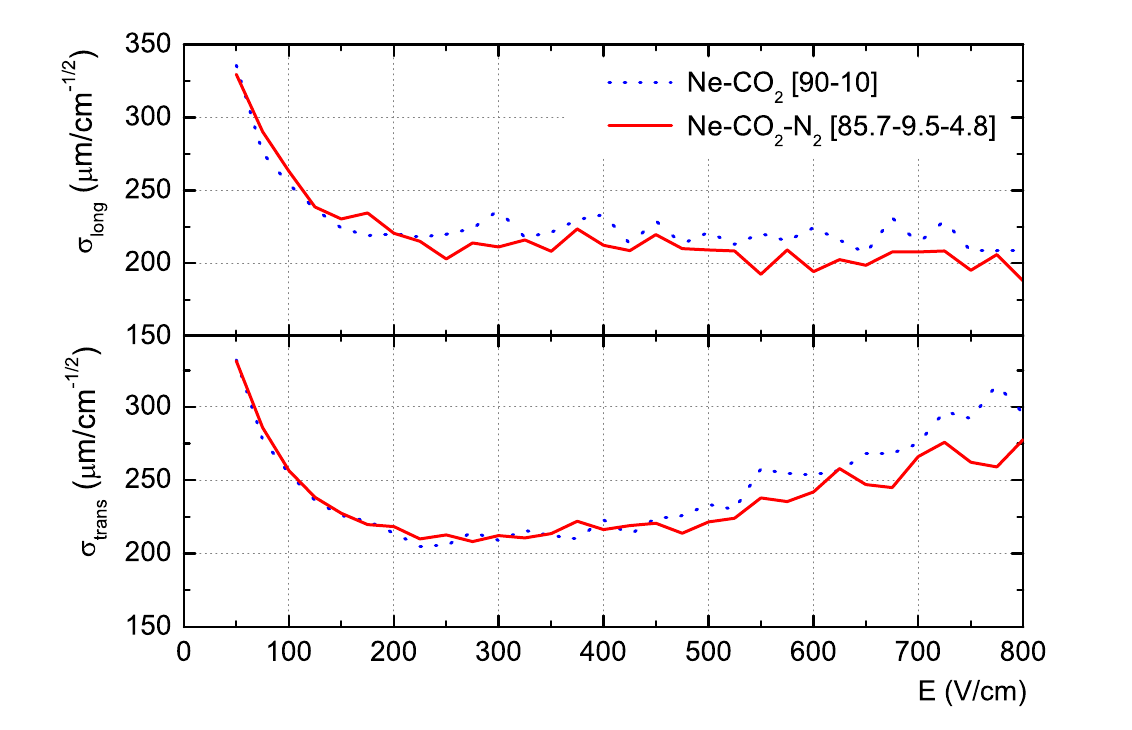} 
\caption{Drift velocity (top) and longitudinal and transverse diffusion coefficients (bottom) as a function of the
electric field for the Ne--CO$_2$ (dashed lines) and the Ne--CO$_2$--N$_2$ (solid lines) mixtures calculated with Magboltz at 750 Torr and $20\,^{\circ}\mathrm{C}$. While the diffusion coefficients do not change with the addition of N$_2$, the drift velocity decreases by 5\% at the nominal field of $\unit{400}{\volt\per\centi\metre}$.}
\label{gas:vd}
\end{figure}

The selection of both the noble gas and the quencher was made by a process of elimination rather than choosing the gas by its merits. Between Ar and Ne, the former, although providing larger primary statistics, was discarded because of material-budget considerations and a slow ion mobility. In particular, Argon would substantially enhance space-charge effects in the drift volume of the TPC due to its relatively slow ion drift velocity. As far as the quencher is concerned, hydrocarbons were excluded due to aging considerations (Malter currents would set in after about 1 year of operation with heavy-ion beams). CF$_4$, on the other hand, presented many concerns with material compatibility at the time of the system design. Therefore, CO$_2$ was chosen. Since the maximum drift field in the field cage was designed to be $\unit{400}{\volt\per\centi\metre}$, and the maximum drift time $\unit{94}{\micro\second}$, the composition results in 10\% CO$_2$ in Ne~\cite{veenhof2003}. During the prototyping phase, a further 5\% of N$_2$ was added to the mixture~\cite{Garabatos2004}. This addition reduces the drift velocity at the nominal field by about 5\%, but it provides a more stable operation of the readout chambers at high gain. This also eliminates the problem that N$_2$ could build up in the gas due to small leaks. The changing nitrogen content would then affect the detector performance since it cannot be removed by the cleaning agents of the gas system. Excited states of neon have energies around $\unit{17}{e\volt}$, for which the quenching capabilities of CO$_2$ are poor. More CO$_2$ in the mixture would rapidly decrease the drift velocity. N$_2$, on the other hand, presents a slightly higher ionization cross section at these energies and so helps to quench the Ne, and affects the drift velocity modestly. The resulting mixture is also less sensitive to the exact composition. The slow proton production due to neutron bombardment of  N$_2$  molecules has been shown in simulations to be reasonably small. The main implication of the choice of a Ne--CO$_2$--N$_2$ gas mixture for the gas system is the necessity of monitoring and controlling a ternary mixture.
Figure~\ref{gas:vd} shows the drift velocity and the longitudinal and transverse diffusion coefficients calculated with Magboltz~\cite{magboltzNIM,magboltzCERN}, as a function of the electric field for the mixtures with and without N$_2$. In both cases, the drift velocity is not saturated at the nominal field, thus making it very sensitive to gas density fluctuations and to the exact electric field. 

\subsubsection{Implications of the gas choice}
\label{gas:Implications}
\begin{figure}[t]
\centering
\includegraphics[width=\linewidth,clip]{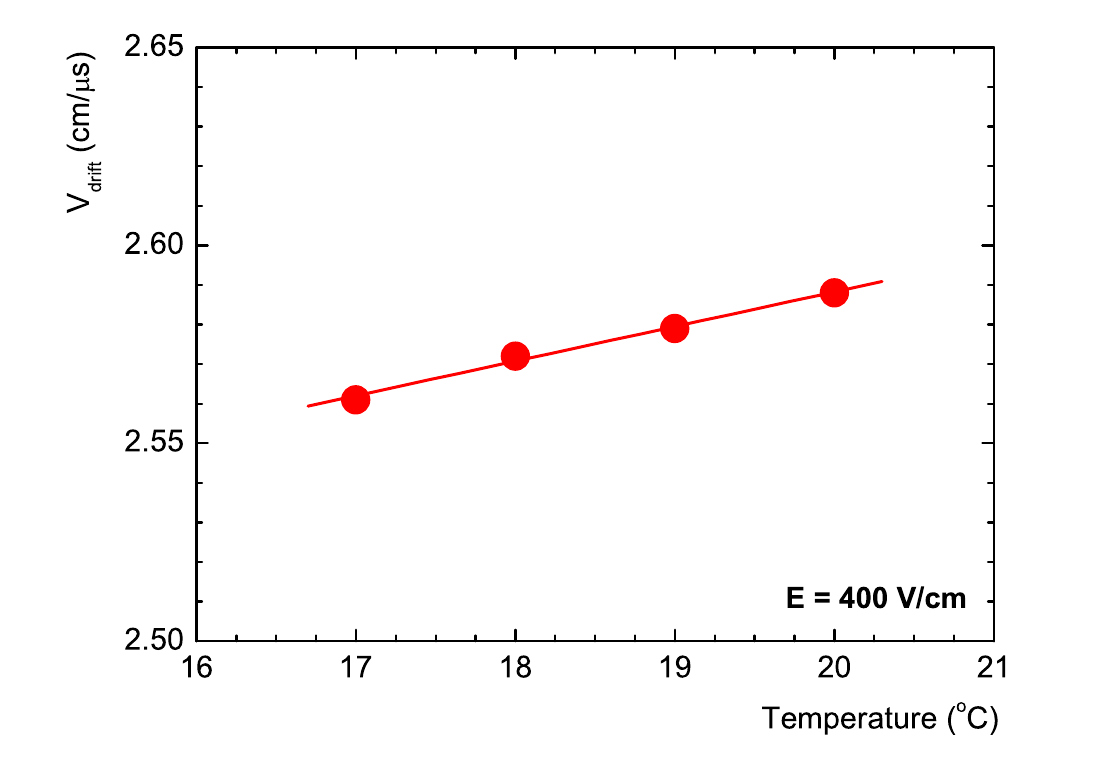} 
\caption{Dependence of the drift velocity of the ternary mixture on the temperature as calculated with the Magboltz simulation package at $\unit{400}{\volt\per\centi\metre}$ and 750 Torr.}
\label{gas:vd_T}
\end{figure}

Figure~\ref{gas:vd_T} shows the drift velocity dependence on temperature at $\unit{400}{\volt\per\centi\metre}$. Since the required position resolution is of order 200\,$\micro$m, the necessary temperature uniformity in the drift volume is 0.1\,K, a stringent requirement which drives the cooling strategy (see Sec.~\ref{cool}). The fluctuations due to ambient pressure variations, which are followed by the pressure regulation of the gas system, are corrected for.

In addition, the drift velocity changes by $-6.4\%$ per \% change in CO$_2$ concentration and by $-1\%$ per \% in N$_2$, while the gain dependence is 15\% per \% change in CO$_2$ and 6\% per \% change in N$_2$. To keep the drift velocity constant at the 10$^{-4}$ level it would be necessary to control the CO$_2$ concentration to better than 0.01\%, which is beyond the precision of current mass-flow controllers. Therefore, in addition to the laser system (see Sec.~\ref{laser}), the gas system is equipped with diagnostic tools to measure the gas composition: a gas chromatograph with a thermal conductivity detector and a high precision drift velocity monitor.
  
The attachment coefficient of electrons to O$_2$ is greatly enhanced in the presence of CO$_2$, reaching 
400 bar$^{-2}\micro$s$^{-1}$. 
This is because negatively ionized excited oxygen molecules rapidly decay into the ground ionic state by energy-transfer collisions with CO$_2$ molecules. An electron drifting over $\unit{2.5}{\metre}$ in this gas, contaminated with 5\,ppm O$_2$, has a 25\% chance to get attached. Therefore, besides the tightness of the detector itself, the gas system and pipe work must be certified to be leak free, and provisions to remove oxygen from the gas are necessary. This issue, combined with the large volume of the system, also determines the total gas flow through the detector. A reasonable choice is to flush the detector volume at a rate close to 5 times a day. The oxygen contamination is monitored by constantly sampling the gas with an oxygen analyser. The water content and the CO$_2$ concentration are also analysed with appropriate sensors in the same chain.

Finally, since neon is a high cost gas, a CO$_2$ absorber system is implemented in the gas system, so that the filling of the detector with the mixture can be done at no waste of the main gas, as explained below.

\subsection{Description of the gas system}
\label{gas:Description}
Most of the gas systems and control programs of the LHC experiments are designed and built under a common modular scheme coordinated by CERN infrastructure groups. Each function of the gas system, such as the gas mixing, the circulation, the cleaning, the analysis, etc., is integrated into a logical module which usually corresponds to one or more racks. A Programmable Logic Controller (PLC) runs the system by executing actions, like opening a valve, or a sequence of actions, regulating devices according to set points, like flows or pressures, reacting to alarms, reading analog and digital values, including Profibus networks, and publishing information to a user interface. The user interface~\cite{Thomas05} allows the operator to change the state of the gas system, for example from fill to run states, and to act on individual modules or on single components of the system. It also logs the various data points into a data base for inspection of trends. A set of recipes, or configuration files, specific for each module, is loaded into the PLC for determination of set points, limits, regulation parameters, timers, etc., which the operator can edit and reload at any time from the user interface.

The racks are distributed in a surface building, in the plug of the shaft (a shielded platform just above the cavern), and in the cavern, according to their functions and specifications.

\subsubsection{Configuration}
\label{gas:Configuration}

\begin{figure}
\begin{center}
\includegraphics[width=\linewidth,angle=-90,clip]{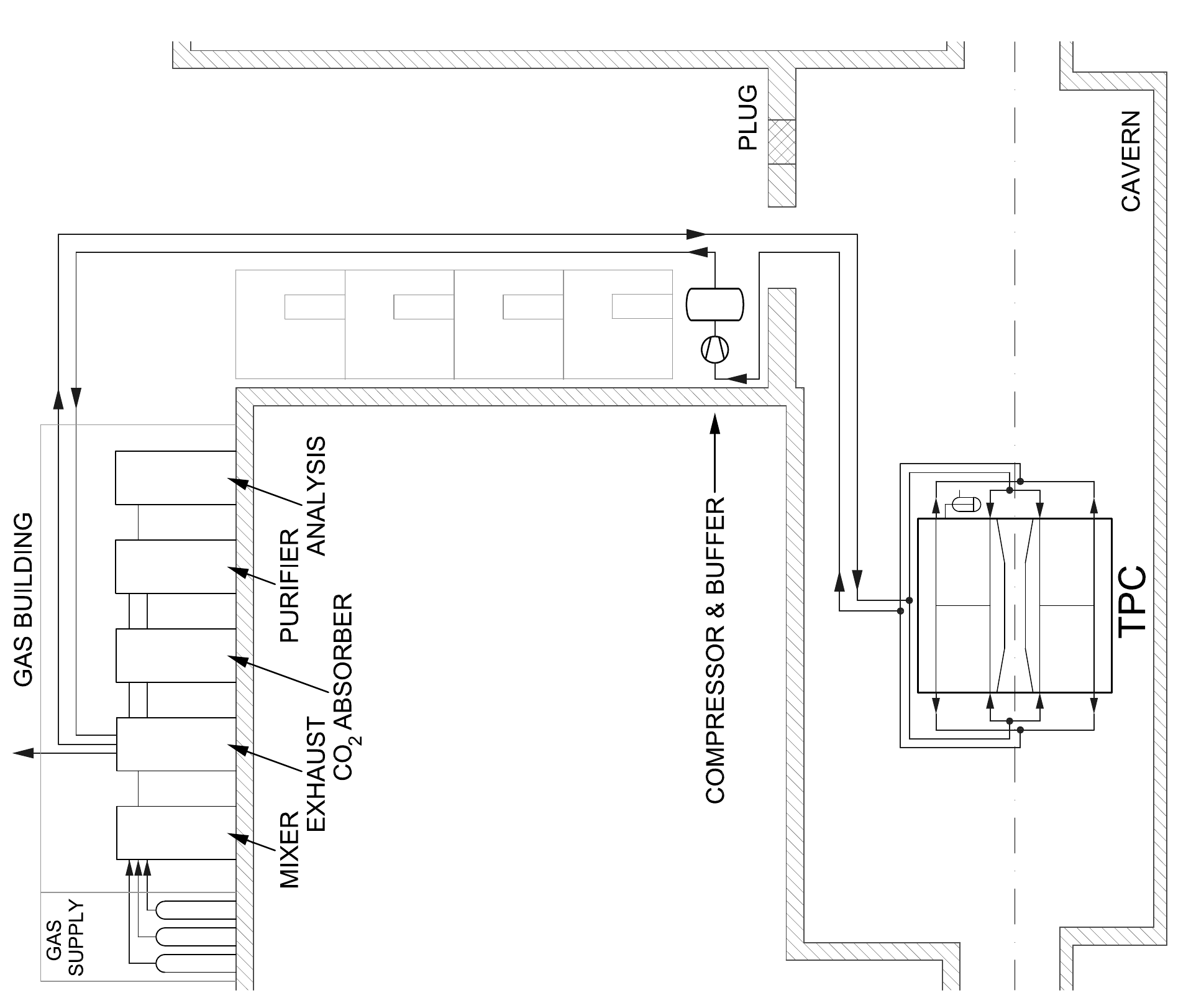} 
\caption{Schematic view of the distribution of the various modules of the TPC gas system on the surface,
on the plug (a platform in the shaft of the experiment, just above the cavern) and in the cavern.}
\label{gas:layout}
\end{center}
\end{figure}

\begin{figure}
\begin{center}
\includegraphics[width=\linewidth,clip]{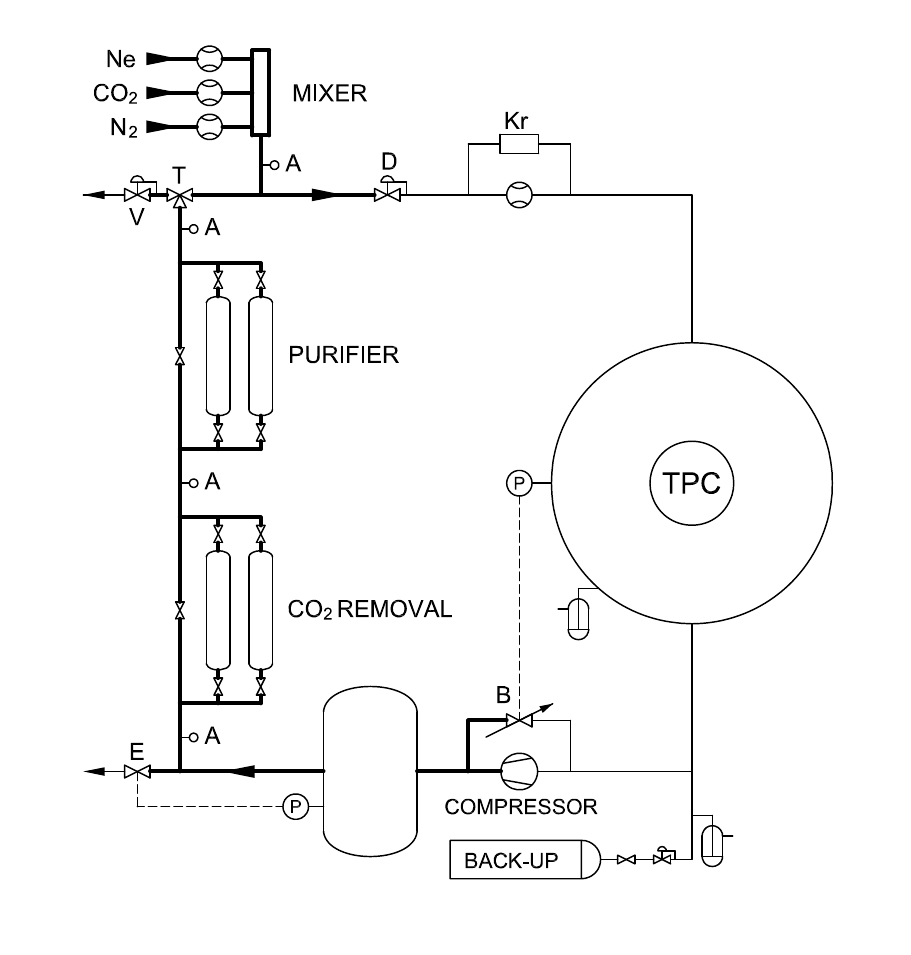} 
\caption{Simplified diagram of the gas system loop, showing its main modules and some of the main elements used for operating and regulating the system in various modes. Bold lines indicate gas pipes that operate at high pressure (about $\unit{3}{\bbar}$).}
\label{gas:flow}
\end{center}
\end{figure}

The gas system is a gas circulation loop with injection, distribution, regulation and other tasks distributed in modules located at different elevations in the ALICE hall (see Fig.~\ref{gas:layout}). The functionality of the loop is schematically depicted in Fig.~\ref{gas:flow}. Gas is circulated through the detector by a compressor module, which extracts the gas from the TPC and fills a high-pressure buffer volume for gas storage. A regulated bypass proportional valve (B in the figure) re-injects part of the compressed gas back into the compressor inlet, as part of a feedback loop to regulate the pressure inside the TPC. This valve is driven by a pressure sensor installed at the detector. which determines the operating pressure set point with respect to atmosphere. The TPC overpressure is thus regulated to $\unit{0.4}{\milli\bbar}$. The high-pressure buffer, a $\unit{1}{\metre\cubed}$ tank at $\unit{2-4}{\bbar}$ overpressure, stores gas that can be delivered to the detector in case of an increase of the atmospheric pressure, or can accept gas from the detector when the ambient pressure decreases. 

Once at a few bars overpressure, the gas, depending on the system mode, flows through several modules, located on the surface, or is directly vented out through a back-pressure regulator (V) if the detector is being flushed with CO$_2$ (purge mode). The purifier, CO$_2$ absorber, the exhaust, the analysis and the mixer modules are installed on the surface. The mixer can be programmed to inject a fixed amount of fresh gas into the loop. Part of the excess gas in the system is exhausted through the analysis lines (A). The remainder of the excess gas in the loop is exhausted through a mass-flow controller (E) regulated against the pressure set-point in the high-pressure buffer. In run mode, the Cu-catalyzer purifier is activated, so that one of its two cartridges continuously removes O$_2$ and H$_2$O from the gas. These cartridges can be regenerated on the spot by a programmed sequence of purging them with hydrogenated gas and heating them.

After the gas injection from the mixer into the gas loop, the pressure is decreased to a fraction of a bar by a high-precision needle valve (D). The resulting pressure will determine the total gas flow (from surface to cavern) through the detector, whose pressure is fixed by the compressor module. This configuration allow for independent settings of the detector pressure and the gas flow, a feature of closed-loop gas systems.


\subsubsection{On-detector distribution}
\label{gas:Distribution}
A final module, the distribution module, is located in the cavern. Here the detector inlet and outlet are distributed and collected. Since the gas is distributed into the drift volume through the strip-holding rods, the inlet and outlet gas pipes have to reach both endplates. For mechanical and access reasons, each gas pipe is split into two in order to service the top and bottom halves of each endplate. Finally, a set of 8 half-circular manifolds is installed at the inner and outer rims of the service support wheels to service the rods through flexible bellows. The diameter and length of these bellows are adjusted to tune the fraction of gas flow for each rod, such that special rods like the HV cable rod, the laser mirror rods and the resistor rods receive a limited flow. In this manner, the gas flows radially from the inner to the outer field cage vessels, and the flow is uniform in the $z$ direction. Therefore, there is no force on the central electrode.

The distribution module also holds a single-pass gas system to supply CO$_2$ to the two insulating volumes surrounding the field cage. The flow in this case is in the $z$ direction, and the gas is exhausted through an extraction system coupled to bubblers installed in the distribution rack. 

\subsubsection{Filling}
\label{gas:Filling}
Before injecting neon, the detector is flushed with CO$_2$ and the return gas is exhausted through the V back-pressure regulator in the high-pressure area (purge mode). Once the air contamination is at the trace level, filling with Ne starts. In this fill mode, the gas is made to recirculate in the loop by switching the three-way valve T (see Fig.~\ref{gas:flow}). Pure neon is injected from the mixer as cartridges filled with molecular sieve alternatively trap CO$_2$ and remove it from the system. The CO$_2$ cartridges are regenerated after saturation, cooled down and put back into service when needed. A thermal conductivity CO$_2$ analyzer and a gas chromatograph are used to measure the composition. The last step is to inject a fixed amount of N$_2$ in order to establish the final composition.

\subsubsection{Running}
\label{gas:Running}
Once the final mixture has been blended, the system is switched to run mode. The high pressure is regulated within precise limits according to the ambient pressure fluctuations. A small amount of fresh gas ($\unit{40}{\litre\per\hour}$) is continuously injected while the excess gas is vented out.

The purifier is activated during normal operation in order to clean residual water and oxygen from the gas. After the gas is decompressed at the surface, a bypass in the gas stream may be used to introduce a Rubidium source, which releases gaseous $^83$Kr isotopes into the gas for detector calibration purposes.

\subsubsection{Back-up system}
\label{gas:Back-up}
 A set of alarms ensures that the system does not produce conditions which are dangerous for the detector integrity, in which case the system stops, i.e. the compressor stops and the high pressures are isolated from the detector via pneumatic valves.
The ultimate safety of the detector for excessive over- or under-pressures is a paraphine oil-bubbler directly connected at the endplate of the TPC. This bubbler has been designed to produce small bubbles and leaves ample space for gas to be exhausted or air to be injected, depending on the pressure conditions.
If the system was stopped for a long time, inevitably air would enter the detector through the safety bubbler when the atmospheric pressure increases. In order to avoid this situation, a back-up supply of gas is put in place as shown in Fig.~\ref{gas:flow}. Premixed gas flows through a line connected to the TPC as soon as the power fails or the system goes to stop mode. This line is equipped with a pressure regulator tuned such that if the pressure in the detector decreases, the differential pressure that the regulator sees increases and therefore opens to the gas stream. The back-up system is located on the platform just above the cavern (i.e. on the plug).

\subsubsection{Analysis}
\label{gas:Analysis}
An analysis module at the surface is connected to various high-pressure points of the system (points A in Fig.~\ref{gas:flow}) through a pressure reducer. The gas to be sampled is extracted from the high-pressure buffer, clean gas from a point after the purifier module, and fresh gas from the mixer outlet.
The analysis module has two single-pass lines (i.e., the outlet gas is vented out). One of them contains the CO$_2$, the H$_2$O, and the O$_2$ analyzers, which operate at a flow rate of a few $\litre\per\hour$. The user interface allows the operator to define which lines to sample and for how long, in a periodic way. The second line contains a drift velocity monitor which also provides a means of calculating the gas composition by measuring, in addition, a gas amplification factor.

\begin{figure}
\centering
\includegraphics[width=\linewidth,clip]{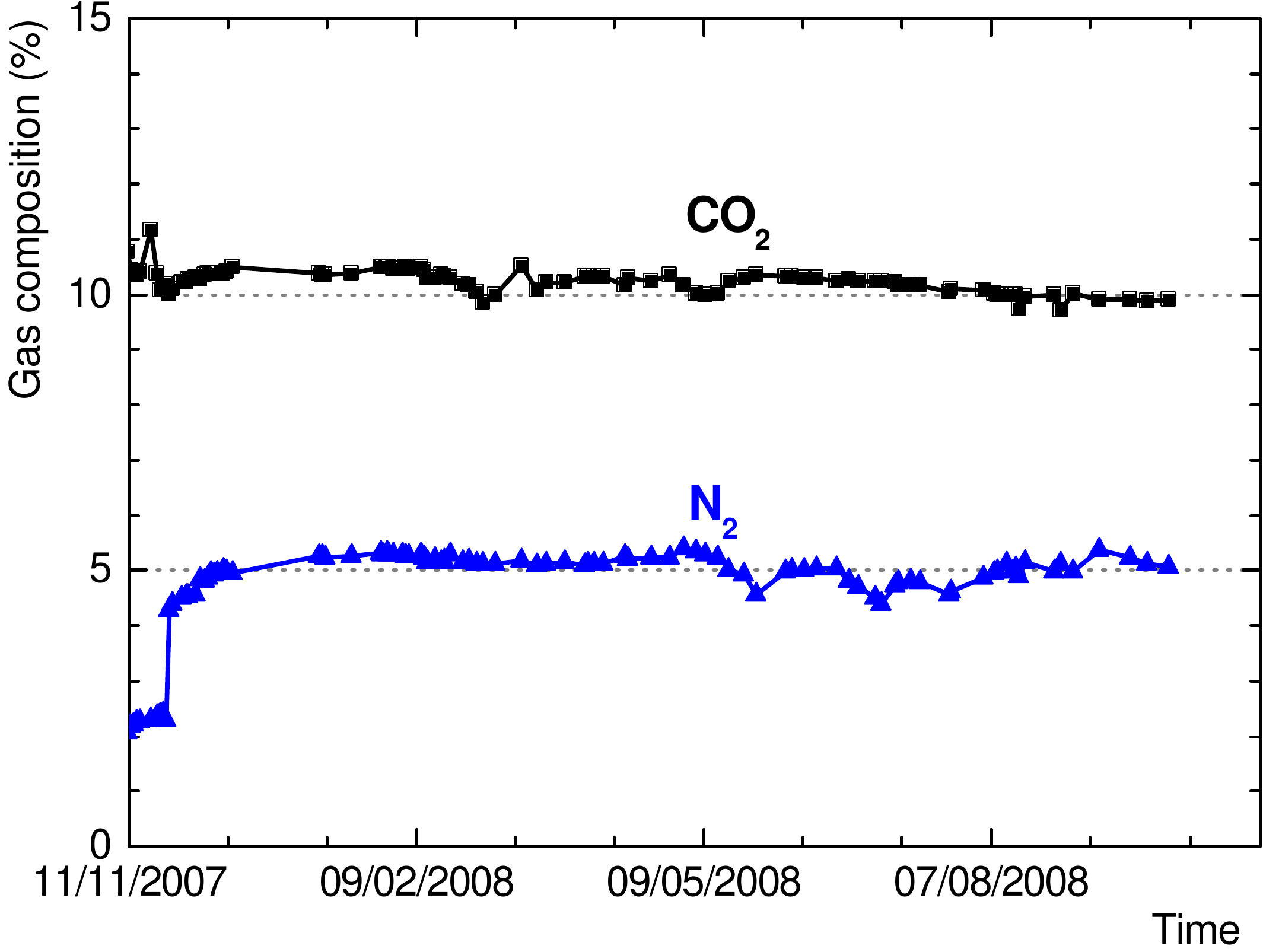}
\caption{CO$_2$ and N$_2$ contents in the TPC during 2008, as measured with a gas chromatograph.}
\label{gas:composition}
\end{figure}

Another line from the gas system is connected to a gas chromatograph through a manual valve. The instrument is equipped with a so-called tandem capillary column which allows for separation of the three gases of the gas mixture as well as the rest of gases that compose the air. The detection of the effluents is carried out by a thermal conductivity cell. In this way, the gas composition and the air contamination can be measured. Figure~\ref{gas:composition} shows the CO$_2$ and N$_2$ concentrations as a function of time as measured with the chromatograph over one year.

\section{Laser system}
\label{laser}

Precise reconstruction of particle tracks in the TPC requires a
thorough understanding of the drift velocity and any inhomogeneities
in the drift field. A non-uniform electron drift can be caused by
mechanical or electrical imperfections in the field cage and
readout chambers, whereas deviations of the electron drift from the
ideal paths inside the gas volume are caused by temperature
variations, relative misalignment of the electrical and magnetic
fields ($E\times B$ effects) and local variations of the electric
field from moving charges (space-charge effects).

To calibrate the drift field parameters against a known standard, a
laser calibration system was built, using a large number of narrow
ultraviolet rays at predefined positions inside the drift volume to
generate tracks. The system was designed to make fast and accurate
measurements of time varying drift velocities. It will run every half hour
interspersed between physics events to measure the drift velocity
and assess space charge effects. The laser system was used extensively
during the detector commissioning for testing of the electronics and
the alignment of the readout chambers and central electrode.

Many features of the ALICE laser calibration system follow the
system built and operated by the STAR experiment at
RHIC~\cite{laser:Lebedev:2002sp, laser:STARlaser}.

\subsection{Requirements}
\label{laser:requirements} The goal is to measure distortions in the TPC drift
field with a relative error of $5 \times 10^{-4}$. Narrow beams of pulsed
UV laser light can be used to generate  tracks in the active
volume of the TPC. If the track positions are well known by
construction, they can be used to calibrate the electron drift
velocity. For a comprehensive review of the use of lasers in gaseous
detectors, see \cite{det_laser_calib}.

To measure the drift velocity on a single event basis to the
required precision, the position of the tracks must be known to a
spatial resolution of $\sigma_{r\varphi} \approx 800~\micro$m and
$\sigma_{z} \approx 1000~\micro$m and the individual laser tracks have
comparable transverse dimensions. The stability of their position must be
assured at the same level. The nature of the laser tracks assures
that the tracks are always straight lines.

We use pulsed monochromatic laser beams of 266~nm wavelength ($E =
h\nu =4.66$~eV) and $\approx 5$~ns pulse duration with approximately
Gaussian cross section with $\sigma \approx 400~\micro$m. The
ionization in the gas volume along the laser path occurs via two
photon absorption by organic impurities with ionization potentials
in the range 5--8 eV. The molecules of the pure Ne--CO$_2$--N$_2$
drift gas have ionization potentials above 10~eV and are not ionized
by the laser.

Because the ionization process is mostly a result of gas impurities,
it is difficult to determine the necessary beam intensity a priori.
Experience from this and other experiments show that energy
densities of approximately $20~\micro$J/mm$^2$ for a 5~ns pulse at
266~nm wavelength are sufficient to obtain an ionization
corresponding to several minimum ionizing particles. We designed our
system to have up to $40~\micro$J/mm$^2$ per pulse.

The aim is to measure the response of the TPC to several hundred
laser tracks generated simultaneously throughout the TPC drift
volume at predefined positions. The laser events can be generated
in special calibration runs or interspersed between physics events.
To obtain the best precision of the measured tracks, the preferred geometry is
one where the tracks have constant drift times and are perpendicular
to the wires. For this configuration, clusters are smallest and the
electronics and reconstruction programs give the best possible
single point resolution. Simultaneously, a extensive coverage of the full
drift volume is desired. This led us to provide tracks in planes
at constant $z$, of which some radiate with approximately constant
$\varphi$. Tracks generated at different $z$ throughout the drift
volume allow easy determination of drift velocities from single
laser events.

Most metallic surfaces have work functions below $4.66$~eV and emit
electrons by photoelectric effect when hit by UV light above this
energy. Being a first order effect in the light intensity, a
considerable amount of low energy electrons are seen from the
diffusely scattered, time correlated UV light produced by
reflections. The signal from the aluminum surface of the central
electrode is used to give a precise picture at the maximum drift
time across the electrode.

\subsection{System overview}
\label{laser:overview}

\begin{figure}[t]
\centering
\includegraphics[width=1.\linewidth,clip]{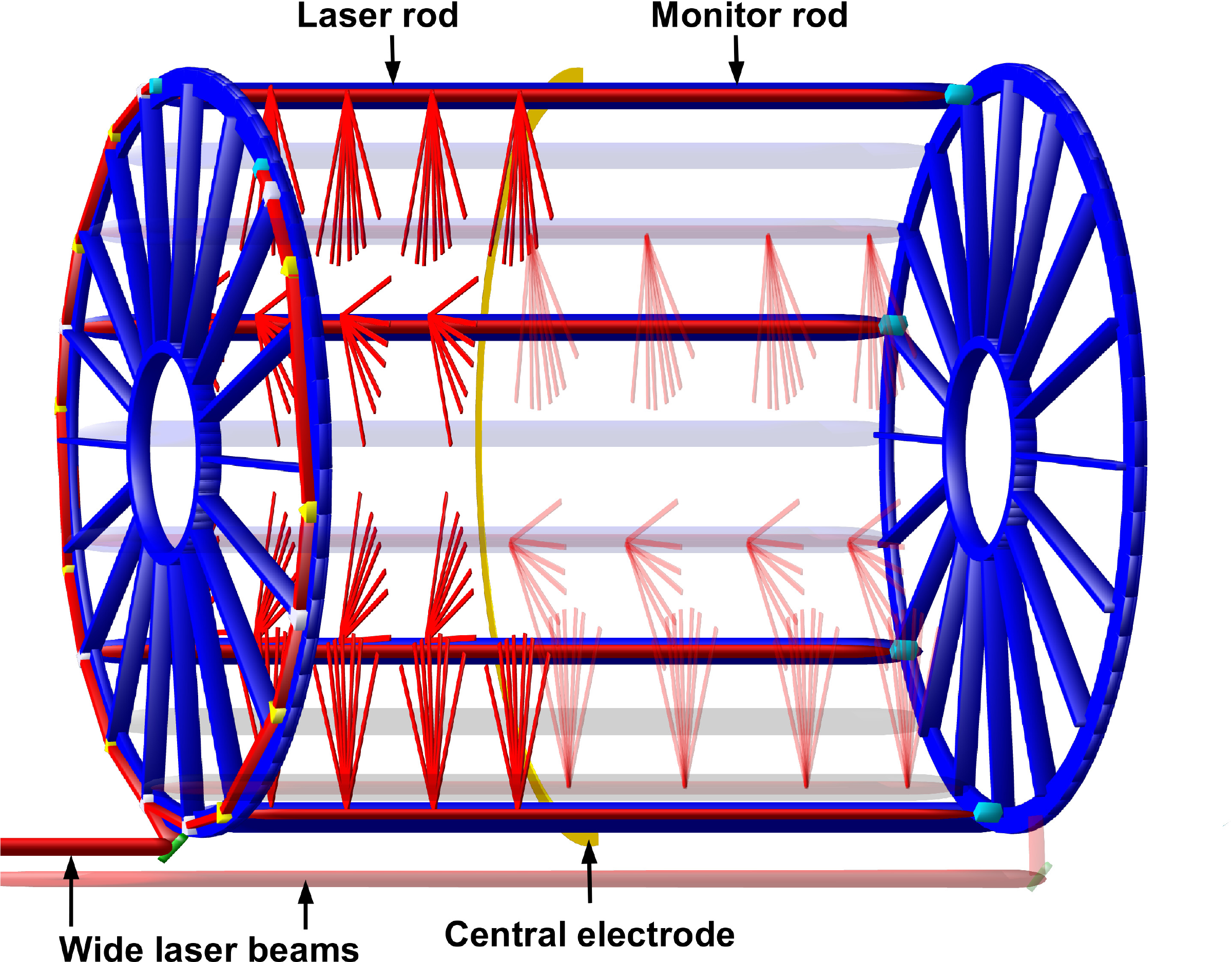}
\caption{Schematic 3D view of the TPC and the laser system. Two
wide pulsed laser beams enter horizontally at the bottom of
the TPC and are guided around the two end-caps by
mirrors, prisms and beam splitters before entering the TPC. 
Bundles of micromirrors in the hollow laser rods intersect the beams 
and generate a large number of thin rays in the TPC drift volume. 
The undeflected part of the beams continue through
the monitor rods to cameras at the far end. All elements are fixed
mechanically, except for the remote controllable entrance
mirrors at the bottom.} 
\label{laser:principle3d}
\end{figure}

The idea of generating hundreds of narrow laser beams
simultaneously was developed for the STAR experiment and was
modified appropriately for ALICE. The basic principle is that the
narrow beams are generated very close to the drift volume by optics
in a mechanically very stable configuration. Figure
\ref{laser:principle3d} shows a sketch of the principle.

\begin{figure*}[t]
\begin{center}
\includegraphics[width=\textwidth,clip]{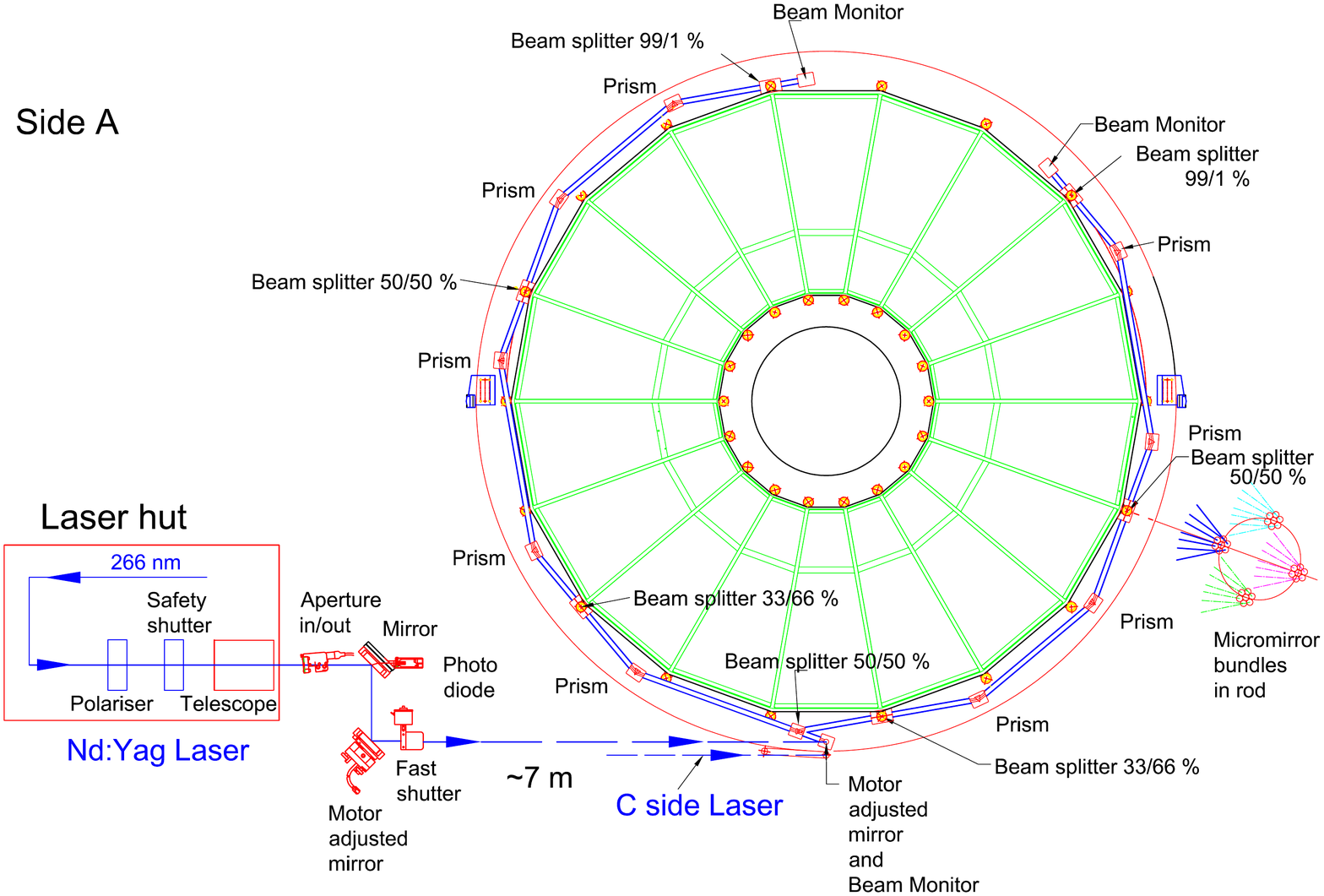}
\end{center}
\caption{Overview of the optical elements to guide the laser beam
from the laser to the entrance windows in the TPC field cage. The A
side system is shown; the C side system is obtained by mirror
symmetry in a vertical plane along the TPC axis.}
\label{laser:beamguidance}
\end{figure*}

A commercial laser outside the TPC generates an energetic pulsed
beam of UV light with 25~mm diameter and very low divergence.
Through an optical system of semitransparent beam splitters, mirrors
and bending prisms, this wide beam is split in several lower
intensity beams and guided into the TPC at different entry points
through quartz windows. The wide beams travel along the inside of
the hollow outer rods of the field cage, used for holding the mylar
strips that define the electric field. Inside the rods, the wide
beams are intersected by a number of very small mirrors (1~mm
diameter) that each deflect a small part of the wide laser beam into
the TPC drift volume. The dimensions, points of origin and
directions of the narrow beams are given by the size, positions and
angles of the micromirrors and only to a very minor degree by the
parameters of the wide beam. The micromirrors are grouped in small
bundles and placed along the length of the rod so that they do not
shadow each other. The undeflected part of the wide beam is used for
position and intensity monitoring by cameras placed at the far end
of the rod. All elements of the optical guidance and splitting
system are static, except for a few remotely controllable mirrors
used to fine tune the beam path.

Six rods in each half of the TPC were equipped with four micromirror
bundles each. Each mirror bundle contains seven small mirrors. The
wide beam originates from one laser for each TPC half and is split
and guided into the six rods. The two lasers are synchronized to
provide simultaneous laser pulses in the full TPC, thus resulting in
a total of 336 simultaneous narrow laser rays in the TPC volume. It
is also possible to operate the system with just one laser for the
full TPC using an additional beam-splitter near the laser.

\subsection{Optical system}
\label{laser:optical_system}

\subsubsection{UV lasers}
\label{laser:lasers}

Energetic pulsed laser light in the UV region is obtained from a
Nd:YAG laser ($\lambda = 1064$~nm) equipped with two frequency
doublers, generating pulses of UV light of 266~nm wavelength. The
same kind of laser was used for the STAR experiment at
RHIC~\cite{laser:STARlaser}, and also in NA49~\cite{NA49}
and CERES/NA45~\cite{laser:CERESlaser} at the CERN SPS and
ALEPH~\cite{ALEPH} at LEP. The typical beam diameter from
this kind of laser is 9--10~mm, but our lasers were fitted with
telescopes to expand the beam diameter to about 25~mm. The power
density of the narrow beams is given by that of the wide beam inside
each rod. A $40~\micro$J/mm$^2$ density in each of the
beams translates into a requirement of the total energy out of the
laser of 100~mJ per pulse.

The laser from one side of the TPC was provided by Spectron Laser
Systems Ltd, model SL805-UPG. Operated in Q-switched mode, it
provides 130~mJ/pulse of $\approx5$~ns duration at 266~nm wavelength
and a repetition rate of 10~Hz. A computer controlled tracking
system continuously optimizes the orientation of the second frequency
doubling crystal to compensate for temperature drifts. Built into
the laser is a beam expanding telescope to enlarge the beam diameter
to 25~mm and reduce the beam divergence to $\approx 0.3$~mrad. Close
to the laser, the beam has a flat intensity profile across the beam
spot which develops smoothly into a Gaussian profile after 20--30~m.

A second laser for the other end of the TPC is a similar Q-switched
Nd:YAG laser from EKSPLA uab, model NL313, similarly fitted with a
computer controlled frequency quadrupling system and beam expanding
telescope. It provides up to 150~mJ pulses at $\lambda = 266$~nm and
3--5~ns duration at 10~Hz repetition rate. After the expander
telescope, the 25~mm diameter beam also has a flat top profile and a
divergence of $<0.5$~mrad.

The lasers are triggered by a fixed rate 10~Hz external clock, such
that their pulses are synchronized to each other and to the readout
clock of the TPC. Both lasers are placed in optically stable
conditions in a hut outside the L3 magnet at $z \approx -10$~m,
2.5~m under the LHC beam line. Together with the actual laser heads
and their power supplies, the hut contains remote adjustable mirrors
to point the wide beams in the correct direction toward the TPC and
remote control electronics for the lasers, monitor cameras and
adjustable mirrors. Each laser beam is deflected through a `knee' of
one fixed and one adjustable mirror before it exits the hut. The hut
ensures personnel safety against UV light in the underground hall.

\subsubsection{Laser beam transport system}
\label{laser:beam_transport_system}

From the laser hut, the two laser beams are guided to the entrance
windows at the outer radius of the TPC field cage by a system of
mirrors, beam splitters and bending prisms, all enclosed in pipes to
ensure personnel safety and stable optical conditions. Both beams
pass through a vertical slit in the L3 magnet. One beam hits the
nearest A-side endplate close to its outer radius, where a mirror
reflects it by $90^\circ$ into a vertical plane parallel to the TPC
endplate. The other beam passes slightly lower, and after another
knee of two $90^\circ$ reflections it enters a tube mounted below
the TPC field cage. It continues in a straight line to the far
C-side endplate where another $90^\circ$ mirror bends it into the
vertical plane parallel to this plate.

Figure~\ref{laser:beamguidance} shows an overview of the optical
elements in the guidance system. High quality fused silica is used
for all optical elements and all surfaces are antireflection coated
for UV light. Dielectric coatings of the mirrors and beam splitters
are designed to divide the beam intensity evenly between the six
rods. First, a 50\% beam splitter directs half of the beam in each
direction around the periphery. Prisms deflect the beams by
$30^\circ$ such that each half of the beam passes over the
prolongation of three of the outer TPC rods. At these points, beam
splitters at $45^\circ$ angles direct equal intensity beams into
each rod along the $z$ axis by deflecting 33\%, 50\% of the
remaining and $\approx$99\% of the then remaining beam through a
$90^\circ$ angle. A small remaining beam is monitored by a camera
and dumped after the last splitter. The beam paths on the two
endplates are virtually identical. By mirror symmetry they are
arranged such that the prolongation of each of the six laser rods at
one TPC end beyond the central electrode corresponds to a hollow rod
(monitor rod) in the opposite end of the TPC.

\begin{figure}
\begin{center}
\includegraphics[width=0.53\linewidth,clip]{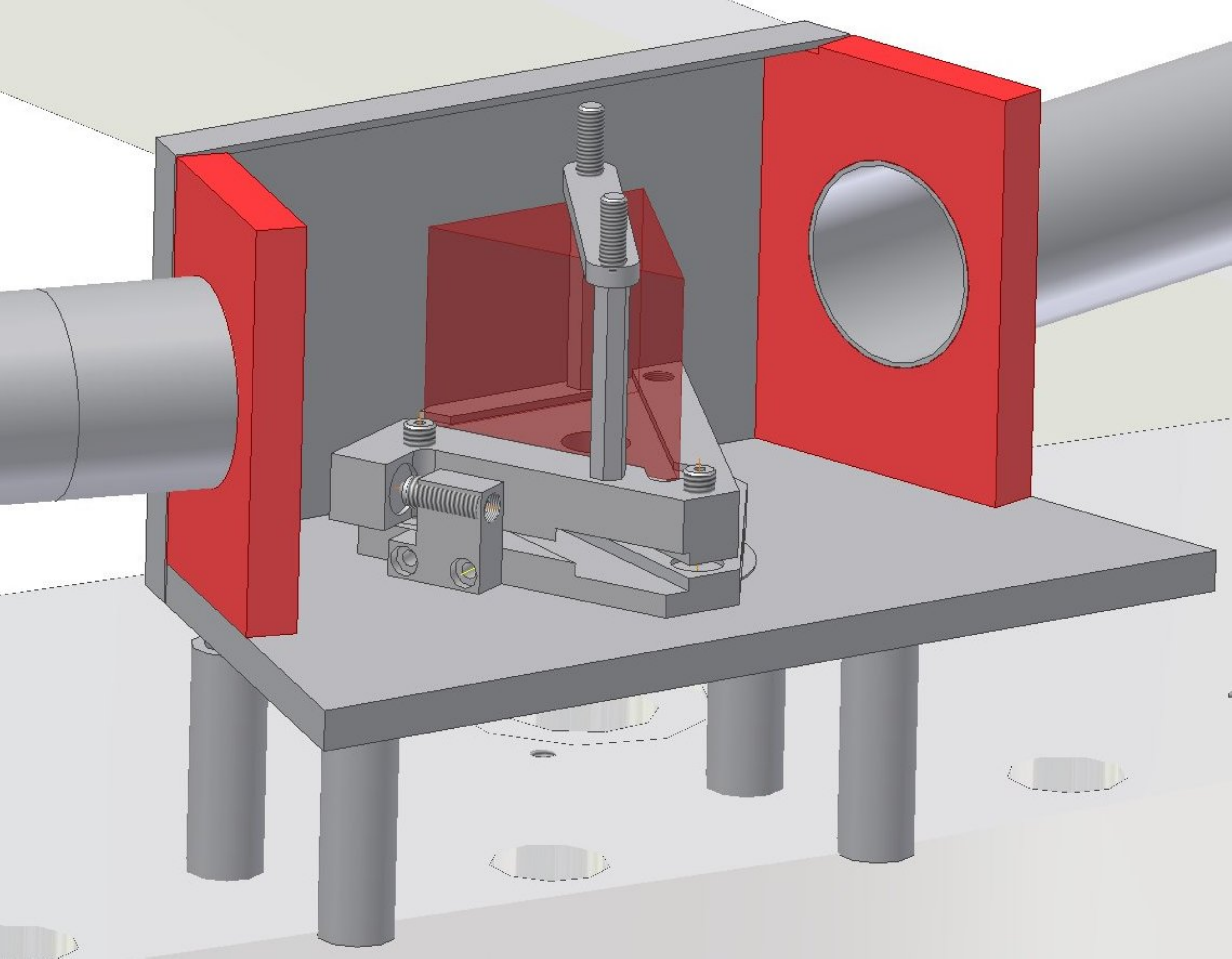}
\hspace{0.05cm}
\includegraphics[width=0.43\linewidth,clip]{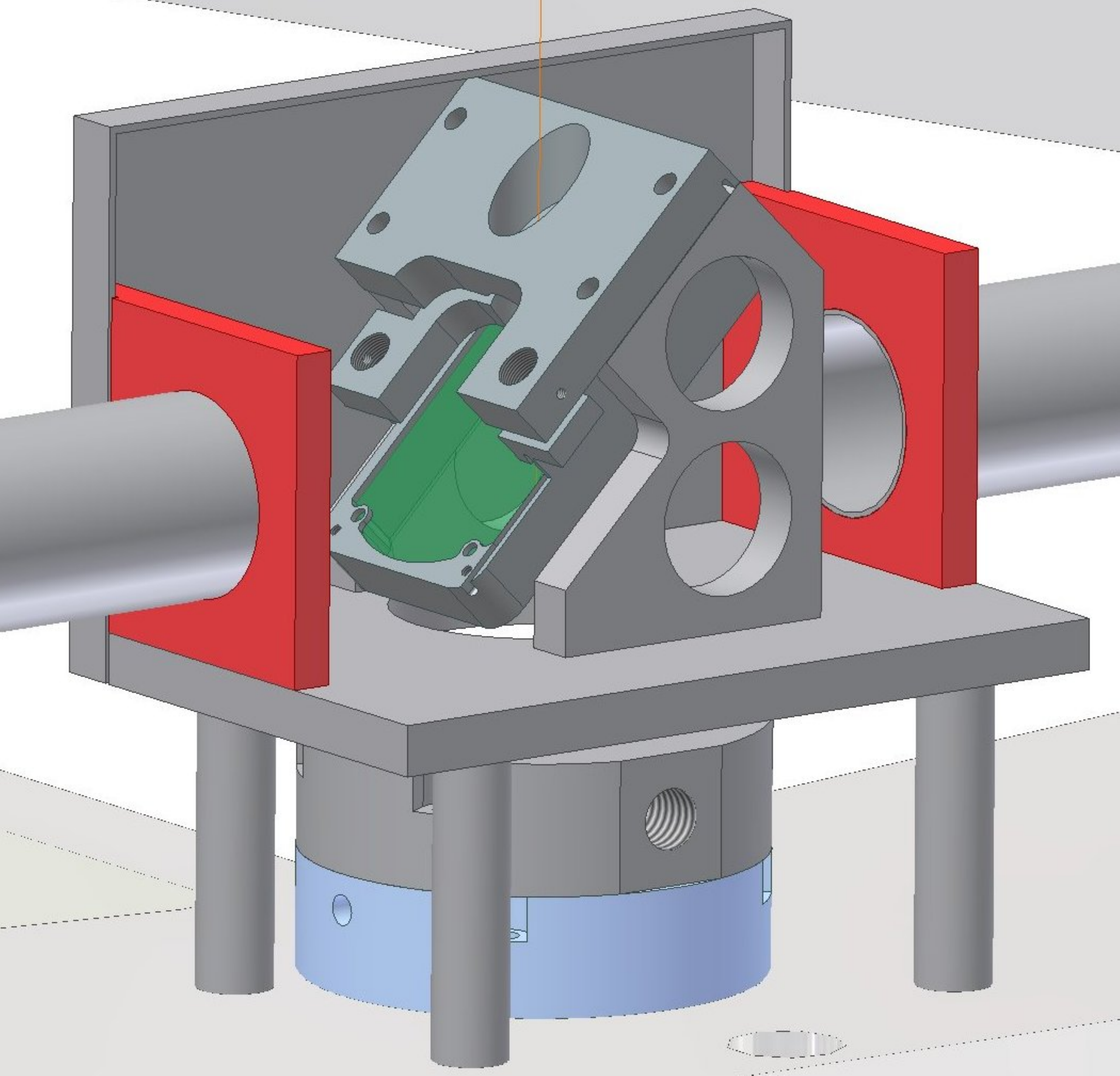}
\end{center}
\caption{Example of the design of the interior of optics boxes
installed on the TPC endplate. The shown boxes contain a $30^\circ$
bending prism and a beam splitter, respectively.}
\label{laser:opticsbox}
\end{figure}

All optical elements on the endplates are placed in small boxes.
Each box is firmly attached to the endplate and the angles of the
optics were fine adjusted manually inside the box after
installation. Figure~\ref{laser:opticsbox} shows examples of the
mechanics in two such boxes.

A few of the mirrors both in the laser hut and at the entrance to
the endplates are remotely adjustable. Together they define the beam
vector at the entry point on the two vertical planes parallel to the
endplates. The rest of the beam guidance system is based on fixed
optics, carefully aligned during the construction and using the
endplates as stable mechanical support.

\subsubsection{Micromirrors and laser rods}
\label{laser:micromirrors}
After entering the six laser rods at each end of the TPC through
sealed quartz windows, the wide laser beams travel along the inside
of the rods as illustrated in Fig.~\ref{laser:principle3d}. They are
intersected by four, roughly equally spaced, micromirror bundles
before arriving at the TPC central electrode. Here, the beam passes
through another sealed quartz window to the hollow rod in the other
half of the TPC and exits at the far end through a third window. At
the far end, the beam position is monitored by a camera before being
dumped.

The generation of narrow beams happens inside the laser rods by
reflecting the wide laser beam off micromirrors at a 45$^\circ$
incidence angle. The mirrors were made from short 1~mm diameter
quartz fibers, cut at a 45$^\circ$ angle at one end. The resulting
elliptical surface was polished and coated for total reflectivity
for 266~nm light. To increase the number of laser tracks, bundles of
micromirrors were assembled with 7 fibers in a unit which generate 7
narrow beams when hit by the wide laser beam at the cut fiber ends.
The rays spread out from the bundle roughly in a plane perpendicular
to the wide laser beam. The fibers were rotated along their axis to
give predefined azimuthal reflection angles: $2.5^\circ$, $\pm
9.2^\circ$, $\pm 16.0^\circ$ and $\pm 31.8^\circ$ relative to the
direction towards the TPC axis. The bundles were constructed with a
tolerance of $1^\circ$ in both azimuth and dip angle, and the angles
of each bundle were measured to a precision of 0.05~mrad.

\begin{figure}[t]
\begin{center}
\includegraphics[width=0.96\linewidth,clip]{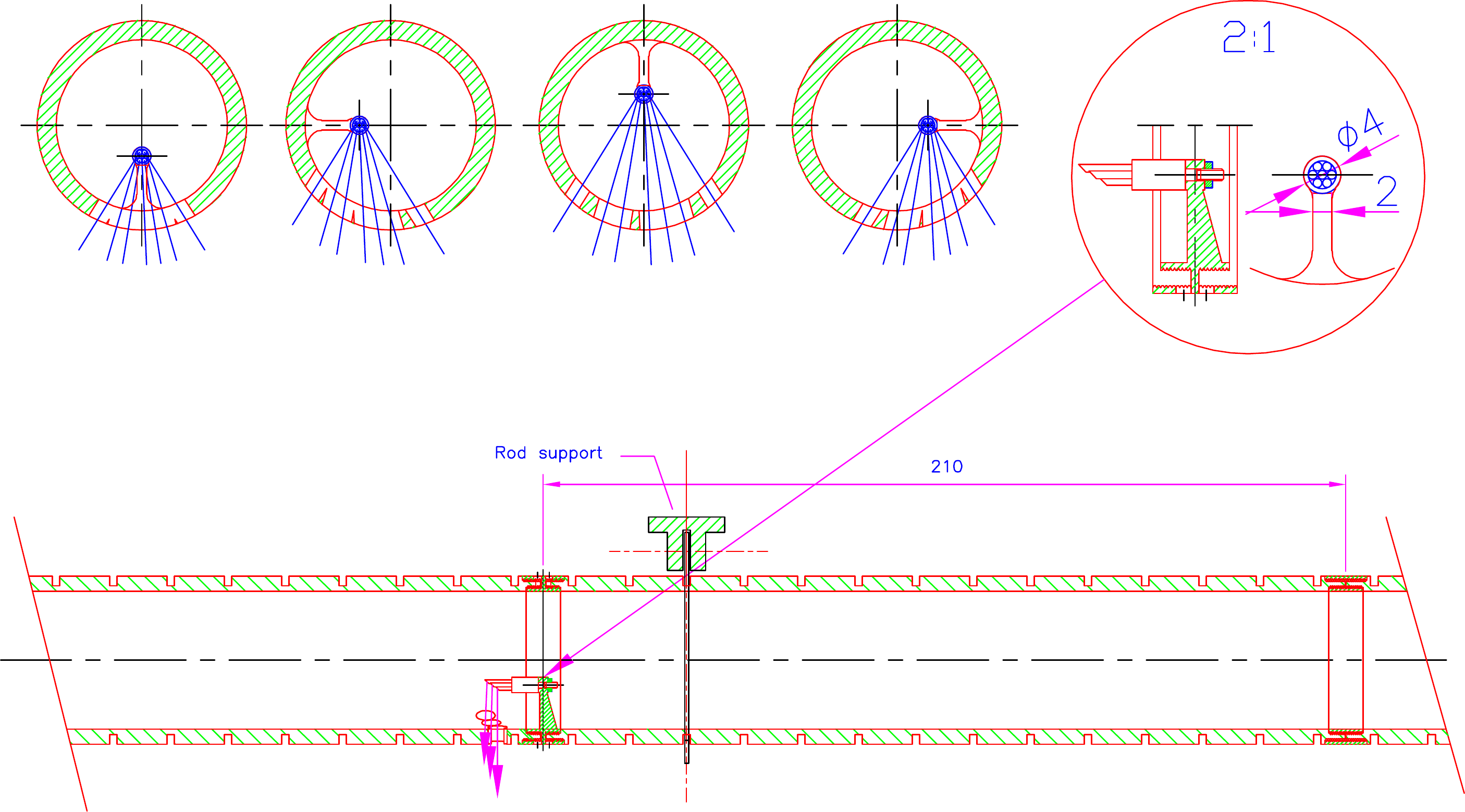}
\end{center}
\caption{Section of a laser rod with a micromirror bundle and its
support. An end view of the tube with the position of the four
mirror bundles is also shown.}
\label{laser:mirrorrod}
\end{figure}

Figure~\ref{laser:mirrorrod} shows the principle of mounting the
micromirrors in the TPC rods. The holders
are integrated into aluminum rings
and glued between the polycarbonate tube pieces that build up the
2.5~m long rod. Holes are drilled in the tube to allow the narrow
beams to exit the rod and enter the drift volume.

Because the position and in particular the mounting angle of the
micromirrors define the narrow beam positions inside the drift
volume, care was taken to assure the mechanical stability of the
mirror holders. The gluing procedure for these rings was specially
adapted to control the position of the mirror holders. The angles of
the reflected beams were measured after the assembly of the 2.5~m
long rod. Also, special care was taken to place the mirror holders
close to the rod supports to the outer field cage to minimize
movements due to mechanical stresses on the rods.

\subsection{Laser beam characteristics and alignment}
\label{laser:beam_characteristics_and_alignment}

\subsubsection{Narrow beam characteristics}
A narrow ray generated by reflection from a circular surface (as
seen along the beam direction) is equivalent to the beam progressing
beyond a screen with a similarly shaped hole. Thus, the narrow beams
in the TPC are approximated by pure Fresnel diffraction of an
infinite planar wave through a circular aperture of 1~mm diameter.

The profile and total energy of the narrow beams, generated by
reflection from each of the micromirror bundles, were measured in
the lab. The reflected beams were measured by a calibrated energy
meter and imaged with a CCD camera as a function of the distance,
$z$, from the mirror bundle. The energies did not vary substantially
for different micromirrors and were stable as a function of time,
reflecting the quality of the coated surfaces and the laser. The
patterns matched qualitatively what one expects from Fresnel
diffraction and the measured FWHM remains at or below 1~mm up to $z
= 200$ cm. For further details, see \cite{TPCint}.

\subsubsection{Narrow beam layout}

The transverse pattern of the narrow beams in the TPC volume follows
from the micromirror angles given in
Sec.~\ref{laser:micromirrors}. In the $z$ direction, the planes
of laser tracks are situated at $z \approx \pm 115, 820, 1660,
2440$~mm. When defining the angles and $z$ positions, we have aimed
at generating beams radiating at constant $z$ that cross sector
boundaries strategically, i.e.\ at points where alignment between
sectors would benefit the most. We have also avoided having too many
tracks with small angles relative to the wires of the readout
chambers.

\begin{figure}
\begin{center}
\includegraphics[width=0.96\linewidth,clip]{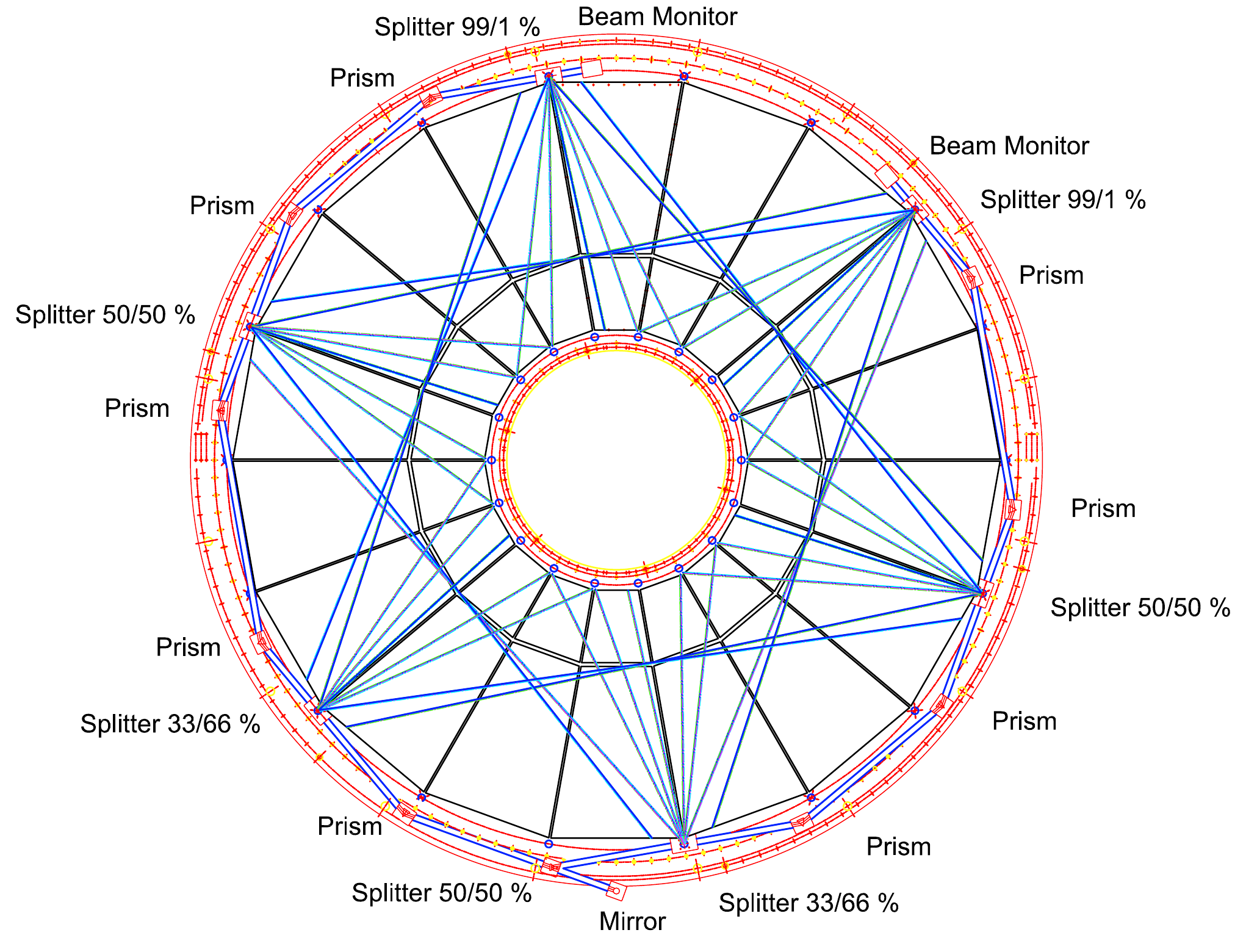}
\end{center}
\caption{Ideal laser tracks in (r,$\varphi$) in the TPC drift volume,
projected to the endcap. The pattern repeats eight times through
the full length of the TPC.} \label{laser:lasertracks}
\end{figure}

Figure~\ref{laser:lasertracks} shows the resulting pattern for the
beams at a single position in $z$. Beams from neighboring laser rods
in $\varphi$ are offset by a few cm in $z$ relative to each other to
avoid most of the apparent beam crossings. The central ray (of the
seven) from each bundle radiates in (r,$\varphi$) in a way very similar
to  tracks from the interaction point while the other angles
were chosen in order to illuminate all sectors and assure beams that
cross over all sector boundaries. In this way, we defined a pattern
optimized for testing and sector alignment as well as for drift
distortion measurements. One should note that, although the position
of the beams was measured to a high precision, the production
tolerance of $1^\circ$ on the angles results in deviations of the
shown paths of up to 40~mm near the inner cylinder.

\subsubsection{Spatial precision and stability}

The TPC calibration would ideally require absolute knowledge with
infinite precision of the spatial position of all laser tracks in
absolute ALICE coordinates. Given the mechanical tolerance, the best
absolute coordinate frame for each half TPC is defined by the plane
of the endplate. All readout chambers and the plane of the central
electrode defining the high voltage surface were aligned and
adjusted relative to the endplates. These surfaces were defined
relative to each other during construction to a precision of
approximately $100~\micro$m with the aim of obtaining a relative
electric drift field error below $5 \times 10^{-4}$.

A final precision goal of $800\,-\,1000~\micro$m for space points translate
into matching requirements on spatial coordinates and
angles of the laser systems:\\
\hspace*{2cm} $(\Delta x, \Delta y, \Delta z)  \leq (800-1000)~\micro$m\\
\hspace*{2cm} $(\Delta \theta, \Delta \varphi ) \leq (0.4-0.5)$~mrad

By far the most important issue in the definition of the laser track
positions is the placement of the micromirrors, both in $(x,y,z)$
and in particular in the angles $(\theta,\varphi)$. The only other
deviation from the ideal rays that matters is the incidence angle of
the wide laser beam on the micromirrors and this is relatively easy
to measure and keep constant because the optical system has long lever
arms.

Even if small movements of the rods and external beam optics cannot
be excluded during and after the assembly of the TPC, the stability
of the finished and installed TPC is well below the requirements.

\subsubsection{Construction and surveys}

The construction errors of the micromirror bundles were specified to
be less than $100~\micro$m in the spatial measures and less than
$1^\circ$ in all reflection angles. The critical surfaces were,
however, measured to remain within a $50~\micro$m tolerance and all the
angles of the reflected beams from the mirror bundles were
subsequently measured to a precision of $0.05$~mrad. The mechanics
of assembly of the rods assured very precise $z$ position of all the
mirror bundles and, after assembly of the rods, a second measurement
of the angles was performed to a precision of $0.1$~mrad, using a
green laser.

Relative $(x,y,z)$ shifts of micromirrors within a rod during the
installation in the TPC are unlikely. However, the absolute position
of the rods relative to the endplates was not guaranteed to a
precision good enough for the laser system, and the mounting of
mylar strips was seen to cause a small bending of the rods between
the support points and result in small rotations of the
micromirrors, especially in the dip angles. In order to monitor such
shifts, the $(\theta,\varphi)$ angles of the central micromirror in
each bundle was remeasured by surveying the intersection of the
narrow laser beam with the inner cylinder using the green laser (see
Fig.~\ref{laser:lasertracks}).

The construction of the TPC required the assembly of the field cage
while it was standing on first one, and then the other end, before
it was finally turned into its final horizontal position. Changes in
the mechanical stresses may have influenced the absolute position of
the rods. However, the stiffness of the rods guarantees a continued
good relative alignment of the micromirrors in the same rod. In this
way, the absolute $(x,y,z)$ of the micromirrors is known to
$100-200~\micro$m and the mirror angles given by the measurements of
the rods prior to installation. Furthermore, the relative angles of
the micromirrors, which are glued together in a bundle, are
determined to a precision of $0.1$~mrad by the lab measurements.

The initial setup of the optics on the endplates needed careful
manual adjustment of all the mirrors and prisms in the system.
Special measurement tools were installed temporarily in the beam
path. Adjustments of the beam paths between the lasers and the
endplates of the TPC must be realigned after interventions, using
the remotely adjustable mirrors and cameras.

\subsubsection{Online and offline alignment}

Given that the readout chambers and the central electrode constitute
the best aligned surfaces during the TPC construction, it is an
advantage to use these surfaces as references also for the absolute
position of the laser tracks. As mentioned, all metallic surfaces
inside the TPC emit electrons by the photoelectric effect,
synchronously with the laser pulse. This is in particular the case
for the central electrode made of a stretched aluminized mylar
foil. We use these electrons to image the whole plane at the end of
the drift volume. The central electrode signal is used to calculate
the full drift time and its $(x,y)$ variations and to obtain an
online measurement of drift velocity variations in time. The full 3D
map of the drift field is obtained in an offline alignment procedure
where the reconstructed tracks are used together with all the survey
information of relative and absolute laser beam positions.

Even without the ultimate, absolute precision of all laser tracks,
much can be learned from a few well determined tracks, in particular
if they span a large lever arm in $z$. Close to the outer TPC
radius, the angular uncertainties of the rays play only a minor
role, and it is possible to obtain a very good drift velocity
measurement near the six rod positions in each half of the TPC.

Furthermore, variations over time are tracked to a very good
accuracy throughout the TPC. An uncertainty in the time variation of
the laser beam positions could come from a possible torsion in the
field cage due to variations in the mechanical loads, magnetic field
and the external temperature. In stable running conditions, these
effects are minimal.

\subsection{Operational aspects}
\label{laser:operational_aspects}

Under the operational conditions of the LHC, it is essential that
all controls and monitoring of the system are done remotely and the
software interfaces are integrated in the Detector Control System
(DCS). The two lasers and the angle adjustable mirrors are
controlled this way. Along the beam path, a number of cameras are
used to look at the laser beam from the control room. Furthermore,
several electronics modules are introduced in the system. One is
used to synchronize the laser pulses to the ALICE trigger and the
TPC readout clock. Another controls various shutters and apertures
that can be inserted into the beam paths by remote control.

\subsubsection{Beam monitoring and steering}
\label{laser:beam_monitoring_and_steering}

The laser heads with their power supplies are equipped with remote
control facilities through RS-232 communication. After conversion to
a network protocol using a Digi terminal server
\cite{digi:PortServer}, all controls are transferred to the online
computers. Both lasers use an active feedback system to optimize the
frequency conversion from 532~nm to 266~nm by very slight adjustment
of the angle of the conversion crystal inside the laser head. The
algorithms delivered by the laser manufacturers were improved and
ported to run on the online computers.

\begin{figure}
\begin{center}
\includegraphics[width=\linewidth]{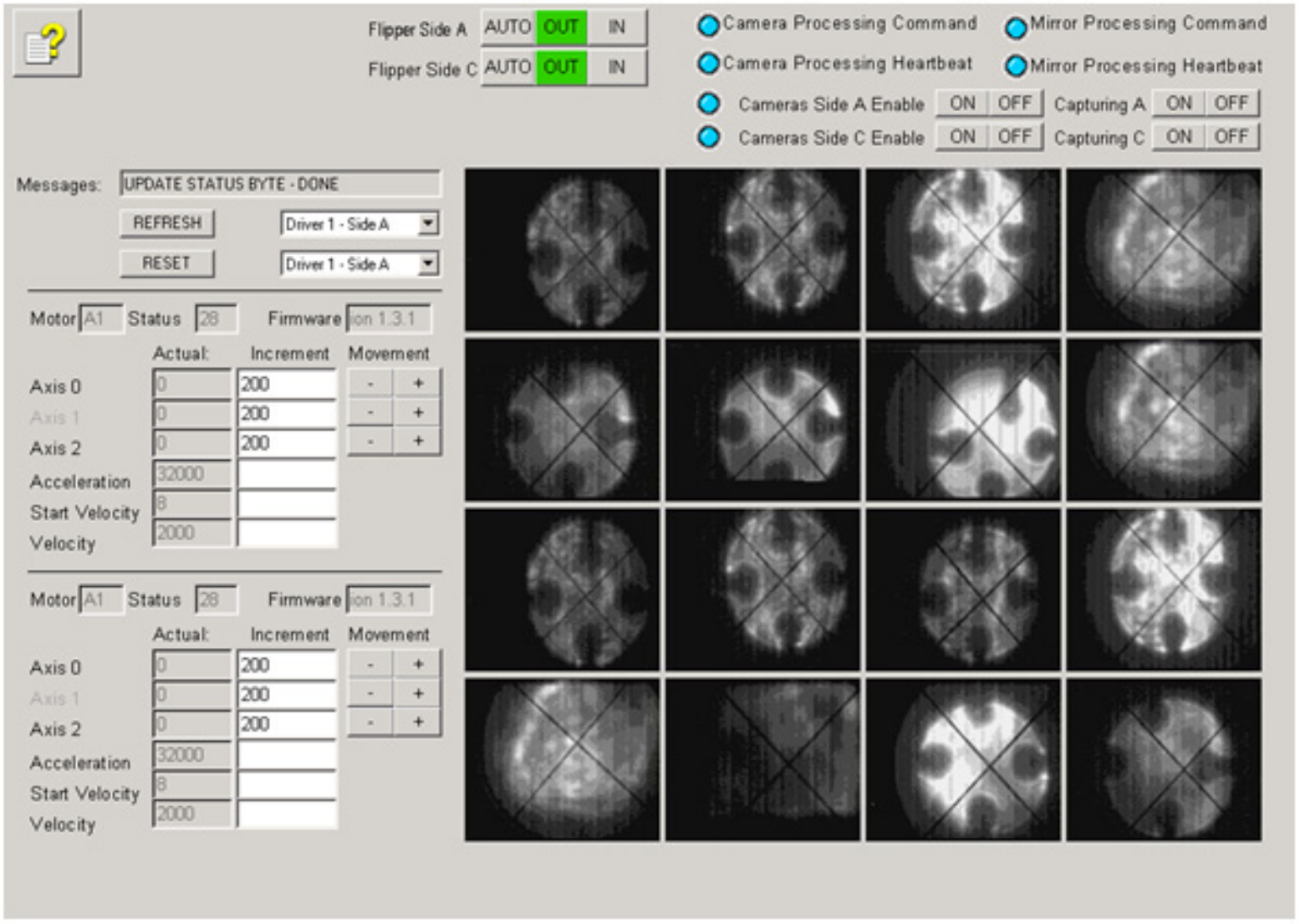}
\end{center}
\caption{User interface panel for the steering of the laser beams
and monitoring by cameras.} \label{laser:beamalignment}
\end{figure}

Eight mirrors were inserted in the beam paths where they make
$90^\circ$ bends. They define the correct alignment of the beams
entering the field cage volume. Five of the mirrors are adjustable
by remote control in two angular dimensions. For the A-side laser
beam, one adjustable mirror is placed in the laser hut and another
at the beam intersection point on the TPC endplate. The C-side laser
beam is guided through the knee at the bottom of the TPC by a third
adjustable mirror. The mechanical movement is driven by the New
Focus Picomotors \cite{iPico} which allow angular steering with
$0.7~\micro$rad resolution, even in the magnetic field.
The direction and quality of the wide laser beams are monitored
online by simple CMOS cameras \cite{quasar}, focused on mat glass
screens, on which the UV pulses are converted to visible (blue)
light. The cameras are interfaced to the computer by coaxial cables
and a frame grabber card \cite{falcon}. In total, there are 18
cameras in the full system, placed at the beam entrance points on
the endplates, at the end of the two paths on the periphery of each
endplate and finally at the far end of the rods, downstream from
each laser rod. Figure~\ref{laser:beamalignment} shows the graphical
user interface with a display of 16 (out of 18) cameras and the
options for beam steering by the five adjustable mirrors. A trained
operator can manipulate the few parameters of the mirrors while
observing the images on the cameras, and the system then remains
stable over months.
To facilitate the adjustment of the mirrors, a 3~mm aperture can be
inserted by remote control in the middle of each of the wide
beams just after the exit from the laser heads, thus reducing the
beam to a thin one useful as a `pointer'.

Finally, two shutters were installed on each laser beam. One set is
internal to the laser head, used as a safety device to block all
light from exiting the head. The other set at the exit from the
laser hut is used in conjunction with the laser synchronization
module (see~Sec.\ref{laser:trigger_and_synchronisation}). They operate
rapidly enough to decide on a pulse-by-pulse level whether to block
the pulse or let it pass.

\subsubsection{Trigger and synchronization}
\label{laser:trigger_and_synchronisation}

Optimal thermal stability of the lasers requires that they are
operated at a constant 10~Hz pulse rate. Equally important for good
calibration is a good synchronization of the laser pulses to
the TPC readout clock. Furthermore, the system is designed to take
data with laser tracks in various run configurations, either as
dedicated calibration runs with one or both lasers running at a
fixed 10~Hz trigger rate or in a mode where the laser events are
interspersed between physics triggers.
A dedicated laser synchronization module was built to handle all
trigger and timing conditions of the laser operation. It is based on
the common RCU module where the optical communication daughter board
was replaced by a dedicated signal driver board. The module provides
programmable timing outputs to control the operation of the lasers
and the shutters, and to interface to the ALICE trigger system.

Based on the LHC 40~MHz clock, the module generates a 10~Hz clock in
phase with the TPC readout clock. Each laser is controlled by
signals from the module to trigger their flash lamps and Q-switches
in order to generate laser pulses synchronous to the 10~Hz base
clock. In case of stand-alone calibration runs, the trigger module
provides an ALICE trigger at the 10~Hz base clock rate. For laser
events interspersed between physics triggers, the module also
generates signals at the 10~Hz rate to fire the laser flash lamps
and generate a calibration event trigger. If this trigger is vetoed
by the central ALICE trigger system, the laser pulse can be
suppressed by vetoing the Q-switch signal to the lasers for this
event. Otherwise, the generation of a Q-switch signal assures a
timely laser pulse from one or both lasers. During LHC collisions,
we foresee to run in a mode where the lasers are first warmed up at
10~Hz with the shutters closed, for as much as one minute until they
are thermally stable. The warm-up is followed by a short burst (less
than one minute) of laser events interleaved between physics
triggers and a period of about one half hour where the lasers are
put into standby mode, without firing the flash lamps. This burst
mode of operation is handled automatically by the detector control
system and is designed to ensure a laser flashlamp lifetime
commensurate with the length of typical LHC runs (of order one
year).

\section{Infrastructure and services}
\label{infra}
The TPC together with the ITS, TRD and TOF detectors of the `central barrel', 
(see Fig.~\ref{intro:ALICElayout}),
are mounted in the so-called
space frame (see Sec.~8 in \cite{TPCint}).

\begin{figure}[t]
	\centering
	\includegraphics[width={\linewidth},clip]{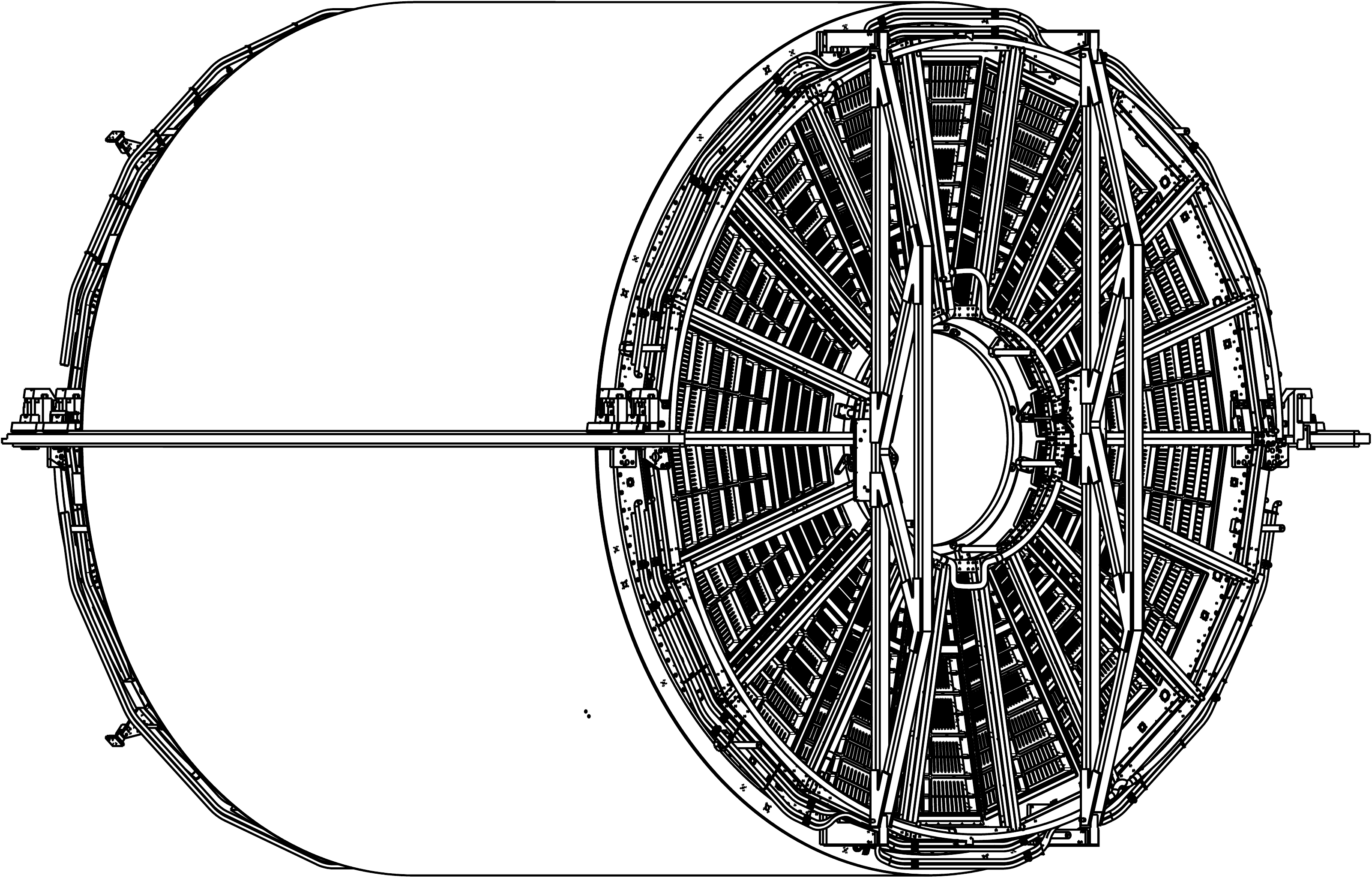} 
	\caption{Overall view of the TPC with the SSW, rails and I-bars.}
	\label{infra:TPC_compl}
\end{figure}

The space frame is supported with four
adjustable feet on two beams traversing the full length of the L3
magnet. The TPC rests on two rails, located in the median plane, in the central
opening of the space frame. The ITS is suspended on two points in the
inner opening of the TPC. An overall view of the TPC together with the Service Support Wheels (SSWs), the 
rails and I-bars (but without the support for the ITS) is shown in Fig.~\ref{infra:TPC_compl}.

The TPC services are organized as follows: the FEE is housed in the SSWs, 
which are
located on the A and C side at a distance of about 20 cm from the TPC
endplates. Connection to the readout chambers is via flexible Kapton
cables. Outgoing services are routed via the so-called baby frame and
back frame on the A and C side, respectively. These frames are short, about
2\,m long extensions of the space frame with nearly the same geometry as
the space frame. They are decoupled mechanically from the space frame
except for the flexible services.

\subsection{Moving the TPC}
\label{infra:Moving_the_TPC}

The TPC can be moved on the rail system.  This was necessary during
installation, and is foreseen in case of future servicing of TPC or
ITS. When the TPC needs to be moved, the rails are extended on the A
side by transfer rails that are connected to a support structure with
rails outside the L3 magnet. For initial installation and later
servicing of the ITS, the TPC is moved to the so-called parking
position, 4.8 m toward the A side. All services, except the 100 kV HV cable and
temporary gas connections have to be removed during movement.

The TPC sits on its rail with 4 feet Teflon-coated gliders which are
adjustable in $x$ and $y$. 
In Fig.~\ref{infra:foot_detail} the detailed design of a pair of feet 
(one for the TPC one for the SSW, see below) with the Teflon padded gliders is shown.

\begin{figure}[t]
	\centering
	\includegraphics[width={0.64\linewidth},clip]{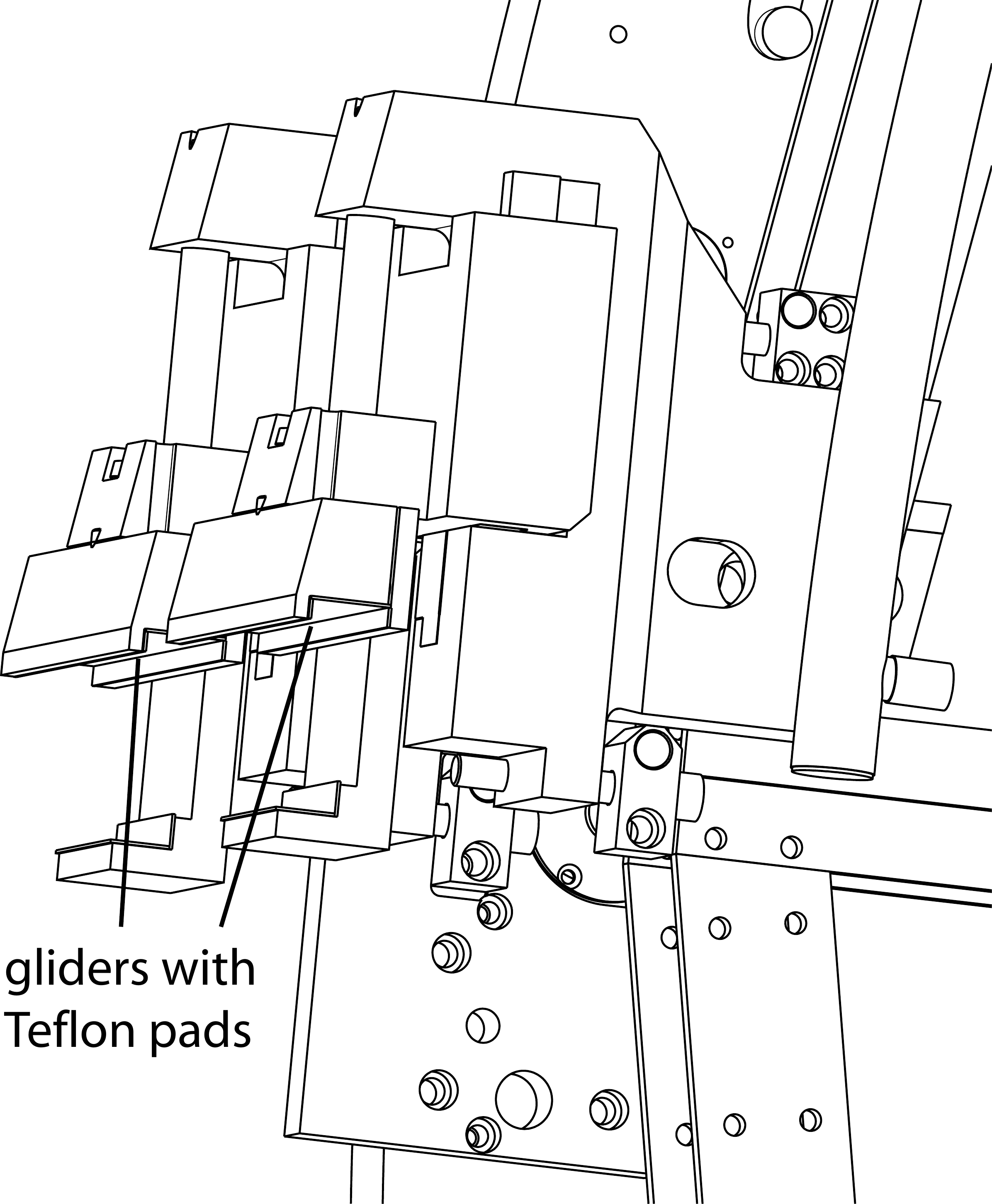} 
	\caption{Design of the feet to support the TPC and the SSW. 
	The surfaces in contact with the rails are covered with Teflon pads.}
	\label{infra:foot_detail}
\end{figure}

The gliders on the
I side have no lateral play, the ones on the O side have $\pm 4$ mm
play.  During movement, two of the feet, either both A or both C side,
are supported  vertically by coupled hydraulic jacks, providing
effectively a 3-point support and completely eliminating torsional
stress on the field cage due to imperfect parallelism of the rails. 
The rails were initially, before loading the space frame with detectors, 
parallel with a tolerance of about 0.3 mm vertically
and about $\pm 1$ mm laterally.  
At rest, the feet are fixed with adjustable screws. The procedure of
letting two feet hydraulically float is also applied during movement of
other heavy detector components that might lead to a deformation of
the space frame at a level of more than a few tenths of a mm.

The two SSWs are equipped with the same type of gliders. During
movement, the TPC and the two SSWs are coupled by 4 so-called track
rods between the feet.  The ensemble, TPC and SSWs, is always moved by
(stepwise) {\em pulling} with two hydraulic jacks attached to the feet
of a SSW and the rail. In order to limit the tensional stress of the
field cage, two steel cords between the A and C side feet on each rail
are pre-tensioned to about 8000~N.  The friction coefficient of the
system Teflon-glider and rail is about 10\%, resulting in a pulling
force $\le 8000$~N on either side, controlled by the hydraulic
pressure. For safety, the hydraulic pressure is limited.

When moving the TPC, the ITS is disengaged from its fixtures at the
TPC and supported by a temporary second rail system, the ITS rails.
These rails are each attached at the hadron absorber
cone (see Fig.~\ref{intro:ALICElayout}) on one end and at a support outside the
L3 magnet on the other end. In between, they have sliding supports at the two TPC
endplates which allows the TPC to move with respect to the ITS. 
The ITS rails are removed when the TPC is at the working
position.

\subsection{Service support wheel}
\label{infra:Service_support_wheel}

\begin{figure}
\begin{center}
	\includegraphics[width=\linewidth,clip]{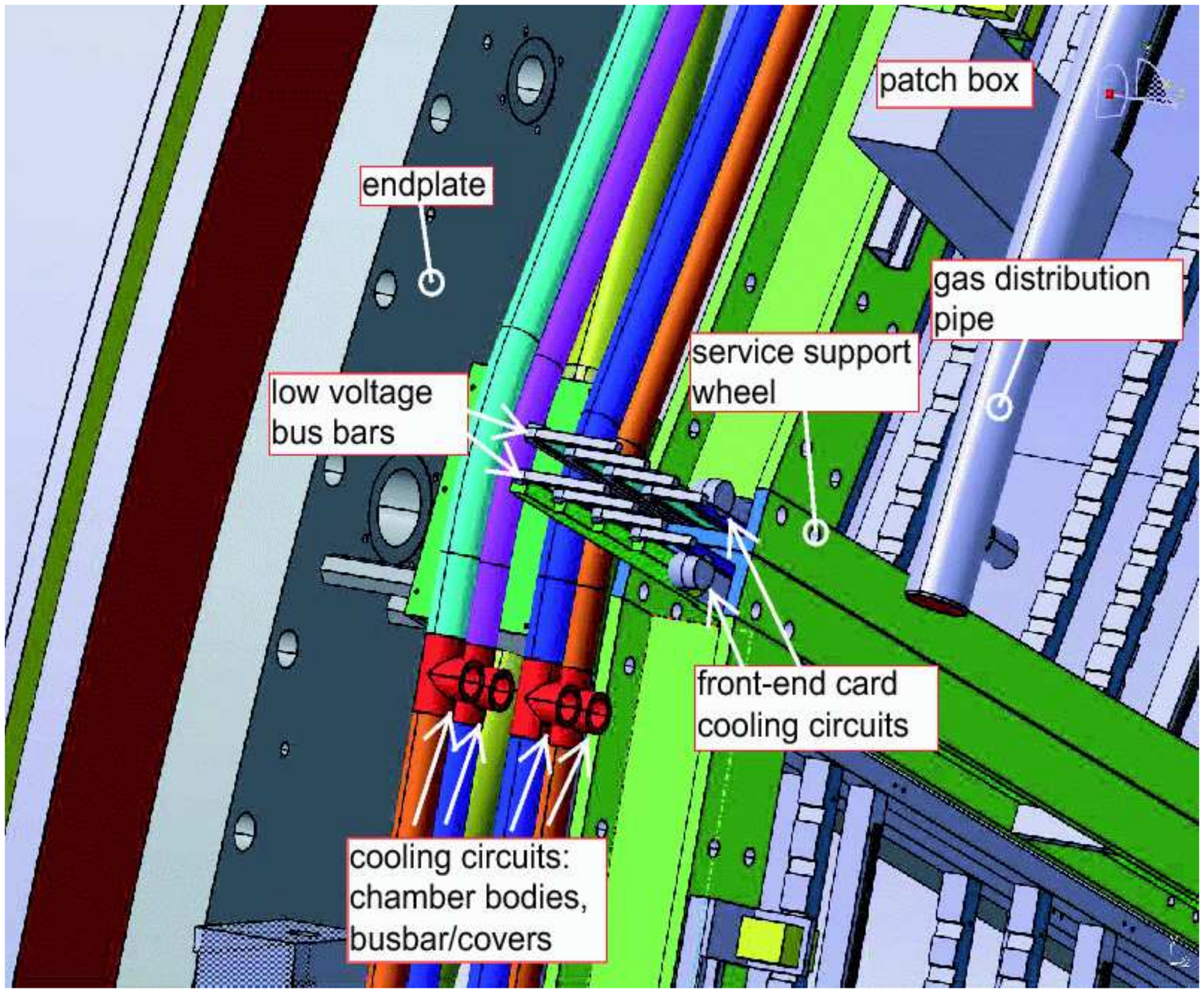} 
	\caption{Closeup of the SSW together with the endplate of the TPC cylinder showing some
	components of the services: cooling lines, gas distribution pipe, a patch box and low 
	voltage bus bars.}
	\label{infra:endplate_SSW_labelled}
\end{center}
\end{figure}

\begin{figure}
\begin{center}
	\includegraphics[width={0.5\linewidth}]{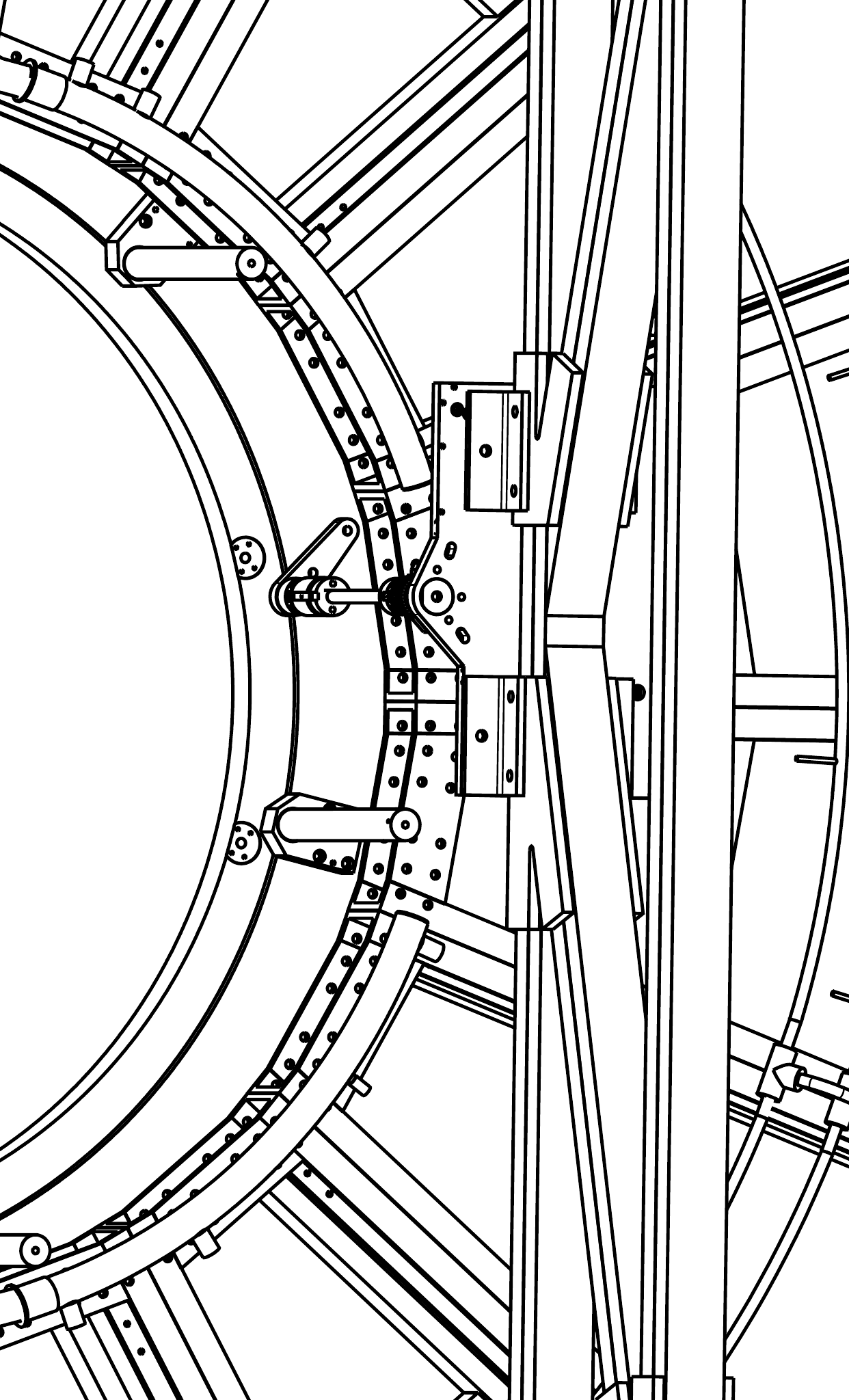} 
	\caption{Details of the connection between I-bar and  endplate.}
	\label{infra:Ibar_detail}
\end{center}
\end{figure}

The SSW houses and supports the front electronics and its services
(LV, DCS, DAQ), the manifolds for the various cooling circuits, and the
drift gas manifolds. 
In Fig.~\ref{infra:endplate_SSW_labelled} details of the SSW are shown in a CAD model. In Fig.~\ref{infra:Ibar_detail} more details of the connection between one I-bar 
and the endplate are shown.

Forces on the endplates and readout chambers
are kept to a minimum. The flexible Kapton cable connections from the
front-end cards to the readout chambers have a maximum play of $\pm
5$~mm. The position of the SSW relative to the respective endplate
has thus to be controlled to much better than this.  While moving, the
constant distance is assured by four so-called tie rods between each
SSW and the corresponding endplate, at the angles of about 45$^\circ$, 
135$^\circ$, 225$^\circ$
and 315$^\circ$ at the outer circumference. The $x-y$ position is tuned
by adjusting the glider feet vertically and laterally. 
In the final position, the support of
the SSW is transferred from the tie rods to fixation points on the
space frame and close to the location of the tie rods, minimizing the
forces on the TPC endplates.


\begin{figure}
	\centering
	\includegraphics[width={1.0\linewidth},clip]{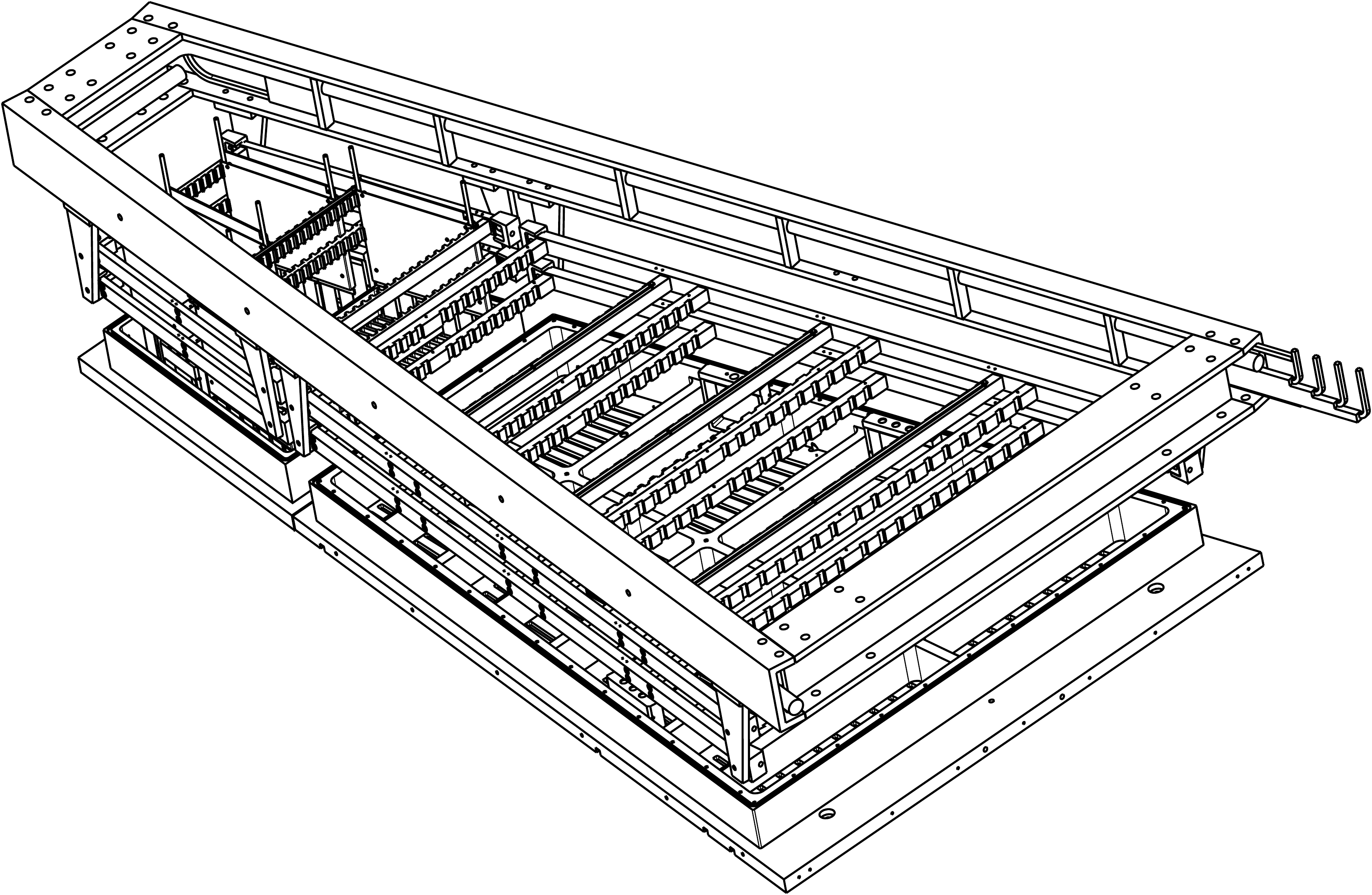} 
	\caption{One frame of the SSW together with the holders for the FECs and 
	the wire chambers. For clarity only a few front-end cards are shown.}
	\label{infra:TPC_SSW_sector}
\end{figure}

The SSW is constructed from 18 essentially identical trapezoidal
frames matching the sectors of the endplate and the readout
chambers. 
In Fig.~\ref{infra:TPC_SSW_sector}, one frame of the SSW with the holder for the 
front-end cards is shown together with an inner and an outer readout chamber.

\begin{figure}
	\centering
	\includegraphics[width={0.8\linewidth},clip]{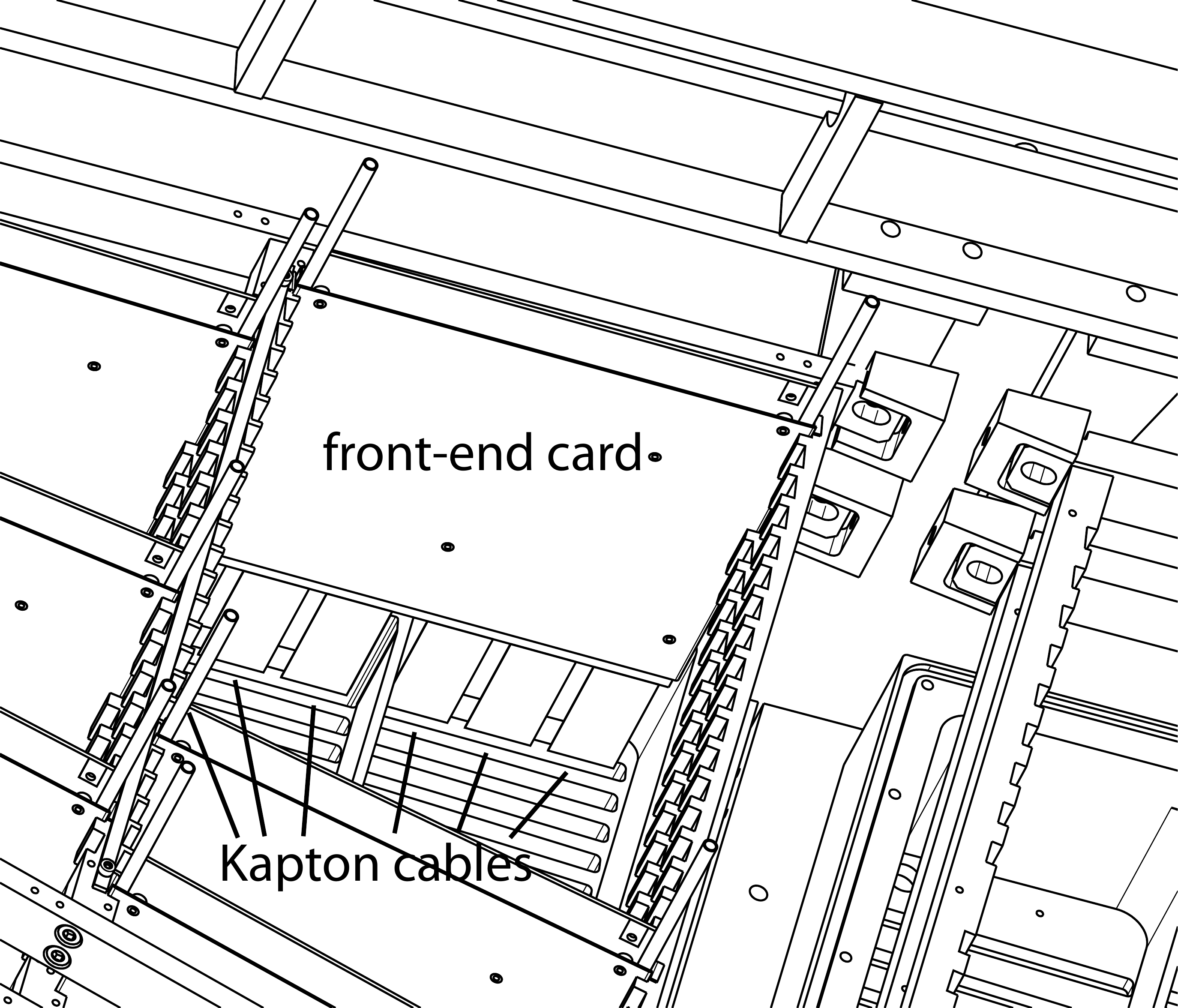} 
	\caption{Construction detail of the holder mechanics for the FECs. The 6 Kapton
	cables connecting to the wire chambers are also shown.}
	\label{infra:SSW_card_holder_detail}
\end{figure}

In Fig.~\ref{infra:SSW_card_holder_detail}, a close up of the mounting 
support for the FECs is given. On the side facing the endplate, front-end-card holders are
attached, adjusted with a precision of about one mm in $x-y$. The LV
services, 12 per sector, are bundled with a cooling circuit and fixed
to one long side of the trapeze. The outer face of each sector is
closed with a cooled cover. A number of distribution boxes is
mounted near the outer circumference of each SSW (see below).

In addition to TPC services, the A side SSW has to serve as the
support for ITS services and beam pipe parts. The carriers of the ITS services, 
two half cones jutting from the SSW to the ITS, are each fixed in 3 points on
the inner circumference of the SSW. Sector 13, pointing downward, carries a
framework which supports a pump and valve of the beam pipe.

 In Tab.~\ref{infra:weights_overview}, an overview of the weights of all components
described in this section is given.

\begin{table}[t]
\begin{center}
\caption{ Weights of all components described in this section.}
\label{infra:weights_overview}

\begin{tabular}{|l|r|r|r|}\hline
Item & weight & units & full TPC \\ 
     & kg &  & kg \\ \hline
\multicolumn{4}{|c|}{Field cage} \\ \hline
Field cage w/o ROC & 6000 & 1 & 6000 \\ \hline
IROC            &  15 & 36 & 540 \\
OROC           &   40 & 36 & 1440 \\ \hline
\multicolumn{3}{|l|}{ROC total}& 1980 \\ \hline
\multicolumn{3}{|l|}{Inner cooling panels}& 100 \\ \hline
gliders        &  80 & 4 & 320 \\ \hline
I-bars C side  &  100 & 2  &   200 \\ \hline
\multicolumn{3}{|l|}{Total field cage}& 8600 \\ \hline

\multicolumn{4}{|c|}{SSW} \\ \hline
SSW frame alone   & 570 & 2 & 1140 \\ \hline 
Frontend cards &  0.56 & 4356 & 2440 \\
RCU            &   0.5  & 216 & 110 \\ 
backplanes per sector & 4 & 36 & 144 \\ \hline
Frontend card supports  & 18 & 36 & 648\\
per sector             &    &    &    \\ \hline
services per sector & 14 & 36 & 504 \\ \hline
cooling covers  & 9 & 36 & 324 \\ \hline
gliders        &  80 & 4 & 320 \\ \hline
total per SSW  & 2800 & 2 & 5600 \\ \hline
\multicolumn{3}{|l|}{TPC total}& 14200 \\ \hline
\end{tabular}
\end{center}
\end{table}

\subsection{Low-voltage distribution}
\label{infra:Low_Voltage_distribution_system}

The FEE Low-Voltage (LV) system is based on modular, low-noise, 
high-efficiency switching power supplies (W-IE-NE-R model PL512 ~\cite{DCS:wiener}).
In each of the 19 water-cooled crates the modules to supply 2 complete 
sectors are housed (rated: 2--7\,V, 2$\times$100\,A, 2$\times$200\,A). Remote control by DCS is provided via a 
network connection (see Sec.~\ref{DCS:Subsystems:LV}).

The power supplies are not connected to ground internally. The ground 
potential is defined at the level 
of the front-end cards. The connection between power supplies and TPC is based on 
large cross section unshielded and uncooled copper cables. Inside each sector the
power is distributed by bus bars running along the spokes of the SSW. Voltage drops along these 
bus bars are about 20~mV. For the distribution of the LV to the individual FECs 12 small distribution boards are connected by screws to the busbars. Up to 15 FECs are supplied by one distribution board. To buffer small voltage variations and also prevent voltage surges which could occur
when the supply lines are interrupted, large capacitors of 10~mF each are mounted on the distribution boards.
In addition smaller capacitors of $\unit{2.2}{\micro\farad}$ for the suppression of HF noise are also mounted.
The sense lines of the power supplies are connected to these 
local bus bars and a dynamic regulation always ensures the correct voltages at the front-end cards. An overview of the parameters of the system is shown in Tab.~\ref{infra:LV_overview}.

\begin{table}[t]
	\begin{center}
\caption{ Characteristics of the low-voltage supply system.
     	The voltages are measured at the power supply.
	The currents and the power dissipation refer to one sector connected to the TPC by
     	40~m long cables.  Due to the longer routing path and additional patch panels the 
	voltage drops and corresponding supply voltages are higher
	on the A side (numbers given in brackets).}
\label{infra:LV_overview}
	\vglue0.2cm
	\begin{tabular}{|l|c|c|} \hline
	\multicolumn{1}{|c|}
{Parameter}  &		Analog supply &		Digital supply		\\ \hline
Supply voltage &	$4.9~(5.2) $~V  & 	$4.1~(4.4) $~V		\\  %
Cable cross section  &	$ 150 $~mm$^2$  &	$300 $~mm$^2$    	\\
Current       &		$ 83$~A 	&	$ 133$~A   		\\
$\Delta V$ in~cables &	0.65~(0.92) V 	&	0.9~(1.2) V 		\\
Total power per sector &  407~(432) W  	&   	545~(585) W        	\\  \hline
\end{tabular}
\end{center}
\end{table}

\subsection{Chamber HV system}
\label{infra:Chamber_High_Voltage_system}

The  high-voltage system for the chamber anode-wire voltages is based on the models EDS 20025p\_204-K1 from 
ISEG, see Ref.~\cite{DCS:iseg}. These modules have 32 output channels grouped on two independent 16 channel boards. 
Each group of 16~channels is supplied by a common HV source with independent control of 
the individual output channels. 
For the auxiliary voltages (edge anode wires, cover electrodes and skirts, see 
Secs~\ref{chamb:Mechanical_structure} and \ref{fcage:skirts}) models EHQ 8005n\_156\_SHV 
and EHQ 8010p\_805\_SHV with 8~independent output channels are used. All power supplies reside in 
one crate and are remotely controlled 
via CAN bus from DCS (see Sec.~\ref{DCS:Subsystems:HV}).
Since the grouping of channels in the power supplies does not match the grouping of wire chambers in the
TPC (16 compared to 18) two types of patch boxes were built. The first type is located directly below the 
HV crate and redistributes the outputs of the
16~channel modules to multi-wire HV cables combining 18~channels plus 2~spares. They connect 
to the second type of patch box located directly on the SSWs of the TPC. From there individual HV 
cables connect to the wire chambers. 

For the supply of the edge anode wires and cover electrodes the single outputs of the HV units are fanned out 
in splitter boxes into
groups of 18~(plus 2 spares) and are also connected via multi-wire HV cables to patch boxes on the SSWs and
from there via individual cables to the chambers. In case of shorts (possible in particular in the last 
anode-wire circuits) individual bridges are installed in the splitter boxes allowing easy disconnection of
the affected circuit.

\subsection{Gate pulser}
\label{infra:Gate_pulser}

A gating grid is installed between the cathode grid and the drift region, to prevent positive ions, generated in the gas-amplification process, to drift back into the drift volume
and create distortions of the drift field  
(see Sec.~\ref{chamb:Mechanical_structure}). The gate is `closed' when a voltage of $V_{\rm G}\pm\Delta V$ is applied to alternating wires. The necessary value of $\Delta$V, given by  the magnetic field, the wire spacing, and drift field \cite{TDR:tpc}, is $\pm$90 V.
A gate pulser system has been devised to enable the rapid transition of the gating grid from the `closed'  to the `open' state upon the receipt of a trigger. 
 
It's general layout is shown in 
Fig.~\ref{infra:gate}. The three main components of the system are:

\begin{figure}[t]
	\centering
	\includegraphics[width={\linewidth},clip]{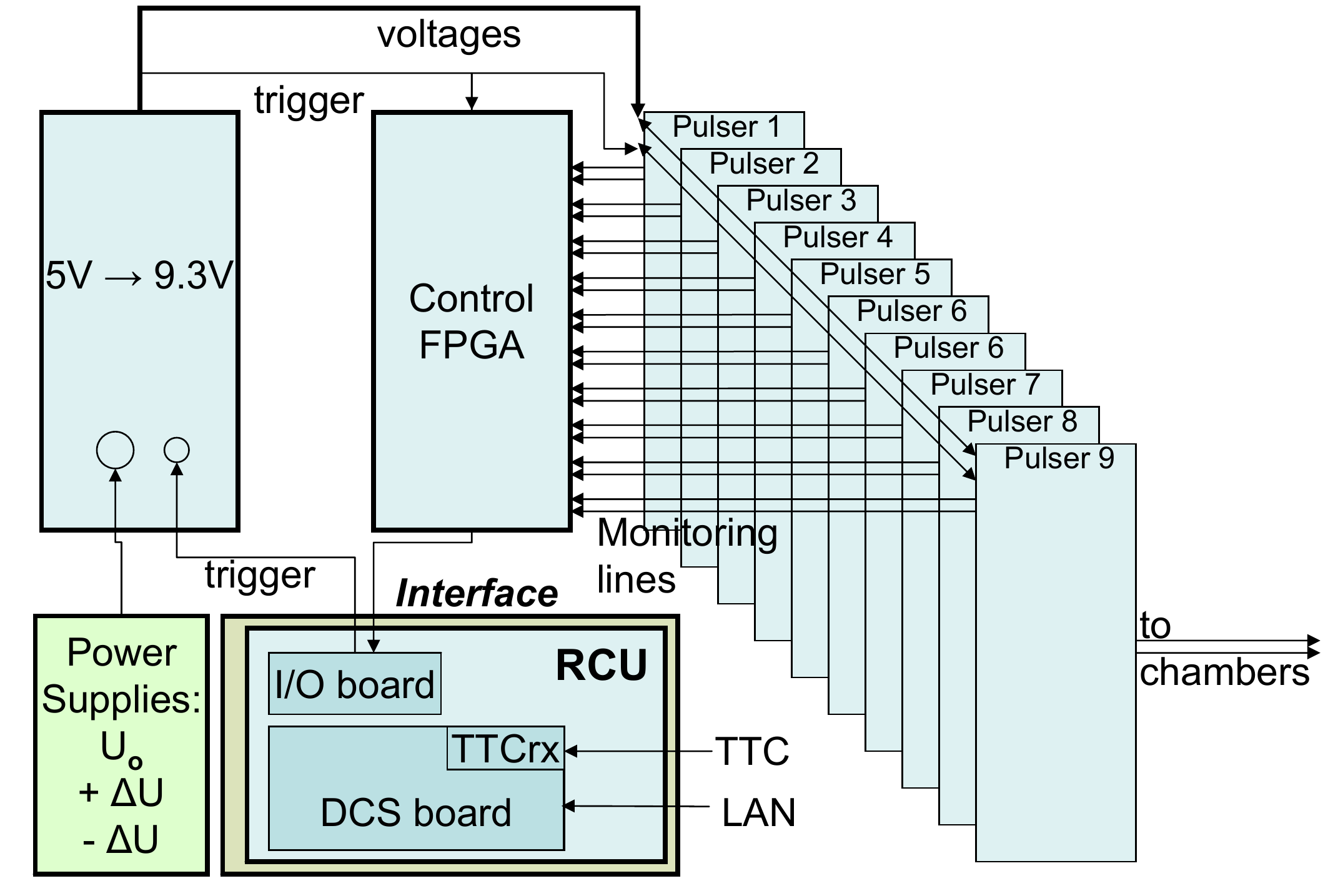} 
	\caption{Schematic setup of the gate pulser system.}
	\label{infra:gate}
\end{figure}

\begin {itemize}
\item
	\textsl{The interface to DCS and trigger system} (see Sec.~\ref{DCS:DCSbased}). 
	This is an RCU board as used in the readout chain with its add-on DCS
	board connecting to the network and the trigger system. The firmware 
	of the RCU is modified to control the state of the gate-pulser system (on/off),  select the trigger (L0 or L1)
	 and adjust the duration of the `open' state of the gating grid  via the DCS system. In addition the actual 
	states of the pulsers are controlled by monitoring  modules and in case of a malfunction an alarm is sent to 	DCS (see below).

\item
	\textsl{The individual pulser modules.}
	A detailed description of the pulser module design is given in \cite{TDR:tpc}. 
	Groups of nine pulsers are housed in 3U Euro crates (six in total) together with a
	monitoring module. This monitoring module is connected via opto-couplers to the  output lines of all pulsers and generates an error flag in case any of the pulsers does not change its state as a function of the trigger.
	In addition, each crate houses a 5~V power supply and a module to receive  the trigger signal and distribute
	it via the backplane.
	This module also receives the voltages $V_{\rm G}$ and $\pm\Delta$V and distributes them via the backplane 
	and, in addition, generates 9.3~V via DC-DC conversion which is needed by the gate-pulser modules. 
\item
	\textsl{Three dual-channel power supplies (Zentro model LD 2$\times$150/1 \cite{DCS:zentro}).}
	The supplies provide 
	$V_{\rm G}$ and $\pm\Delta$V. 
	IROCs and OROCs are supplied by separate channels. The power supplies are remotely controlled via RS232 	connections from the DCS (see Sec.~\ref{DCS:Subsystems:Gate}).
\end{itemize}

The gate-pulser system consists of two subsystems, one per TPC side.
The gating circuit has to put large voltage swings $\Delta$V on the
gating grid as fast as possible with minimal pick-up on the readout
electronics resulting from the transients.  To achieve this a three
step procedure is followed. First, the FET switches used in the pulser
output stage are grouped according to their switching time. Then, FETs
with similar characteristics are mounted together on the same
board. In the last step, the relative timing of the positive and
negative side of a pulser is tuned by a potentiometer such that a
minimum in residual signal is reached when summing the two outputs and
monitoring them on an oscilloscope.
 
For $\Delta$V = $\pm$90~V the amplitude of the pickup signal on the TPC reaches 80 to 100 ADC counts for the IROCs, 
70 to 90 ADC counts for the inner and 110 to 130 ADC counts for the outer part of the OROCs respectively. 
The duration
of the pickup signal is about 10 time bins and 21 time bins for IROCs and OROCs respectively.
Since these pickup signals are constant in time they are subtracted out in the pedestal subtraction procedure. 
The time between trigger and full transparency of the gating grid was measured to be $\unit{1.5}{\micro\second}$ \cite{Frankenfeld2002}.
The measured gating efficiency against positive ion feedback is described in Sec.~\ref{chamb:Tests}.

\subsection{Calibration pulser}
\label{infra:Calibration_pulser}

A calibration pulser system is used to measure the gain and timing calibrations of all readout channels. In addition, the calibration pulser can be used to make 
general tests of the readout chain such as to identify dead channels.
To generate a signal in the readout electronics, charge is injected onto the 
pads by pulsing the cathode wire plane. For homogeneous gas amplification and good position resolution, the mechanical tolerances  of the chambers (pad sizes and distances between wire 
and pad-planes) are quite narrow. 
Therefore the variation of charge injected onto different pads
is quite small and should, in the worst case, reflect only large-scale variations
across the chamber surface.
In addition, by controlling the cable length between calibration pulser and 
wire chambers, the timing of the readout can be calibrated.

The working principle of the calibration pulser system is described in \cite{TDR:tpc}.
 A step function is generated by an 
arbitrary waveform generator. Due to the capacitive coupling between the cathode-wire plane and the pad-plane the signal is differentiated and a narrow spike is 
detected by the readout electronics. The falling edge of the signal is 
always kept outside the readout time window of  $\unit{100}{\micro\second}$ to avoid unnecessary data flow.

\begin{figure}
	\centering
	\includegraphics[width={\linewidth},clip]{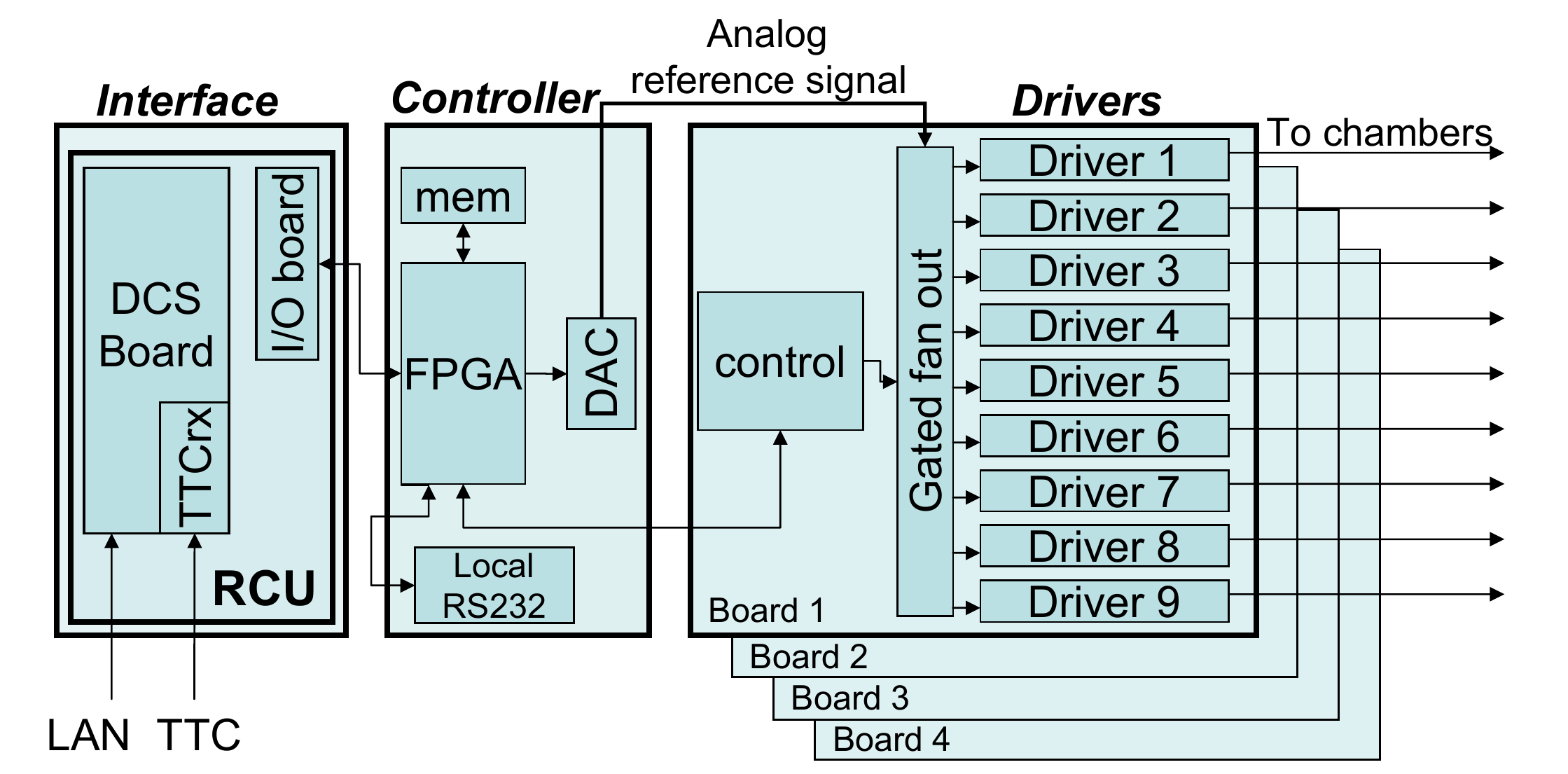} 
	\caption{Schematic setup of the calibration pulser system.
	Each side of the TPC is equipped with one such system.}
	\label{infra:calib}
\end{figure}

The schematic layout of the calibration-pulser system is shown in 
Fig.~\ref{infra:calib}. It comprises 3 main components:
\begin {itemize}

\item
	\textsl{The interface to DCS and the trigger system} (see Sec.~\ref{DCS:DCSbased}). 
	This is an RCU as described before for the gate-pulser system. The firmware 
	of the RCU  is modified and includes a 
	communication module with the control FPGA of the pulser.
\item
	\textsl{The controller module.}
	Its main components is a Field Programmable Gate Array (FPGA) acting as the controller for the pulser system. 
	Different pulse shapes can be stored as a sequence of amplitudes 
	in its memory. On command, the FPGA is set to an active pulse generating 
	state taking into account the desired amplitude, the pulse delay relative 
	to the trigger and the 
	pulse shape. In addition, the FPGA controls which driver channels are 
	activated. Besides a simple step function, generating one single signal per
	readout cycle, a sequence 
	of signals with identical amplitudes or a ramp (signals with increasing amplitudes) can be generated.
	A special fine delay unit shifts the position of the output signal in 1~ns steps.
	The standard step size is 50~ns given by the 20~MHz internal clock of the pulser.
	 
	\item
	\textsl{The output drivers.}
	They are connected via a high-impedance input to
	a DAC which is controlled by the FPGA. Their function is to drive the 50~$\Omega$ cables 
	connected to the individual TPC readout chambers. The 18 sectors of one side of the TPC are supplied by
	four modules with nine channels each.
	An additional control circuit on the driver cards allows the activation/deactivation of
	individual channels.
\end {itemize}

 There is one such system (6U Euro crate) per TPC side.

The pulser cables are terminated by
50 $\Omega$ resistors on the readout chamber side.
Due to the large capacitance of the wire plane this termination is far from ideal.
To avoid reflections back to the chambers the output stage of the calibration pulser has
 a 50~$\Omega$ impedance. This has the disadvantage that 
the effective output amplitude of the driver stage has to be twice the 
amplitude desired at the chambers (7.8~V, see below). Since the pulse 
width is about $\unit{110}{\micro\second}$ in order to keep the falling edge 
outside the readout time window, a considerable power is dissipated. 
As a consequence, each of the four~driver modules  has its own 18~V power 
supply.

\begin{figure}[t]
	\centering
	\includegraphics[width={0.95\linewidth},clip]{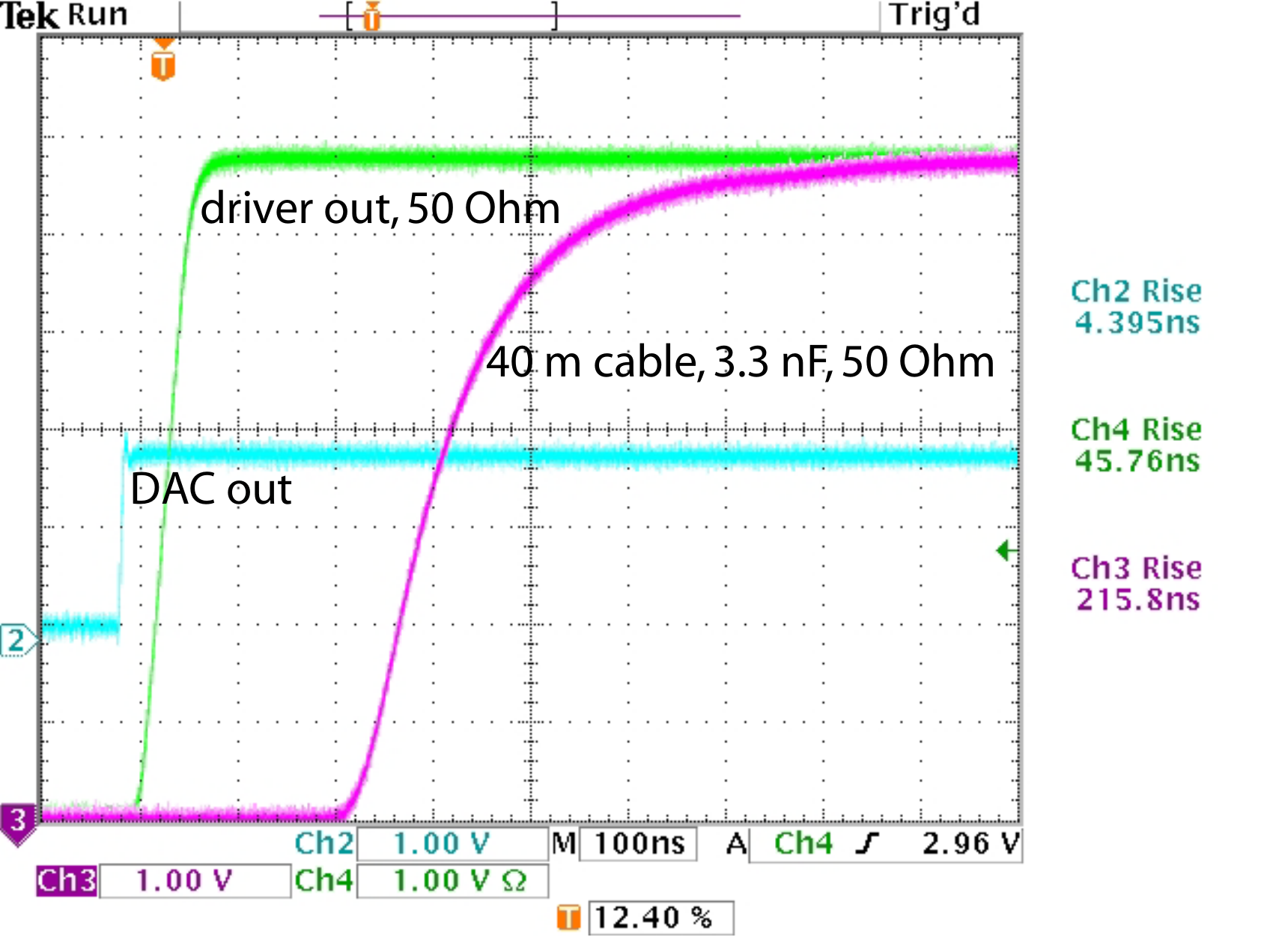} 
        \caption{Signal shapes in the calibration-pulser system: output of 
	the DAC (channel 2); output of the driver, terminated 
	by 50 $\Omega$~(channel 4);  
	test setup simulating a chamber using a 40~m cable with 3.3~nF and 
	50~$\Omega$~termination (channel~3). On the right side the measured rise times 
	are indicated.}
	\label{infra:pulser_signals}
\end{figure}

In Fig.~\ref{infra:pulser_signals} the shapes of the rising edges at 
various stages of the signal generation in the calibration pulser are 
shown. The simulation of an OROC chamber, with their rather large capacitive 
load, shows that the rise time and the width of the induced signal
is given by the charging process and much 
less by the shaping time of the preamplifier.

The output amplitudes of the drivers are equalized to $<1\%$.
The measured nonlinearity is $<1\%$.
The remote control of the amplitudes allows to check the linearity of 
individual readout channels. The maximum possible amplitude  at the 
chambers is 7.8 V. This corresponds roughly  to the dynamic range of the readout chain. 
The rather high output voltage is necessary since a considerable part of the signal strength arriving
at the cathode-wire plane is 
absorbed by the anode wire plane: the buffer capacities of 4.7~nF installed to stabilize the anode wire 
voltages represent an effective AC connection to ground.

\section{Detector Control System (DCS)}
\label{DCS}

The primary task of the Detector Control System (DCS) is to ensure
safe and reliable operation of the TPC. It provides remote control
and monitoring of all detector equipment in such a way that the
TPC can be operated from a single workplace (the ALICE
experimental control room at LHC Point 2) through a set of
operator interface panels. The system is intended to provide
optimal operational conditions so that the data taken by the TPC
is of the highest quality. More information about the TPC DCS can
be found in \cite{TPCint}.

\subsection{Overview}
\label{DCS:Overview}

\begin{figure}
  \centering
  \includegraphics[width=.96\linewidth]{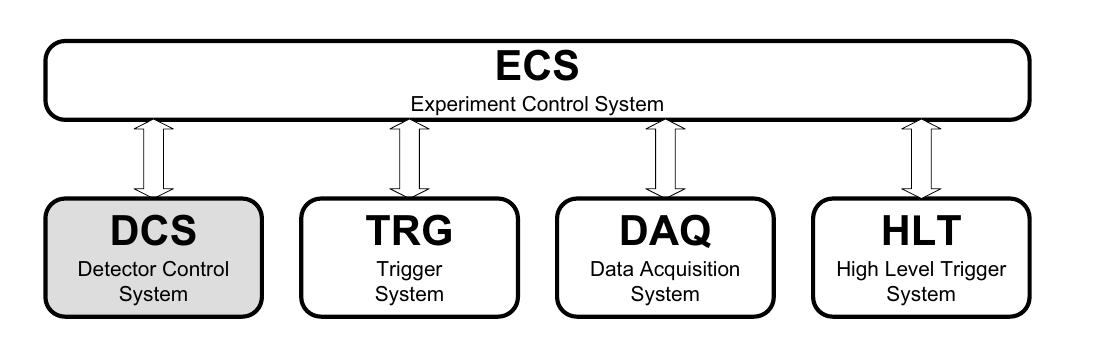}
  \caption{Overview of the ALICE online systems. The DCS interfaces
    to the other systems through the ECS.}
  \label{DCS:Fig:Online}
\end{figure}

The TPC control system is part of the ALICE DCS \cite{DCS:Augustinus2008}.
Like the other three ALICE online systems~\cite{TDR:daq} (Data Acquisition
system (DAQ), Trigger system (TRG) and High Level Trigger system (HLT)),
the ALICE DCS is controlled by the Experiment Control System (ECS), where the
ECS is responsible for the synchronization between the four systems. This is
schematically shown in Fig.~\ref{DCS:Fig:Online}.

\subsubsection{Hardware architecture}
\label{DCS:Overview:HWarch}

The hardware architecture of the TPC DCS can be divided in three
functional layers. The {\sl field layer} contains the actual
hardware to be controlled (power supplies, FEE ...). The {\sl
control layer} consists of devices for collecting and processing
information from the field layer and making it available to the
{\sl supervisory layer}. At the same time the devices of the
control layer receive commands from the supervisory layer to be
processed and distributed to the field layer. The equipment in the
{\sl supervisory layer} consists of personal computers, providing
the user interfaces and connecting to disk servers holding
databases for archiving data, etc. The three layers interface
mainly through a Local Area Network (LAN).

\subsubsection{Software architecture}
\label{DCS:Overview:SWarch}

\begin{figure}[t]
  \centering
  \includegraphics[width=.97\linewidth]{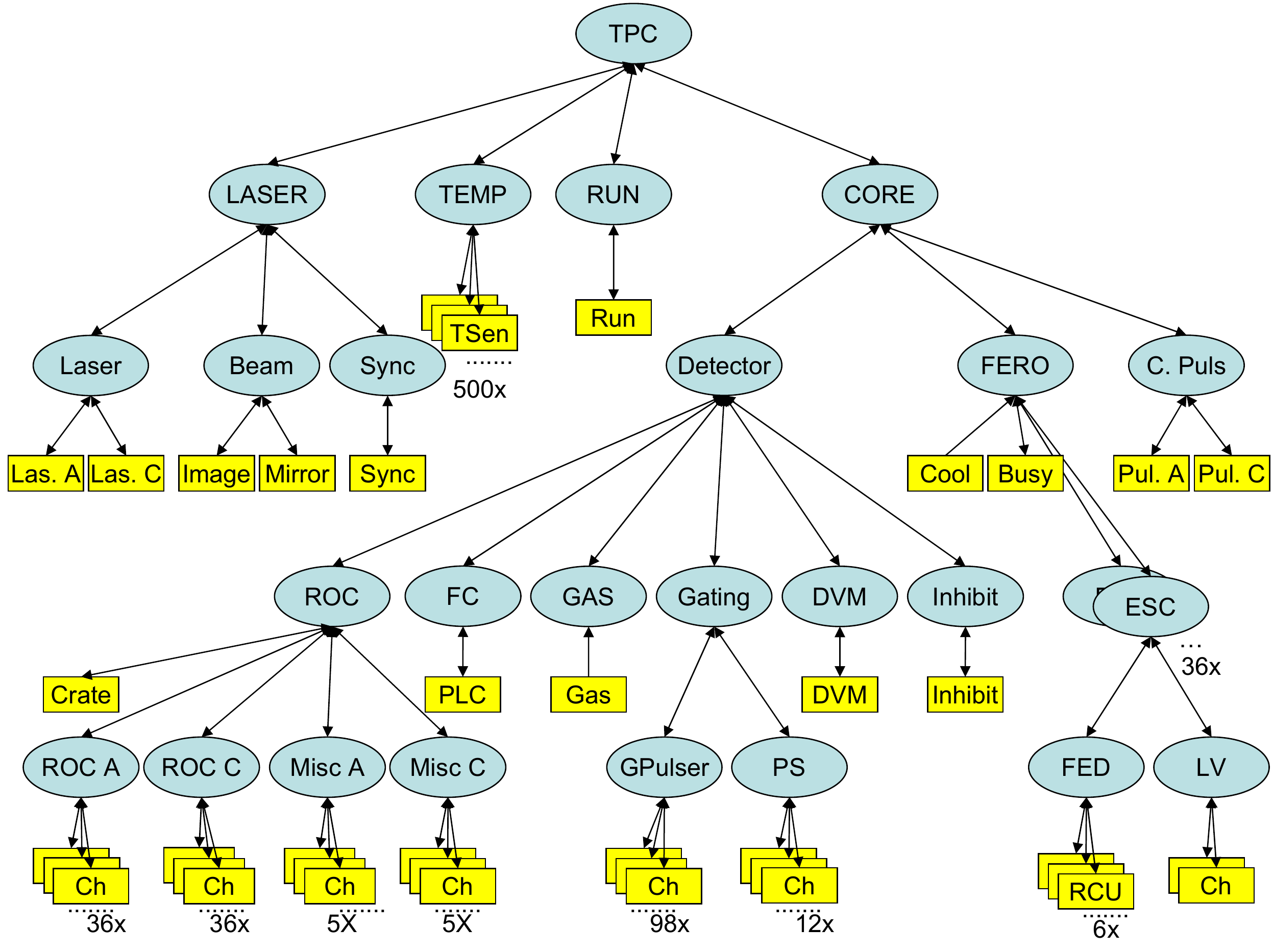}
  \caption{Overview of the software architecture of the DCS. The tree
    structure is build out of Device Units (boxes) and Control
    Units (ellipses).}
  \label{DCS:Fig:fsm}
\end{figure}

The software architecture is a tree structure that represents the
structure of the TPC, its sub-systems and devices. The structure,
as shown in Fig.~\ref{DCS:Fig:fsm}, is composed of nodes, each
having a single `parent', except for the top node called the `root
node'. Nodes may have zero, one or more children. There are two
types of nodes, the parent nodes are called Control Units (CU) and
the leaf nodes are called Device Units (DU). The control unit
controls the sub-tree below it, and the device unit `drives' a
device. The behavior and functionality of each control unit is
implemented as a finite state machine.

The control system is built using a `controls framework' that is flexible and
allows for easy integration of separately developed
components~\cite{DCS:Schmeling2006,DCS:SMI}. This framework includes drivers for
different types of hardware, communication protocols, and configurable components
for commonly used applications such as high or low voltage power supplies.
The framework also contains many other utilities such as interfaces to the
various databases (configuration, archiving), visualization tools, access
control, alarm configuration and reporting, etc.

\subsubsection{System implementation}

\label{DCS:Overview:impl}

The core software of the control system is the commercial SCADA
(Supervisory Controls And Data Acquisition) system PVSSII (Prozess
Visualisierungs und Steuerungs System) from the company
ETM~\cite{DCS:etm}. PVSSII is an object-oriented process
visualization and control system that is used in industry and
research as well as by the four LHC experiments. PVSSII is
event-driven and has a highly distributed architecture. The SCADA
System for the TPC is distributed over twelve Computers.

\subsubsection{Interfaces to devices}
\label{DCS:Interfaces}
\label{DCS:Subsystems}
\label{DCS:Subsystems:HV} \label{DCS:Subsystems:Gate}
\label{DCS:Subsystems:DriftHV} \label{DCS:Subsystems:LV}
\label{DCS:Subsystems:Temp} \label{DCS:Subsystems:Cool}
\label{DCS:Subsystems:Gas} \label{DCS:Subsystems:Laser}

Where possible, commercial servers using the OPC standard of
process control are used to interface the SCADA system to devices
like power supplies. OPC servers interface the  ROC high
voltage, the field cage high voltage, the front-end electronics
low voltage and the temperature monitoring system~\cite{TPCint}.
The power supplies for the gating grid of the readout chambers and
the lasers are controlled via a TCP/IP to RS232 bridges and RS232
interfaces. For non-commercial hardware the communication has been
developed based on the communication framework Distributed
Information Management (DIM\,\cite{DCS:DIM}); it is used in the
laser system, in the drift velocity monitor, the electronics
control and the pulser control.

\subsubsection{Interlock}
\label{DCS:InterLock}

The safety of the equipment and the detector is based on three layers of
interlocks:

\begin{itemize}
\item {\sl Internal interlock}. The internal mechanism of devices
(e.g.~power supply trip) are used wherever applicable to guarantee
the highest level of reliability and security. The threshold and
status of these interlocks are controlled by the SCADA system, but
their function is independent of the communication between
hardware and software. \item {\sl Cross system interlock}. The
interlocks between different subsystems are realized by open loop
contacts. Programmable Logical Device (PLC) systems are used to
delay the signals or to give the possibility to enable or disable
these interlocks. \item {\sl Software interlock}. The software
interlocks are realized in the supervisory layer. They rely on the
communication between the hardware and the SCADA system. Therefore
they are only used to prevent the system from unwanted but not
harmful events like switching off the power supplies under full
load. The safety of the equipment does not rely on the software
interlocks.
\end{itemize}
Internal interlocks are used for the ROC high voltage, the field
cage high voltage, the front-end electronics low voltage, the
cooling and the gas system. External interlocks are implemented
for the field cage high voltage, the front-end electronics low
voltage and the cooling system. Software interlocks are used for
the ROC high voltage, the front-end electronics low voltage and
the front-end electronics \cite{TPCint}.

In addition to the interlocks the alert system of the SCADA system
is set-up to inform the shifter of unusual or potentially
dangerous situations.

\subsection{Electronics control}
\label{DCS:Electr} \label{DCS:DCSbased}

\begin{figure}[t]
  \centering
  \includegraphics[width=.97\linewidth]{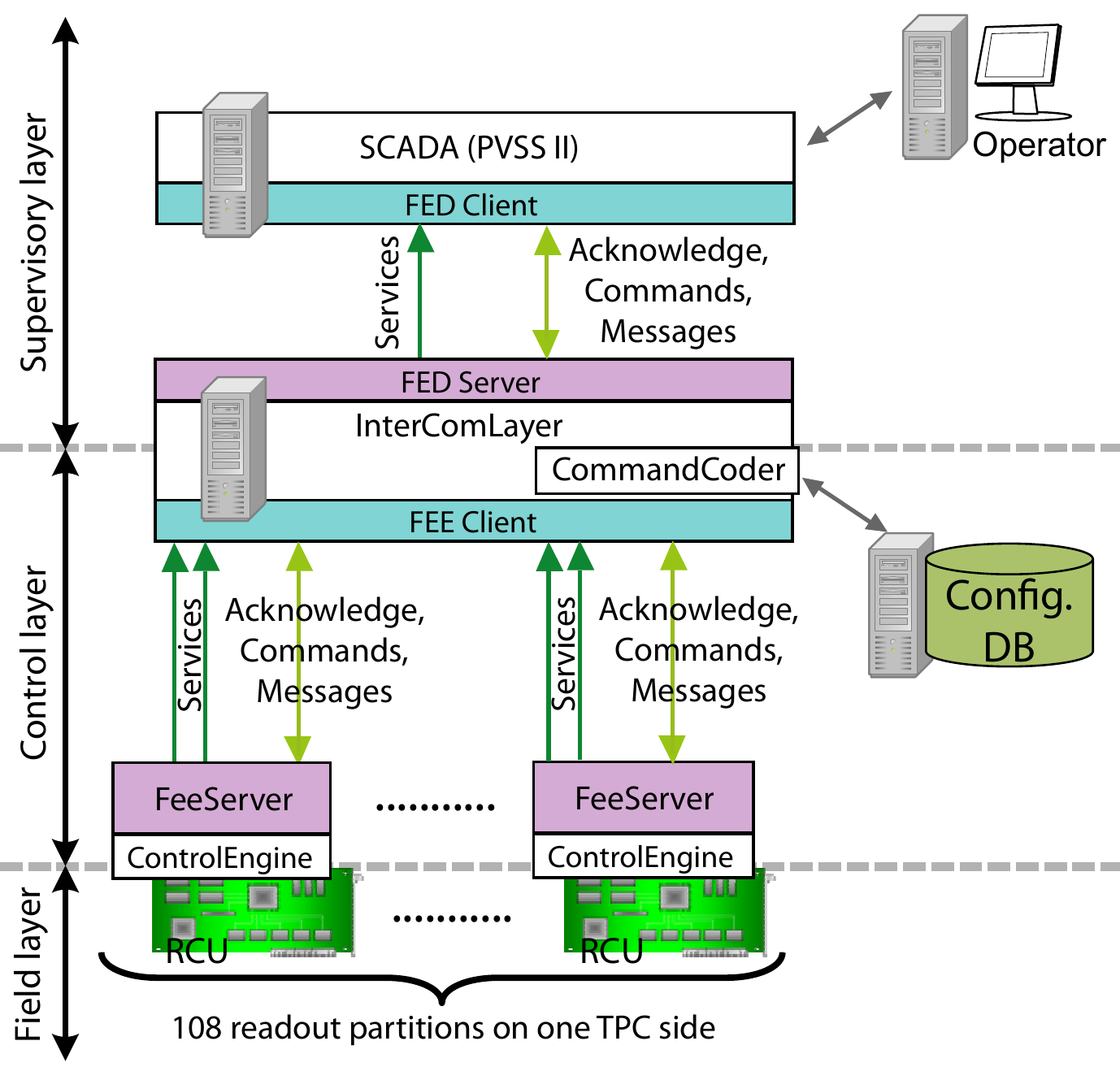}
  \caption{Schematic setup of the DCS for the TPC FEE. The operator in the
    ALICE control room can open graphical user interfaces to display the
    monitoring values as they are provided by two PVSS systems, corresponding
    to the two sides of the TPC. The operator can also send commands to the
    control and field layer.}
  \label{DCS:Fig:icl}
\end{figure}

The three functional levels (as described in
Sec.~\ref{DCS:Overview}) for the DCS for the FEE are shown in
Fig.~\ref{DCS:Fig:icl}. The 216 RCUs and the FECs with the ALTRO
chips (Sec.~\ref{elect}) are part of the field layer. The
interface between the different layers relies on communication
using DIM. Independent, but identical setups are used for the two
sides of the TPC.

The FEE communication software for the TPC is described in \cite{Bablok:2006zj}
and is used for many detectors in the ALICE experiment. A front-end electronics
server (FeeServer) interfaces the hardware, publishes monitoring data and receives
commands and configuration data. The software can be used for different hardware
devices, since it has been divided into a device-independent core and an
interface to the hardware dependent functionality ({\em ControlEngine}).

The TPC FeeServer runs on each of the 216 DCS boards (Sec.~\ref{elect})
that control the corresponding RCU and FECs. It reads temperatures, voltages and
currents on the FECs and a number of status and configuration registers on the RCU
and publishes the acknowledge and message channels. For one readout partition with 25
FECs (for the full TPC) about 165 (30\,000) services are published.

The InterComLayer gathers, buffers and bundles monitoring data from the 108
FeeServers and redistributes them to the SCADA system. The services of each
FeeServer are grouped in dedicated service channels, reducing the number of
services to which the SCADA system has to subscribe to 216 per TPC side. Since
the performance of the DIM client in the SCADA system is limited to less than
1\,kHz, an algorithm is implemented in the InterComLayer that effectively reduces
the network traffic to below that limit, while still ensuring that the latest
monitoring information is made available. The implementation of the hardware
dependent functionality is moved to a configuration database and a separate
software component that retrieves the configuration data from the database
({\em CommandCoder}).

\subsubsection{Front-end monitoring}
\label{DCS:Electr:FEE:Moni}

The SCADA system implements the graphical user interfaces to
display monitoring data. For the TPC FEE this data is mainly the
temperatures, voltages and currents from the FECs and status
information of the FEE. In the commissioning phase this
functionality allowed the system to identify unwanted voltage
drops and insufficiently cooled FECs.

\subsubsection{Front-end configuration and control}
\label{DCS:Electr:FEE:Conf}

There are about 5 million configurable parameters for each TPC configuration.
The structure of the TPC configuration database follows the structure of the
hardware. The parameters for the ALTROs, FECs and RCUs are collected in
dedicated tables which are linked via relational tables.

The configurations for the FEE are subdivided in 216 blocks (one block per readout
partition). The configurations change over time due to disfunctional or replaced
hardware or due to changing hardware behavior. One complete configuration needs
$\approx$\,30\,MByte of space in the configuration database, not including the
ALTRO pedestal memories. It is expected that the TPC will need 20 to 100 different
FEE configurations to be stored in the database.

Only simple configuration commands containing a parameter which
describes the configuration type are passed from the supervisory
layer to the control layer. The actual complex configuration
process takes place mainly in the control layer. The CommandCoder
assembles queries to the configuration database, which are
executed serially and retrieve the configuration data for each
readout partition. The InterComLayer sends the data to the
FeeServers, which execute the commands that are contained in the
configuration data block. The time to compile, send and execute
the configurations is of the order of 20\,s for 108 readout
partitions. A new version of the InterComLayer will make use of
parallel invocation of the CommandCoder, which should effectively
speed up the configuration process.

The FeeServer stores the latest retrieved configuration data locally on the DCS
board making possible a very fast reconfiguration of the FEE, without
the overhead of querying the configuration database.

The TPC FEE can also be configured via the optical Detector Data Link (DDL),
bypassing the DCS and the FeeServer.
This method is used for configuring the ALTRO pedestal memories, where the
volume of the configuration data is about 700\,MByte for the full TPC.
Freshly calculated pedestal and noise values can be used to configure the FEE
directly after pedestal data has been acquired.

\subsection{Interfaces to experiment control and offline}
\label{DCS:IF}

The DCS interfaces to  ECS, which steers the whole experiment. It also
interfaces to the offline data analysis framework, since for a proper
interpretation of the recorded data the configuration and status of the FEE
and the environmental conditions in the cavern and on/inside the detector
have to be known.

A run type defines how the FEE will be configured and which
subsystems are to be activated at the start of the run. For the
TPC, four run types are of relevance. A {\em Physics} run is the
general run type for recording data with beam--beam collisions. The
remaining run types are used for the calibration of the detector.
For a {\em Pedestal} run the FEE is configured to read out black
events (no zero suppression) in order to analyze the pedestal and
noise values in all channels. For {\em Laser} ({\em Pulser}) runs
the TPC laser system (the calibration pulser system) is activated.
The run type is propagated from ECS to DCS, together with the run
number and the list of readout partitions which will be read out
in the upcoming data taking. This is needed in order to properly
configure the BusyBox.

\section{Commissioning and calibration}
\label{calcom}

\subsection{Calibration requirements}
\label{calcom.requirements}

The main goal for the calibration procedures is to provide the information needed
for the offline software to reconstruct the position and energy of
clusters with sufficient precision so that the design performance can be
achieved (see Sec.~\ref{perform}).

To cope with the huge amount of raw data (about $\unit{750}{\mega\byte\per event}$), zero
suppression is performed on the level of the FEE. The first
step in the calibration chain is to obtain the parameters that are
uploaded to the FEE and used to process the raw data online. Because
the online zero suppression uses a threshold for removing noise, the noise design value is included in the calibration
requirements. The requirements are:
\begin{itemize}
\item \textsl{Noise.} The TPC and FEE were designed to have an
  overall Equivalent Noise Charge (ENC) better than 1000\,$e$ in the
  whole TPC, corresponding to $\approx1$\,ADC channel. With the nominal
  gain ($2\times 10^{4}$) this should give a signal\footnote{Here signal means the
  maximum value of the charge for a pad-timebin cell in a cluster
  (often denoted $Q_{\text{max}}$).} to noise ratio of about 20 in the
  IROCs and 30 in the OROCs for minimum ionizing particles. This still
  leaves a large (order of 30) dynamic range for the larger ionization
  of low-energy tracks.
\item \textsl{Gain homogeneity.} The gain has to be calibrated to
  better than 1.5\,\% over all pads. The time dependence of the residual
  gain variations can then be obtained from precision measurements of
  temperature and pressure variations.
\item \textsl{Space-point resolution.} The systematic contribution of
  each of the following effects to the space-point resolution has to
  be kept below $\unit{200}{\micro\metre}$:
  \begin{itemize}
  \item \textit{Drift velocity.} The drift velocity has to be known to a
    precision better than $10^{-4}$. This results from considering 
    a resolution of $\unit{200}{\micro\metre}$ over the full drift length ($\unit{250}{\centi\metre}$).
  \item \textit{Alignment.} The residual rotation (translation) after alignment
    has to be kept below 0.1\,mrad ($\unit{200}{\micro\metre}$).
  \item \textit{$E\times B$ effects.} The $E\times B$ effect was estimated from simulations using the
    measured magnetic field to be of the order of $\unit{8}{\milli\metre}$ for the
    full drift length. The precision of the correction therefore has
    to be on the level of 2\%.
    \end{itemize}
\end{itemize}

The alignment and $E\times B$ calibration will be described in detail in a future
publication dedicated to the calibration and performance of the TPC.

\subsection{Commissioning}
\label{calcom.commissioning}

\subsubsection{Commissioning phases}

After the TPC was assembled, the commissioning activities began.
The main activities in the commissioning procedures were:
\begin{itemize}
\item \textsl{Phase 1:} First tests at the surface (2006). For these tests
  limited services were available so that only 2 sectors could be
  powered at a time. No zero suppression was applied to the
  data (black events). 
\item \textsl{Phase 2:} First commissioning in the ALICE experimental area underground (Dec.\ 2007). 
  A full side (side C) was successfully operated with final services attached and online zero
  suppression.
\item \textsl{Phase 3:} Commissioning under final running conditions
  (Mar.\ to Sept.\ 2008). The TPC was operated stably over several months. Different run types (see Sec.~\ref{calcom.commissioning.runs}) were
  successfully implemented, and extensive calibration data was taken.
\end{itemize}
The bulk of the calibration data was taken in phase~3, and most of the
results shown in the following sections are from this phase. 

\subsubsection{Data sets}
\label{calcom.commissioning.runs}

\begin{itemize}
\item {\sl Pedestal runs.} This run type, where the zero suppression
  is switched off, is used to determine the
  pedestal and noise for all readout channels. The extracted values
  are used to perform zero suppression with the FEE.
\item {\sl Calibration pulser runs.} The cathode wire grids of the ROCs can be pulsed to determine the response of the electronics chain (see
  Sec.~\ref{infra:Calibration_pulser}). The extracted data is used
  for drift-time calibration as well as for the detection of dead
  channels and floating wires.
\item {\sl Laser runs.} The TPC laser system (see Sec.~\ref{laser}) provides
  well defined straight tracks within the TPC volume. In addition,
  scattered laser light creates photo-electrons on the central
  electrode. Both sets of data are used to determine the drift velocity
  and study the inter-chamber alignment.
\item {\sl Cosmic runs.} Cosmic runs were taken primarily with the
  ACORDE scintillator array~\cite{ALICEjinst} as trigger. 
\item {\sl Krypton runs.} To determine the gain of each individual 
readout channel with high precision, krypton gas was released into the
TPC gas system. Krypton data were accumulated with
cosmic-triggered events for three different anode wire voltage settings (gains).

\end{itemize}

\subsection{Electronics calibration}
\label{calcom.fee.calib}
\subsubsection{Pedestal and noise determination}

\begin{figure}
  \centering
  \includegraphics[width=0.8\columnwidth]{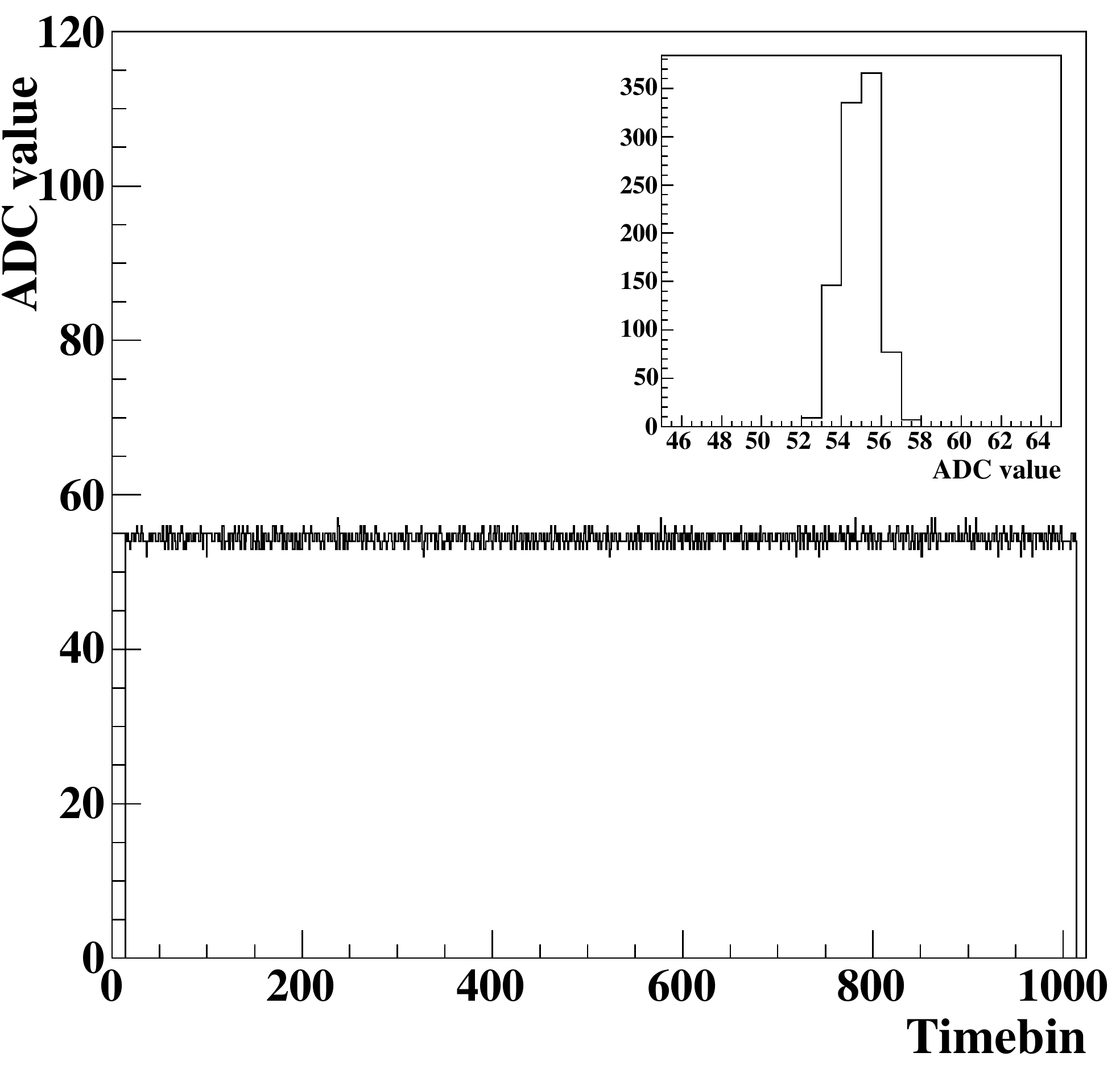}
  \caption{Typical electronic baseline of one channel. The insert shows its distribution.}
  \label{fig:calcom.pedestal}
\end{figure}

In pedestal runs the
electronics baseline (pedestal) and its width (noise) are determined.
The typical baseline of one electronic channel is displayed in
Fig.~\ref{fig:calcom.pedestal}. The insert shows its
distribution. The Gaussian mean defines the pedestal value, while the
sigma corresponds to the noise.  The measured pedestal and noise
values are stored in the OCDB~\cite{ALICEjinst} for offline reconstruction usage, as
well as on the LDC machines to be uploaded into the FEE. The
zero-suppression threshold at this stage is set at three sigma of the baseline
distribution.

\paragraph{Results of noise measurements}

\begin{figure}[t]
   \includegraphics[width=0.8\columnwidth]{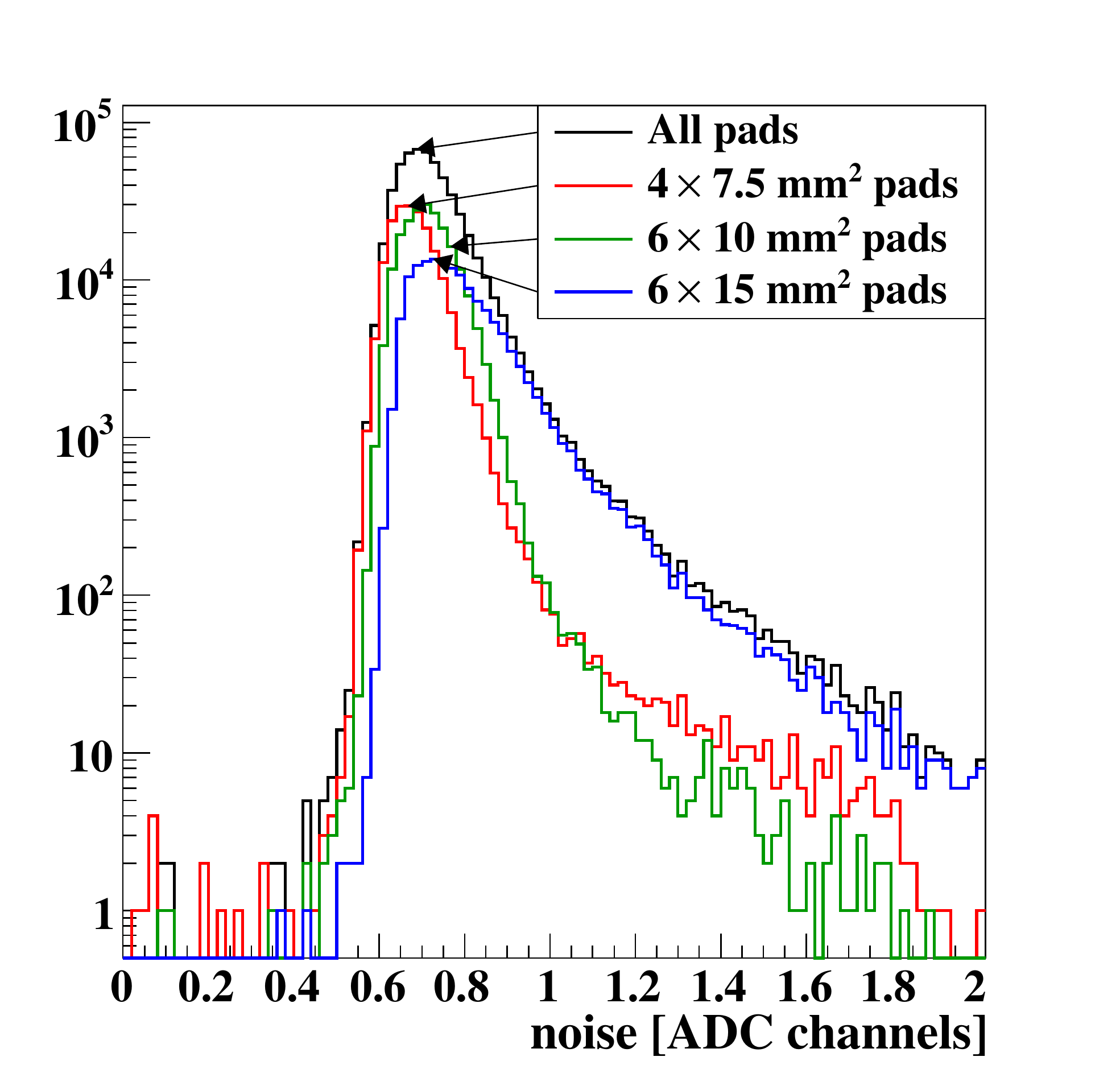}
   \caption{Noise distribution; for all chambers and separately for the different pad sizes.}
   \label{fig:calcom.noise_dist}
\end{figure}

In Fig.~\ref{fig:calcom.noise_dist} the histogrammed noise 
distributions obtained during the final stage of commissioning 
are shown for all pads and for the different pad sizes
separately. Truncated mean values in the range
$0-2$~ADC channels and the corresponding RMS of the distributions are
summarized in Tab.~\ref{tab:calcom.noise}. 

\begin{table}[t]
  \centering
  \caption[Truncated mean and its RMS of the noise distribution for the different pad sizes]{Truncated mean and RMS of the noise distributions in Fig.~\ref{fig:calcom.noise_dist} (see text). Values are given separately for the different pad sizes. In addition, the fraction of pads above 1\,(1.5)\,ADC channels is shown.}
  \label{tab:calcom.noise}
  \begin{tabular}{|c|c|c|c|}
    \hline
    {pad size [mm$^2$]} & {mean} & {RMS} & {$>$ 1 (1.5) \,ADC [\%]}\\
    \hline
    4 $\times$ 7.5 & 0.686 & 0.068 & 0.419 (0.072)\\
    6 $\times$ 10  & 0.719 & 0.064 & 0.244 (0.015)\\
    6 $\times$ 15  & 0.792 & 0.127 & 5.692 (0.420)\\
    \hline
  \end{tabular}
\end{table}

Only 1.7\,(0.14)\,\% of the channels show noise values above
1\,(1.5)\,ADC channels. The largest pads with 5.7\,(0.4)\,\%
contribute most to this value. The mean ENC in the TPC is about
730\,$e$. 

\begin{figure}[t]
  \centering
  \includegraphics[width=0.8\columnwidth]{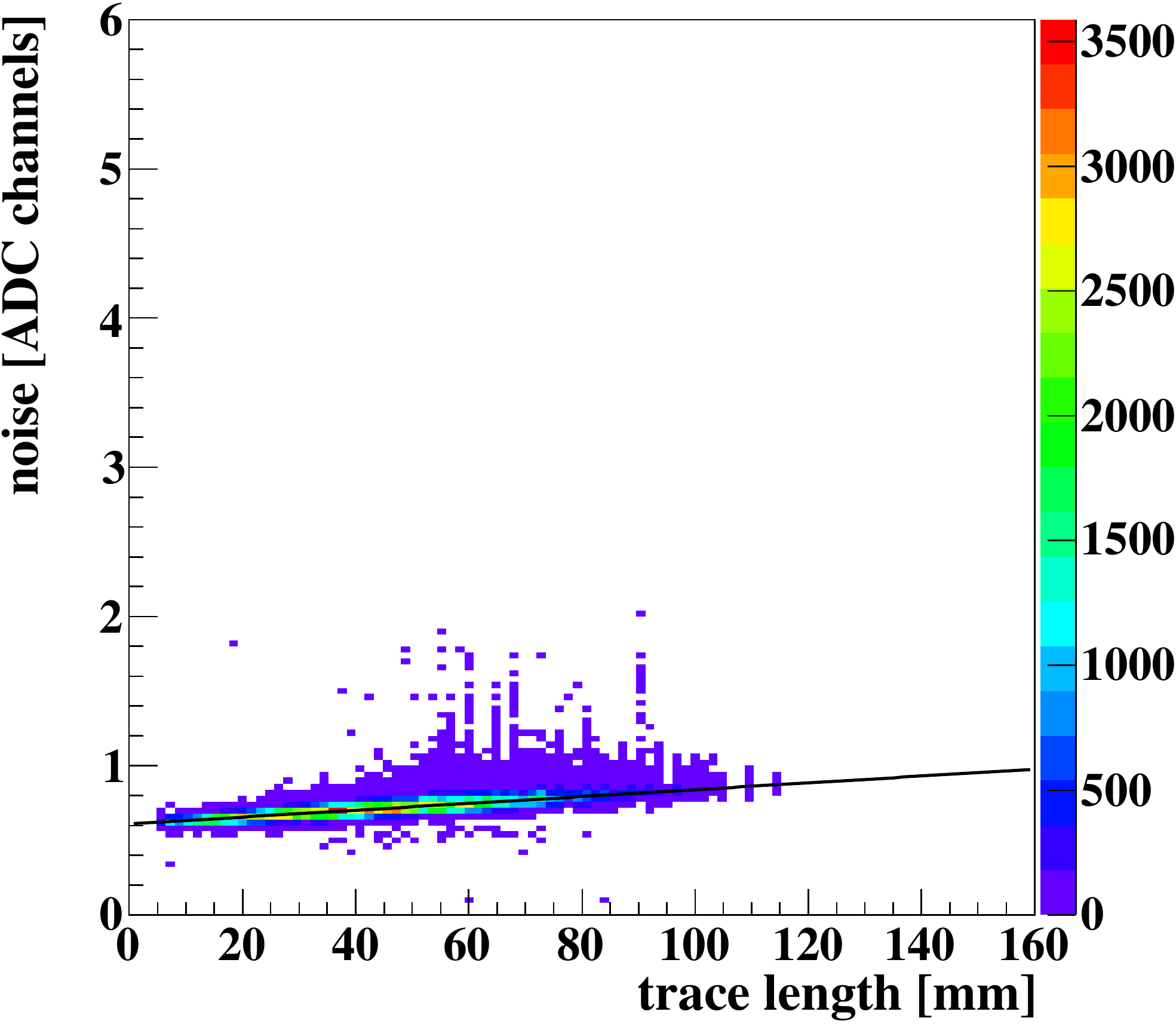}\
  \caption{Correlation of noise and the trace length on the pad-plane PCB board. A straight line was fit to the data.}
  \label{fig:calcom.noise_tracklength}
\end{figure}

A systematic variation in the noise level is observed increasing from the
center of each readout partition towards its edges~\cite{Wiechula2009}.
This variation is directly related to the variation in length of the traces on the
pad-plane PCB board, which connects the pads with the connectors on
its back side. With the trace length the capacitance at the input of
the charge-sensitive preamplifier/shaping chips (PASAs) rises and
hence the noise level rises. Figure~\ref{fig:calcom.noise_tracklength} shows
the dependence of the noise on the trace length for the medium-sized
pads.

A detailed discussion on noise measurements in the TPC can be found in~\cite{Wiechula2009}.

\paragraph{Improvements to decrease the noise level}
Measurements during the first commissioning in 2006 showed that a
large fraction of pads ($\approx$10\,\%) had noise values above 1\,ADC
channel. In the largest pads this fraction was $\approx$24\,\%. Two
modifications reduced the noise to the desired level. First, the start of readout for groups of channels was desynchronized to minimize the peak current drawn by the FEE (reduce ground-bounce
effects). Secondly, the grounding scheme for the FECs was revised and
optimized in terms of the noise behavior.

\subsubsection{Tail-cancellation filter parameter extraction}

The signal from a gas detector with a MWPC readout is often
characterized by a long tail with a rather complex shape. Detailed
simulations can be found
in~\cite{Mota:2004kq,Ross09}.

A Tail-Cancellation Filter (TCF) (see Sec.~\ref{elect:ALTRO}) is implemented in
the ALTRO chip for filtering the digital signal after the initial
baseline subtraction so that zero suppression can be applied in an
efficient way. The TCF is
based on the approximation of the tail by the sum of exponential
functions. The parameters for the TCF were extracted from non-zero
suppressed cosmic data. The method is described in detail in~\cite[chap.3]{Rossegger:1217595}.

\begin{figure}[t]
  \centering
  \includegraphics[width=0.9\columnwidth]{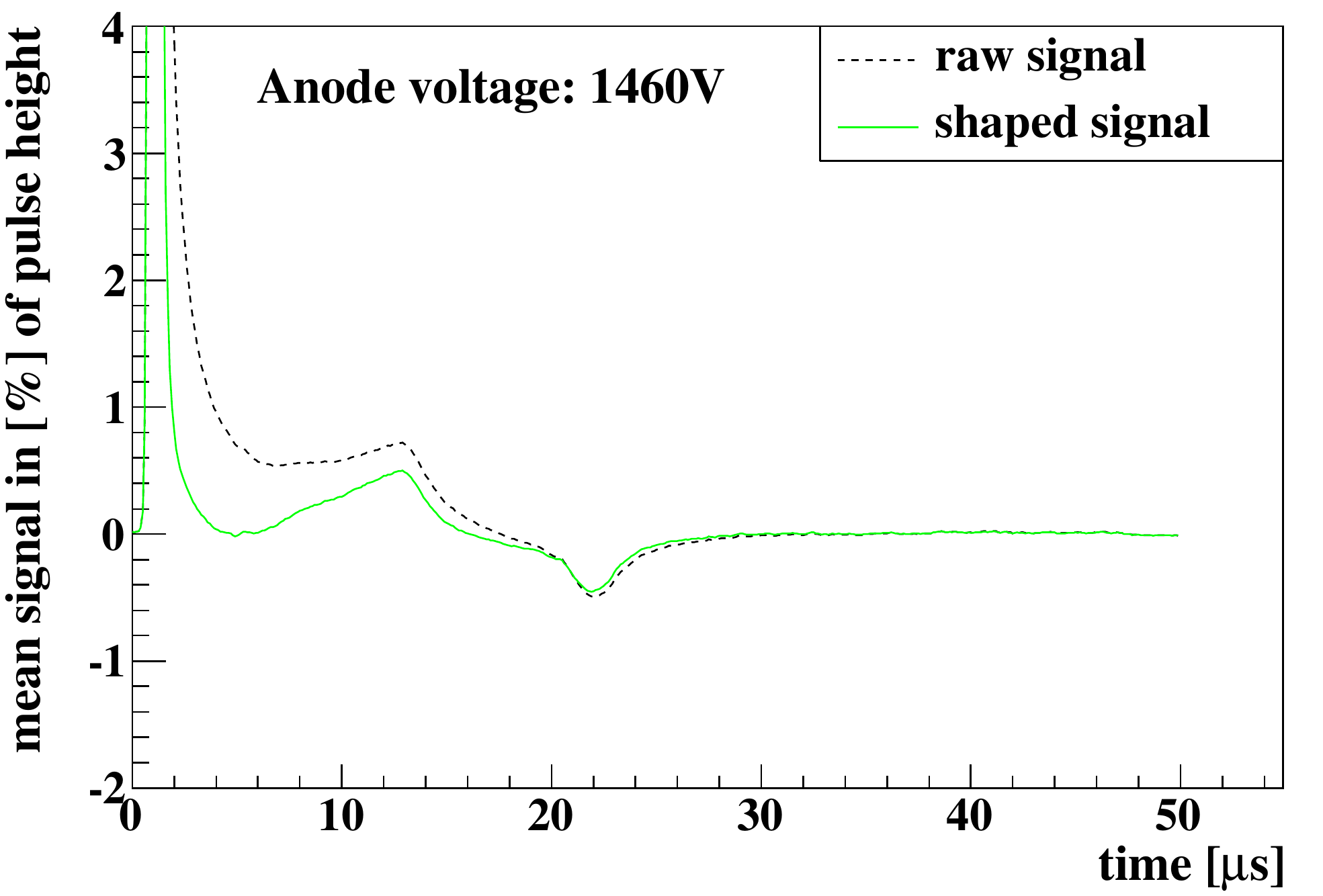}
  \includegraphics[width=0.9\columnwidth]{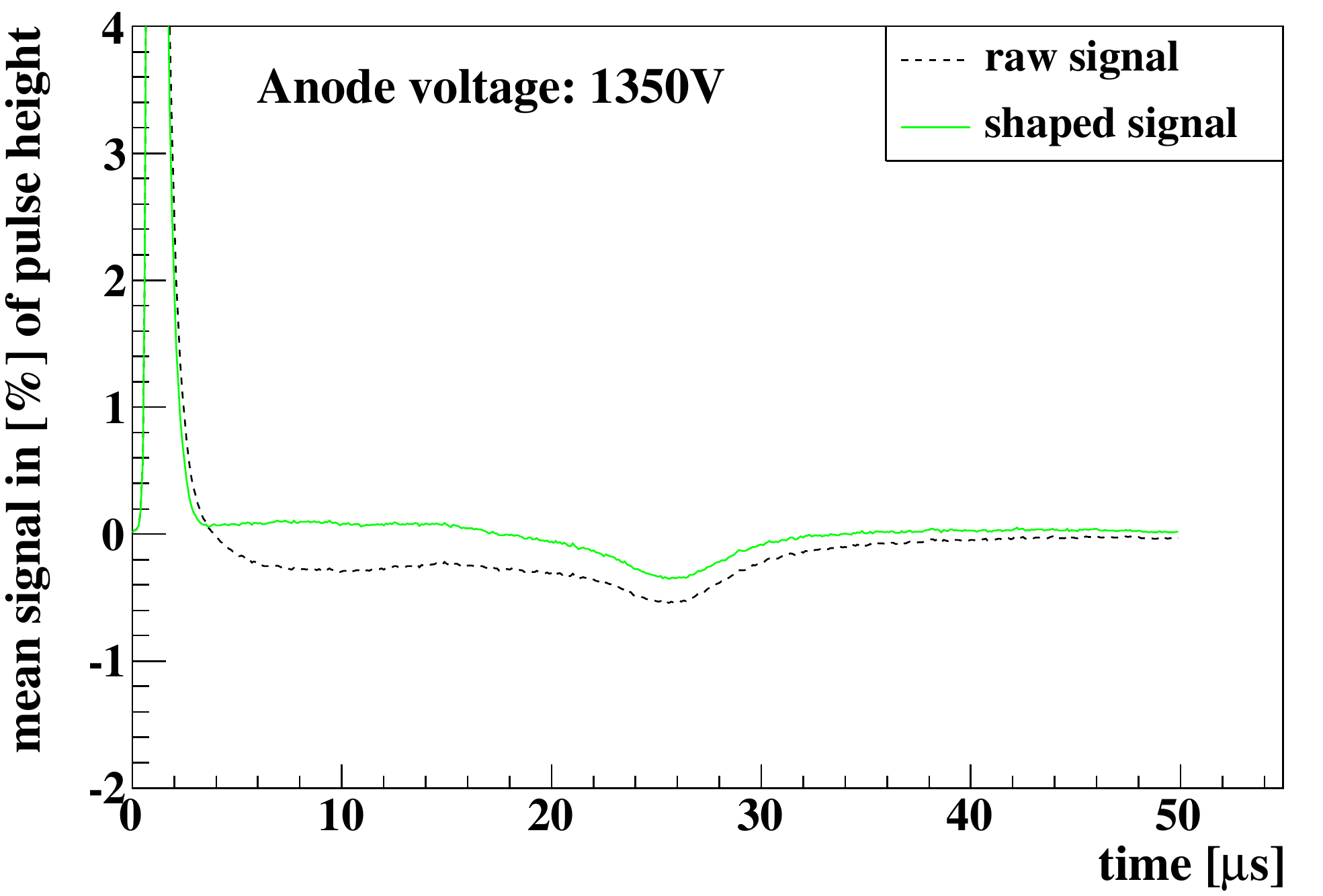}
  \caption{Mean pulses in the IROC at two different gains, before
  (raw) and after (shaped) the tail-cancellation filter was
  applied.}

  \label{fig:calcom.TCF.gain}
\end{figure}

Figure~\ref{fig:calcom.TCF.gain} shows an example of mean pulses in the IROCs at
two different gains before and after the TCF was
applied. Different gain (anode voltage) settings were applied to achieve a detailed
ion-tail characterization of the MWPCs\footnote{With
  increasing gain, the avalanche size around the multiplication wire
  increases. Therefore, the ratio between the number of ions, which go
  to the pad, and the number of ions, which go to the cathode,
  changes~\cite{Ross09}.}. At an anode wire voltage of 1350\,V the ion tail
reveals an immediate undershoot after the main peak of the signal
(Fig.~\ref{fig:calcom.TCF.gain} (bottom)).
 
The pulse-by-pulse fluctuations were found to have a bigger impact
than the pad-per-pad systematics. Differences between medium and long
pad sizes in the OROC were not significant either. The geometrical
differences between IROC and OROC as well as the different gain
settings were found to have the biggest impact on the ion-tail
shape. Therefore, shaping parameters for the different sectors were
found to be sufficient. Since small gas composition changes do not
have a major influence on the shape of the tail either, it is foreseen
to redo the shaping-parameter-finding-procedure just once a year in
order to study the influence of long-term variations like aging
effects on the chambers.

The TCF was not applied for any of the results presented in the
following sections. When it is employed, in the future, we anticipate that we will make the time signal more
symmetric.

\subsection{Gain calibration}
\label{calcom.gain}

\subsubsection{Krypton calibration}
\label{sec:krypton}
The krypton calibration method was developed by the ALEPH
\cite{ALEPH,Blum_alephKr} and DELPHI \cite{calcom:delphiKr}
collaborations. It was also successfully applied by the NA49 collaboration
\cite{NA49}. The advantage of this method is that it provides
an absolute calibration of the total gain (electronics and gas
amplification) for each pad.

The pad-gain factors are measured from decay clusters of radio-active
krypton ($^{83}_{36}$Kr) which is released into the TPC gas. Dedicated krypton
data taking is planned once a year.

\paragraph{Analysis}

Data samples were collected for three different gains (anode voltage settings) over one week of data taking. 
In the following paragraphs only results for the case of 1350\,V and 1550\,V applied to the IROCs and OROCs anode wires, respectively (quoted as `nominal voltage setting') are presented. 
The gain for the nominal voltage setting is about 6500--7500. A gain curve can be found in~\cite{TPCint}.
11.3 million events were collected where each event had roughly 80 krypton clusters.

A dedicated krypton-cluster finder is used to reconstruct the krypton decays.
The reconstructed cluster charge is associated with the pad with the maximum amplitude.
For each pad the charge spectrum is accumulated. The pad gain is then defined as the mean
of a Gaussian fit to the main peak (41.6\,keV) of the charge spectrum. The error on the mean 
obtained from the fit is of the
order 0.2\,\% on the single pad level, which is well below the requirements of 1.5\,\% specified
in Sec.~\ref{calcom.requirements}. 

\begin{figure}[t]
  \begin{center}
    \includegraphics[width=0.8\columnwidth]{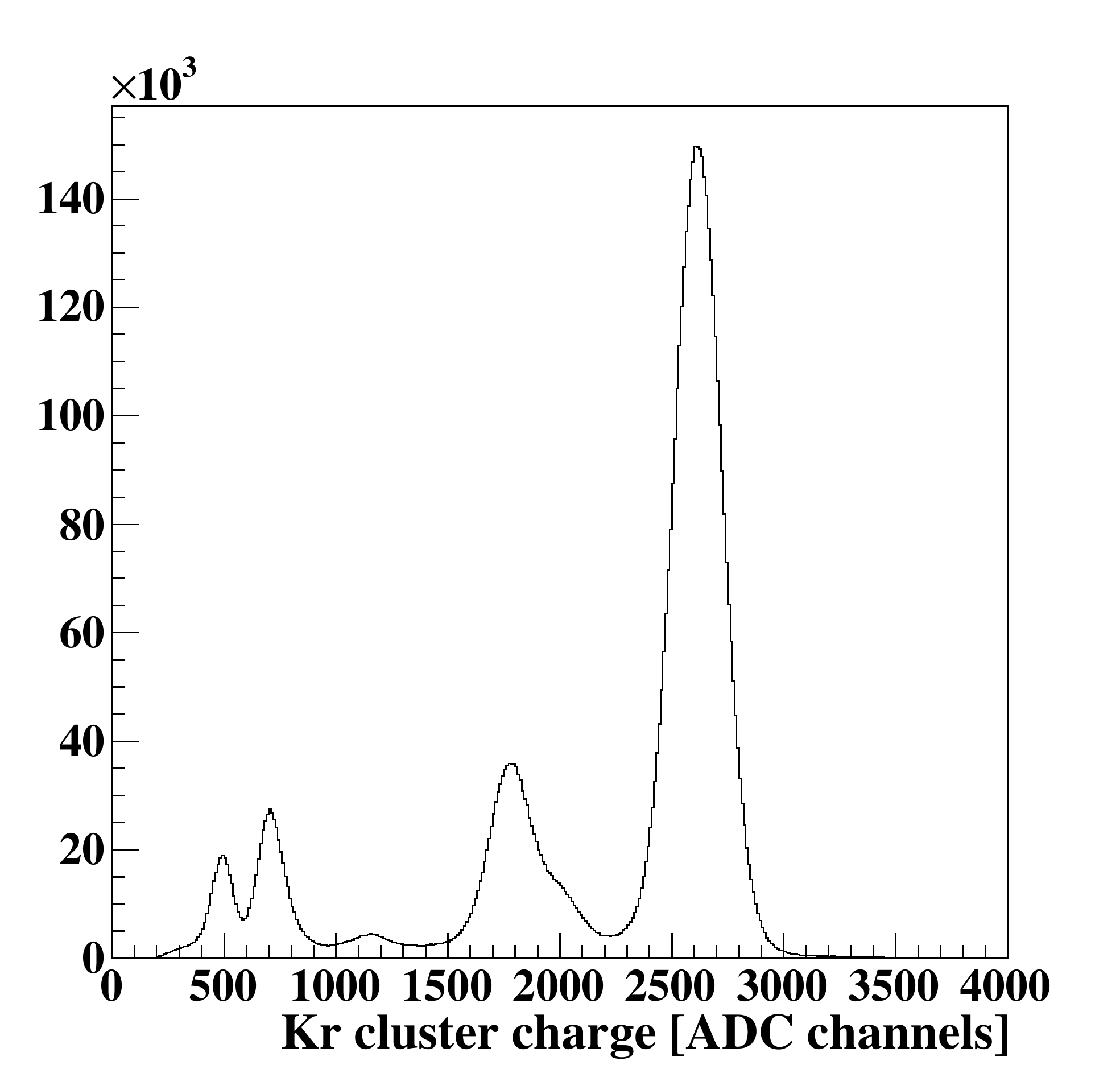}
  \end{center}
   \caption[Krypton spectrum]{Krypton spectrum of all OROCs at nominal gain.} 
  \label{fig:calcom.krspectrum}
\end{figure}

Figure~\ref{fig:calcom.krspectrum} shows an example of the accumulated charge spectrum of all OROCs, 
corrected for the pad-by-pad variations.
The resolution of the main peak
for inner and outer ROCs is of the same magnitude, $4.0\,\%$ for
IROCs and $4.3\,\%$ for OROCs.

\begin{figure}[t]
  \centering
  \includegraphics[width=\columnwidth]{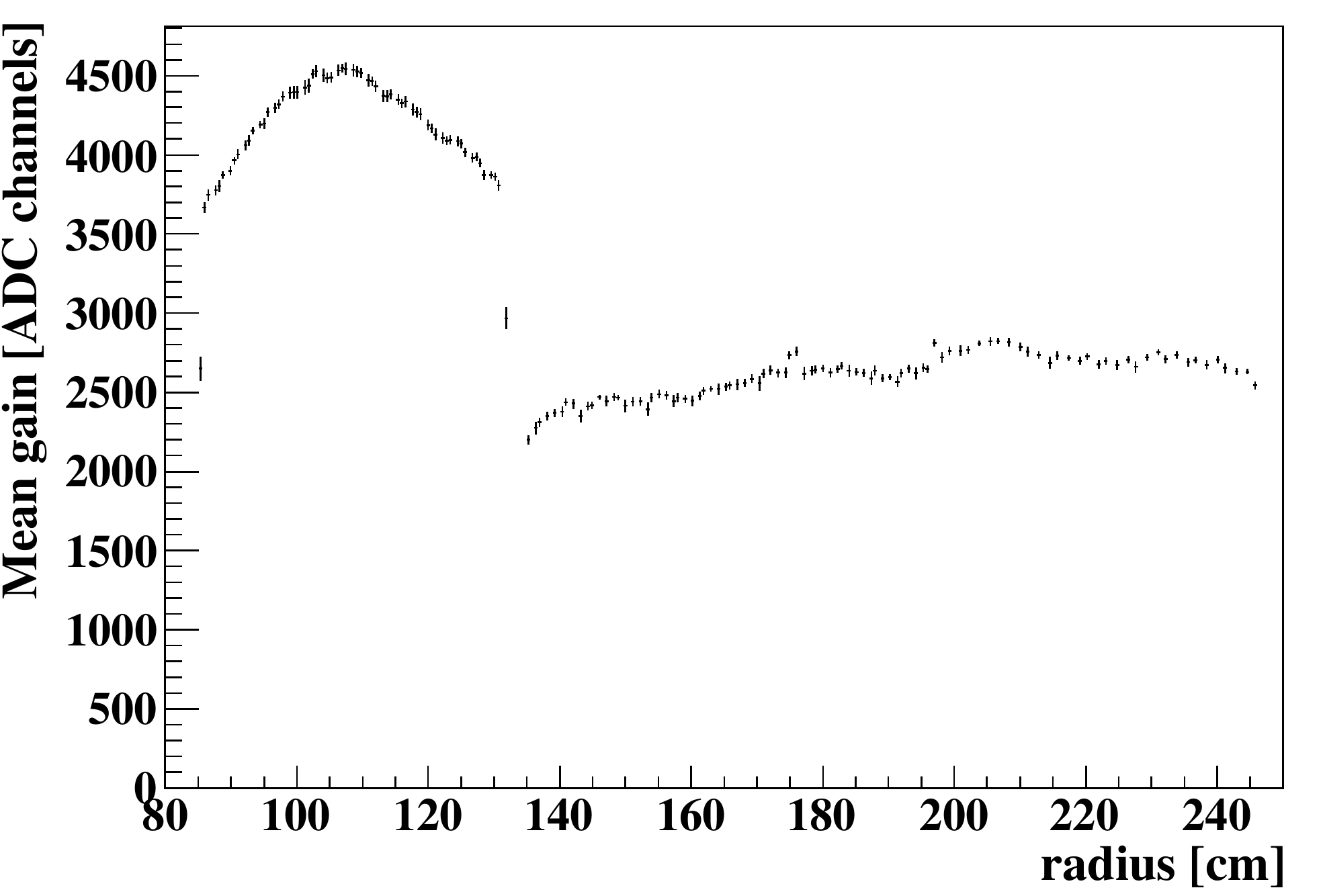}
  \includegraphics[width=\columnwidth]{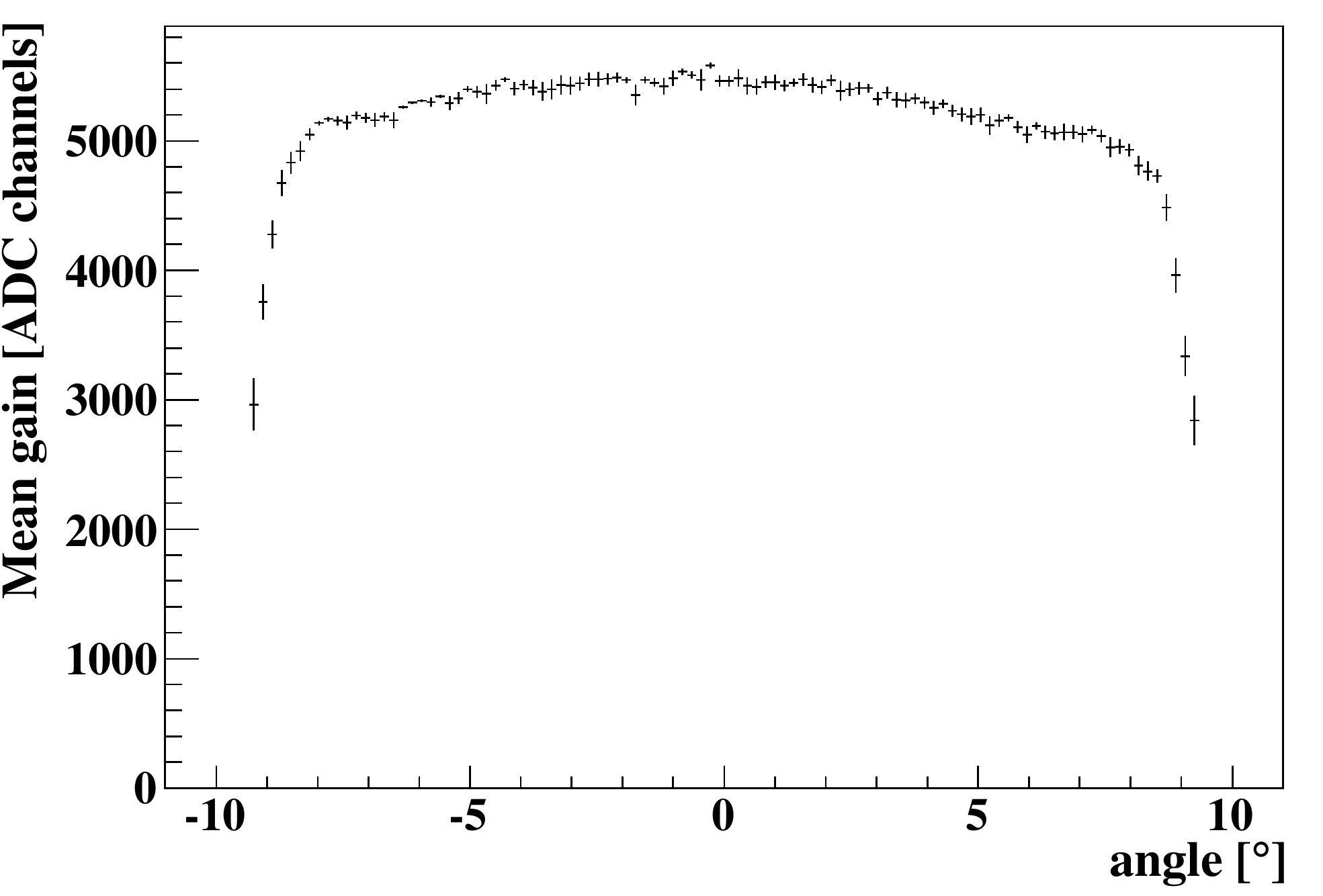}
  \caption{Radial (top) and azimuthal (bottom) systematics of gain variations 
within one sector. The step observed in the radial variations is due to the IROC-OROC transition.}
  \label{fig:calcom.projections}
\end{figure}

\paragraph{Results}

The results of the krypton calibration are used to produce pad-by-pad
calibration constants which reflect the gain topology.

Gain variations within a chamber reflect mechanical deformations and imperfections (see \cite{NA49} and
\cite{Rybicki:604807}). The geometrical characteristics of a sector are visible in
the radial and azimuthal projections. Figure~\ref{fig:calcom.projections} shows the average
gain variations in radial (top) and azimuthal (bottom) direction over all chambers.

Within a single sector, sizable systematics are observed in the radial direction, 
typically reaching up to $18$\,\%,
$23$\,\% and $11$\,\% for IROCs, OROCs short pads and OROCs long
pads, respectively. 
The maximal variations in the azimuthal direction are $27$\,\% for IROCs and $22$\,\%
for OROCs.
A decrease of gain on the edges is visible especially in the azimuthal
direction. It is related to the fact that on the edges the full
krypton cluster cannot be reconstructed. For this reason a parabolic
extrapolation of the gain is used for gain correction in these regions.

\subsection{Drift-time calibration}
\label{calcom.driftvelocity}

\subsubsection{Shaping variations in the FEE}
\label{calcom.pulser}
To determine the shaping characteristics of the front-end electronics, a pulse is injected on the cathode-wire plane of the readout chambers (see Sec.~\ref{infra:Calibration_pulser}). This induces a signal on the pads, without gas amplification. Due to manufacturing tolerances of the PASA chips the shaping of the signal is expected to vary, resulting in the detection of different arrival times and integrated charges.

In order to correct for these effects, calibration pulser runs (see
Sec.~\ref{calcom.commissioning.runs}) will be taken on a regular
basis to monitor the chip characteristics. The resulting correction
values are stored in the OCDB and used in the offline reconstruction.

\begin{figure}[t]
  \begin{center}
    \includegraphics[width=\columnwidth]{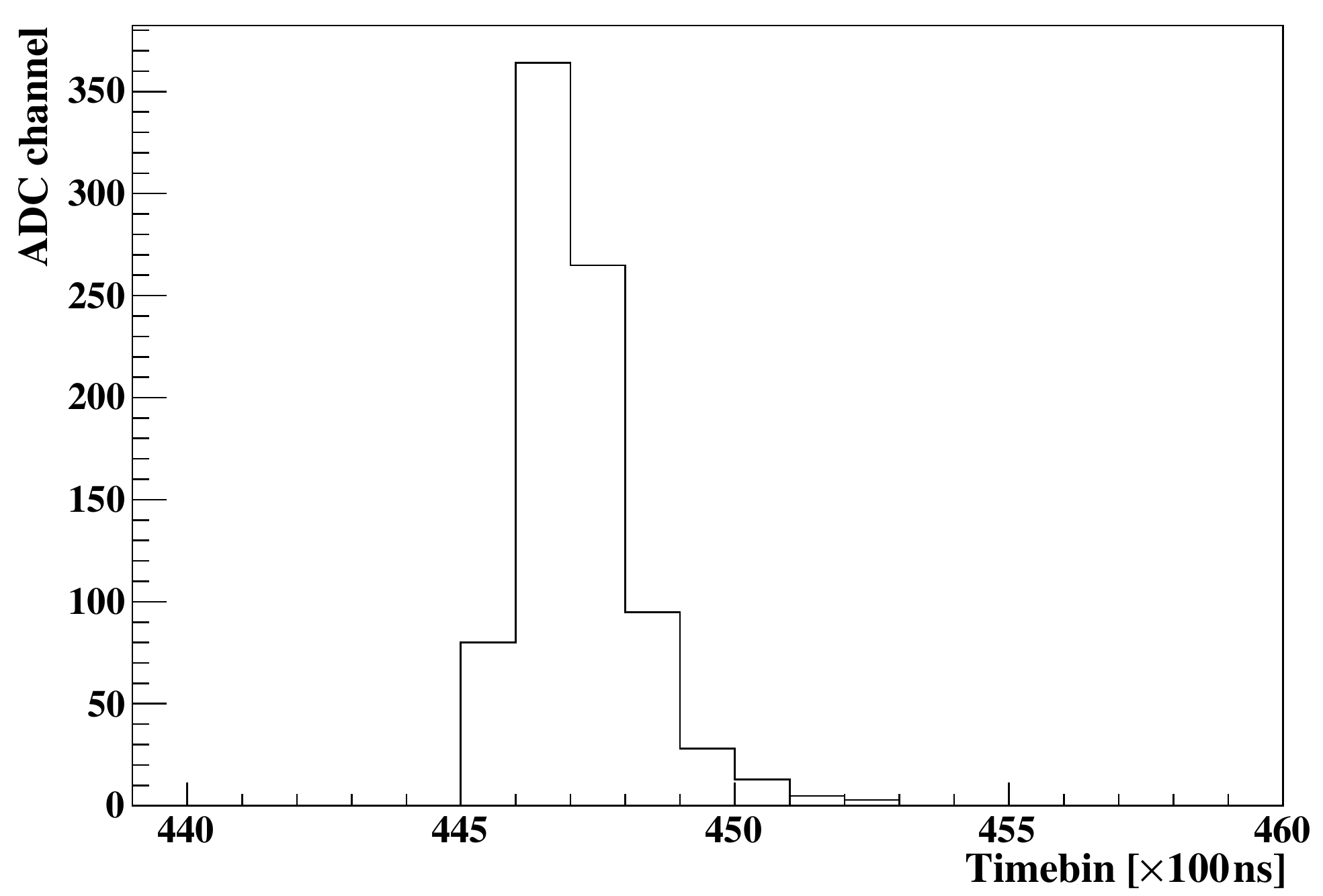}
  \end{center}
   \caption{Typical calibration pulser signal in one readout channel.}
  \label{fig:calcom.pulsersignal}
\end{figure}

Figure~\ref{fig:calcom.pulsersignal} shows a typical pulser signal of one channel. A calibration algorithm accumulates a number of pulser events and calculates the position (center of gravity), width (RMS) and area (integral) for each pad signal. The signal analysis is done in a window of minus two to plus two time bins around the maximum bin, as used in the offline cluster finder.

\begin{figure}[h!]
  \includegraphics[width=0.8\columnwidth]{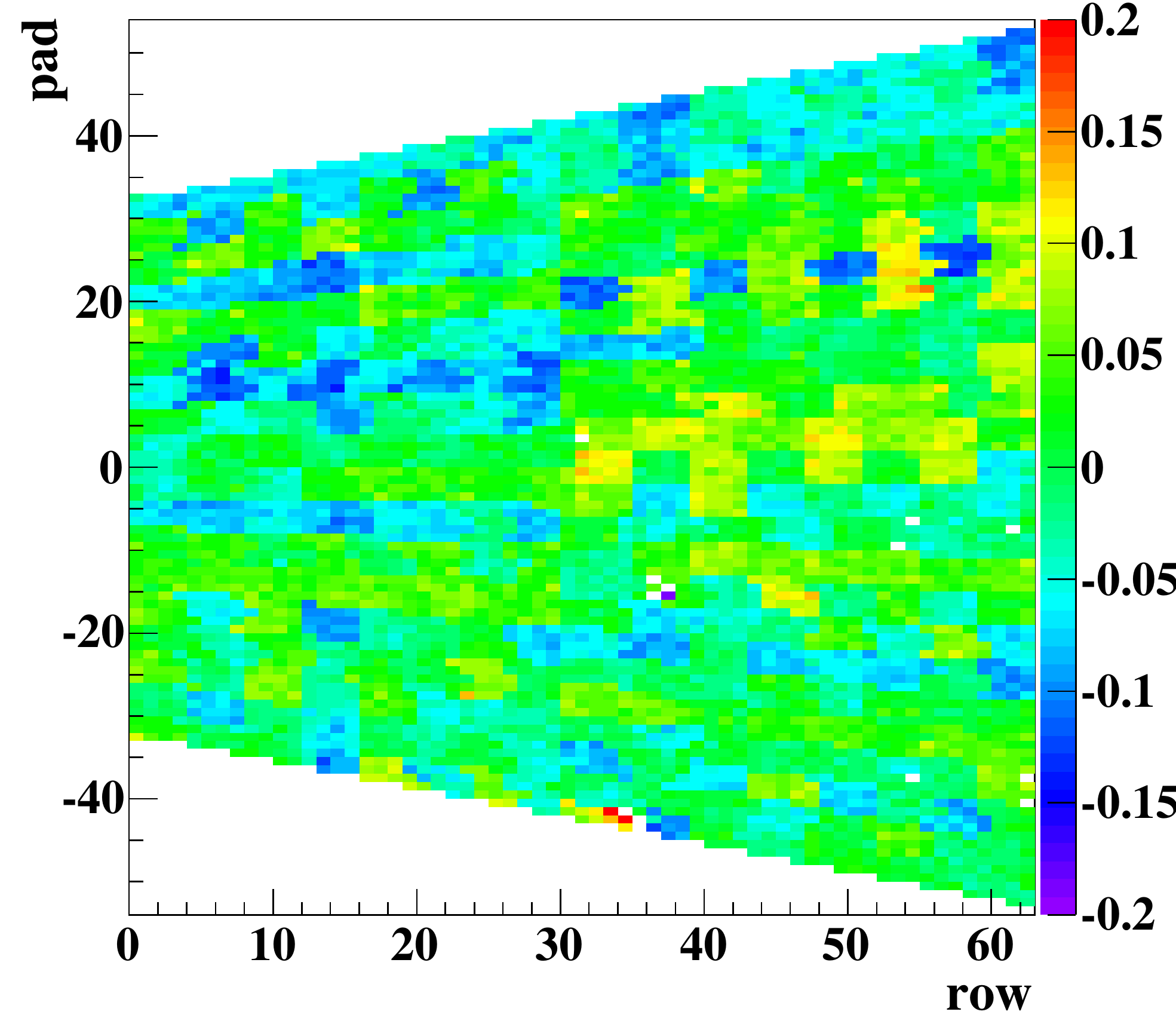}
  \caption{Topology of pulser-timing variations in one IROC.}
  \label{fig:calcom.pulTdist.2D}
\end{figure}

Figure~\ref{fig:calcom.pulTdist.2D} shows the timing differences within one IROC. Clear patterns can be seen: groups of 16 pads are found, showing nearly the same values. Differences between the groups can be larger. Each of the groups corresponds to one PASA chip. The variations result from manufacturing tolerances.

\begin{figure}[h!]
  \centering
  \includegraphics[width=0.8\columnwidth]{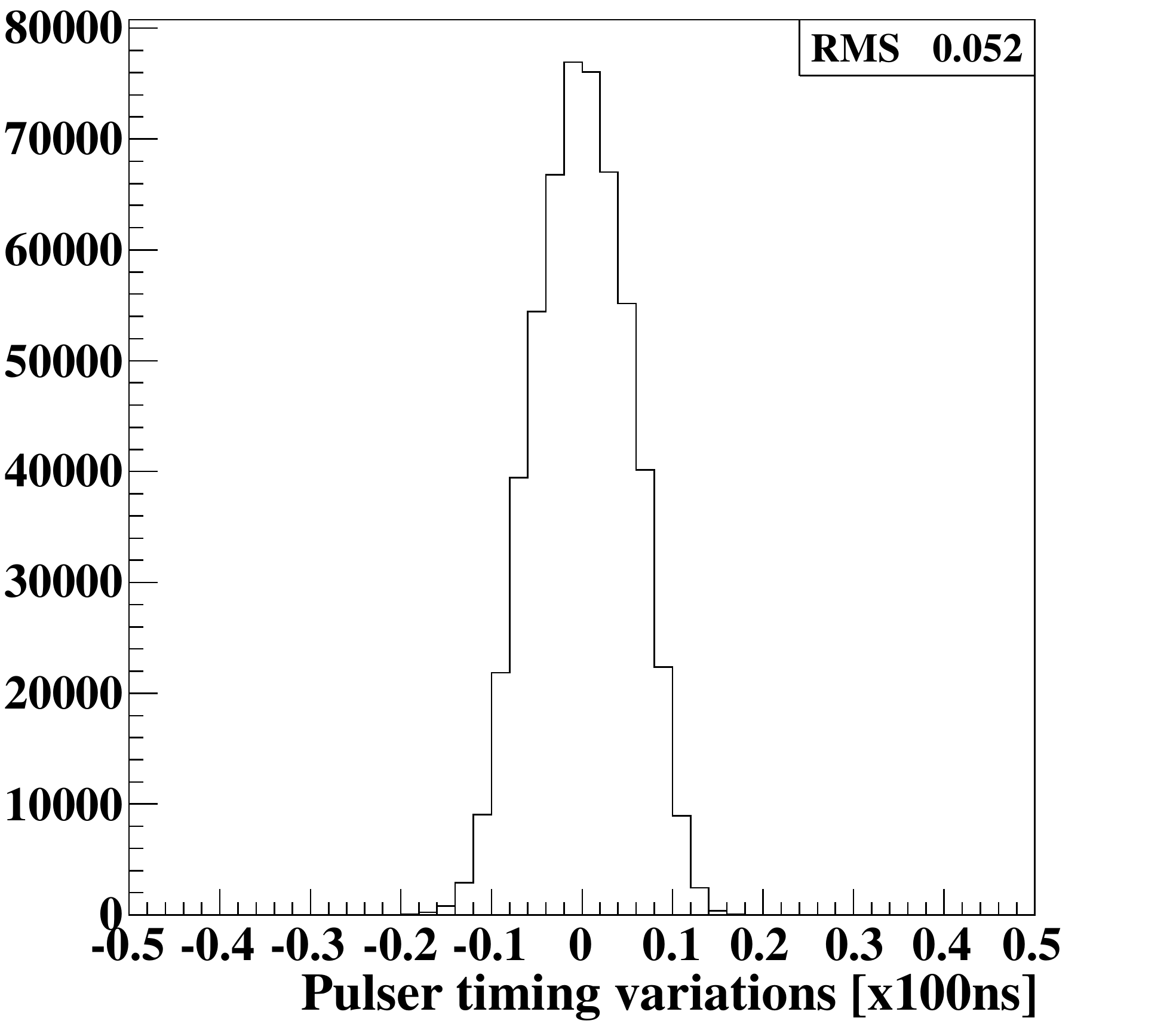}
  \caption{Distribution of pulser-timing variations of all pads.}
  \label{fig:calcom.pulTdist}
\end{figure}

In Fig.~\ref{fig:calcom.pulTdist} the distribution of the timing variations in the complete TPC is shown. The RMS of the distribution is 0.052, corresponding to 5.2\,ns. Considering a drift velocity of about 2.65\,cm/\textmu s, this would yield a systematic error in the cluster-position resolution of about 140\,\textmu m. 
Compared to the intrinsic cluster-position resolution of 300--800\,\textmu m, given by the diffusion and therefore depending on the $z$-position, this is a second order effect. 

Details on the pad-by-pad shaping variations are discussed in~\cite{Wiechula2009}.

\subsubsection{Drift velocity}

The drift velocity is a function of the field (electric, magnetic) and the mobility  \cite{blumrolandi}. The mobility depends on the gas density which is a function of the environment variables as well as the gas composition. The drift velocity is therefore a function of many parameters:
\begin{equation}
v_d = v_d(E,B,N(P,T),C_{CO_2},C_{N_2}),
\end{equation} 
where $E$ and $B$ are the field values (electric, magnetic), $N$ is
the gas density, $P$ is the atmospheric pressure, $T$ is the
temperature inside the TPC and $C_{CO_2}$ and $C_{N_2}$ are two
concentrations out of three components of the drift gas \gas (\mixture). 
We assume that these parameters, especially
the environment variables, will vary in time within a reasonable
range. According to
 \textsl{Magboltz-2} \cite{magboltzNIM,magboltzCERN}
simulations, a first order Taylor expansion of the dependencies around
the nominal values, $\Delta{v_d}=v_d-v_{d0}$, is sufficient.

Within the TPC volume, the parameters in the expansion are changing
with different time constants. A significant change of the drift
velocity due to changes in the gas composition as well as $E$ and $B$
field variations has a time constant of several hours, while 
the changes due to pressure and
temperature variations have to be corrected on the level of minutes.

In the following we will therefore disentangle the two time scales and summarize
the long term variations under the term $k_0(t)$:
\begin{equation}
x=\frac{\Delta{v_d}}{v_{d0}}= k_0(t)+k_{N}\frac{\Delta{N(P,T)}}{N_0(P,T)} 
 = k_0(t)+k_{P/T}\frac{\Delta{(P/T)}}{(P/T)_0}.
\end{equation}

The correction factor $x$ can be measured using different methods:
\begin{itemize}
\item matching laser tracks with the surveyed mirror positions;
\item matching with tracks from the Inner Tracking System (ITS);
\item matching of the TPC primary vertices from the two halves of the TPC;
\item matching tracks from two halves of the TPC using cosmic tracks.
\end{itemize}

The unknown parameters $k_0(t)$ and $k_{P/T}$ can be determined using a
Kalman filter approach.

\begin{figure}[t]
\centering
\includegraphics[width=\columnwidth]{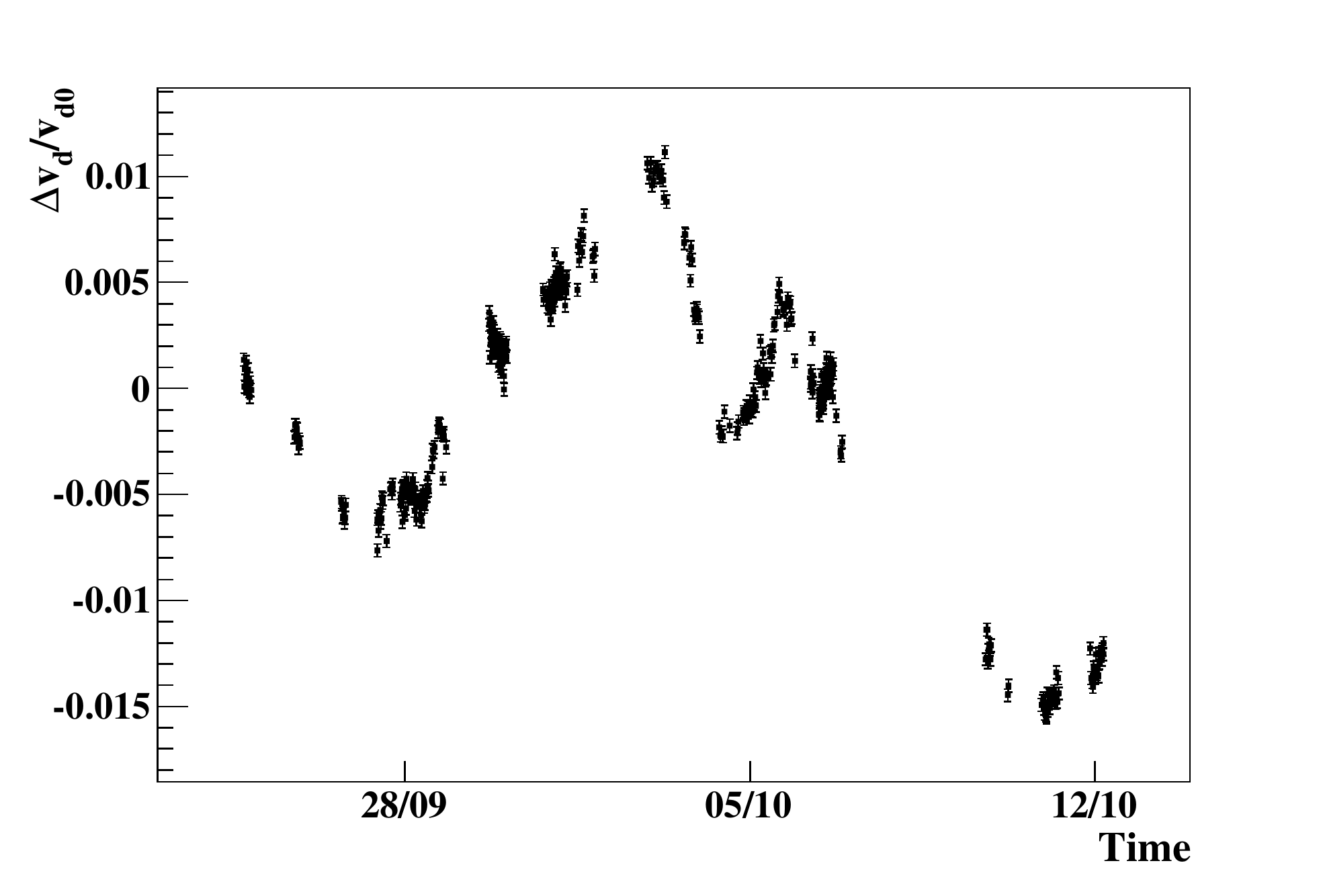}
\includegraphics[width=\columnwidth]{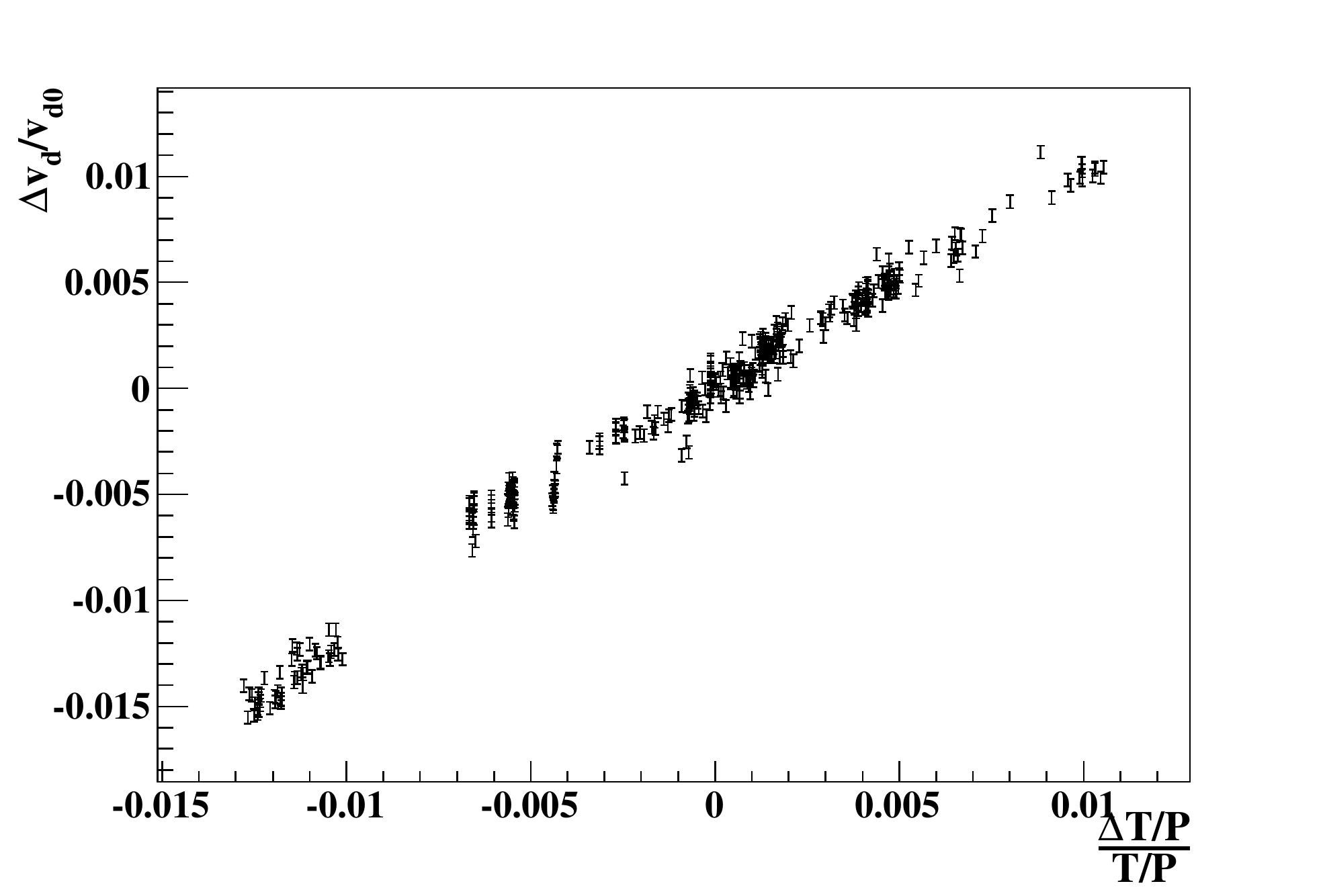}
\caption{Drift velocity as function of time (top) and as a function of $\Delta(T/P)$ (bottom).
} 
\label{figVDrift}
\end{figure}

\paragraph{Precision of the correction}

The precision of the drift velocity correction is proportional
to the  precision of the pressure and temperature measurement and to the length of the time
interval 
\begin{equation}
    \sigma^2_x=\frac{\partial\sigma^2_{k_0(t)}}{\partial t}\Delta{t}+\frac{k^2_{P/T}}{(P/T)_0^2}\sigma^2_{P/T} = \dot{\sigma}^2_{k_0(t)}\Delta{t}+k'_{P/T}\sigma^2_{P/T}.
\label{eq:sigmaX}
\end{equation}

The typical relative resolution of the pressure and temperature
measurement is on the level of $6\times10^{-5}$ and $1\times10^{-5}$
respectively~\footnote{This is the resolution of the pressure sensor
used in 2008. As this resolution was close to the required drift
velocity precision (see Sec.~\ref{calcom.requirements}) it has now
been replaced with a sensor with higher precision.}. For a cool gas
the coefficient $k_N$ is close to one. The contribution of the P/T
correction to the drift velocity uncertainty is approximately
$6.1\times10^{-5}$ (150~$\micro$m for the full drift length of 250~cm)

\begin{figure}[t]
\centering
\includegraphics[width=\columnwidth]{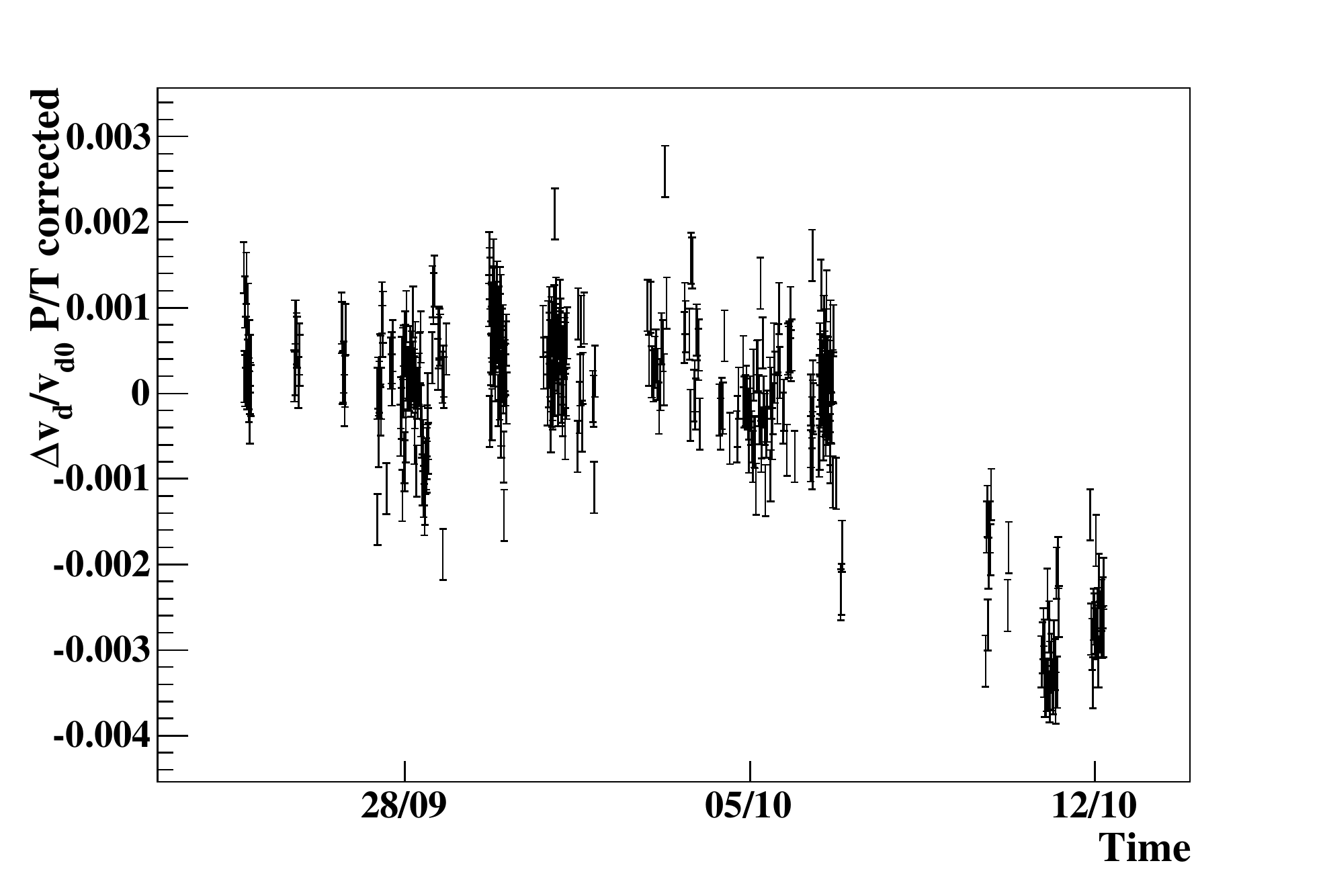}
\includegraphics[width=\columnwidth]{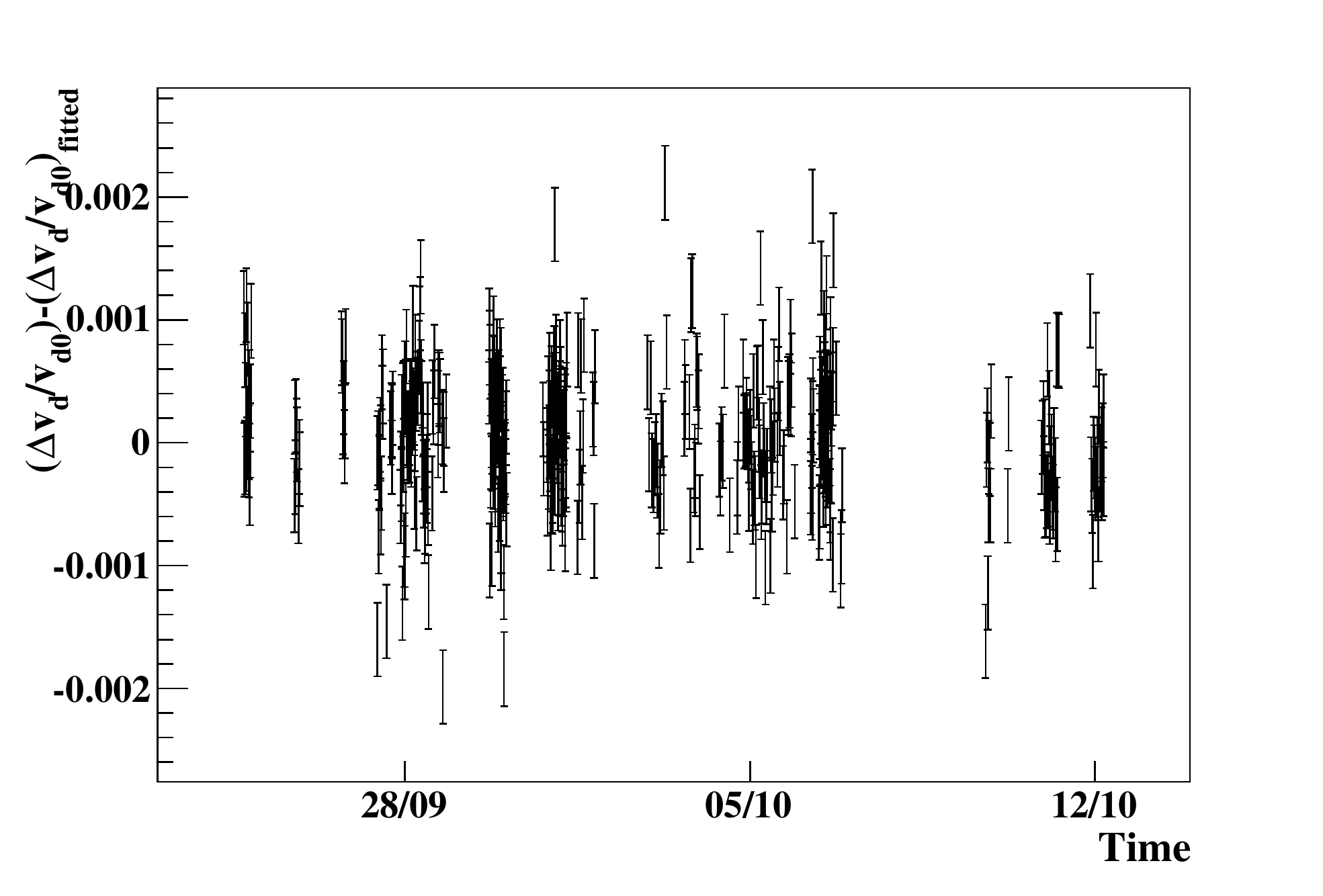}
\caption{
	Drift velocity corrected for P/T variations as function of time
	(top).  In the lower plot the correction for time-dependent
	offset is also applied. }
\label{figVDriftCorrected}
\end{figure}

Figure~\ref{figVDrift} shows the input to the Kalman filter and
Fig.~\ref{figVDriftCorrected} shows the results after drift velocity
correction.

A similar method will be applied to correct for the time variation of the gain.

The $\dot{\sigma}_{k_0(t)}$ from Eq.~\ref{eq:sigmaX} was  estimated from Fig.~\ref{figVDriftCorrected} and is on the level of 0.001 in a four day period. This estimate was obtained for the period of largest change in the present data sample. Further studies will be performed for extended time periods.

For the TPC drift velocity determination, the required relative resolution is on the level of $6\times10^{-5}$.
Entering the observed sigmas into Eq.~\ref{eq:sigmaX} the minimal frequency of the drift velocity updates were estimated to be about 1 hour.

\section{Performance}
\label{perform}
The ALICE TPC is the main tracking device of the experiment, therefore its performance is 
a crucial issue. In this section we discuss the space-point resolution, which mainly 
determines the tracking performance (momentum and angular resolutions), the track matching performance 
and the particle identification performance.

\subsection{Space-point resolution}
In general, the space-point resolution ($\sigma_{{\mathrm{COG}}}$) depends on many 
parameters; namely, the readout geometry, the gas composition and the track characteristics.
Here we discuss those, which are dominant for the present detector, namely:
\begin{itemize} 
\item the drift length ($L_{\mathrm{Drift}}$); 
\item the track inclination angle ($\alpha$); 
\item the charge deposited on the anode wire ($Q$).
\end{itemize}
The space-point resolutions presented in this section were determined as a function of 
these values.

For further studies, it is convenient to parametrize the space-point resolution
as a function of the parameters mentioned above.
We obtained the parametrization of the space-point resolution
by fitting parameters $p_{\mathrm{0}}$, $p_{\mathrm{L}}$ and 
$p_{\mathrm{A}}$ in the formula below to data with cosmic ray measurements.
\begin{eqnarray}\
     \sigma^2_{{\mathrm{COG}}} \propto p^2_0+p^2_{L}L_{\mathrm{Drift}}+p^2_{A}\tan^2\alpha,  
     \nonumber\\
     p^2_L \propto  \frac{\sigma^2_DG_{\mathrm{g}}}{N_{\mathrm{ch}}},
     \nonumber\\
     p^2_A \propto \frac{L_{\rm{pad}}^2G_{\rm{Lfactor}}}{N_{\mathrm{eprim}}}. 
\label{eqResCOG0}
\end{eqnarray}
where $N_{\mathrm{ch}}$ is the number of electrons created during the amplification process, $L_{\mathrm{pad}}$ is the pad length, and 
$N_{\mathrm{eprim}}$ is the number of primary electrons per pad. 
There are two main factors which degrade the space-point resolution, namely 
the gas gain fluctuations (factor $G_{\mathrm{g}} \approx 2$)
and the Landau fluctuations of the ionization energy loss (factor $G_{\rm{Lfactor}}$).
One should note that:
\begin{align}
     N_{\rm{ch}}&\propto {L_{\rm{pad}}}, \nonumber \\
     N_{\rm{eprim}}&\propto {L_{\rm{pad}}}, 
\end{align}
and therefore
\begin{align}\
     p_L &\propto \frac{1}{\sqrt{L_{\rm{pad}}}},
     \nonumber\\
     p_A &\propto \sqrt{L_{\rm{pad}}} .
\label{eq:ResolScaling}	
\end{align}
Analysis of cosmic ray data determines the resolution parameters  $p_{\mathrm{0}}$,
$p_{\mathrm{L}}$ and $p_{\mathrm{A}}$. Results from a fit which are scaled according to Eq.~\ref{eq:ResolScaling} are shown in 
Tab.~\ref{table:PointResolFitParam}. 
The fit was done separately in the $z$ (drift) and the $r\varphi$
directions, denoted further as $z$ and $y$ accordingly.

\begin{table}
\begin{center}
\caption{Values of parameters describing the space-point resolution.}
\label{table:PointResolFitParam}
\begin{tabular}{|l|ccc|}
\hline
Pad size 		     & $\unit{0.75\times0.4}{\centi\metre\squared}$ 
                             & $\unit{1.0 \times0.6}{\centi\metre\squared}$
                             & $\unit{1.5 \times0.6}{\centi\metre\squared}$  \\
\hline
$p_{0y}$                     & $\unit{0.026}{\centi\metre}$ 
                             & $\unit{0.031}{\centi\metre}$
                             & $\unit{0.023}{\centi\metre}$ \\  
$p_{0z}$                     & $\unit{0.032}{\centi\metre}$
                             & $\unit{0.032}{\centi\metre}$ 
                             & $\unit{0.028}{\centi\metre}$ \\  
$p_{Ly}\sqrt{L_\text{pad}}$  & $\unit{0.0051}{\centi\metre}$
                             & $\unit{0.0060}{\centi\metre}$
                             & $\unit{0.0059}{\centi\metre}$ \\   
$p_{Lz}\sqrt{L_\text{pad}}$  & $\unit{0.0056}{\centi\metre}$
                             & $\unit{0.0056}{\centi\metre}$
                             & $\unit{0.0059}{\centi\metre}$ \\
$p_{Ay}/\sqrt{L_\text{pad}}$ & $\unit{0.13}{\centi\metre^{1/2}}$
                             & $\unit{0.15}{\centi\metre^{1/2}}$
                             & $\unit{0.15}{\centi\metre^{1/2}}$  \\  
$p_{Az}/\sqrt{L_{\mathrm{pad}}}$ & $\unit{0.15}{\centi\metre^{1/2}}$
                                 & $\unit{0.16}{\centi\metre^{1/2}}$
                                 & $\unit{0.17}{\centi\metre^{1/2}}$ \\
\hline 
\end{tabular}
\end{center}
\end{table}

\begin{figure}[t]
  \centering
  \includegraphics[width=\linewidth]{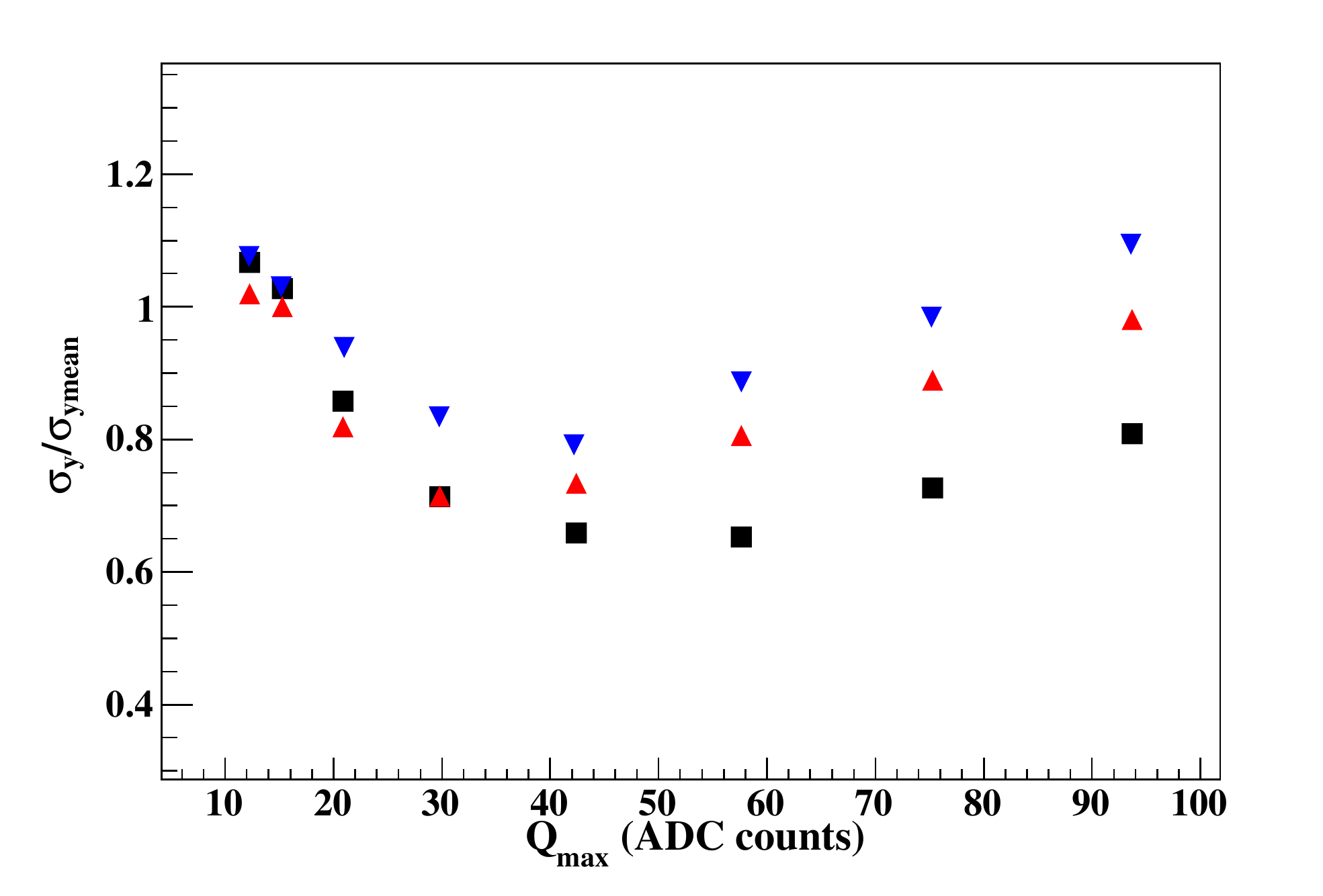}
  \centering
    \includegraphics[width=\linewidth]{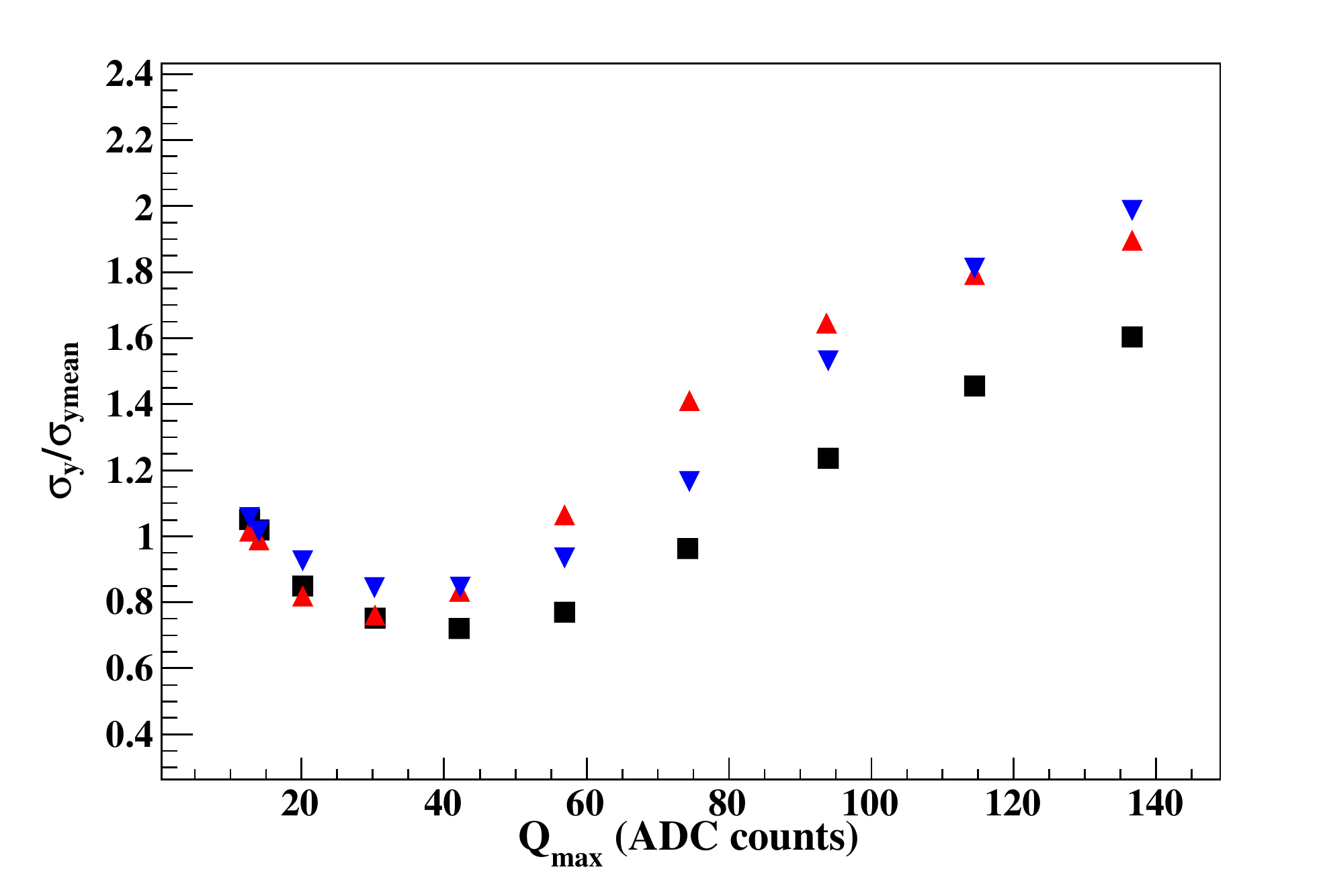}
\caption{Space-point resolution in $y$ ($r\varphi$) direction  as a function of the maximum of
deposited charge within a cluster $Q_{\mathrm{max}}$, with (upper panel) and without (lower panel) magnetic field.
The different curves correspond to the three pad regions short (squares), medium (triangles), and long (inverted triangles).}
 \label{figPointResolYQ}
\end{figure}

In the ALICE TPC three different pad geometries are used, thus the  space-point 
resolution was obtained for each of them separately. One should note that the scaled values of 
$p_{Ly}\sqrt{L_{\mathrm{pad}}}$, $p_{Lz}\sqrt{L_{\mathrm{pad}}}$ and, separately, $p_{Ay}/\sqrt{L_{\mathrm{pad}}}$,
$p_{Az}/\sqrt{L_{\mathrm{pad}}}$ are equal to within 10\%, as expected.

\begin{figure}[t]
  \centering
  \includegraphics[width=\linewidth]{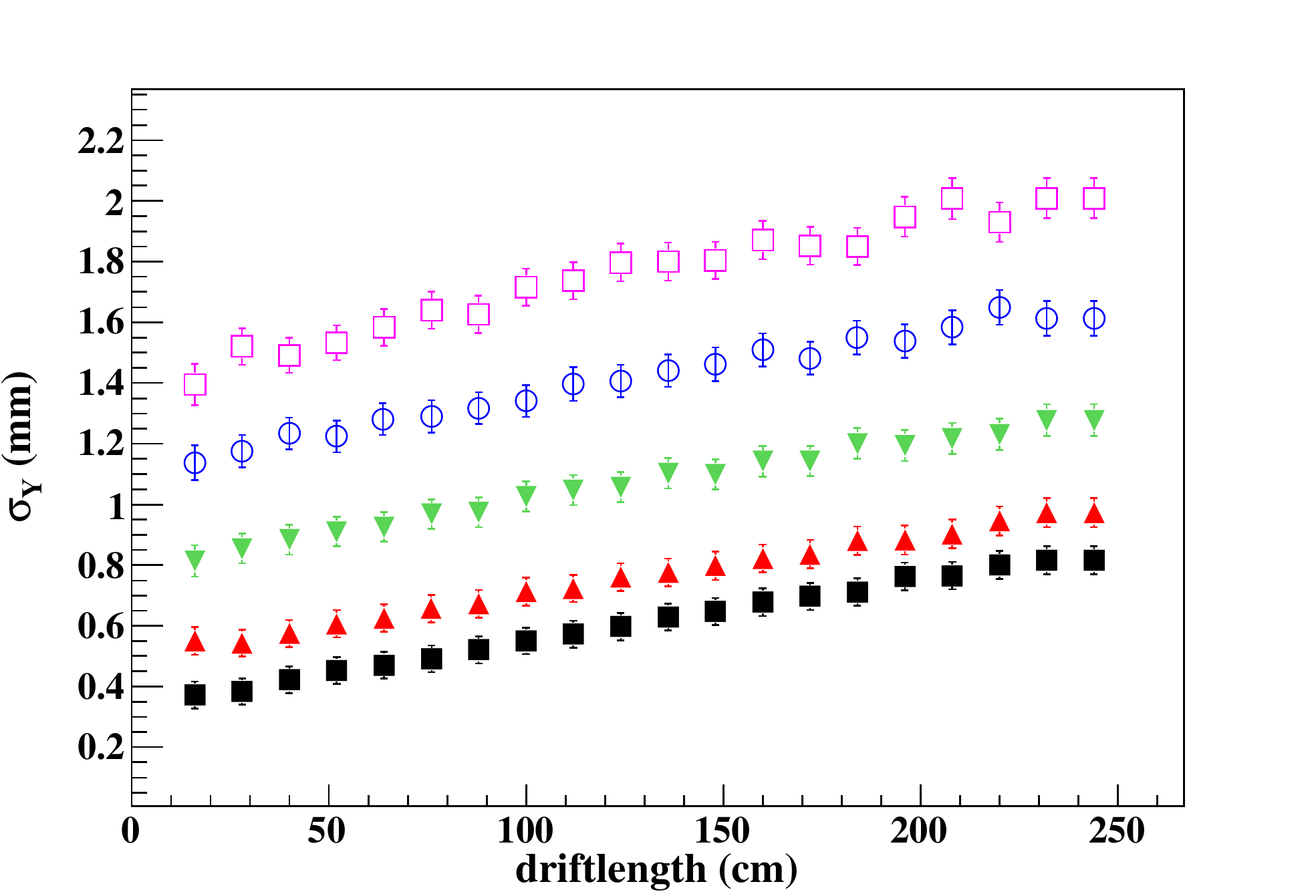} 
  \centering
  \includegraphics[width=\linewidth]{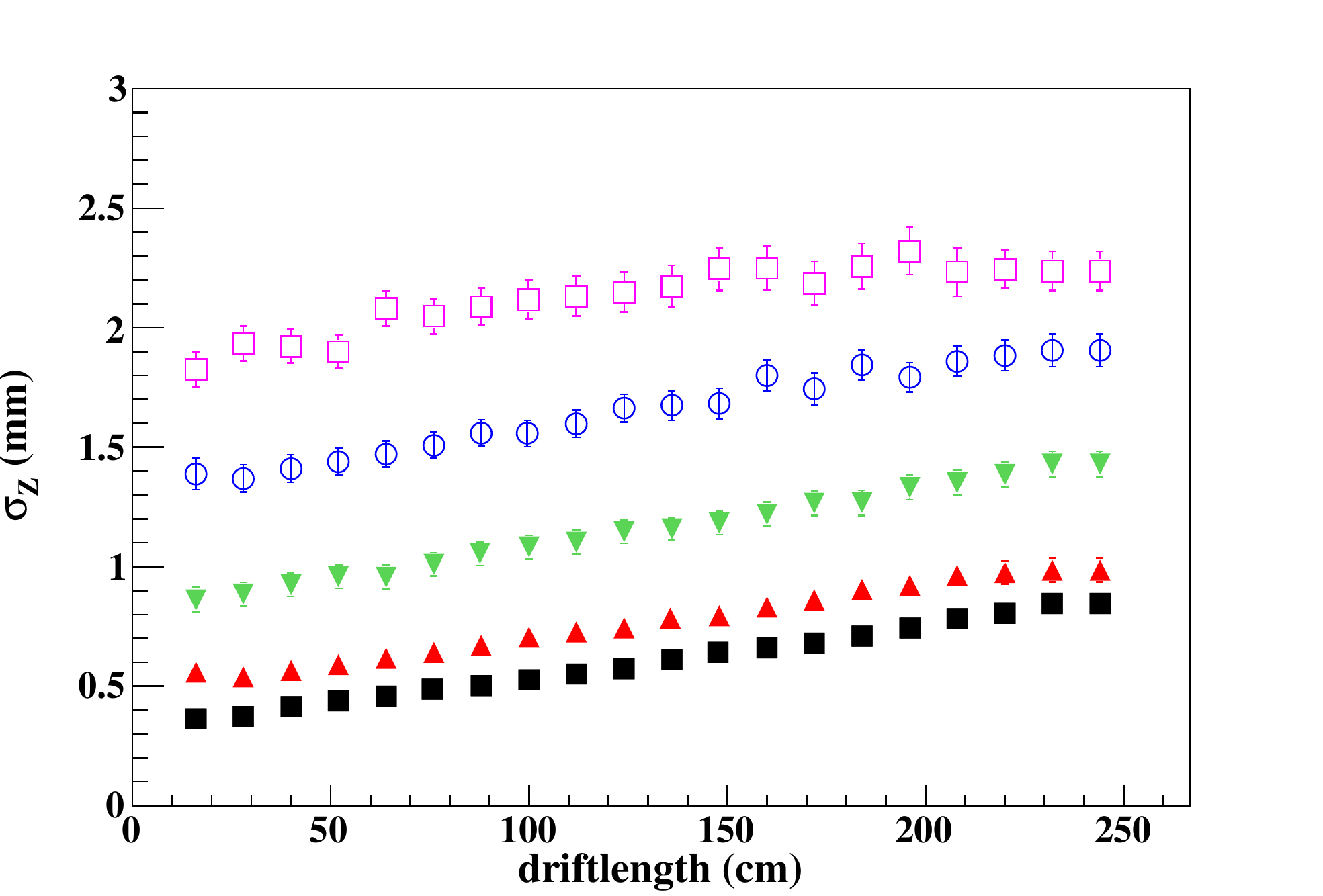}
  \caption{Space-point resolution in $y$ ($r\varphi$) and $z$ (drift) direction as a function of the drift length. The different symbols correspond to inclinations from $\tan(\alpha)=0$ (full squares) to  $\tan(\alpha)=0.92$ (open squares) in steps of 0.23.}
  \label{figPointResolDRTAN}
\end{figure}

In the previous formula we assumed that all electrons created in the ionization process 
contribute
to the measured signal. In a real experiment, because of the applied zero suppression, 
part of the signal is lost. The fraction of the signal below the threshold is proportional to 
the width of the response function and increases with the drift length and
the track inclination angle. 
We have corrected for these effects, replacing p$_{\mathrm{L}}$ and p$_{\mathrm{A}}$ from  
Eq.~\ref{eqResCOG0}:
\begin{align}
  p_{\mathrm{L}}'&\propto p_{\mathrm{L}} \cdot p_{\mathrm{LC}}=p_{\mathrm{L}}\cdot
  (1+p_{\mathrm{L1}}\cdot L_{\mathrm{Drift}}+p_{\mathrm{L2}}\tan^2\alpha),
  \nonumber\\
  p_{\mathrm{A}}'&\propto p_{\mathrm{A}} \cdot p_{\mathrm{AC}}=p_{\mathrm{A}}\cdot
  (1+p_{\mathrm{A1}}\cdot L_{\mathrm{Drift}}+p_{\mathrm{A2}}\tan^2\alpha),
\label{eq:PointResolFitCorrection}
\end{align}
where p$_{\mathrm{L1}}$, p$_{\mathrm{L2}}$, p$_{\mathrm{A1}}$ and p$_{\mathrm{A2}}$ are
parameters to be fitted.
We also added to the Eq.~\ref{eq:PointResolFitCorrection} terms proportional
to $1/Q$, where $Q$ is a total charge of the cluster, to account for the number of 
electrons which contribute to the signal.
However, the space-point resolution improves only 
slightly, and after a certain value of $Q$ deteriorates (see Fig.~\ref{figPointResolYQ}).
This is due to the production of $\delta$-electrons. 
The influence of $\delta$-electrons is much smaller in presence of the magnetic 
field because of their smaller effective range.

The measured space-point resolution in $y$ ($r\varphi$) and $z$ (drift) directions are shown in 
Fig.~\ref{figPointResolDRTAN}. The parametrization (see Eq.~\ref{eq:PointResolFitCorrection}) describes 
the data within 2\%.

The observed dependence of the space-point resolution on drift length and inclination
angle is, for these cosmic ray tracks, mostly determined by geometrical factors.
The values observed for small inclination angles are close to those specified in the 
TPC TDR~\cite{TDR:tpc}. Note that small inclination angles are common for tracks 
originating from the collision vertex of an LHC pp or Pb--Pb event.

\subsection{Momentum resolution}
\begin{figure}[t]
 \centering
 \includegraphics[width=\linewidth]{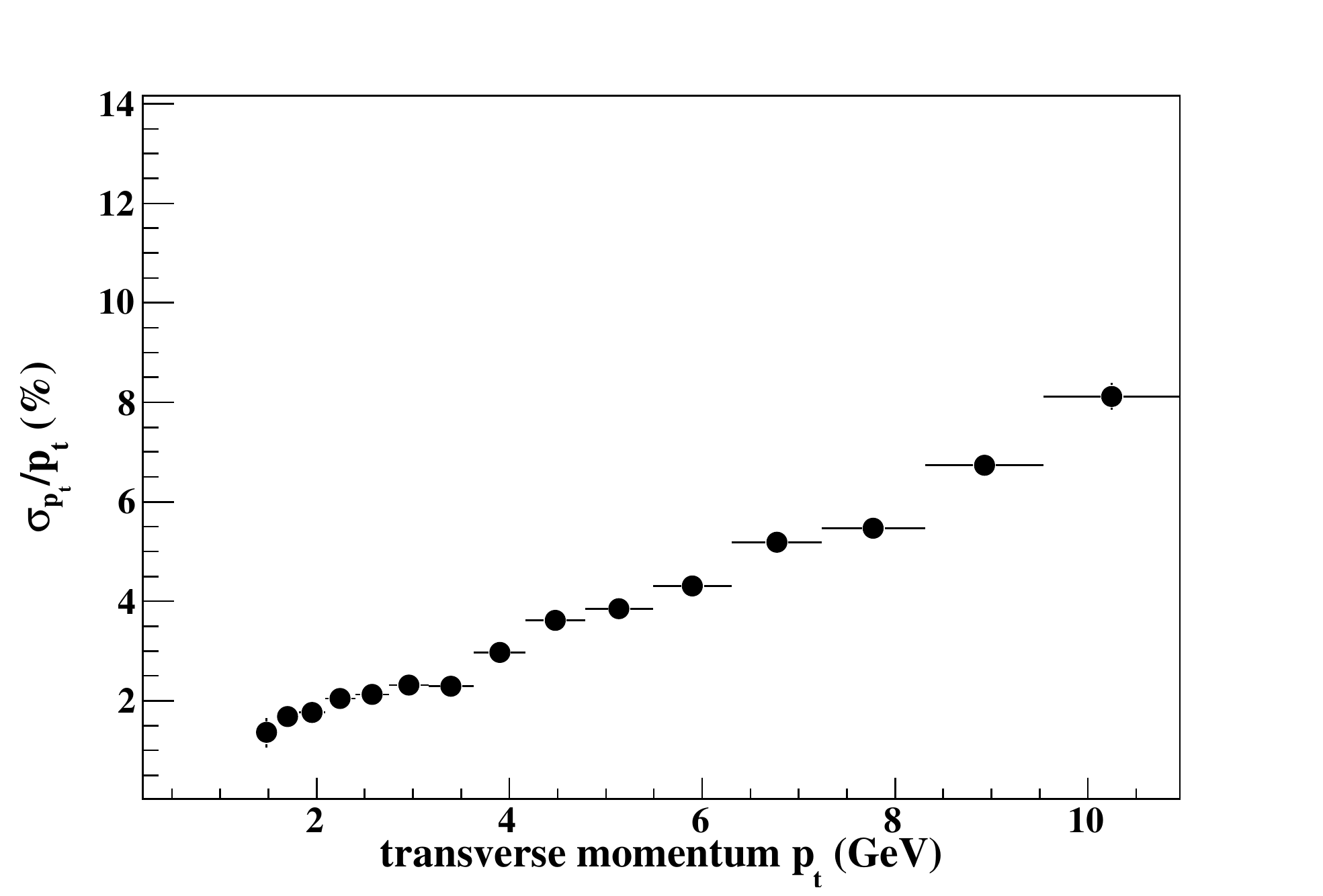}
 \caption{Transverse momentum resolution measured with cosmic rays.}
 \label{figMomentumResolution}
\end{figure}

The momentum resolution achievable with the ALICE TPC used as a stand-alone detector can be determined by using cosmic ray tracks passing through the center of the TPC. Comparison of the momenta for the first and second half of each track yields the momentum resolution curve depicted in Fig.~\ref{figMomentumResolution}. At the current stage of calibrations a momentum resolution of better than 7\% is reached at 10 GeV close to the value listed in the TDR~\cite{TDR:tpc}. We are currently continuing to improve the correction of various (small) distortions and further improvements are expected.

\subsection{Particle identification performance}
The simultaneous measurement of the momentum $p$ of  a particle and its
specific ionization loss in the TPC gas provides particle identification
over a wide momentum range. In practice, only relative values of the
ionization need to be 
known to distinguish between different particle species. The d$E$/d$x$ information 
for a given track must be extracted from the $n_{\mathrm{cl}}$ clusters 
($50 \; < \; n_{\mathrm{cl}} \; < \;160$), which are assigned to the track. 
For each cluster its maximal charge $Q_{max}$ (the highest ADC value) and its total 
charge $Q_{\mathrm{tot}}$ can be obtained. The question of whether the d$E$/d$x$ information 
should be extracted from $Q_{\mathrm{max}}$ or $Q_{\mathrm{tot}}$ is still under discussion. 
Results shown here are based on evaluations from $Q_{\mathrm{tot}}$.

Because of the long tail towards higher energy losses in the straggling function, the average energy 
loss is not a good estimator as it would be for a Gaussian distribution. Therefore, the so-called 
truncated mean is used. The truncated mean $<S>_{\eta}$, called also the TPC signal, is defined as 
the average over 
$m$ lowest values, which correspond to the $\eta$-fraction of the whole sample, 

\begin{equation}
 <S>_{\eta} = {1 \over m} \sum_{i=0}^{m} Q_{i} \; ,
\end{equation}

\noindent where $i = 0,\ldots,n$ and $Q_{i-1} \leq Q_{i}$ for all $i$.   
Values of $<S>_{\eta}$ follow an almost perfect Gaussian distribution. At present, 
the value of $\eta$ is set to 0.7,  as the result of an optimization process, 
but will be a subject of further investigation.
The d$E$/d$x$ or the energy loss resolution $\sigma_{\mathrm{d}E/\mathrm{d}x}$ is given by the 
variance of the Gaussian distribution of $<S>_{\eta}$.

\begin{figure}[t]
 \centering
 \includegraphics[width=\linewidth]{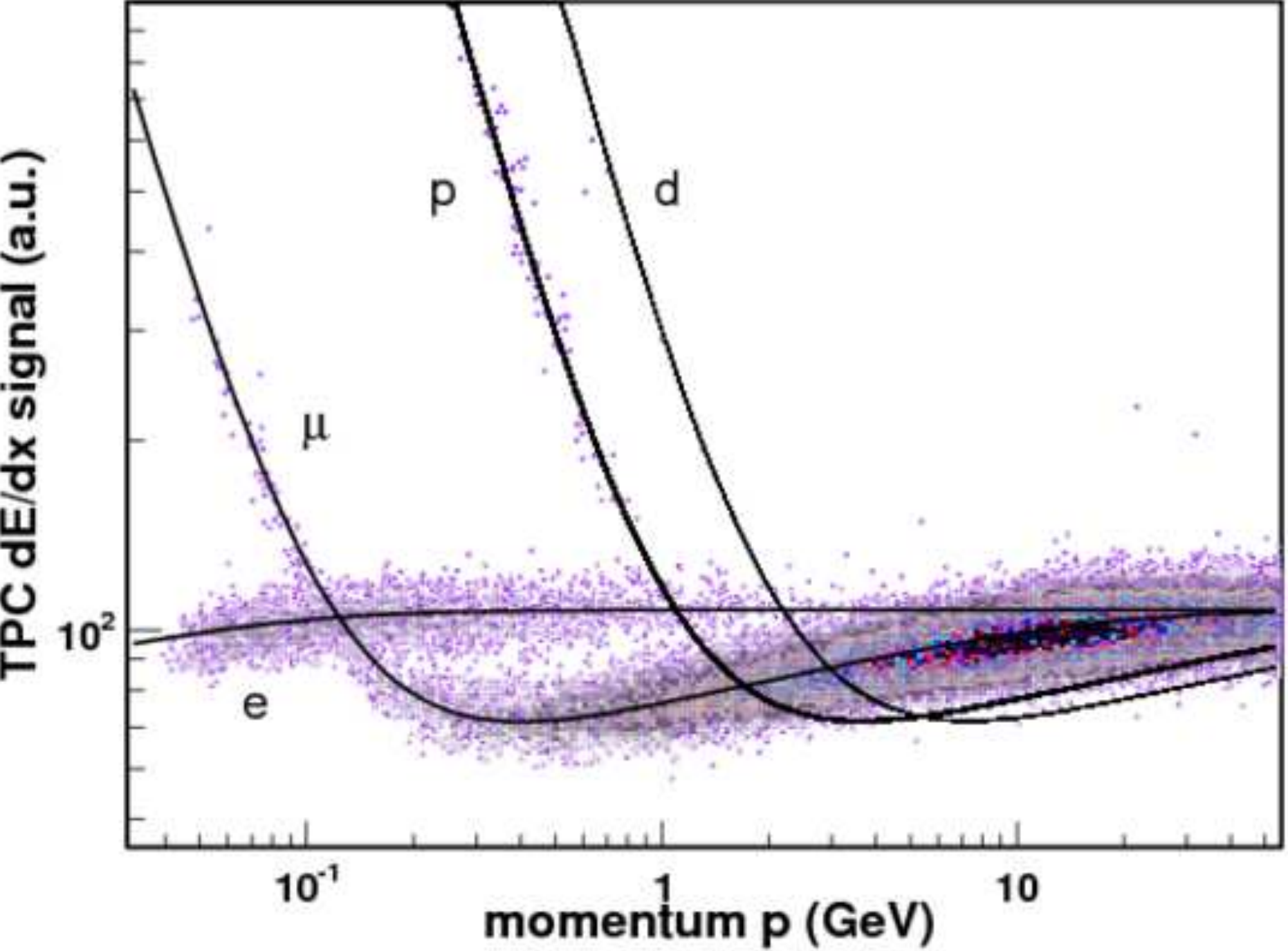}
 \caption{d$E$/d$x$ spectrum of cosmic rays.}
 \label{figDeDxSpectrum}
\end{figure}

Figure~\ref{figDeDxSpectrum} shows the TPC signal of cosmic tracks versus their momentum, from 
8.3 million events. For these data, the maximal inclination angle of $\tan (\alpha) <1$ and at least 
120 out of 160 possible TPC points per track were required.
Characteristic bands for various particles (electrons, muons, protons, 
deuterons) are clearly visible. 
The energy loss is described by the  Bethe-Bloch function:
\begin{equation}
 \langle{dE\over dx}\rangle = {4\pi N e^{4} \over mc^{2}} {Z^{2} \over \beta^{2}} \Bigl(\ln{2mc^{2}\beta^{2}\gamma^{2} \over I} - \beta^{2} - {\delta(\beta) \over 2} \Bigr) \; ,
 \label{eq.:Bethe}
\end{equation}
\noindent where $mc^{2}$ is the rest energy of the electron, $Z$ the charge of the projectile, $N$ the number density of electrons in the traversed matter, $e$ the elementary charge, $\beta$ the velocity of the projectile and $I$ is the mean excitation energy of the atom.  
In the analysis of experimental data, other
parameterizations than the Bethe-Bloch function are often used. Here we
use the parameterization proposed by the ALEPH experiment of the form~\cite{blumrolandi}:
\begin{equation}
f(\beta \gamma) = {P_{1} \over \beta^{P_{4}}} \Bigl(P_{2} - \beta^{P_{4}} - \ln(P_{3} + {1 \over (\beta \gamma)^{P_{5}}})\Bigr)   \; .
 \label{eq.:AlephParametrization}
\end{equation}
They are shown as lines in  Fig.~\ref{figDeDxSpectrum}.

The decisive quantity for particle identification is the resolution $\sigma_{\mathrm{d}E/\mathrm{d}x}$ of 
the d$E$/d$x$-measurement. Assuming a perfect gain calibration, it depends on the number of samples $n$, 
the pad size $x$ and the gas pressure $p$. In a given gas cell, the energy loss distribution depends 
only on the cluster-size distribution and on the number of primary interactions in the gas. 
This implies that the ionization distribution varies with $p$ in the same way as it does with $x$ 
and therefore the width of the distribution scales inversely proportional with the product $xp$.

For the remaining dependence on $n$ we expect a statistical scaling according to the 
law $\sigma_{\mathrm{d}E/\mathrm{d}x} \propto 1/\sqrt{n}$. In addition to this, the measurement of 
the energy loss is influenced by systematic uncertainties $\sigma_{syst}$. Therefore, the overall 
resolution is assumed to be of the form

\begin{equation}
 \sigma_{\mathrm{d}E/\mathrm{d}x}^{2}(n) = \sigma_{\mathrm{syst}}^{2} + {\sigma_{\mathrm{stat}}^{2} \over n} \; .
\end{equation}

\begin{figure}
 \centering
 \includegraphics[width=\linewidth]{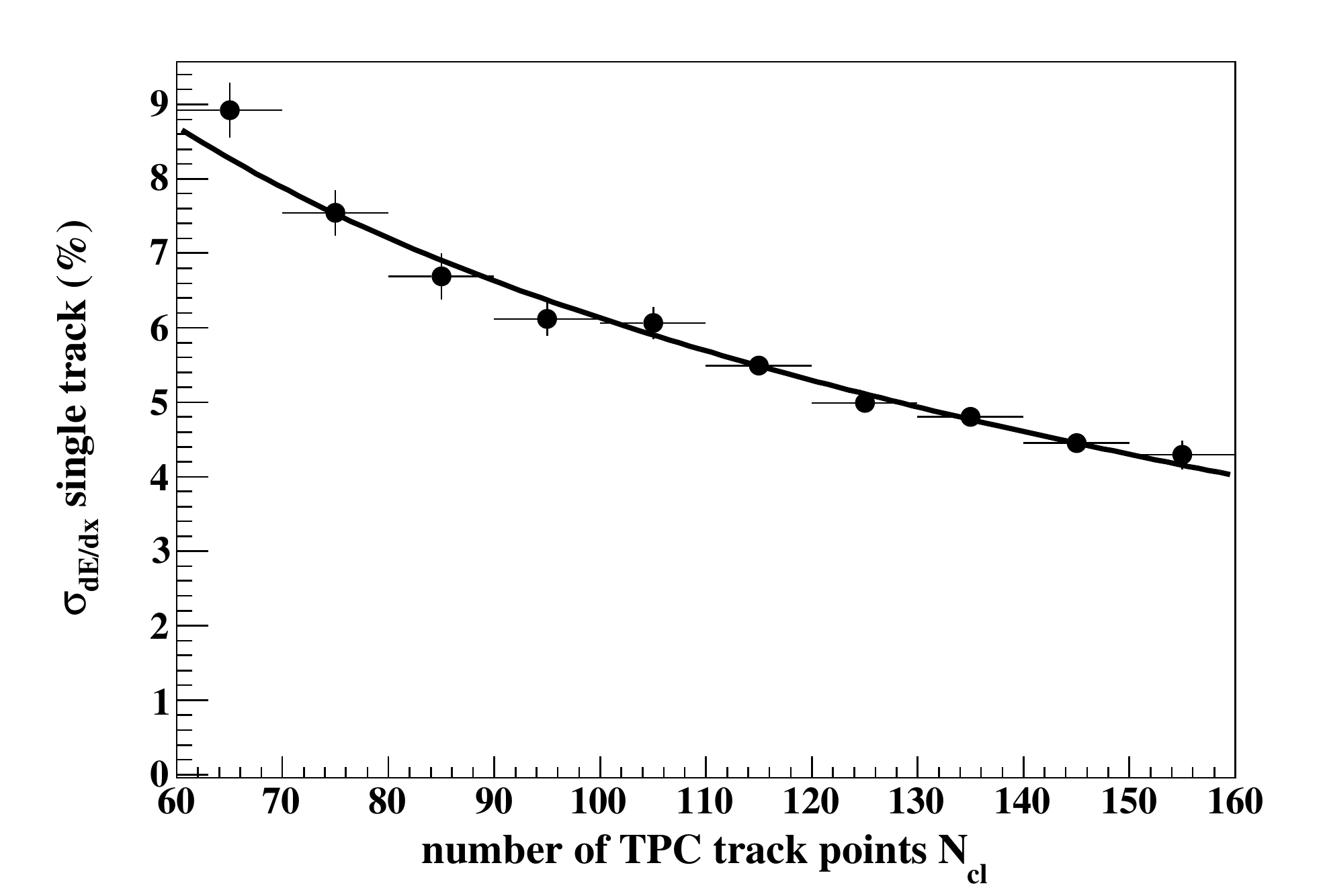}
 \caption{Dependence of the d$E$/d$x$-resolution on the number of TPC track points measured with cosmic tracks.}
 \label{figNclSingleFit}
\end{figure}

\noindent A measurement of this dependence with cosmic tracks is shown in Fig.~\ref{figNclSingleFit}.

The results demonstrate that the energy loss resolution reaches 5\% 
for cosmic tracks with 160 clusters,
(corresponding to about 1.5 times the minimum ionizing energy loss)
which is close to and actually slightly better than the design value.
In summary, space-point and {$\mathrm{d}E/\mathrm{d}x$} resolutions 
as specified in the TPC TDR~\cite{TDR:tpc} have been reached with 
the ALICE TPC.

\section{Conclusions}
\label{concl}

The ALICE TPC has been constructed over a period of five years between 2002 and
2006 with most of the assembly taking place in a large clean room located above ground near the ALICE experiment at CERN. 
In January 2007
it was transported into the ALICE underground cavern and installed into the
ALICE experiment. In late 2007 and 2008 there were extensive campaigns to test
all components and the full system, using pulsers, laser beams, and cosmic
rays. 
In September 2008 the TPC was ready for first collisions. After the LHC
incident \cite{LHCincident08} a major effort started in late 2008 to improve accessibility of
the TPC electronics with all other ALICE detectors installed. 
Since August 2009
the TPC is in full operating mode. As described in this paper, all systems
perform close to or even exceed specifications and the calibration scheme is
sufficiently advanced that momentum resolution of better than 6\% at 10 GeV
and d$E$/d$x$ resolution of better than 5\% are reached with the TPC
alone. Calibration is further continuing with cosmic rays and the TPC team
very much looks forward to taking first data with proton--proton and Pb--Pb
collisions at high rate (around 1400 and $\unit{300}{\hertz}$, respectively) in the coming 
runs.

\begingroup
\setlength{\parindent}{0mm}
\setlength{\parskip}{0.pt}      %
\newcommand{\eoinst}{\newline\nobreak}
\newcommand{\eonames}{\newline\vskip0.1cm plus0.1cm minus0.1cm}
\def\lab#1,#2,#3:#4\par{\goodbreak\textbf{#1, #2,} #3:\par#4\goodbreak\bigskip}
\begin{flushleft}

\end{flushleft}

\endgroup

\noindent{\bf Acknowledgements}

\vspace{0.3cm}

\noindent

\noindent
The ALICE TPC Collaboration wishes to acknowledge significant contributions 
to the TPC project given by:

G.~Augustinski,
L.~Bozyk, 
B.~Cantin,
H.-C.~Christensen,
H.W.~Daues,
A.~Ferrand,
F.~Formenti, 
D.~Fraissard, %
S.~G\"artner,
J.~Hehner,
S.~Lang, 
J.~Lien,
A.~Martinssen,
T.~Morhardt, 
A.~Przybyla, 
A.~Rosseboe,
A.~Stangeland,
R.~Toft-Petersen,
T.~Weber.

\noindent
In particular we wish to thank J.H.~Thomas for his very helpful suggestions and careful proofreading. 

\noindent
We acknowledge the support of the following funding agencies:

\begin{itemize}
\item{}
CERN;

\item{}
Danish National Science Research Council and the Carlsberg Foundation;

\item{}
German BMBF and the Helmholtz Association;

\item{}
Research Council of Norway;

\item{}
Polish Ministry of Science and Higher Education;

\item{}
Ministry of Education of the Slovak Republic;

\item{}
Swedish Research Council (VR) and Knut $\&$ Alice Wallenberg Foundation (KAW).

\end{itemize}

\bibliography{general}

\begin{thebibliography}{10}

\bibitem{TP:alice}
ALICE{ Collaboration},
\newblock \href{https://edms.cern.ch/document/316077/1}{{\em Technical Proposal
  for A Large Ion Collider Experiment at the CERN LHC}},
\newblock CERN/LHCC 95-71, 1995.

\bibitem{ALICEjinst}
{ALICE Collaboration},
\newblock \href{http://stacks.iop.org/1748-0221/3/S08002}{{\em The ALICE
  experiment at the CERN LHC}},
\newblock Journal of Instrumentation {\bf 3}, S08002 (2008),
\newblock \href {http://dx.doi.org/10.1088/1748-0221/3/08/S08002}
  {\path{doi:10.1088/1748-0221/3/08/S08002}}.

\bibitem{TDR:tpc}
ALICE{ Collaboration},
\newblock \href{https://edms.cern.ch/document/398930/1}{{\em Technical Design
  Report of the Time Projection Chamber}},
\newblock CERN/LHCC 2000-001, 2000.

\bibitem{lhc_predictions}
N.~Armesto~(ed.) {\em et~al.},
\newblock {\em {Heavy Ion Collisions at the LHC - Last Call for Predictions}},
\newblock J. Phys. {\bf G35}, 054001 (2008),
\newblock \href {http://arxiv.org/abs/0711.0974} {\path{arXiv:0711.0974}},
\newblock \href {http://dx.doi.org/10.1088/0954-3899/35/5/054001}
  {\path{doi:10.1088/0954-3899/35/5/054001}}.

\bibitem{ppr1}
ALICE{ Collaboration},
\newblock {\em ALICE: Physics Performance Report, Volume I},
\newblock J. Phys. {\bf G30}, 1517 (2004),
\newblock \href {http://dx.doi.org/10.1088/0954-3899/30/11/001}
  {\path{doi:10.1088/0954-3899/30/11/001}}.

\bibitem{ppr2}
ALICE{ Collaboration},
\newblock {\em ALICE: Physics Performance Report, Volume II},
\newblock J. Phys. {\bf G32}, 1295 (2006),
\newblock \href {http://dx.doi.org/10.1088/0954-3899/32/10/001}
  {\path{doi:10.1088/0954-3899/32/10/001}}.

\bibitem{NA49}
{NA49 Collaboration}, S.~Afanasev {\em et~al.},
\newblock {\em {The NA49 large acceptance hadron detector}},
\newblock Nucl. Instrum. Methods Phys. Res. {\bf A430}, 210 (1999),
\newblock \href {http://dx.doi.org/10.1016/S0168-9002(99)00239-9}
  {\path{doi:10.1016/S0168-9002(99)00239-9}}.

\bibitem{Vranic1997}
D.~Vrani\'{c},
\newblock \href{https://edms.cern.ch/document/106879/1.0}{{\em Drift
  Distortions in Alice TPC Field Cage}},
\newblock ALICE-INT-1997-22, CERN-ALICE-INT-1997-22, 1997.

\bibitem{Wieman1998}
H.~Wieman {\em et~al.},
\newblock {\em {Recent developments on the STAR detector system at RHIC}},
\newblock Nucl. Phys. {\bf A638}, 559 (1998),
\newblock \href {http://dx.doi.org/10.1016/S0375-9474(98)00385-6}
  {\path{doi:10.1016/S0375-9474(98)00385-6}}.

\bibitem{Janik2009}
R.~Janik, M.~Pikna, B.~Sitar, P.~Strmen and I.~Szarka,
\newblock {\em TPC cathode read-out with C-pads},
\newblock Nucl. Instrum. Methods Phys. Res. {\bf A598}, 681 (2009).

\bibitem{Sauli2004}
F.~Sauli,
\newblock Nucl. Instrum. Methods Phys. Res. {\bf A522}, 93 (2004).

\bibitem{blumrolandi}
W.~Blum, W.~Riegler and L.~Rolandi,
\newblock {\em Particle Detection with Drift Chambers}, 2nd ed.
  (Springer-Verlag, 2008).

\bibitem{Frankenfeld2002}
U.~Frankenfeld {\em et~al.},
\newblock {\em The ALICE TPC Inner Readout Chamber: Results of Beam and Laser
  Tests},
\newblock ALICE-INT-2002-030, 2002.

\bibitem{Stelzer2003}
H.~Stelzer {\em et~al.},
\newblock {\em The ALICE TPC Readout Chamber: From Prototypes to Series
  Production},
\newblock ALICE-INT-2003-017, 2003.

\bibitem{Christiansen2005}
{P. Christiansen \em {et al.}},
\newblock {\em Performance Test of the ALICE TPC IROC Module at the PS},
\newblock ALICE-INT-2005-029  (2005).

\bibitem{Mota2004500}
B.~Mota {\em et~al.},
\newblock
  \href{http://www.sciencedirect.com/science/article/B6TJM-4D3WCPJ-9/2/d83e743%
cf933ac919b2c1fc6e00847a3}{{\em Performance of the ALTRO chip on data acquired
  on an ALICE TPC prototype}},
\newblock Nucl. Instrum. Methods Phys. Res. {\bf A535}, 500  (2004),
\newblock Proceedings of the 10th International Vienna Conference on
  Instrumentation,
\newblock \href {http://dx.doi.org/DOI: 10.1016/j.nima.2004.07.179}
  {\path{doi:DOI: 10.1016/j.nima.2004.07.179}}.

\bibitem{Ross09}
S.~Rossegger and W.~Riegler,
\newblock \href{https://edms.cern.ch/document/1054231/1}{{\em Signal Shapes in
  a TPC Wire Chamber}},
\newblock ALICE-INT-2009-038, 2009.

\bibitem{elect:Soltveitpaper}
H.~K. Soltveit,
\newblock \href{https://edms.cern.ch/document/1054232/1}{{\em {The
  Preamplifier-Shaper for the ALICE TPC-Detector}}},
\newblock ALICE-INT-2009-039, 2009.

\bibitem{EsteveBosch:2003bj}
R.~Esteve~Bosch, A.~Jimenez~de Parga, B.~Mota and L.~Musa,
\newblock {\em {The ALTRO chip: A 16-channel A/D converter and digital
  processor for gas detectors}},
\newblock IEEE Trans. Nucl. Sci. {\bf 50}, 2460 (2003),
\newblock \href {http://dx.doi.org/10.1109/TNS.2003.820629}
  {\path{doi:10.1109/TNS.2003.820629}}.

\bibitem{elect:STM_ADC}
STMicroelectronics,
\newblock {\em {TSA1001 product information and data sheet}},
\newblock (2001),
\newblock http://www.st.com/stonline/books/pdf/docs/7333.pdf.

\bibitem{elect:DDL}
G.~Rybun and C.~So{\'{o}}s,
\newblock
  \href{https://edms.cern.ch/file/112129/1.4/ALICE-INT-1998-21.pdf}{{\em ALICE
  DDL -- Hardware Guide for the Front-end Designers}},
\newblock ALICE-INT-1998-21, 2007.

\bibitem{Xilinx:2007a}
Xilinx Inc.,
\newblock {\em {Virtex-II Pro and Virtex-II Pro X Platform FPGAs: Complete Data
  Sheet, DS083 (v4.7)}}, 2007,
\newblock Available from: http://www.xilinx.com.

\bibitem{Xilinx:2007b}
Xilinx Inc.,
\newblock {\em {Virtex-II Pro and Virtex-II Pro X FPGA User Guide, UG012
  (v4.2)}}, 2007,
\newblock Available from: http://www.xilinx.com.

\bibitem{elect:TTC}
B.~Taylor,
\newblock {\em Timing Distribution at the LHC},
\newblock in {\em Eighth Workshop on Electronics for LHC, Colmar, 2002}, pp.
  63--74, 2002.

\bibitem{Actel:2007}
Actel Corporation,
\newblock {\em {Actel ProASICPLUS Flash Family FPGAs, v5.5}}, 2007,
\newblock Available from: http://www.actel.com.

\bibitem{Excalibur:2002}
Altera,
\newblock {\em {Excalibur Devices, Hardware Reference Manual, Version 3.1}},
  2002,
\newblock Available from: http://www.altera.com.

\bibitem{Krawutschke:2008}
T.~Krawutschke,
\newblock {\em {Reliability and Redundancy of an Embedded System used in the
  Detector Control System of the ALICE Experiment}},
\newblock PhD thesis, University of Applied Sciences Cologne - University of
  Heidelberg, Germany, 2008.

\bibitem{elect:CTP}
A.~Bhasin {\em et~al.},
\newblock {\em Implementation of the ALICE Trigger System},
\newblock in {\em Real-Time Conference, 2007 15th IEEE-NPSS}, pp. 1--8, 2007,
\newblock \href {http://dx.doi.org/10.1109/RTC.2007.4382861}
  {\path{doi:10.1109/RTC.2007.4382861}}.

\bibitem{elect:ALICE-INT-2002-028}
A.~Morsch and B.~Pastir\v{c}\'ak,
\newblock \href{https://edms.cern.ch/file/358706/1/ALICE-INT-2002-028.pdf}{{\em
  Radiation in ALICE Detectors and Electronic Racks}}, 2002,
\newblock ALICE-INT-2002-028.

\bibitem{Roed:2009}
K.~Roed,
\newblock {\em {Single Event Upsets in SRAM FPGA Based Readout Electronics for
  the Time Projection Chamber in the ALICE Experiment}},
\newblock PhD thesis, University of Bergen, 2009.

\bibitem{TPCint}
{ALICE TPC Collaboration},
\newblock \href{https://edms.cern.ch/document/1049095/1}{{\em The ALICE TPC, a
  Large 3-Dimensional Tracking Device with Fast Read-out for Ultra-high
  Multiplicity Events}},
\newblock ALICE-INT-2009-034, 2009.

\bibitem{Wiechula2005}
J.~Wiechula {\em et~al.},
\newblock
  \href{http://www.sciencedirect.com/science/article/B6TJM-4GFV296-3/2/1c0fc05%
c48c1d3c76a7b65a0f1b2ec1d}{{\em High-precision measurement of the electron
  drift velocity in Ne-CO2}},
\newblock Nucl. Instrum. Methods Phys. Res. {\bf A548}, 582 (2005),
\newblock \href {http://dx.doi.org/10.1016/j.nima.2005.05.031}
  {\path{doi:10.1016/j.nima.2005.05.031}}.

\bibitem{cool:Wiechula2004}
J.~Wiechula,
\newblock {\em Pr\"azisionsmessung der Elektronen-Driftgeschwindigkeit in
  $NeCO_2$},
\newblock Master's thesis, Universit\"at Frankfurt, 2004.

\bibitem{cool:Mueller2004}
A.~M\"uller,
\newblock CERN TS-Department Report No. 528133, 2004 (unpublished).

\bibitem{Bonneau:1999xi}
P.~Bonneau and M.~Bosteels,
\newblock {\em Liquid cooling systems (LCS2) for LHC detectors},
\newblock (1999),
\newblock Prepared for $5^{th}$ Workshop on Electronics for the LHC Experiments
  (LEB 99), Snowmass, Colorado, 20--24 Sep 1999.

\bibitem{Pimenta2003}
M.~Pimenta~dos Santos,
\newblock \href{https://edms.cern.ch/document/490540/1}{{\em ALICE TPC Readout
  Chambers Cooling System}},
\newblock CERN-ST/CV-2003-490540, 2003.

\bibitem{cool:Popescu2005}
S.~Popescu, U.~Frankenfeld and H.~R. Schmidt,
\newblock {\em Thermal influences of the front-end electronics on the ALICE TPC
  readout chamber},
\newblock IEEE Trans. Nucl. Sci. {\bf 52}, 2879 (2005),
\newblock \href {http://dx.doi.org/10.1109/TNS.2005.862794}
  {\path{doi:10.1109/TNS.2005.862794}}.

\bibitem{Frankenfeld2005_cooling}
U.~Frankenfeld, S.~Popescu and H.~Schmidt,
\newblock \href{https://edms.cern.ch/document/537240/1}{{\em Experimental
  Evaluation of the ALICE TPC Front-End Electronics Cooling Strategy}},
\newblock ALICE-INT-2005-001, 2005.

\bibitem{Frankenfeld2005_temp_system}
U.~Frankenfeld, S.~Popescu and H.~Schmidt,
\newblock \href{https://edms.cern.ch/document/544269/1}{{\em Temperature
  Monitoring System for the ALICE TPC}},
\newblock ALICE-EN-2005-001, 2005.

\bibitem{veenhof2003}
R.~Veenhof,
\newblock \href{https://edms.cern.ch/document/404406/1}{{\em Choosing a Gas
  Mixture for the ALICE TPC}},
\newblock ALICE-INT-2003-29, 2003.

\bibitem{Garabatos2004}
C.~Garabatos,
\newblock
  \href{http://www.sciencedirect.com/science/article/B6TJM-4D2X8RC-M/2/fce5ba1%
18ed40946e6f6488f6e48954b}{{\em The ALICE TPC}},
\newblock Nucl. Instrum. Methods Phys. Res. {\bf A535}, 197 (2004),
\newblock \href {http://dx.doi.org/10.1016/j.nima.2004.07.127}
  {\path{doi:10.1016/j.nima.2004.07.127}}.

\bibitem{magboltzNIM}
S.~F. Biagi,
\newblock
  \href{http://www.sciencedirect.com/science/article/B6TJM-418RG6K-S/2/57aaef3%
72781e8596dce3f6dd5c2c550}{{\em Monte Carlo simulation of electron drift and
  diffusion in counting gases under the influence of electric and magnetic
  fields}},
\newblock Nucl. Instrum. Metheods Phys. Res. {\bf A421}, 234  (1999),
\newblock \href {http://dx.doi.org/10.1016/S0168-9002(98)01233-9}
  {\path{doi:10.1016/S0168-9002(98)01233-9}}.

\bibitem{magboltzCERN}
S.~Biagi,
\newblock \href{http://consult.cern.ch/writeup/magboltz}{{\em Magboltz-2,
  Transport of Electrons in Gas Mixtures}},
\newblock http://consult.cern.ch/writeup/magboltz.

\bibitem{Thomas05}
G.~Thomas {\em et~al.},
\newblock CERN Report No. CERN-OPEN-2005-031, 2005 (unpublished).

\bibitem{laser:Lebedev:2002sp}
A.~Lebedev,
\newblock {\em {A laser calibration system for the STAR TPC}},
\newblock Nucl. Instrum. Methods Phys. Res. {\bf A478}, 163 (2002),
\newblock \href {http://dx.doi.org/10.1016/S0168-9002(01)01747-8}
  {\path{doi:10.1016/S0168-9002(01)01747-8}}.

\bibitem{laser:STARlaser}
{STAR Collaboration}, J.~Abele {\em et~al.},
\newblock {\em {The laser system for the STAR time projection chamber}},
\newblock Nucl. Instrum. Methods Phys. Res. {\bf A499}, 692 (2003),
\newblock \href {http://dx.doi.org/10.1016/S0168-9002(02)01966-6}
  {\path{doi:10.1016/S0168-9002(02)01966-6}}.

\bibitem{det_laser_calib}
H.~J. Hilke,
\newblock {\em Detector calibration with lasers -- A review},
\newblock Nucl. Instrum. Methods Phys. Res. {\bf A252}, 169 (1986),
\newblock \href {http://dx.doi.org/10.1016/0168-9002(86)91177-0}
  {\path{doi:10.1016/0168-9002(86)91177-0}}.

\bibitem{laser:CERESlaser}
D.~Miskowiec and P.~Braun-Munzinger,
\newblock {\em {Laser calibration system for the CERES Time Projection
  Chamber}},
\newblock Nucl. Instrum. Methods Phys. Res. {\bf A593}, 188 (2008),
\newblock \href {http://arxiv.org/abs/0801.4920} {\path{arXiv:0801.4920}},
\newblock \href {http://dx.doi.org/10.1016/j.nima.2008.02.034}
  {\path{doi:10.1016/j.nima.2008.02.034}}.

\bibitem{ALEPH}
D.~Decamp {\em et~al.},
\newblock
  \href{http://www.sciencedirect.com/science/article/B6TJM-4729K35-D/2/c1dc188%
e79c8a64c213a31e1f6c2b778}{{\em ALEPH: A detector for electron-positron
  annihilations at LEP}},
\newblock Nucl. Instrum. Methods Phys. Res. {\bf A294}, 121  (1990),
\newblock \href {http://dx.doi.org/10.1016/0168-9002(90)91831-U}
  {\path{doi:10.1016/0168-9002(90)91831-U}}.

\bibitem{digi:PortServer}
\href{http://www.digi.com}{{\em {Digi PortServer TS 16, Digi International
  Inc., 11001 Bren Road East, Minnetonka, MN 55343, USA}}}.

\bibitem{iPico}
{\em {Intelligent Picomotor (iPico), New Focus, 2584 Junction Avenue, San Jose,
  CA 95134, USA, www.newfocus.com}}.

\bibitem{quasar}
{\em {M3185A B/W Camera Module, Quasar Electronics Ltd, PO Box 6935, Bishops
  Stortford, CM23 4WP, UK, www.quasarelectronics.com}}.

\bibitem{falcon}
{\em {FALCONquattro Express, Imaging Development Systems GmbH, Dimbacher
  Strasse 6, 74182 Obersulm, Germany, www.ids-imaging.com}}.

\bibitem{DCS:wiener}
\href{http://www.wiener-d.com/}{{\em W-IE-NE-R, Plein \& Baus GmbH,
  M\"ullersbaum 20, D - 51399 Burscheid, Germany}}.

\bibitem{DCS:iseg}
\href{http://iseg-hv.de}{{\em iseg Spezialelektronik GmbH, Bautzner Landstraße
  23, D-01454 Radeberg, Germany}}.

\bibitem{DCS:zentro}
\href{http://www.zentro-elektrik.de}{{\em Zentro-Elektrik GmbH KG, Sandweg 20,
  D-75179 Pforzheim, Germany}}.

\bibitem{DCS:Augustinus2008}
A.~Augustinus {\em et~al.},
\newblock {\em {The ALICE control system}},
\newblock ICFA Beam Dyn. Newslett. {\bf 47}, 90 (2008).

\bibitem{TDR:daq}
ALICE{ Collaboration},
\newblock \href{https://edms.cern.ch/file/456354/2}{{\em Technical Design
  Report of the Trigger, Data Acquisition, High Level Triggerand Control
  System}},
\newblock CERN-LHCC-2003-062, 2004.

\bibitem{DCS:Schmeling2006}
S.~Schmeling,
\newblock {\em {Common tools for large experiment controls: A common approach
  for deployment, maintenance, and support}},
\newblock IEEE Trans. Nucl. Sci. {\bf 53}, 970 (2006),
\newblock \href {http://dx.doi.org/10.1109/TNS.2006.873706}
  {\path{doi:10.1109/TNS.2006.873706}}.

\bibitem{DCS:SMI}
B.~Franek and C.~Gaspar,
\newblock {\em SMI++: An Object Oriented Framework for Designing Distributed
  Control Systems},
\newblock IEEE Trans. Nucl. Sci. {\bf 45}, 1946 (1998),
\newblock \href {http://dx.doi.org/10.1109/23.710969}
  {\path{doi:10.1109/23.710969}}.

\bibitem{DCS:etm}
\href{http://www.etm.at}{{\em ETM professional control GmbH, A Siemens Company,
  Kasernenstraße 29 A-7000, Eisenstadt, Austria}}.

\bibitem{DCS:DIM}
C.~Gaspar, M.~D\"onszelmann and P.~Charpentier,
\newblock \href{http://dim.web.cern.ch/dim/papers/CHEP/DIM.PDF}{{\em DIM, a
  portable, light-weight package for information publishing, data transfer and
  inter-process communication}},
\newblock Computer Phys. Communi. {\bf 140}, 102 (2001).

\bibitem{Bablok:2006zj}
S.~Bablok {\em et~al.},
\newblock {\em {Front-end-electronics communication software for multiple
  detectors in the ALICE experiment}},
\newblock Nucl. Instrum. Methods Phys. Res. {\bf A557}, 631 (2006),
\newblock \href {http://dx.doi.org/10.1016/j.nima.2005.11.208}
  {\path{doi:10.1016/j.nima.2005.11.208}}.

\bibitem{Wiechula2009}
J.~Wiechula,
\newblock {\em Commissioning and Calibration of the ALICE-TPC},
\newblock PhD thesis, Goethe-Universit\"at Frankfurt am Main, 2009.

\bibitem{Mota:2004kq}
B.~Mota {\em et~al.},
\newblock {\em {Performance of the ALTRO chip on data acquired on an ALICE TPC
  prototype}},
\newblock Nucl. Instrum. Methods Phys. Res. {\bf A535}, 500 (2004),
\newblock \href {http://dx.doi.org/10.1016/j.nima.2004.07.179}
  {\path{doi:10.1016/j.nima.2004.07.179}}.

\bibitem{Rossegger:1217595}
S.~Rossegger,
\newblock {\em Simulation and Calibration of the ALICE TPC Including Innovative
  Space Charge Calculations.},
\newblock PhD thesis, University of Technology, Graz, 2009,
\newblock CERN-THESIS-2009-124.

\bibitem{Blum_alephKr}
W.~Blum,
\newblock {\em The ALEPH Handbook} (CERN, Geneva, 1989),
\newblock CERN-ALEPH-89-077.

\bibitem{calcom:delphiKr}
A.~De~Min {\em et~al.},
\newblock {\em {Performance of the HPC calorimeter in DELPHI}},
\newblock IEEE Trans. Nucl. Sci. {\bf 42}, 491 (1995),
\newblock \href {http://dx.doi.org/10.1109/23.467923}
  {\path{doi:10.1109/23.467923}}.

\bibitem{Rybicki:604807}
A.~Rybicki,
\newblock {\em Charged Hadron Production in Elementary and Nuclear Collisions
  at 158 GeV/c},
\newblock PhD thesis, H. Niewodniczanski Institute of Nuclear Physics, Polish
  Academy of Sciences, Krakow, 2002.

\bibitem{LHCincident08}
\href{http://cdsweb.cern.ch/record/1135729/}{{\em Summary of the analysis of
  the 19 September 2008 incident at the LHC.}},
\newblock http://cdsweb.cern.ch/record/1135729/, 2008.

\end{thebibliography}

\section{Acronyms}
\subsection*{A}
\vspace{-6mm}
\begin{tabular}{ll}
~~~~~~~~~~~~~~~~~~~~~~~ & ~~~~~~~~~~~~~~~~~~~~~~~~~~~~~~~~~~~~~~~~~~~~~~~~~ \\

ADC	 	& Analog to Digital Converter \\
ACORDE	& ALICE COsmic Ray DEtector \\
ALEPH    & Apparatus for LEP PHysics                \\ 
ALICE	  & A Large Ion Collider Experiment \\
ALTRO 	& ALICE TPC ReadOut chip \\
\end{tabular}

\subsection*{B}

\subsection*{C}
\vspace{-6mm}
\begin{tabular}{ll}
~~~~~~~~~~~~~~~~~~~~~~~ & ~~~~~~~~~~~~~~~~~~~~~~~~~~~~~~~~~~~~~~~~~~~~~~~~~ \\
CERN    	& Conseil Europ\'een pour la Recherche Nucl\'eaire \\
                & (European Organization for Nuclear Research) \\
CFD             & Computational Fluid Dynamics \\
CMOS 	& Complementary Metal-Oxide-Semiconductor \\
CSA		& Charge Sensitive Amplifier \\
CTP     	& Central Trigger Processor \\
CU		& Control Unit \\

\end{tabular}

\subsection*{D}
\vspace{-6mm}
\begin{tabular}{ll}
~~~~~~~~~~~~~~~~~~~~~~~ & ~~~~~~~~~~~~~~~~~~~~~~~~~~~~~~~~~~~~~~~~~~~~~~~~~ \\
 DAC     	& Digital to Analog Converter  \\
 DAQ            & Data Acquisition System \\
 DC &           Direct Current\\ 
 DCS    	& Detector Control System \\
 DDL     	& Detector Data Link \\
 DIM      	& Distributed Information Management system \\
 DNL  & Differential Non-Linearity \\
 D-RORC   	& DAQ RORC Data ReadOut Receiver Card \\
 DU		& Device Unit \\

\end{tabular}

\subsection*{E}
\vspace{-6mm}
\begin{tabular}{ll}
~~~~~~~~~~~~~~~~~~~~~~~ & ~~~~~~~~~~~~~~~~~~~~~~~~~~~~~~~~~~~~~~~~~~~~~~~~~ \\
 ECS  	& Experiment Control System \\
 ELMB		& Embedded Local Monitor Board \\
 EMCAL 	& ElectroMagnetic CALorimeter \\
 ENC 		& Equivalent Noise Charge \\
 ENOB& Equivalent Number Of Bits \\

\end{tabular}

\subsection*{F}
\vspace{-6mm}
\begin{tabular}{ll}
~~~~~~~~~~~~~~~~~~~~~~~ & ~~~~~~~~~~~~~~~~~~~~~~~~~~~~~~~~~~~~~~~~~~~~~~~~~ \\
 FEC		& Front-End Card \\
 FEE	 	& Frond-End Electronics  \\
 FEM        & Finite Element Method\\
 FET & Field-Effect Transistor \\ 
 FMD		& Forward Multiplicity Detector \\
 FPGA    	& Field Programmable Gate Array \\
 FSM     	& Finite State Machine \\
 FWHM		& Full Width Half Maximum \\

\end{tabular}

\subsection*{G}
\vspace{-6mm}
\begin{tabular}{ll}
~~~~~~~~~~~~~~~~~~~~~~~ & ~~~~~~~~~~~~~~~~~~~~~~~~~~~~~~~~~~~~~~~~~~~~~~~~~ \\
 GEM &  Gas Electron Multipliers \\
 GTL	 	&  Gunning Transistor Logic (FEE-bus \\
     & technology) \\

\end{tabular}

\subsection*{H}
\vspace{-6mm}
\begin{tabular}{ll}
~~~~~~~~~~~~~~~~~~~~~~~ & ~~~~~~~~~~~~~~~~~~~~~~~~~~~~~~~~~~~~~~~~~~~~~~~~~ \\
 HMPID 	& High Momentum Particle Identification Detector \\
HCMOS   &     High-speed CMOS         \\
 HV    	& High Voltage \\

\end{tabular}

\subsection*{I}
\vspace{-6mm}
\begin{tabular}{ll}
~~~~~~~~~~~~~~~~~~~~~~~ & ~~~~~~~~~~~~~~~~~~~~~~~~~~~~~~~~~~~~~~~~~~~~~~~~~ \\

 INL &  Integral Non-Linearity \\
 IROC&  Inner ReadOut Chamber \\
 ITS		& Inner Tracking System \\
\end{tabular}

\subsection*{J}
\vspace{-6mm}
\begin{tabular}{ll}
~~~~~~~~~~~~~~~~~~~~~~~ & ~~~~~~~~~~~~~~~~~~~~~~~~~~~~~~~~~~~~~~~~~~~~~~~~~ \\

\end{tabular}

\subsection*{L}
\vspace{-6mm}
\begin{tabular}{ll}
~~~~~~~~~~~~~~~~~~~~~~~ & ~~~~~~~~~~~~~~~~~~~~~~~~~~~~~~~~~~~~~~~~~~~~~~~~~ \\
 L0 		& Level 0 trigger \\
 L1		& Level 1 trigger \\
 L2 		& Level 2 trigger \\
 L3     & Magnet used by LEP-L3 experiment\\
 LDC  	& Local Data Concentrator \\
 LEP		& Large Electron Positron collider \\
 LHC  	& Large Hadron Collider \\

 LSB	 	& Least Significant Bit \\

 LV		& Low Voltage \\
 LVCMOS   &     Low-Voltage CMOS\\

\end{tabular}

\subsection*{M}
\vspace{-6mm}
\begin{tabular}{ll}
~~~~~~~~~~~~~~~~~~~~~~~ & ~~~~~~~~~~~~~~~~~~~~~~~~~~~~~~~~~~~~~~~~~~~~~~~~~ \\
MSPS & Mega-Samples Per Second \\
MWPC & Multi-Wire Proportional Chamber \\
 
\end{tabular}

\subsection*{N}
\vspace{-6mm}
\begin{tabular}{ll}
~~~~~~~~~~~~~~~~~~~~~~~ & ~~~~~~~~~~~~~~~~~~~~~~~~~~~~~~~~~~~~~~~~~~~~~~~~~ \\

NMOS & N-type Metal-Oxide-Semiconductor field \\
     & effect transistors \\
\end{tabular}

\subsection*{O}
\vspace{-6mm}
\begin{tabular}{ll}
~~~~~~~~~~~~~~~~~~~~~~~ & ~~~~~~~~~~~~~~~~~~~~~~~~~~~~~~~~~~~~~~~~~~~~~~~~~ \\
 OCDB		& Offline Conditions Data Base \\ 
 OLE		& Object Linking and Embedding \\
 OPC		& OLE for Process Control \\  
 OROC&  Outer ReadOut Chamber \\
\end{tabular}

\subsection*{P}
\vspace{-6mm}
\begin{tabular}{ll}
~~~~~~~~~~~~~~~~~~~~~~~ & ~~~~~~~~~~~~~~~~~~~~~~~~~~~~~~~~~~~~~~~~~~~~~~~~~ \\
PASA & PreAmplifier ShAper  \\
PC   	&  Personal Computer \\
PCB & Printed Circuit Board \\
PHOS		& PHOton Spectrometer \\
PEEK& Polyaryl-Ether-Ether-Ketone\\
PID  &   Proportional-Integral-Derivative \\
PLC     	& Programmable Logic Controller  \\
PMOS &  P-type Metal-Oxide-Semiconductor field \\
     & effect transistors \\
PS     	& Power Supply \\ 

 PMD 		& Photon Multiplicity Detector \\
PVSS    	&  Prozessvisualisierungs- und Steuerungs-\\
     & System  \\

\end{tabular}

\subsection*{Q}
\vspace{-6mm}
\begin{tabular}{ll}
~~~~~~~~~~~~~~~~~~~~~~~ & ~~~~~~~~~~~~~~~~~~~~~~~~~~~~~~~~~~~~~~~~~~~~~~~~~ \\

\end{tabular}

\subsection*{R}
\vspace{-6mm}
\begin{tabular}{ll}
~~~~~~~~~~~~~~~~~~~~~~~ & ~~~~~~~~~~~~~~~~~~~~~~~~~~~~~~~~~~~~~~~~~~~~~~~~~ \\

 RCC & Ring Cathode Chamber \\
 RCU	 	& Readout Control Unit \\
 RHIC	 	& Relativistic Heavy Ion Collider \\
 RMS & Root Mean Square         \\ 
 ROC & ReadOut Chamber    \\

\end{tabular}

\subsection*{S}
\vspace{-6mm}
\begin{tabular}{ll}
~~~~~~~~~~~~~~~~~~~~~~~ & ~~~~~~~~~~~~~~~~~~~~~~~~~~~~~~~~~~~~~~~~~~~~~~~~~ \\
 SCADA   	& Supervisory Controls And Data Acquisition  \\
SEL  & Single Event Latchup \\
 SEU     	& Single Event Upset \\
SFDR & Spurious-Free Dynamic Range \\ 

  SIU 		& System Interface Unit	\\
 S/N  & Signal-to-Noise ratio  \\
SRAM &  Static Random Access Memory \\

 SSW & Service Support Wheel\\

   STAR	 	& Solenoidal Tracker At RHIC \\

\end{tabular}

\subsection*{T}
\vspace{-6mm}
\begin{tabular}{ll}
~~~~~~~~~~~~~~~~~~~~~~~ & ~~~~~~~~~~~~~~~~~~~~~~~~~~~~~~~~~~~~~~~~~~~~~~~~~ \\
  TCF& Tail Cancellation Filter \\ 
  TCP/IP	& Transmission Control Protocol/Internet \\
     & Protocol \\
 TDR & Technical Design Report\\

 TOF 		& Time-Of-Flight detector \\

 TPC 		& Time Projection Chamber \\
 TQFP& Thin Quad Flat Pack (chip package) \\

 TRD 		& Transition Radiation Detector \\

 TTC 		& Timing, Trigger and Control \\

\end{tabular}

\subsection*{U}
\vspace{-6mm}
\begin{tabular}{ll}
~~~~~~~~~~~~~~~~~~~~~~~ & ~~~~~~~~~~~~~~~~~~~~~~~~~~~~~~~~~~~~~~~~~~~~~~~~~ \\
 UPS  	& Uninteruptible Power Supply \\

 UV      	&  Ultra Violet  \\

\end{tabular}

\subsection*{V}
\vspace{-6mm}
\begin{tabular}{ll}
~~~~~~~~~~~~~~~~~~~~~~~ & ~~~~~~~~~~~~~~~~~~~~~~~~~~~~~~~~~~~~~~~~~~~~~~~~~ \\

\end{tabular}

\subsection*{W}
\vspace{-6mm}
\begin{tabular}{ll}
~~~~~~~~~~~~~~~~~~~~~~~ & ~~~~~~~~~~~~~~~~~~~~~~~~~~~~~~~~~~~~~~~~~~~~~~~~~ \\

\end{tabular}

\subsection*{Z}
\vspace{-6mm}
\begin{tabular}{ll}
~~~~~~~~~~~~~~~~~~~~~~~ & ~~~~~~~~~~~~~~~~~~~~~~~~~~~~~~~~~~~~~~~~~~~~~~~~~ \\
 ZDC		& Zero Degree Calorimeter \\
\end{tabular}

\end{document}